\documentclass[usegraphicx,onecolumn,usenatbib]{mn2e}

\usepackage{graphicx}	
\usepackage{amsmath}	
\usepackage{amssymb}	
\usepackage{bm}		
\usepackage{pdflscape}	
\usepackage{newtxtext,newtxmath}

\usepackage[T1]{fontenc}
\usepackage{ae,aecompl}
\usepackage{amsmath,amssymb}
\usepackage{graphicx}
\usepackage{latexsym,color}
\usepackage{hyperref}
\usepackage{cleveref}
\usepackage[utf8]{inputenc}
\usepackage{mathtools,cuted}
\usepackage{natbib}

\usepackage{graphicx}	
\usepackage{amsmath}	
\usepackage{amssymb}	

\title[Lense-Thirring effect on accretion flow]{Lense-Thirring effect on accretion flow from   counter-rotating tori }

\author[D. Pugliese et al.]{
D. Pugliese,\thanks{E-mail: d.pugliese.physics@gmail.com}
Z. Stuchl\'{\i}k
\\
Research Centre for Theoretical Physics and Astrophysics, Institute of Physics,
  Silesian University in Opava,\\
 Bezru\v{c}ovo n\'{a}m\v{e}st\'{i} 13, CZ-74601 Opava, Czech Republic
}

\date{Accepted XXX. Received YYY; in original form ZZZ}

\pubyear{2022}

\begin{document}
\label{firstpage}
\pagerange{\pageref{firstpage}--\pageref{lastpage}}
\maketitle
\count\footins = 1000
\begin{abstract}
We study the accretion flow from a counter-rotating torus orbiting  a central Kerr  black hole (\textbf{BH}).
We  characterize the flow  properties  at the turning point  of  the accreting matter flow from the orbiting torus, defined by the  condition $u^{\phi}=0$ on the flow torodial velocity.   The  counter-rotating accretion  flow  and  jet-like flow turning point location along  \textbf{BH}  rotational axis is given. Some properties of the  counter-rotating flow thickness and counter-rotating  tori energetics are studied. The maximum amount of matter swallowed by the \textbf{BH} from the counter-rotating tori is  determined  by the background properties.  The fast spinning  \textbf{BH} energetics   depends mostly   on  \textbf{BH} spin  rather than on   the properties of the counter-rotating fluids or  the tori masses. The  turning point is  located in  a narrow orbital corona (\emph{spherical shell}), for photons and matter flow constituents,  surrounding the \textbf{BH} stationary limit (outer ergosurface), depending  on the \textbf{BH}   spin--mass ratio and  the fluid initial momentum only.   The  turning corona for jet-like-flow has larger thickness,   it is separated from the  torus flow turning  corona    and  it is closer to the \textbf{BH}  stationary limit.  Turning  points of   matter accreting  from torus  and from jets are independent explicitly   of the   details of the accretion and tori model.  The turning corona could be observable due to  an increase of flow  luminosity and temperature. The corona is larger on the  \textbf{BH} equatorial plane,   where it is the farthest from the central attractor, and narrower on the \textbf{BH} poles. \end{abstract}




\def\be{\begin{equation}}
\def\ee{\end{equation}}
\def\bea{\begin{eqnarray}}
\newcommand{\cc}{\mathrm{C}}
\newcommand{\jj}{\mathrm{J}}
\def\eea{\end{eqnarray}}
\newcommand{\tb}[1]{\textbf{{#1}}}
\newcommand{\ttb}[1]{\textbf{#1}}
\newcommand{\rtb}[1]{\textcolor[rgb]{1.00,0.00,0.00}{\tb{#1}}}
\newcommand{\btb}[1]{\textcolor[rgb]{0.00,0.00,1.00}{\tb{#1}}}
\newcommand{\otb}[1]{\textcolor[rgb]{1.00,0.47,0.22}{\tb{#1}}}
\newcommand{\ptb}[1]{\textcolor[rgb]{0.57,0.14,0.58}{\tb{#1}}}
\newcommand{\gtb}[1]{\textcolor[rgb]{0.00,0.47,0.24}{\tb{#1}}}
\newcommand{\pp}{\textbf{()}}
\newcommand{\non}[1]{{\LARGE{\not}}{#1}}
\newcommand{\mso}{\mathrm{mso}}
\newcommand{\Qa}{\mathcal{Q}}
\newcommand{\mbo}{\mathrm{mbo}}
\newcommand{\il}{~}
\newcommand{\rc}{\rho_{\ti{C}}}
\newcommand{\dd}{\mathcal{D}}
\newcommand{\Sie}{\mathcal{S}}
\newcommand{\Sa}{\mathcal{S}}
\newcommand{\mnras}{MNRAS}
\newcommand{\aap}{A\&A}
\newcommand{\prd}{Phys. Rev. D}
\newcommand{\actaa}{Acta Astron.}
\newcommand{\Fa}{\mathcal{F}}
\newcommand{\Mie}{\mathcal{M}}
\newcommand{\Em}{\mathcal{E}}
\newcommand{\La}{\mathcal{L}}
\newcommand{\Ta}{{\mbox{\scriptsize  \textbf{\textsf{T}}}}}
\newcommand{\apjs}{APJS}
\newcommand{\apj}{APJ}

    \newcommand{\oo}{\mathrm{O}}

\section{Introduction}
Counter-rotating accreting tori orbiting a central attractor,   particularly a Kerr    black hole (\textbf{BH}), are a   known possibility  of the \textbf{BH} Astrophysics \citep{Murray,Kuznetsov,Violette,Ensslin:2002gn,Beckert:2001az,Kim:2016qsr,Barrabes:1995ue,2010ApJ...710..859E,
Christodoulou,Nixonx,Garofalo,2010ASPC..427....3V,Volonteri:2002vz,2012MNRAS.422.2547N,Dyda:2014pia,
2016A&A...591A.114A,Cui,Rao,Reis,Middleton-miller-jones,Morningstar,Cowperthwaite}. In  Active Galactic Nucleai   (\textbf{AGNs}),  corotating and counter-rotating tori, or strongly misaligned disks, as related  to the central Kerr {\textbf{BH}} spin, can report traces of the \textbf{AGNs} evolution. Chaotical, discontinuous   accretion   can  produce accretion disks  with different rotation orientations with respect to the central Kerr \textbf{BH} where aggregates of  corotating and counter-rotating  toroids can be mixed \citep{Dyda:2014pia,Aligetal(2013),Carmona-Loaiza:2015fqa,Lovelace:1996kx,Romanova,ringed,Multy,long}. Eventually, misaligned disks  with respect to the central \textbf{BH} spin may characterize these strong attractors \citep{Nixon:2013qfa,2015MNRAS.449.1251D,Bonnerot:2015ara,Aly:2015vqa}.

Counter-rotating accretion disks can form in  transient systems due to tidal disruption events and
 by
episodic or
prolonged phases  of accretion, by galaxies merging, or binary  (multiple) system components    merging, or also  from  stellar   formation in a counter-rotating cloud environment--see also \cite{2022MNRAS.511.3795N,2017MNRAS.469.4483T,Wongetal,2021MNRAS.502.2023P,Zhangetal}.
Phenomena related to counter--rotation play very  important role around  \textbf{BHs} in the center of \textbf{AGNs}  due to their possible complex accretion history; both co-rotating and counter-rotating accretion and  equilibrium toroidal structures can orbit the central \textbf{BH}, being sometimes endowed with related jets \citep{ella-correlation,ella-jet,proto-jets}.
Counter-rotating accretion structures are possible occasionally in the  binary systems containing stellar mass \textbf{BH}, as  in  \textbf{3C120} \citep{Kataoka,Cowperthwaite},
the Galactic binary \textbf{BHs} \citep{Cui,Reis}, \textbf{BH} binary system  (\citep{Morningstar} or \citep{Christodoulou})\footnote{
They were also connected to
\textbf{SWIFT J1910.2 0546}   \citep{Reis},
and   the faint luminosity of
\textbf{IGR J17091--3624} (a transient X-ray source  believed to be a Galactic \textbf{BH} candidate)  \citep{Rao}.}. %
The counter-rotation phenomena
and a wide range of their demonstration were discussed in a large variety of papers.
Disk counter-rotation may   also  distinguish \textbf{BHs} with or without jets
\citep{Ensslin:2002gn,Beckert:2001az}.
Counter-rotating tori and jets were studied   in relation to
radio--loud \textbf{AGN} and   double radio source associated  with galactic nucleus \citep{2010ApJ...710..859E,GarofaloEvans}.
{Observational evidence of counter-rotating disks has been   provided  by  M87, observed by the Event Horizon Telescope\cite{M87}\footnote{The rotation orientation of  the jet and funnel wall, governed  by the \textbf{BH} spin, was studied
in dependence of the  relative  large-scale jet  and  \textbf{BH}  spin axis orientation,   and   the relative  disk--\textbf{BH} rotation orientation.
 The effect of \textbf{BH}  and disk spin  on ring  (image) asymmetry  produced
 (from emission generated in the funnel wall) was studied.
The  rotation  orientation of both the jet and funnel wall are controlled by the \textbf{BH} spin.
It was found that the image may correspond to a counter-rotating disk,  with the  emission region  very close to the central \textbf{BH}, corresponding  to gas   being  forced to  co-rotate due to the \textbf{BH} dragging effect\cite{M87}.}}.
In \cite{Middleton-miller-jones},
counter-rotation of the  extragalactic microquasars have been investigated  as
engines for  jet emission powered by Blandford--Znajek  processes\footnote{
Jets may be produced by sweeping
the magnetic flux in the "plunging region" on to the \textbf{BH}. This  region, bounded by the
marginally stable circular orbit, $r_{mso}^+$,  and the \textbf{BH} horizon,  is larger for counter-rotating tori, increasing with the \textbf{BH} spin in magnitude. Consequently  the magnetic flux trapped on the \textbf{BH} can  be enhanced \citep{Garofalo}.}.

\textbf{BHs} may  accrete from disks having  alternately corotating and
counter-rotating  senses of rotation
\citep{Murray}. Less massive
\textbf{BH} in counter-rotating configurations may "flip" to corotating
configurations (this effect  has been related  to
radio-loud systems  turning  into radio-quiet systems)\footnote{
 Reversals
in the rotation direction of an accretion disk have  been
considered  to explain state transitions \citep{Cui}.
In  X-ray binary, the  \textbf{BH} binaries with no detectable
ultrasoft component above 1-2 keV in their high luminosity state may contain a fast-spinning retrograde \textbf{BH}, and
   the  spectral state transitions can correspond to a temporary "flip-flop" phase of  disk reversal,
showing the characteristics of both
counter-rotating   and  corotating   systems, switching from one state to another
(the hard X-ray luminosity of a corotating  system is generally much lower than that of a counter-rotating  system)--\citep{Cui}.
Counter-rotating  tori  have  been modelled also as a counter-rotating gas layer on the
surface of a corotating disk. The matter  interface  in these configurations is a mix of the two components
 with zero net angular momentum which tends to free-fall towards the  center.
In   \cite{Dyda:2014pia}  a  high-resolution axisymmetric hydrodynamic simulation of viscous counter-rotating
disks was presented for the cases where the two components are vertically separated and radially separated.
The accretion rates are increased
over that for  corotating  disks--see also  \cite{Kuznetsov}.
A time-dependent, axisymmetric  hydrodynamic simulation of complicated composite counter-rotating  accretion disks is in \cite{Kuznetsov}, where the disks
consist of  combined counter-rotating  and corotating components.}.
Counter-rotating tori have been studied  in  more complex structures, for example featuring accreting disks/\textbf{BH} rotation  "flip", i.e., alternate phases of cororation and counter-rotation accretion, or
 with the presence of relative counter-rotating  layers in the same torus, vertically separated corotating-counter-rotating tori, or  finally agglomerates of corotating and counter-rotating  tori centered on one central Kerr \textbf{BH} orbiting on its equatorial plane \citep{ringed,open,dsystem}.
However, there is  a crucial exception in the close vicinity of the \textbf{BH} horizon in the ergoregion where the accreting or non accreting, e.g. jets, matter must be co-rotating  with the Kerr \textbf{BH}, from the point of view of distant observers, due to spacetime dragging \citep{dragged}. Therefore, it is of high relevance to study the  turning of the accretion matter originally being in  counter-rotation to the co--rotating motion, and its astrophysical consequences. In the present paper we study the turning effect  in connection to matter accreting  from counter-rotating toroidal structures, or the jets.

The toroidal structures are considered in the framework of the Ringed Accretion Disks model (\textbf{RAD}) developed in \cite{pugtot,ringed,open,dsystem} and widely discussed in subsequent papers\citep{letter,long,Multy,proto-jets,Fi-Ringed,ella-correlation,mnras2,cqg2020}.
Evidences of the presence of a cluster composed by an inner corotating torus and outer counter-rotating  torus has been provided by
\textbf{A}tacama
\textbf{L}arge \textbf{M}illimeter/submillimeter \textbf{A}rray (\textbf{ALMA}).
 In \cite{Violette} counter-rotation and high-velocity outflow in the \textbf{NGC1068} galaxy  molecular torus
were studied.
\textbf{NGC 1068}  center hosts a super--massive \textbf{BH}  within a
thick dust and gas doughnut-shaped cloud.  \textbf{ALMA} showed evidence that the molecular torus
consists of counter-rotating and misaligned disks on parsec scales which can explain   the \textbf{BH} rapid growth.
From the observation of gas motion around the \textbf{BH}
inner orbits, the presence of two disks of gas rotating in
opposite directions was pointed out.  It has been assumed that the outer disk could have been  formed in a recent times from molecular gas falling. The
inner disks follows the rotation of the galaxy, whereas the outer
disk rotates (in stable orbit) the opposite way.
The interaction between counter-rotating
disks may enhance the accretion rate with a rapid multiple-phases of accretion.

This  orbiting structure could be interpreted as a special \textbf{RAD},  composed by  an
  outer disk counter-rotating  relative to  inner
disk.  This double structure with counter-rotating outer disk has been studied in particular in   \cite{dsystem,letter,Fi-Ringed}.
These couples are generally stabilized (for tori collision)
  for high spins of the \textbf{BHs}, where the distance  between the two   tori  can be  very   large, the inner co-rotating disk can be in the ergoregion and  the  two tori can be both in accretion phase, or the outer or the inner torus of the couple be in the accretion with a quiescent  pair  component.
  (The case of  an orbiting pair of tori with an outer counter-rotating torus,  differs strongly from  the couple  of  an inner  counter-rotating  torus, limiting strongly the possibility of  simultaneous accreting phase, mostly inner accretion counter-rotating  torus  and outer  quiescent corotating torus, being possible only for slowly  spinning attractors \citep{letter,dsystem}.)

Definition of torus counter-rotation   is grounded on the hypothesis  that the torus shares its symmetry plane and equatorial plane with  the central stationary  attractor and, within proper  assumptions  on the flow direction, the torus corotation or counter-rotation  is a well defined property.
In a more general frame, including the  misaligned (or tilted) tori, these configuration  symmetries   do not hold--\citep{ella-correlation,mnras2,cqg2020}. In this context we expect  that, because of   the  background  frame--dragging, combined eventually with magnetic field or viscosity, and depending on tori inclination angle, the orbiting tilted torus  can split  in an  inner part, forming eventually  an equatorial co-rotating torus  and an outer torus, producing a multiple  disks structure  composed by two orbiting tori centered on the \textbf{BH} with different relative  rotation orientation, affecting the \textbf{BH} spin and mass. This complex phenomenon depends   on several tori characteristic as its   geometrical thickness,  symmetries, maximum density points.
In this context, the Lense--Thirring effect can express, being  combined with the vertical stresses in the tori and the polar gradient of  pressure,  in  the Bardeen--Petterson effect on the  originally  misaligned torus,  broken due to the  frame dragging and  other factors   as the fluids viscosity, in an inner  corotating torus  and an outer torus which may also  be counter-rotating,  where the
\textbf{BH} spin  can change  under the action of the tori  torques \citep{BP75,Nealon:2015jya,Martin:2014wja,King:2018mgw,2012ApJ...757L..24N,2012MNRAS.422.2547N,2006MNRAS.368.1196L,Feiler,King2005}.
The  frame-dragging can   affect the accretion process,  in particular for  counter-rotating tori   acting on the matter and photons flow from the accreting tori.
The flow, having an initial counter-rotating component, due to the Lense--Thirring effect tends to reverse the rotation direction (toroidal component of the velocity in the proper frame)  along its trajectory  from the counter--rotating orbiting torus  towards the \textbf{BH}. The flow, assumed to be free-falling into the central attractor,  inherits some properties of the accreting   configurations.
Its trajectory  is characterized by the presence of  flow turning point, defined by the condition  $u^{\phi}=\Omega=0$ on the   axial component of the flow velocity and relativistic angular velocity relate to the distant static observer.

From methodological view--point we consider one-particle-species counter-rotating, geometrically thick toroids centered on the equatorial plane of a  Kerr \textbf{BH}, considering "disk--driven" free--falling accretion flow constituted by matter and photons. We use a full  GRHD  Polish doughnut  (PD) model \citep{abrafra}, considering also the case of  "proto-jets (or jets)  driven" flows. (For the toroids influenced by the dark energy, or relict cosmological constant see \cite{2005MPLA...20..561S,2009CQGra..26u5013S,2020Univ....6...26S,2016EPJC...76...32S,2021Univ....7..416S}) Proto-jets are open HD  toroidal configurations, with matter funnels along the \textbf{BH} rotational axis, associated to PD models, and emerging under special conditions on the fluid  forces balance.  Toroidal surfaces are the closed, and closed cusped PD solutions, proto-jets are the open cusped solutions of the PD model.

{In this work we focus on the conditions for the existence of  the  turning point and the flow properties at this point. We discuss   properties of the flow at the turning points distinguishing photon from matter components in the flow, and proto-jets driven and tori driven accreting flows.
The turning point  could  be   remarkably active  part of the accreting flux of matter and photons,    and we consider here particularly the region of the  \textbf{BH} poles the equatorial plane,  eventually characterized  by an increase of the flow luminosity and temperature.
However,  we expect that the  observational properties  in  this region could   depend strongly on the  processes timescales (related to  the  time flow reaches the turning points).}

\medskip

The paper is organized as follows:

In Sec.\il(\ref{Sec:first}) we define the problem setup. In
Sec.\il(\ref{Sec:quaconsta}) equations and constants of motion are introduced.
Details on tori models are in Sec.\il(\ref{Sec:tori-models}).
Characteristics of the fluids at the flow  turning point are the focus of Sec.\il(\ref{Sec:fluidsturing}).
In Sec.\il(\ref{Sec:turning}) there is the analysis of the  flow turning point, where
definition of the turning point radius and plane is provided in Sec.\il(\ref{Sec:fishing-particles}).
The analysis of the  extreme values of  the turning point radius and plane is in Sec.\il(\ref{Sec:extreme-turning-box}).
Fluid velocity at the turning point is studied in Sec.\il(\ref{Sec:velocity}).
In Sec.\il(\ref{Sec:basis-eq}) we specialize the investigation to the
  equatorial plane case, distinguishing the cases
of flow turning point  located on the equatorial plane of the central attractor in Sec.\il(\ref{Sec:argument}), with the discussion of the
conditions on the counter-rotating flows with Carter constant $\Qa=0$ in Sec.\il(\ref{Sec:basisCQu0}). Then the case of
flow off the equatorial plane and general considerations  on initial configurations are explored in Sec.\il(\ref{Sec:point-actua}).
Turning points of the  counter-rotating proto-jet  driven flows are studied in Sec.\il(\ref{Sec:proto-jets-driven}).
  Verticality of the  counter-rotating flow turning point (location along the \textbf{BH} rotational axis) is discussed in Sec.\il(\ref{Sec:vertical-z}).
Flow thickness and counter-rotating tori energetics are investigated in Sec.\il(\ref{Sec:flow}).
In Sec.\il(\ref{Sec:accelerations-fluids})  there are some considerations on the
    fluids  at the turning point.
    Discussion and concluding  remarks follow in Sec.\il(\ref{Sec:discussion}).

\section{Counter-rotating  accreting tori orbiting   Kerr black holes}\label{Sec:first}
\subsection{Equations and constants of geodesic  motion}\label{Sec:quaconsta}
We consider    counter-rotating  toroidal   configurations orbiting a central Kerr \textbf{BH} having      spin $a=J/M\in]0,M]$, total angular momentum  $J$  and  the  gravitational mass parameter $M$.
The non-rotating   case $a=0$ is the   Schwarzschild \textbf{BH} solution while the extreme Kerr \textbf{BH}  has dimensionless spin $a/M=1$.
The background metric,
in the Boyer-Lindquist (BL)  coordinates
\( \{t,r,\theta ,\phi \}\),  is\footnote{We adopt the
geometrical  units $c=1=G$ and  the $(-,+,+,+)$ signature, Latin indices run in $\{0,1,2,3\}$.  The radius $r$ has unit of
mass $[M]$, and the angular momentum  units of $[M]^2$, the velocities  $[u^t]=[u^r]=1$
and $[u^{\phi}]=[u^{\theta}]=[M]^{-1}$ with $[u^{\phi}/u^{t}]=[M]^{-1}$ and
$[u_{\phi}/u_{t}]=[M]$. For the seek of convenience, we always consider the
dimensionless  energy and effective potential $[V_{eff}]=1$ and an angular momentum per
unit of mass $[L]/[M]=[M]$.}:
%
\bea \label{alai}&& ds^2=-\left(1-\frac{2Mr}{\Sigma}\right)dt^2+\frac{\Sigma}{\Delta}dr^2+\Sigma
d\theta^2+\left[(r^2+a^2)+\frac{2M r a^2}{\Sigma}\sin^2\theta\right]\sin^2\theta
d\phi^2-\frac{4rMa}{\Sigma} \sin^2\theta  dt d\phi,
\eea
where
\bea
\Delta\equiv a^2+r^2-2 rM;\quad \Sigma\equiv a^2 \cos ^2\theta+r^2.
\eea
 In the following, where more convenient, we  use dimensionless units where $M=1$.
  The horizons $r_-<r_+$ and the outer  and inner stationary  limits $r_{\epsilon}^\pm$ (ergosurfaces) are respectively given by
\bea
r_{\pm}\equiv M\pm\sqrt{M^2-a^2};\quad r_{\epsilon}^{\pm}\equiv M\pm\sqrt{M^2- a^2 \cos\theta^2};
\eea
where $r_+<r_{\epsilon}^+$ on   $\theta\neq0$  and  $r_{\epsilon}^+=2M$  in the equatorial plane $\theta=\pi/2$. The equatorial plane   is a metric symmetry plane  and   the  equatorial  (circular) trajectories are confined on the equatorial  plane as a consequence of the metric tensor symmetry under reflection through  the plane $\theta=\pi/2$.

The constants of geodesic motions are
\bea&&\label{Eq:EmLdef}
\Em=-(g_{t\phi} \dot{\phi}+g_{tt} \dot{t}),\quad \La=g_{\phi\phi} \dot{\phi}+g_{t\phi} \dot{t},\quad  g_{ab}u^a u^b=-\mu^2,
\\
&&\label{Eq:eich}
\Qa=(\cos\theta)^2 \left[a^2 \left(\mu^2-\Em^2\right)+\left(\frac{\La}{\sin\theta}\right)^2\right]+(g_{\theta\theta} \dot{\theta})^2;
\eea
with   $u^a\equiv\{ \dot{t},\dot{r},\dot{\theta},\dot{\phi}\}$, where
$\dot{q}$ indicates the derivative of any quantity $q$  with respect  the proper time (for  $\mu=1$) or  a properly defined  affine parameter for the light-like orbits (for $\mu=0$).
In Eqs\il(\ref{Eq:EmLdef}) quantities  $\Em$ and $\La$ are defined  from   the Kerr geometry  rotational  Killing field   $\xi_{\phi}=\partial_{\phi}$,
     and  the Killing field  $\xi_{t}=\partial_{t}$
representing the stationarity of the  background. The constant $\La$ in Eq.\il(\ref{Eq:EmLdef}) may be interpreted       as the axial component of the angular momentum  of a test    particle following
timelike geodesics and $\Em$  represents the total energy of the test particle
 coming from radial infinity, as measured  by  a static observer at infinity,
  while  $\Qa$ in Eq.\il(\ref{Eq:eich})  is known as Carter constant. If $a>0$, then   particles counter-rotation  (corotation) is defined by $\La a<0$ ($\La a>0$).

From the constants of motion $(\Em, \La)$ we obtain the relations for the velocity components $(u^t,u^\phi)$:
\bea\label{Eq:ufidottdotinconst}
\dot{t}= \frac{g_{\phi\phi} \Em+g_{t\phi} \La}{g_{t\phi}^2-g_{\phi\phi} g_{tt}},\quad \dot{\phi}= -\frac{g_{t\phi} \Em+g_{tt} \La}{g_{t\phi}^2-g_{\phi\phi} g_{tt}}.
\eea
The relativistic angular velocity and the specific  angular momentum are respectively
 \bea&&\nonumber\label{Eq:flo-adding}
\Omega \equiv\frac{u^\phi}{u^{t}}=-\frac{\Em g_{\phi t}+g_{tt} \La}{\Em g_{\phi \phi}+g_{\phi t} \La}= -\frac{g_{t\phi}+g_{tt} \ell}{g_{\phi\phi}+g_{t\phi} \ell},\quad
\ell\equiv\frac{\La}{\Em}=-\frac{u_\phi}{u_{t}}=-\frac{g_{\phi\phi}u^\phi  +g_{\phi t} u^t }{g_{tt} u^t +g_{\phi t} u^\phi } =-\frac{g_{t\phi}+g_{\phi\phi} \Omega }{g_{tt}+g_{t\phi} \Omega}.
\eea
 If $a>0$ the  fluid  counter-rotation  (corotation) is defined by $\ell a<0$ ($\ell a>0$)\footnote{In this case we assume $\Em>0$. This condition  for corotating fluids in the ergoregion has to be discussed further. In the ergoregion  particles can also have $\La=0$, associated to fluids with  $\ell=0$. However this condition characterizing the ergoregion  is not associated to geodesic  circular motion in the \textbf{BH} spacetimes,  while  it is a well known feature of some  Kerr naked singularities (\textbf{NSs}) ($a>M$),  where there are  also circular geodesic with $\Em\leq0$ or   $\La\leq0$ \citep{ergon,observers,2013CQGra..30g5012S,1980BAICz..31..129S,2016PhRvD..94h6006B,2011CQGra..28o5017S,2012CQGra..29f5002S,Pu:Kerr}).}.
 (Static  observers, with  four-velocity   $\dot{\theta}=\dot{r}=\dot{\phi}=0$
cannot exist inside the ergoregion, then trajectories   $\dot{r}\geq0$, including particles  crossing the stationary  limit and escaping outside
in the region $r\geq r_{\epsilon}^+$ are possible.).

%
 For convenience  we summarize  the  Carter equations  of motion as follows  (see\cite{Carter}):
\bea&&\label{Eq:eqCarter-full}
 \dot{t}=\frac{1}{\Sigma}\left[\frac{P \left(a^2+r^2\right)}{\Delta}-{a \left[a \Em (\sin\theta)^2-\La\right]}\right],\quad
\dot{r}=\pm \frac{\sqrt{R}}{\Sigma};\quad \dot{\theta}=\pm \frac{\sqrt{T}}{\Sigma},\quad \dot{\phi}=\frac{1}{\Sigma}\left[\frac{a P}{\Delta}-\left[{a \Em-\frac{\La}{(\sin\theta)^2}}\right]\right];
\eea
where
\bea\nonumber
&& P\equiv \Em \left(a^2+r^2\right)-a \La,\quad R\equiv P^2-\Delta \left[(\La-a \Em)^2+\mu^2 r^2+\Qa\right],\quad  T\equiv \Qa-
(\cos\theta)^2 \left[a^2 \left(\mu^2-\Em^2\right)+\left(\frac{\La}{\sin\theta}\right)^2\right].
\eea
\subsection{Details on tori models}\label{Sec:tori-models}
\begin{figure*}
\centering
    \includegraphics[width=6.5cm]{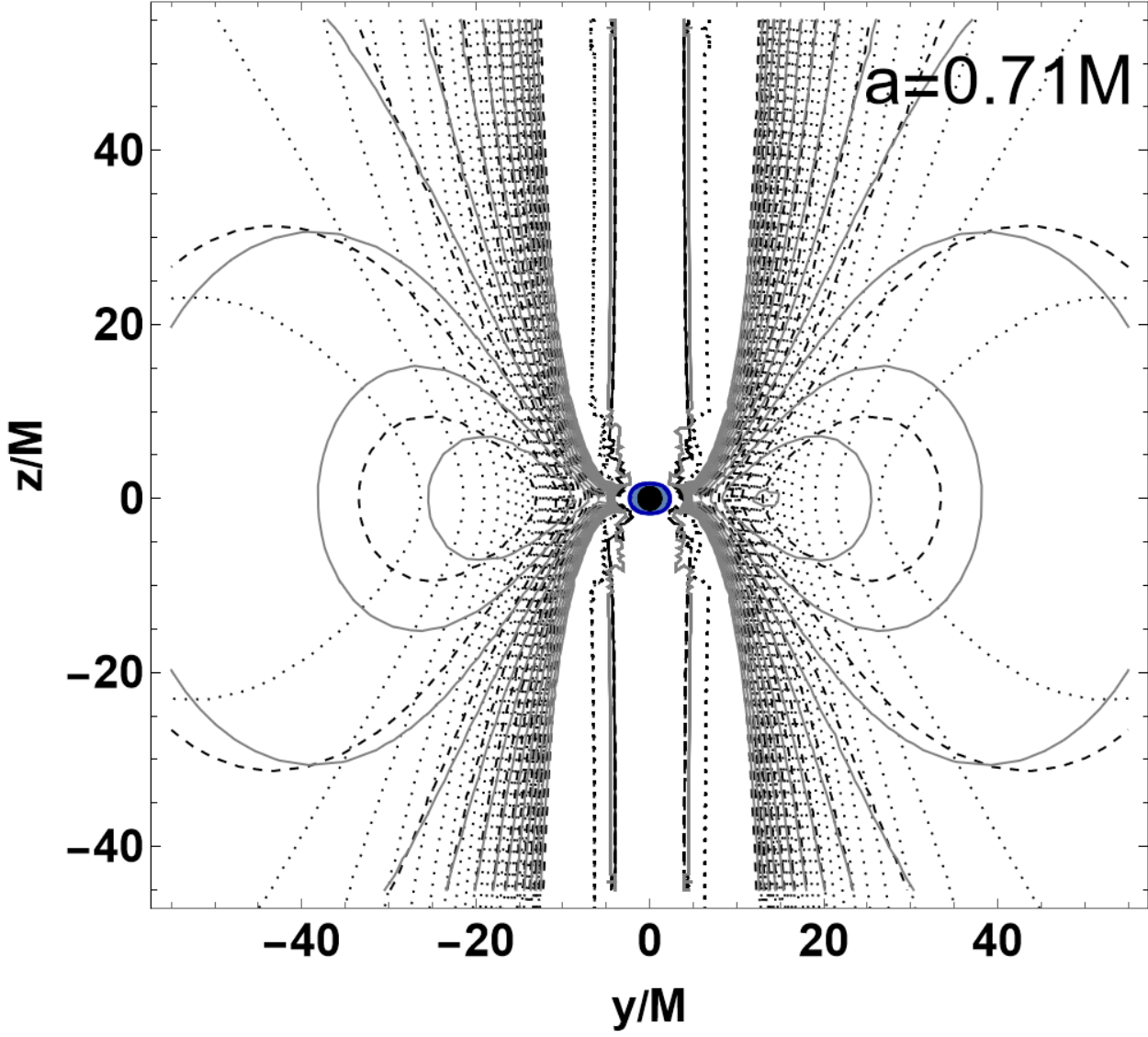}
    \includegraphics[width=6.5cm]{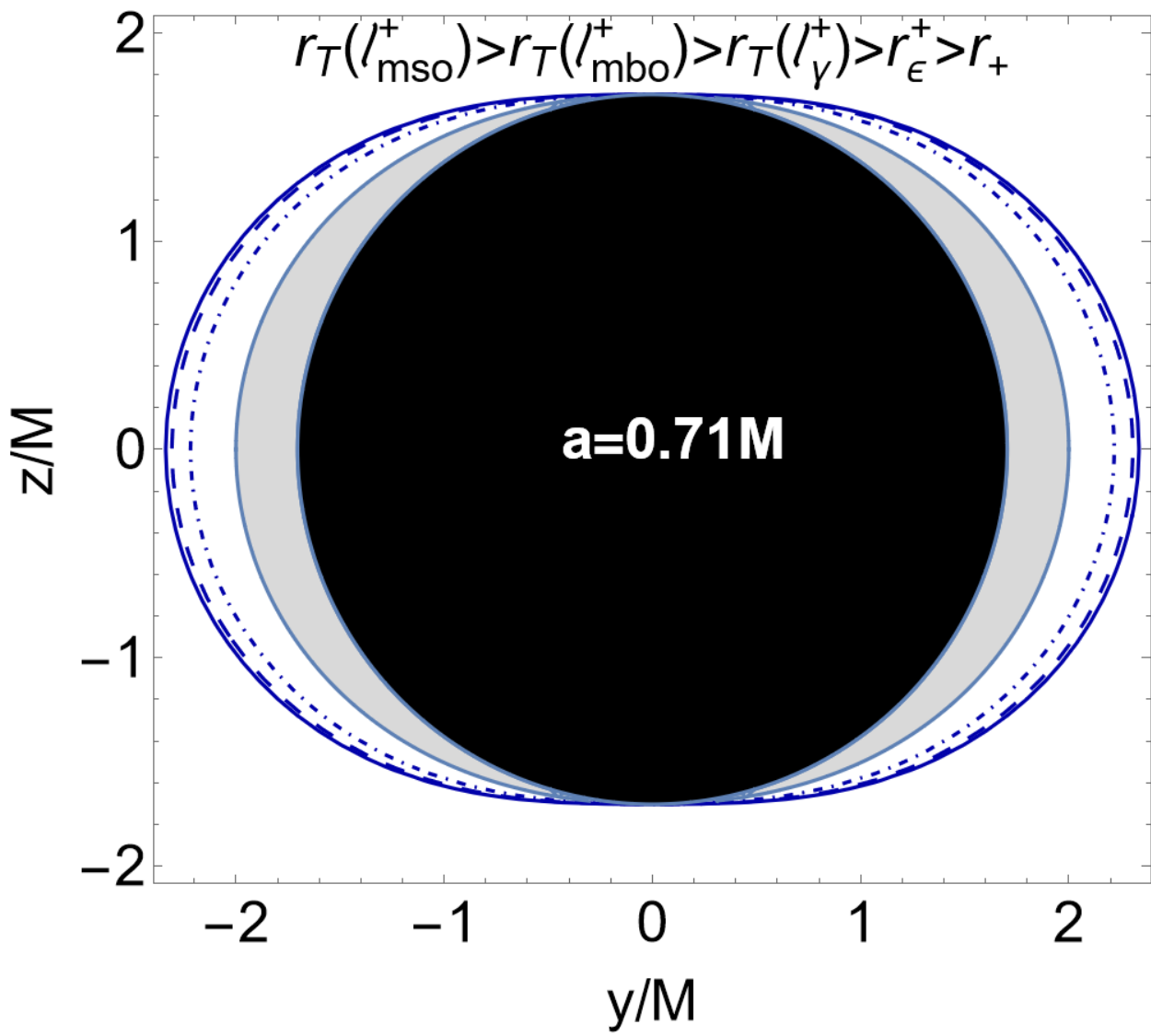}
   \caption{Counter-rotating flows.   There is $r= \sqrt{y^2+z^2}$ and $\theta=\arccos({z}/r)$. Black central region is the central \textbf{BH}, gray region is the outer ergoregion.  The  \textbf{BH} spin-mass ratio is $a/M=0.71$. In this spacetime there are the limiting specific angular momenta $ \{\ell_{mso}^+=-4.21319,\ell_{mbo}^+=-4.61534,\ell_{\gamma}^+=-6.50767\}$,   defined in Eqs\il(\ref{Eq:def-nota-ell}). Left panel shows  the  toroidal fluid equi-pressure (equi-density) surfaces  evaluated at $\ell=-6.6$ (dotted-curves), $\ell=-5$ (dashed curves), $\ell=-4.5$ (gray curves). Closed toroids and the proto-jets (open funnels of matter parallel the \textbf{BH} rotational axis) are shown. Right panel is a zoom  in the region close to the \textbf{BH}, $r_+$ is the \textbf{BH} horizon,  and  $r_{\epsilon}^+$ is  the outer ergosurface radius.  The counter-rotating
flow turning points, $r_{\Ta}$  of Eq.\il(\ref{Eq:turning-point-radius}) are shown,  evaluated for fluid specific angular momenta $\ell_{mso}^+$ (plain curve), $\ell_{mbo}^+$  (dashed curve) and $ \ell_{\gamma}^+$  (dotted-dashed curve).  The corona defined by the
   radii $(r_\Ta(\ell_{mso}^+)-r_\Ta(\ell_{mbo}^+))$ ($(r_\Ta(\ell_{\gamma}^+)-,r_\Ta(\ell_{mbo}^+))$) is  the range of  the  counter-rotating flow turning points location from cusped tori (proto-jets) driven flows. Radii reach the  maximum at the equatorial plane $(z=0)$--see Eqs\il(\ref{Eq:max-cusp-eq-exte1}),  Eqs\il(\ref{Eq:max-cusp-eq-exte})  and Figs\il(\ref{Fig:bluecurvehereq}). }\label{Fig:PlotlongVie}
\end{figure*}

We specialize our analysis to   GRHD toroidal  configurations centered on the Kerr \textbf{BH}  equatorial plane, which is  coincident with  the tori equatorial symmetry  plane. Tori are  composed by a  one particle-specie   perfect fluid, with  constant fluid specific angular momentum $\ell$ \citep{abrafra,pugtot,ringed,review},
 total energy density $\varrho$  and
pressure   $p$, as measured by an observer comoving with the fluid with velocity $u^{a}$--Figs\il(\ref{Fig:PlotlongVie}).
We  assume $\partial_t \mathbf{q}=0$ and
$\partial_{\varphi} \mathbf{q}=0$,  with $\mathbf{q}$ being a generic spacetime tensor.
The  continuity equation
is  identically satisfied and the  fluid dynamics  is  governed by the Euler equation only. 
Assuming  a barotropic equation of state ($p=p(\varrho)$), and orbital motion with  $u^{\theta}=0$ and
$u^r=0$, 
and by setting  $\ell=$constant  as a torus parameter fixing  the maximum density points in the disk,
the pressure gradients are  regulated by  the gradients of an effective potential function for the fluid  $V_{eff}(r;\ell,a)$,
which is  invariant under the mutual transformation of  the parameters
$(a,\ell)\rightarrow(-a,-\ell)$.

The fluid  effective potential, emerging from the GRHD  constrain equation for the  pressure, reads \cite{abrafra,ringed}
\bea
V_{eff}^2=\left(\frac{\Em}{\mu}\right)^2={\frac{g_{t\phi}^2-g_{\phi\phi} g_{tt}}{g_{\phi\phi}+2 g_{t\phi} \ell +g_{tt} \ell ^2}},
\eea
 assuming at the initial data $\dot{r}=\dot{\theta}=0$ and using   the definitions of constants of motions $(\mu, \Em, \La)$ of Eqs\il(\ref{Eq:EmLdef}) and  $\ell$ in Eqs\il(\ref{Eq:flo-adding}).
The extremes of  the pressure in Eq.\il(\ref{Eq:toricenter-inner}) are therefore regulated by  the angular momentum distributions $\ell(r,\theta;a):\partial_r V_{eff}=0$ which, on the equatorial plane $\theta=\pi/2$,  is
\bea
\ell^{\mp}\equiv\frac{a^3\mp r^{3/2} \Delta-a (4-3 r) r}{a^2-(r-2)^2 r}
\eea
for corotating $(-)$  and counter-rotating $(+)$  fluids respectively.
Fluid effective potential defines  the function $K(r)=V_{eff}(\ell(r))$. Cusped tori have parameter $K=K_\times\equiv K(r_{\times})\in]K_{center}, 1[\subset ]K_{mso}, 1[$, where $K_{center}\equiv K(r_{center})$. (More in general we adopt the notation $q_{\bullet}\equiv q(r_{\bullet})$ for any quantity $q$ evaluated on a radius $r_{\bullet}$.).
Super-critical tori have parameter $K=K_s\in ]K_{\times},1[$ and they are  characterized by  an accretion throat (opening of the cusp), considered  in more details  in Sec.\il(\ref{Sec:flow}).

 Torus cusp $r_\times$ is the  minimum point of pressure and density in the torus  corresponding  to the maximum point of the  fluid effective potential. The torus center $r_{center}$ is the maximum point of pressure and density in the torus,  corresponding  to the minimum point of the fluid  effective potential. For the  cusped co-rotating and  counter-rotating tori, there is:
\bea&&\label{Eq:toricenter-inner}
r_{center}=r_{\Upsilon}^{+},\quad r_{\times}=r_{\Upsilon}^{-},\quad\mbox{with}\quad
r_{\Upsilon}^{\pm }(a,\ell)\equiv \frac{\lambda_d\pm\lambda_e}{12},\quad\mbox{where}
\\\nonumber
&&
\lambda _a\equiv (a-2 \ell) (a-\ell);\quad \lambda _b\equiv\left(2 a^2-3 a \ell+\ell^2\right)^2,\\
&&\nonumber \lambda _{c_1}\equiv27 a^2 \ell^4 (a-\ell)^2-144 a^2 \lambda _a (a-\ell)^2+16 \lambda _a^3-72 \ell^2 \lambda _a (a-\ell)^2+432 (a-\ell)^4,
\\
&&\nonumber \lambda_{c_2}\equiv\left[16 \lambda _a^3-72 \lambda_a (a-\ell)^2 \left(2 a^2+\ell^2\right)+27 (a-\ell)^2 \left[a^2 \left(\ell^4+16\right)-32 a \ell+16 \ell^2\right]\right]{}^2-256 \left(2 a^2-3 a \ell+\ell^2\right)^6,
\\
&&\nonumber \lambda _c\equiv 6 \sqrt[3]{\lambda _{c_1}+\sqrt{\lambda _{c_2}}},
\\
&&\nonumber \lambda _d\equiv \sqrt{\frac{288 \sqrt[3]{2} \lambda _b}{\lambda _c}+2^{2/3} \lambda _c+9 \ell^4-48 \lambda _a}+3 \ell^2,
\\
&&\nonumber\lambda _e\equiv 6 \sqrt{\frac{3 \left[32 (a-\ell)^2+\ell^2 (\ell^4-8 \lambda _a)\right]}{2 \left(\lambda _d-3 \ell^2\right)}-\frac{8 \lambda _a}{3}-\frac{8 \sqrt[3]{2} \lambda _b}{\lambda _c}-\frac{\lambda _c}{18 \sqrt[3]{2}}+\frac{\ell^4}{2}}.
\eea
The matter outflows as consequence of the violation of mechanical equilibrium  in the balance of the gravitational and inertial forces and the pressure gradients in the tori  regulated by the fluid effective potential (Paczynski-Wiita (P-W)  hydro-gravitational instability  mechanism  \citep{Pac-Wii}). At the cusp ($r\leq r_\times$) the fluid may be considered pressure-free.

\begin{figure*}
\centering
    \includegraphics[width=5.75cm]{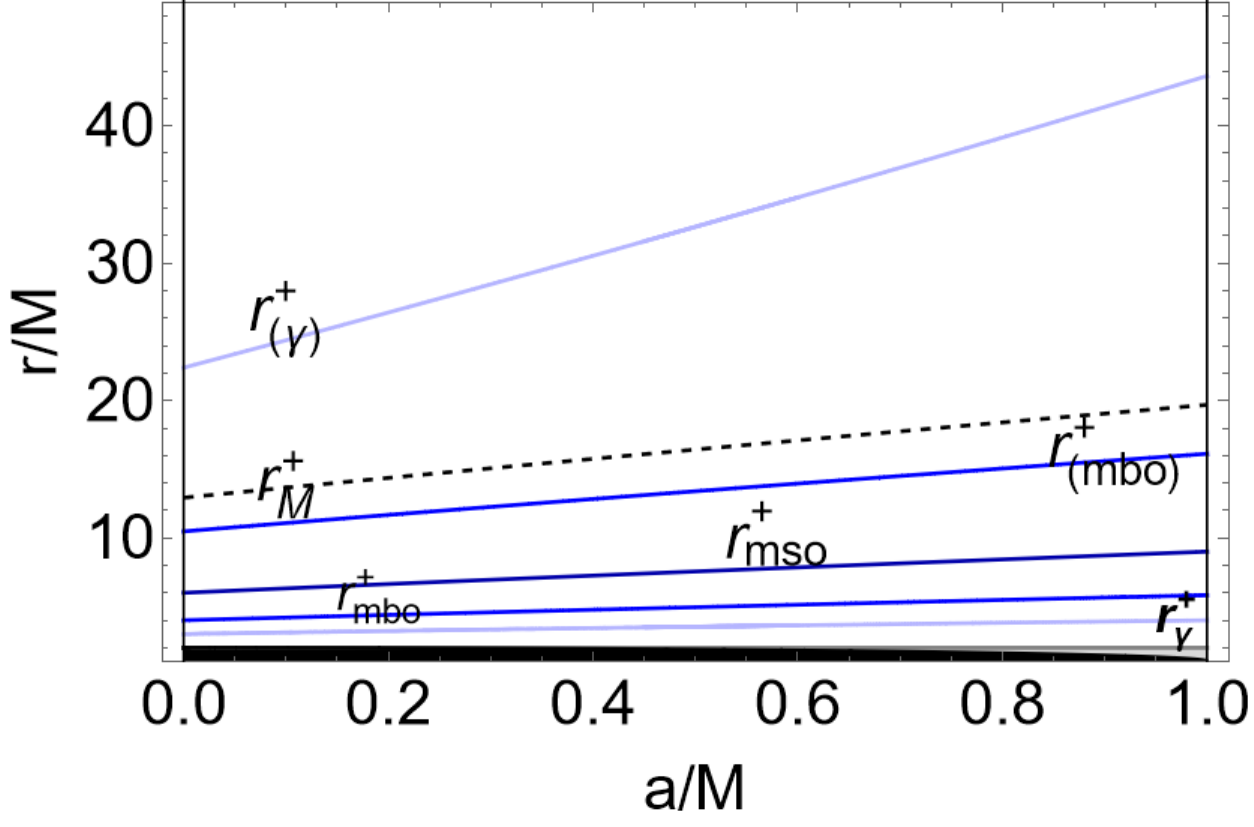}
  \includegraphics[width=5.75cm]{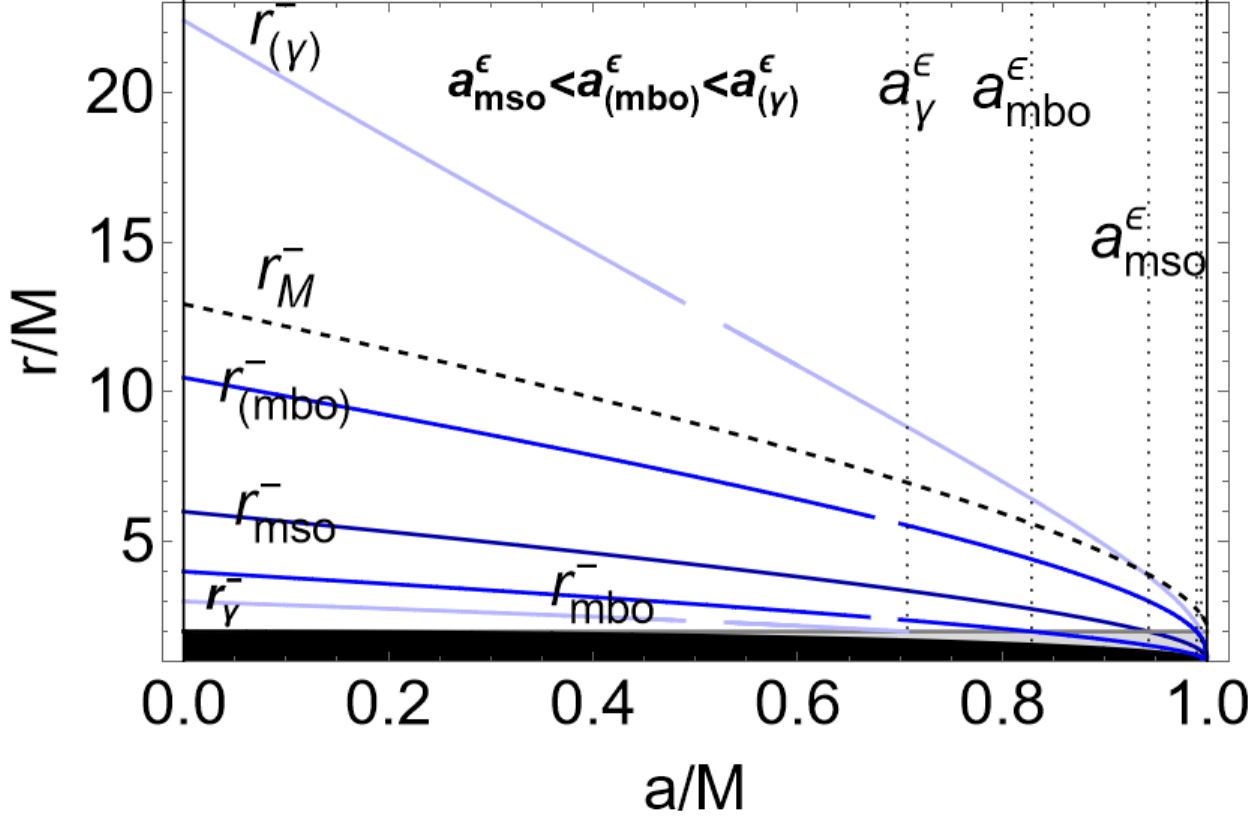}
   \includegraphics[width=5.75cm]{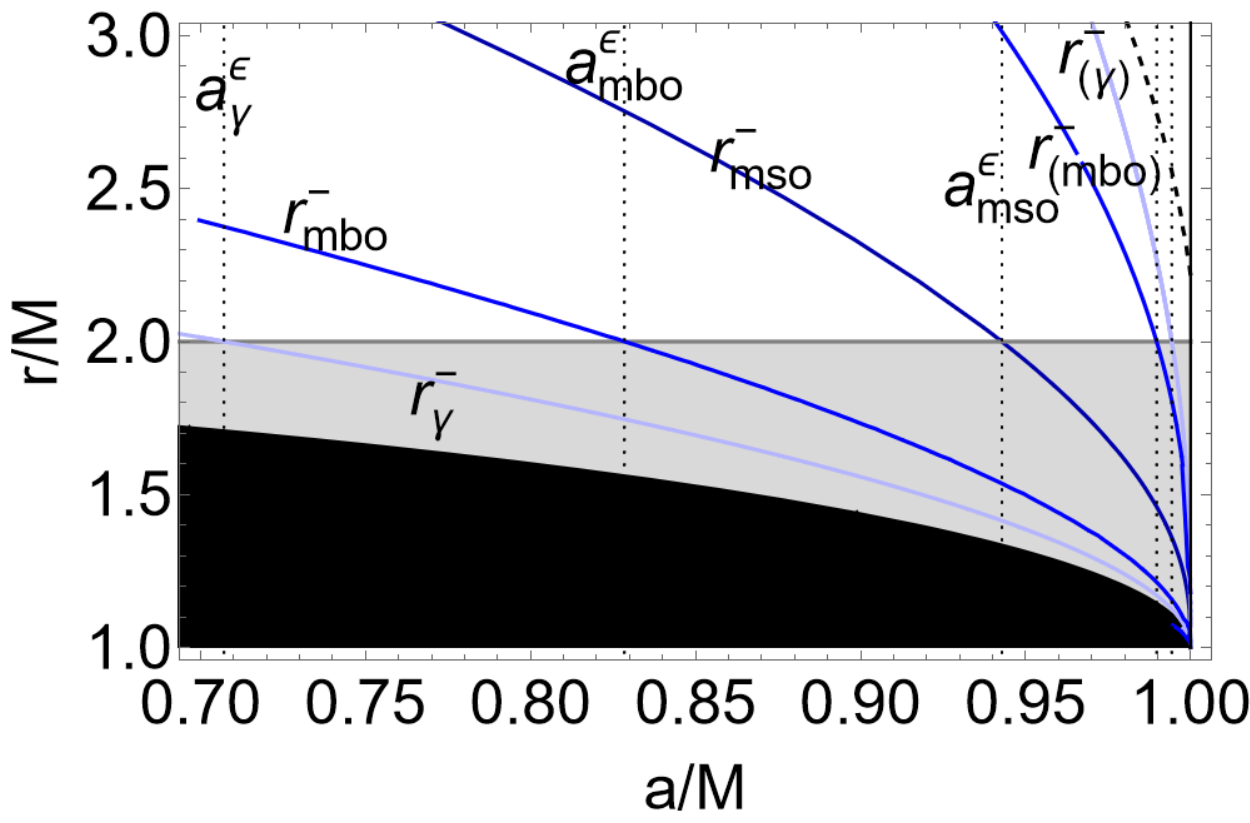}
  \caption{Kerr spacetime geodesic structure   of Eqs\il(\ref{Eq:def-nota-ell})  for corotating $(-)$ (center and right panels) and counter-rotating orbits $(+)$ (left panel)  as functions of \textbf{BH} spin-mass ratio $a/M$, where black region is the \textbf{BH} at $r<r_+$, where $r_+$ is the outer \textbf{BH} horizon, gray region is the outer ergoregion $]r_{+},r_{\epsilon}^+[$ of  the Kerr spacetime on the equatorial plane. Spins $\{a_{\gamma}^\epsilon\equiv 0.707107M, a_{mbo}^\epsilon\equiv 0.828427M, a_{mso}^\epsilon\equiv 0.942809M, a_{(mbo)}^\epsilon\equiv 0.989711M,a_{(\gamma)}^\epsilon\equiv 0.994298M\}$ are the geometries where  the  corotating geodesic structure radii coincide with the outer ergosurface $r_{\epsilon}^+=2M$ on the equatorial plane.
  }\label{Fig:Plotbalchocasep}
\end{figure*}

 Accretion  disk   physics is regulated by the Kerr background  circular geodesic structure   constituted by the marginally  circular orbit for timelike particles  $r_{\gamma}^{\pm}$,  which is also the photon circular  orbit, the marginally  bounded orbit, $r_{mbo}^{\pm}$, and the marginally stable circular orbit, $r_{mso}^{\pm}$ for corotating and counter-rotating motion.
 We consider  also the radius $r_{\mathcal{M}}^{\pm}:\partial_r^2\ell=0$, and
the set of radii  $r_{(mbo)}^{\pm}$ and $r_{(\gamma)}^{\pm}$  and $r_{(\mathcal{M})}^{\pm}$ defined from
\bea&&\nonumber
r_{\mathrm{(mbo)}}^{\pm}:\;\ell^{\pm}(r_{\mathrm{mbo}}^{\pm})=
 \ell^{\pm}(r_{\mathrm{(mbo)}}^{\pm})\equiv \mathbf{\ell_{\mathrm{mbo}}^{\pm}},\quad
  r_{(\gamma)}^{\pm}: \ell^{\pm}(r_{\gamma}^{\pm})=
  \ell^{\pm}(r_{(\gamma)}^{\pm})\equiv \mathbf{\ell_{\gamma}^{\pm}},\quad  r_{(\mathcal{M})}^{\pm}:\ell^{\pm}(r_\Mie^{\pm})\equiv \ell_\Mie^{\pm}=\ell^{\pm}(r_\Mie^{\pm}),
  \\&&\label{Eq:def-nota-ell}
\mbox{where}\quad r_\gamma^{\pm}<r_{\mathrm{mbo}}^{\pm}<r_{\mathrm{mso}}^{\pm}<
 r_{\mathrm{(mbo)}}^{\pm}<
  r_{(\gamma)}^{\pm}
  \eea
  see Figs\il(\ref{Fig:Plotbalchocasep}).
Ranges  $(\mathbf{L_1,L_2,L_3})$  of fluids specific angular momentum  $\ell$ govern the  tori    topology, according to the geodesic structure of Eqs\il(\ref{Eq:def-nota-ell}),  as follows:
\begin{description}
\item[\textbf{$ \mathbf{L_1} $}:] for $\ell\in \mathbf{L_1} $ there are  quiescent (i.e. not cusped)  and cusped tori--where there is $
\mp \mathbf{L_1}^{\pm}\equiv[\mp \ell_{mso}^{\pm},\mp\ell_{mbo}^{\pm}[$.
  The cusp is   $r^{\pm}_{\times}\in]r^{\pm}_{mbo},r^{\pm}_{mso}]$ (with $K_{\times}^{\pm}<1$)) and  the center with maximum pressure in $r^{\pm}_{center}\in]r^{\pm}_{mso},r^{\pm}_{(mbo)}]$.
\item[\textbf{$\mathbf{L_2}$:}] for $\ell\in \mathbf{L_2}$ there are  quiescent  tori and proto-jets (open-configurations) --where there is $\mp \mathbf{L_2}^{\pm}\equiv[\mp \ell_{mbo}^{\pm},\mp\ell_{\gamma}^{\pm}[ $.
The    cusp  $r_{\times}^{\pm}\in]r_{\gamma}^{\pm},r_{mbo}^{\pm}]$  is associated to the proto-jets,  with $K_{\times}>1$,  and the  center with maximum pressure is in $r_{center}^{\pm}\in]r_{(mbo)}^{\pm},r_{(\gamma)}^{\pm}]$;
\item[\textbf{ $\mathbf{L_3}$:}] for $\ell\in \mathbf{L_3}$ there are only quiescent  tori where there is   $\mp \mathbf{L_3}^{\pm}\equiv\mp \ell \geq\mp\ell_{\gamma}^{\pm}$
and the torus center is at  $r^{\pm}_{center}>r_{(\gamma)}^{\pm}$,
\end{description}
--see Figs\il(\ref{Fig:PlotlongVie}). Configurations   with momentum in      $\mathbf{L_2}$  range  and $K>1$ are   associated to (not-collimated) open  structures,   proto-jets, with matter funnels along the \textbf{BH} rotational axis--see Figs\il(\ref{Fig:PlotlongVie})--\citep{pugtot,open,proto-jets,ella-jet}.  Tori--driven and  proto-jets-driven  flows are flows originated  from  tori or proto-jet configurations    respectively.
In Figs\il(\ref{Fig:PlotlongVie}) examples of different   orbiting  configurations  and  torus driven and proto-jets driven flow   turning points are shown.  Conditions for the turning points from proto-jets driven flows are discussed   in more details in Sec.\il(\ref{Sec:proto-jets-driven}).
\begin{figure*}
\centering
 \includegraphics[width=4.5cm]{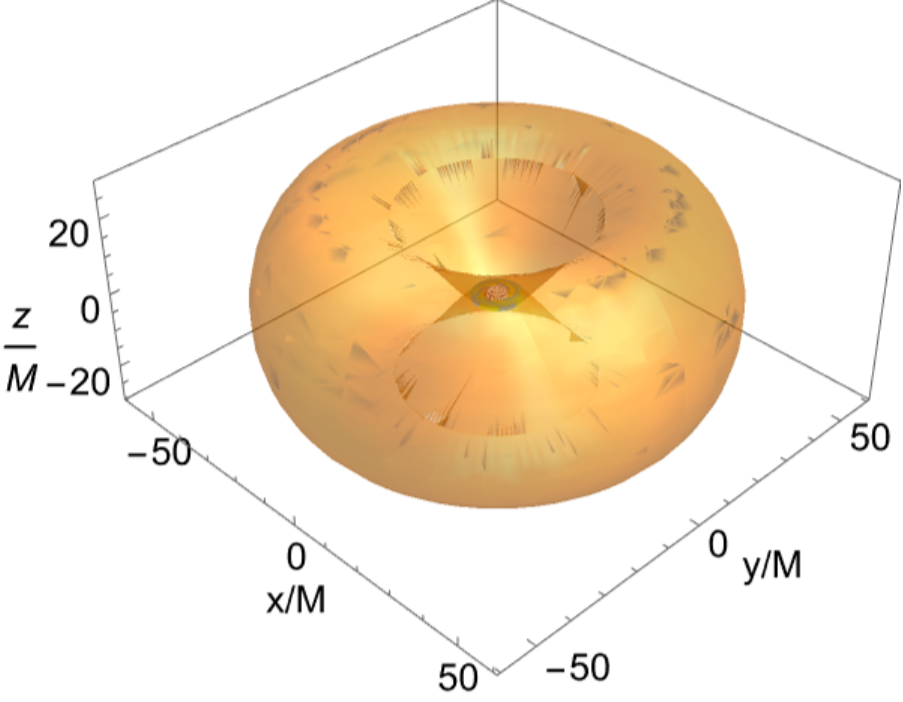}
    \includegraphics[width=5.5cm]{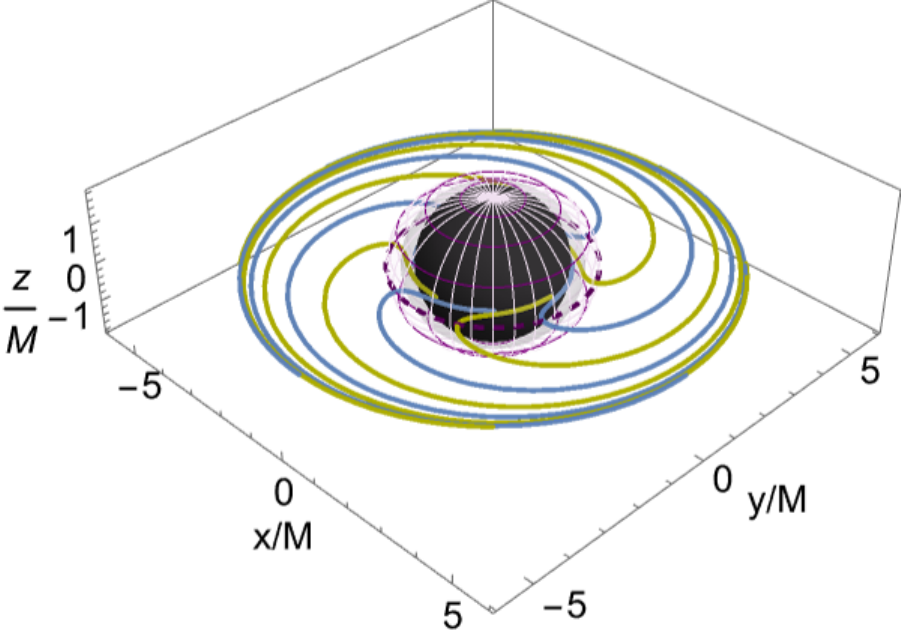}
    \includegraphics[width=5cm]{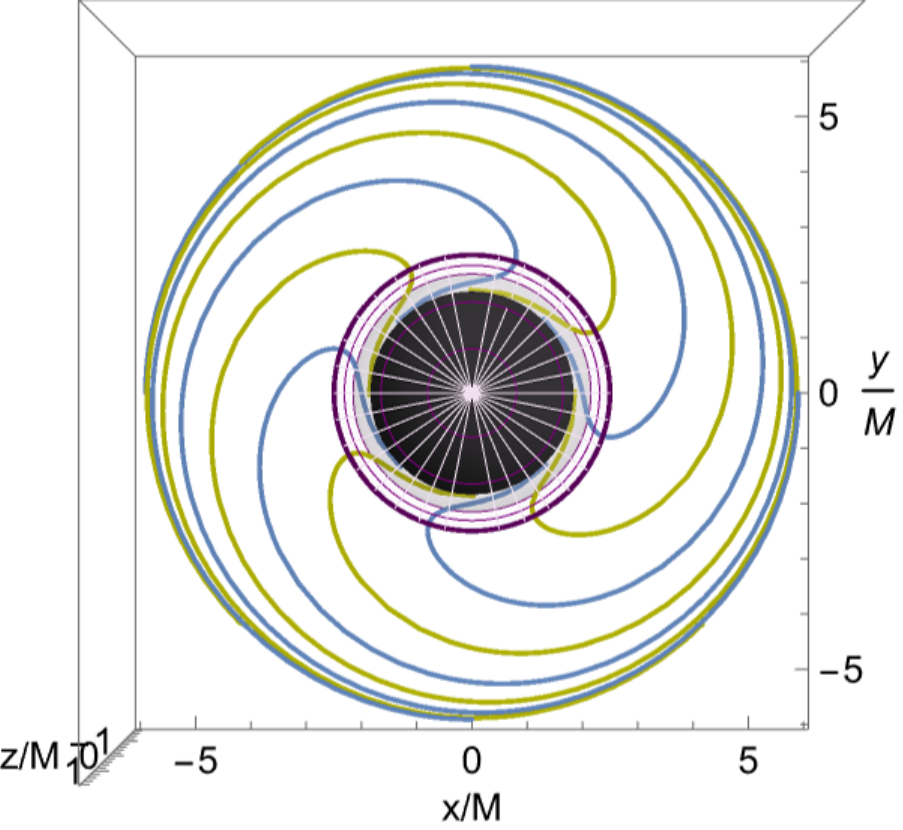}
   \caption{Tori driven counter-rotating flows turning points in the \textbf{BH} spacetime with spin-mass ratio  $a/M=0.71$ and cusped  tori with  the fluid specific angular momentum   $\ell=-4.5$,  where is $\{z=r \cos\theta,y=r \sin \theta \sin\phi,x=r \sin\theta \cos\phi\}$ in dimensionless units. The limiting fluid specific angular momenta, defined  in Eqs\il(\ref{Eq:def-nota-ell}), are  $ \{\ell_{mso}^+=-4.21319,\ell_{mbo}^+=-4.61534,\ell_{\gamma}^+=-6.50767\}$--see also  Figs\il(\ref{Fig:PlotlongVie}). Left panel:  the cusped  torus    orbiting the equatorial plane of the central \textbf{BH}. Central and right panels show a  front and above view  of the  counter-rotating  flow stream section on the equatorial plane from the torus inner edge (cusp)  to the central \textbf{BH}. Black region is the  central  \textbf{BH}  (region $r<r_+$, radius $r_+$ is the outer horizon). Flow turning point $r_{\Ta}=2.31556M$ of Eqs\il(\ref{Eq:turning-point-radius}),(\ref{Eq:max-cusp-eq-exte1}), (\ref{Eq:rte-second-sigma1})  is plotted as the deep-purple curve. Radius $r_{\Ta}$
   lies  in the  turning corona defined by the
   range $(r_\Ta(\ell_{mso}^+)-r_\Ta(\ell_{mbo}^+))$. Gray region is the outer ergosurface, light-purple shaded region is the region $r< r_\Ta(\sigma_\Ta)$ (where $\sigma\equiv\sin^2\theta$)--
see Figs\il(\ref{Fig:Plotrte4}). The analysis for photons is in Figs\il(\ref{Fig:bluecurvehereq})}\label{Fig:bluecurvehereq}
\end{figure*}
\section{Fluids at the  turning point of the azimuthal motion}\label{Sec:fluidsturing}
\subsection{Flow turning points}\label{Sec:turning}
\subsubsection{Definition of the turning point radius and plane}\label{Sec:fishing-particles}
The flow turning point  is defined by the condition   $u^{\phi}=0$--see Figs\il(\ref{Fig:bluecurvehereq}). We thus obtain equation relating the motion constants of the infalling matter to the orbital turning point given generally by coordinates $(r_\Ta,\theta_\Ta)$.
The value of the constant of motion $\ell=\Em/\La$ is determined  by the values of specific angular momentum $\ell$ at the cusp of the accreting torus that is assumed uniform accross the torus. In the following  we use the notation $q_\Ta$ or $q(\Ta)$ for any quantity $q$ considered at the turning point, and $q_0=q(0)$ for  any  quantity $q$ evaluated at the initial point of the free-falling flow trajectory.  In the special case where  the initial flow  particles  location is  coincident  with the torus cusp, we use notation $q_{\times}$.
In the following it will be useful to use the variable $\sigma\equiv \sin^2\theta$.

Conditions at the flow  turning point can be found from  the Carter equations  of motion in Eqs\il(\ref{Eq:eqCarter-full}), within the condition  $u^{\phi}=0$ and  using the   constants of motion Eqs\il(\ref{Eq:EmLdef}) and Eqs\il(\ref{Eq:eich}).

From the definition of constant $\ell$,  fixed by the torus initial data, and  turning point definition   we obtain:
\bea\label{Eq:lLE}
&&\ell=\left.-\frac{g_{t\phi}}{g_{tt}}\right|_\Ta=\frac{2 a r_{\Ta} \sigma_{\Ta} }{a^2 (\sigma_\Ta -1)-(r_\Ta -2) r_\Ta},
\eea
as on the turning point there is
\bea\label{Eq:lLE1}
\Em = -g_ {tt}(\Ta) \dot {t}_\Ta,\quad \La = g_ {t\phi}(\Ta)\dot{t}_\Ta.
\eea
Quantities $(\Em,\La)$ are constants of motion, and could be found as $\Em=\Em_\Ta$ and $\La=\La_\Ta$ at the initial point where (for timelike particles on the equatorial plane)  it could be assumed $V_{eff}(r_\times)=\Em$ at the cusp of  the accreting torus, corresponding to an unstable circular geodesic. The parameters $(\Em,\La)$ are thus  the energy and axial angular momentum of the circular geodesic of the cusp in the equatorial plane of the Kerr geometry.
We also note the independence of the turning point definition on the Carter constant $Q$, affecting the off-equatorial motion. We shall see that definition Eq.\il(\ref{Eq:lLE}) defines, for fixed $\ell$, a spherical region surrounding the \textbf{BH}.  For the turning point the crucial role is played by the constant specific angular momentum $\ell$ which is assumed uniform across the torus. To describe the more general situation then we mainly consider here $\ell$ fixed by the torus, and $(\Em,\La)$ evaluated at the turning point as in  Eqs.\il(\ref{Eq:lLE1}) within  the  (sign) constraint provided by the torus.
(Note that  there is $\La\ell<0$ with $\ell<0$ if $ g_ {t\phi}(\Ta)<0$ where   $\dot{t}_\Ta>0$ which is the natural condition for the future-oriented particle motion \citep{1989BAICz..40..133B},  while  there is $ \Em < 0$ where  $\dot{t}_\Ta > 0$ if   $g_ {tt}(\Ta)>0$ in the  ergoregion).

The flow  turning point is located at a radius $r_{\Ta}$ on a plane $\sigma_\Ta$, related as follows:
\bea&&\label{Eq:add-equa-sigmata}
\sigma_{\Ta}(r_{\Ta})=\frac{\ell  \Delta_{\Ta}}{a (a \ell -2 r_{\Ta})};
\\\label{Eq:turning-point-radius}
&&r_{\Ta}(\sigma_\Ta)=\sqrt{a^2 \left(\sigma_\Ta +\frac{\sigma_\Ta ^2}{\ell ^2}-1\right)-\frac{2 a \sigma_\Ta }{\ell }+1}-\frac{a \sigma_\Ta }{\ell }+1.
\eea
Note that $\sigma_\Ta (r_\Ta)$ and  $r_{\Ta}(\sigma_{\Ta})$ depend on constant of motion  $\ell$ only\footnote{Turning radius $r_\Ta$ and plane $\sigma_\Ta$ of Eqs\il(\ref{Eq:add-equa-sigmata}) and Eqs\il(\ref{Eq:turning-point-radius}) are not independent variables, and they can be found solving the equation of motion or using   further assumptions at any other point of the fluid trajectory.},
holding for matter and photons, not depending  explicitly from  the normalization condition. Quantities   $r_{\Ta}$ and  $\sigma_{\Ta}$ are independent from the initial velocity  $\dot{\sigma}_{\Ta}$ or  the constant  $\Qa$, therefore their dependence on the tori models and accretion process is limited to the dependence on the fluid specific angular momentum $\ell$ and the results considered here are adaptable to a variety of different general relativistic accretion models.
\footnote{{At fixed $\ell$, function $\sigma_\Ta(r_\Ta)$ (or radius $r_\Ta(\sigma_\Ta)$) defines a spherical surface surrounding the central attractor. The  point $(r_\Ta,\theta_\Ta,\phi_\Ta)$ on the sphere can be determined by the set of equations (\ref{Eq:eqCarter-full}) which also  relates $(r_\Ta,\sigma_\Ta)$ to the initial values $(r_0, \sigma_0)$,  obviously depending  on the single particle trajectory. We address this aspect in part in Sec.\il(\ref{Sec:point-actua})}.}.

By using  Eq.\il(\ref{Eq:turning-point-radius}) in Eqs\il(\ref{Eq:lLE}), particles  energy and  angular momentum   at the turning point are:
\bea\label{Eq:tdotemLem}
\Em=\frac{a \sigma  \dot{t}_\Ta}{a \sigma_\Ta -\ell },\quad \La=\frac{a \sigma  \dot{t}_\Ta \ell }{a \sigma_\Ta -\ell },\quad \mbox{with}\quad
\dot{t}_\Ta=\La\left(\frac{1}{\ell }- \frac{1}{a \sigma_\Ta }\right)
\eea
(and there is $\Em\gtrless 0$ for $\ell\lessgtr a\sigma_\Ta$, while there is $\Em>0$  and  $\ell>0$ for $\ell\in]0,a\sigma[$, assuming $\dot{t}_\Ta>0$ which implies $\La\gtrless 0,\ell\lessgtr a\sigma_\Ta$).

There is a turning point $(u^{\phi}=0)$   from Eq.\il(\ref{Eq:lLE}), within  the following  conditions
\bea
&&\label{Eq:limellcoro}
a\in ]0,1], \quad \ell <0\cup \ell >\ell_{lim}^->0,\quad\mbox{where}
\quad \ell_{lim}^-\equiv 2 \left(\frac{1}{a}+\sqrt{\frac{1}{a^2}-1}\right)>\ell_{\gamma}^->0,
\eea
respectively, where
the following limits hold
\bea\label{Eq:protre-spess-comunci}
&&
\lim\limits_{\ell\rightarrow \pm\infty} r_{\Ta}(\sigma_\Ta)=r_{\epsilon}^+,\quad \lim\limits_{a\rightarrow 0} r_{\Ta}(\sigma_\Ta)=r_{+}=2M,\quad \lim\limits_{\sigma\rightarrow 0} r_{\Ta}(\sigma_\Ta)=r_{+},
\eea
see --Figs\il(\ref{Fig:Plotrte}) and Figs\il(\ref{Fig:Plotrte4}).
{It should be stressed that there are no (time-like and photon-like) co-rotating turning points,  solutions $u^\phi=0$ with  the conditions $\ell>0$ with $\Em>0$ and $\La>0$ ($u^t>0$), where $(\ell,\Em,\La)$ are in Eqs\il(\ref{Eq:lLE}).  (Conditions (\ref{Eq:limellcoro}) and limits  (\ref{Eq:protre-spess-comunci})  take into account only  function  $\ell_\Ta$ in  Eq.\il(\ref{Eq:lLE}) providing a more general solution in  $r>r_+$, dependent only on the condition  $\ell=$constant, not necessarily related to the orbiting tori, and without considering the further constraints of  $(\Em,\La)=$constant). We will detail this aspect in Sec.\il(\ref{Sec:turning-sign-existence}). }

From Eqs\il(\ref{Eq:protre-spess-comunci}) we note that
asymptotically,  for  $\ell$ very large in magnitude,  function $r_\Ta(\ell)$ approaches the ergosurface  $r_{\epsilon}^+$ for any plane $\sigma_\Ta$, from the region $r_\Ta>r_{\epsilon}^+$ for counter-rotating flows, and $r_\Ta<r_{\epsilon}^+$ for $\ell>0$---Figs\il(\ref{Fig:Plotrte4}). (Radius   $r_\Ta$  for $\ell>0$ must be in the ergoregion at any $\sigma$, while  the counter-rotating fluids turning point must located  out of the ergoregion.)
 At the \textbf{BH} poles, in the limit  $\sigma\rightarrow 0$, the flow turning points coincide  (according  to the  adopted  coordinate frame)  with  the \textbf{BH} horizon.
(This condition eventually holds also for the limit of static background where the eventual  flow turning point is not  induced by the frame-dragging.).

As pointed out in Sec.\il(\ref{Sec:tori-models}), a very large magnitude of $\ell$, explored in  Eqs\il(\ref{Eq:protre-spess-comunci}),   corresponds  to  quiescent tori  with $(\ell^\pm)^2\gg(\ell^\pm_\gamma)^2$, which can be  very large and located   far from the central \textbf{BH}  (i.e. $r_{center}^\pm>r_{(\gamma)}^\pm$), for counter-rotating tori, and   very close to the central attractor for co-rotating  tori orbiting  fast  spinning \textbf{BHs}--Figs\il(\ref{Fig:Plotbalchocasep}).  As     there is  $ \ell_{lim}^->\ell_{\gamma}^-$,  this condition holds  for very large centrifugal component of the co-rotating quiescent torus force balance, for initial tori centered at $r>r_{(\gamma)}^-$, such radius is very far from the attractor for slower rotating \textbf{BHs}, and  located in the ergoregion for fast spinning \textbf{BH}, i.e.  for attractors with spins $a\geq a_{(\gamma)}^{\epsilon}\approx 0.994M$--Figs\il(\ref{Fig:Plotbalchocasep}).
The lower bound  $ \ell_{lim}^-$ in   Eqs\il(\ref{Eq:limellcoro}) for  $\ell>0$, is independent from $\sigma$  and  from the system initial data (initial fluid velocity and  location)  being a function of the \textbf{BH} spin $a/M$ only, and therefore it is   independent from the tori models.

For the  counter-rotating fluid turning points $\sigma_\Ta(r_{\Ta})$  there is
\bea\label{Eq:max-cusp-eq-exte1}
(\ell<0)\quad
r_{\Ta}\in ]r_{\epsilon}^+,r_{\Ta}^e]\quad \mbox{with}\quad r_{\Ta}^e\equiv \left. r_\Ta\right|_{\sigma=1},
\eea
where
$r_{\Ta}^e$ is the turning point  $r_{\Ta}$ for $\sigma_\Ta=1$-- the  turning point is located on the torus and the central attractor equatorial plane. (The role of  equatorial plane in this problem is detailed in Sec.\il (\ref{Sec:basis-eq}).).
This implies that   the turning point  reaches  its  maximum value  $r_\Ta=2.45455M$ on the \textbf{BH} equatorial plane for the  extreme Kerr \textbf{BH}   spacetime.
Remarkably, radius  $r_{\Ta}(\sigma_\Ta)$ in Eqs\il(\ref{Eq:turning-point-radius}) and (\ref{Eq:add-equa-sigmata}), depending on   $\ell$ only,  has  no explicit dependence on the flow initial data.
Therefore, at any plane  $\sigma_\Ta\in [0,1[$, the   turning point radius $r_\Ta$ is located in  a range $r_\Ta/M\in]2,2.45455[$, independently from other flow initial data.

\begin{figure*}
\centering
    \includegraphics[width=7.75cm]{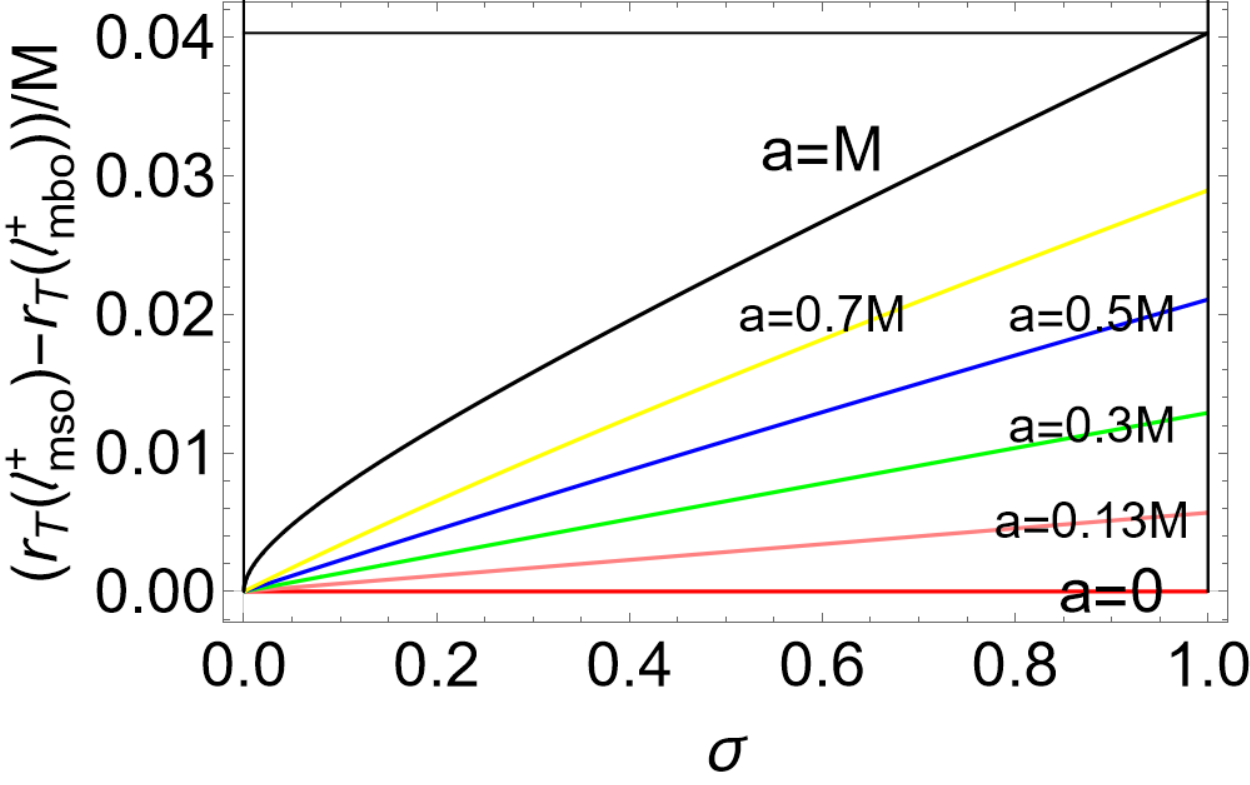}
  \includegraphics[width=7.75cm]{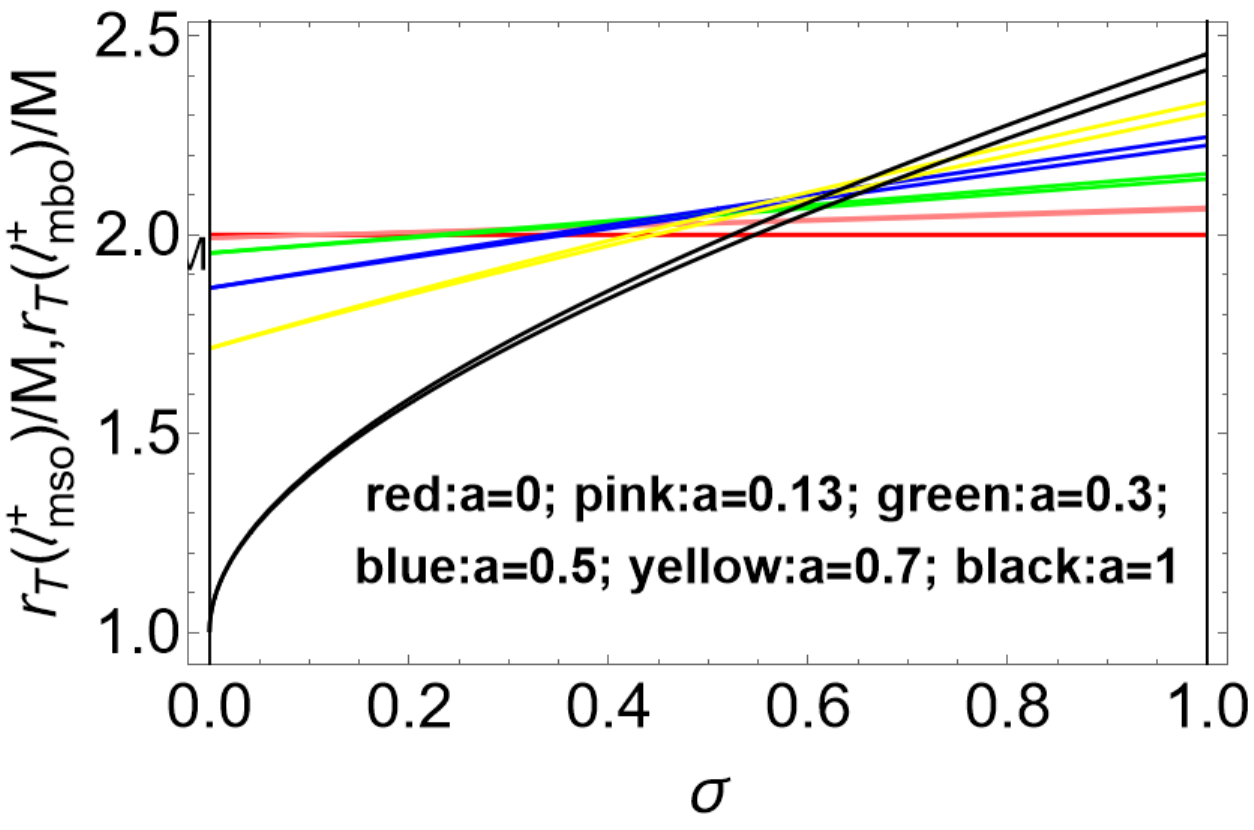}
   \includegraphics[width=7.75cm]{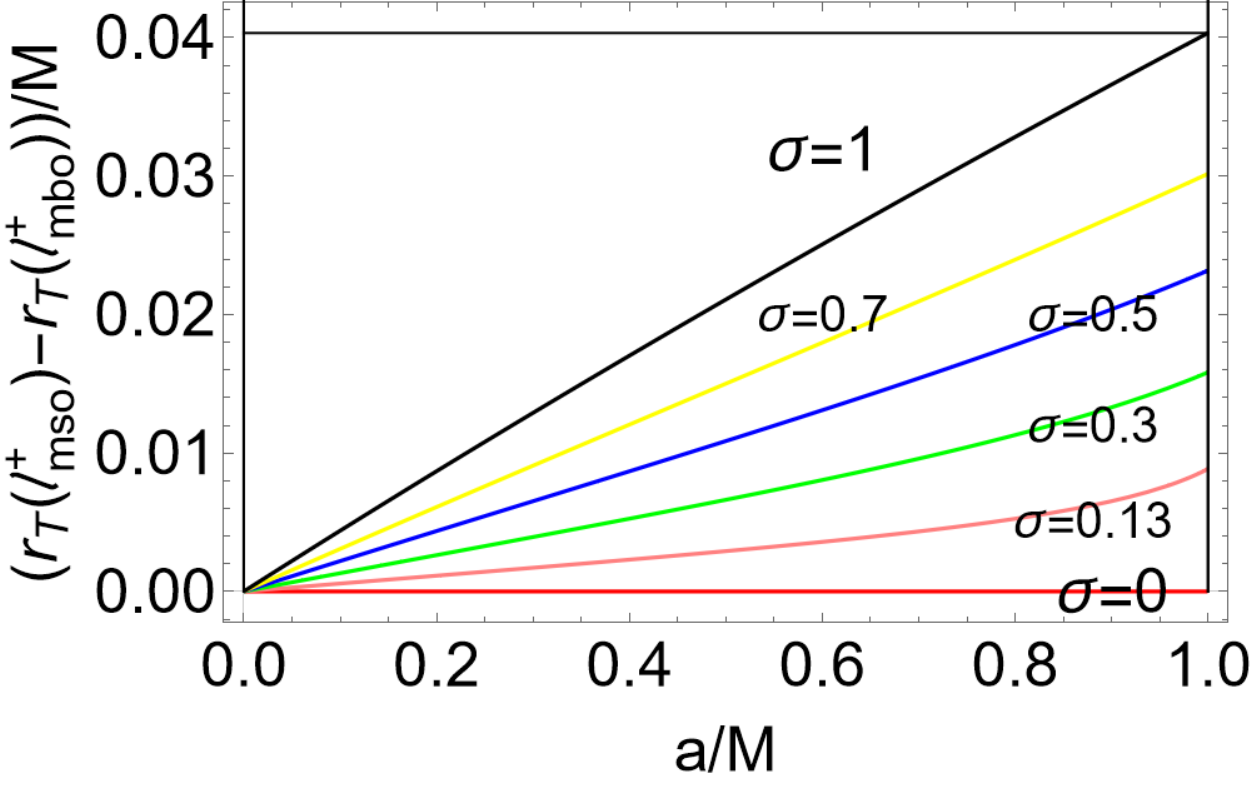}
    \includegraphics[width=7.75cm]{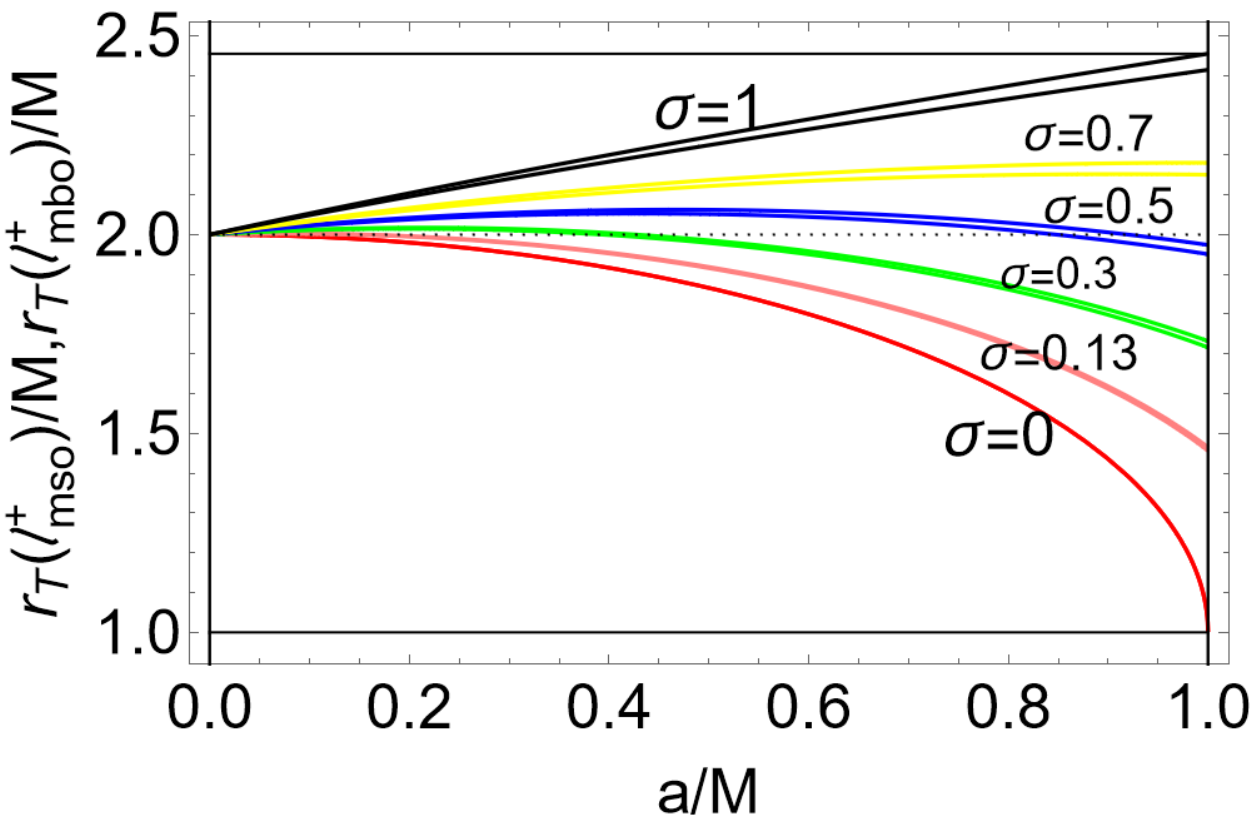}
  \caption{Turning radius $r_{\Ta}$ of the tori driven counter-rotating  flows, at the boundaries of  the turning corona, is shown. The boundary radii are evaluated at the specific angular momenta $\ell=\ell_{mso}^+$ and $\ell=\ell_{mbo}^+$, defined in  Eqs\il(\ref{Eq:def-nota-ell}). Upper left panel: Corona  radius, difference  $(r_{\Ta}(\ell_{mso}^+)-r_{\Ta}(\ell_{mbo}^+)$, as function of the plane $\sigma\equiv \sin^2\theta$ is shown  for different \textbf{BH} spin $a/M$ signed on the curves  (upper left panel). and as function of $a/M$ for different plane $\sigma$  signed on the curve (bottom left panel).  Radii  $r_{\Ta}(\ell_{mso}^+)>r_{\Ta}(\ell_{mbo}^+)$ as functions of the plane $\sigma$,  for different \textbf{BH} spins $a/M$ are shown in  the upper right  panel,  and as functions of $a/M$, for different planes $\sigma$  in bottom right  panel. In Figs\il(\ref{Fig:Plotrteag}) the analysis is repeated  for the counter-rotating proto-jets driven flows.}\label{Fig:Plotrtea}
\end{figure*}

\begin{figure*}
\centering
    \includegraphics[width=6.75cm]{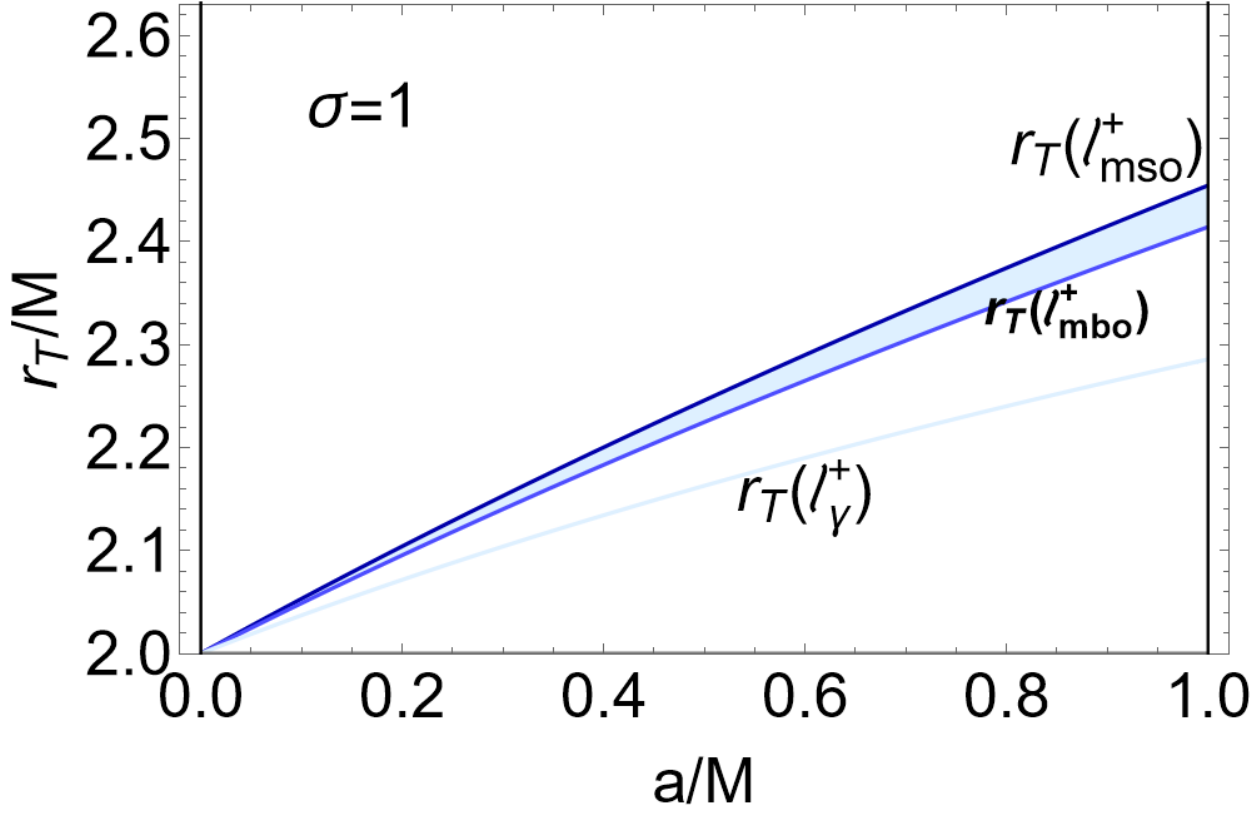}
  \includegraphics[width=6.75cm]{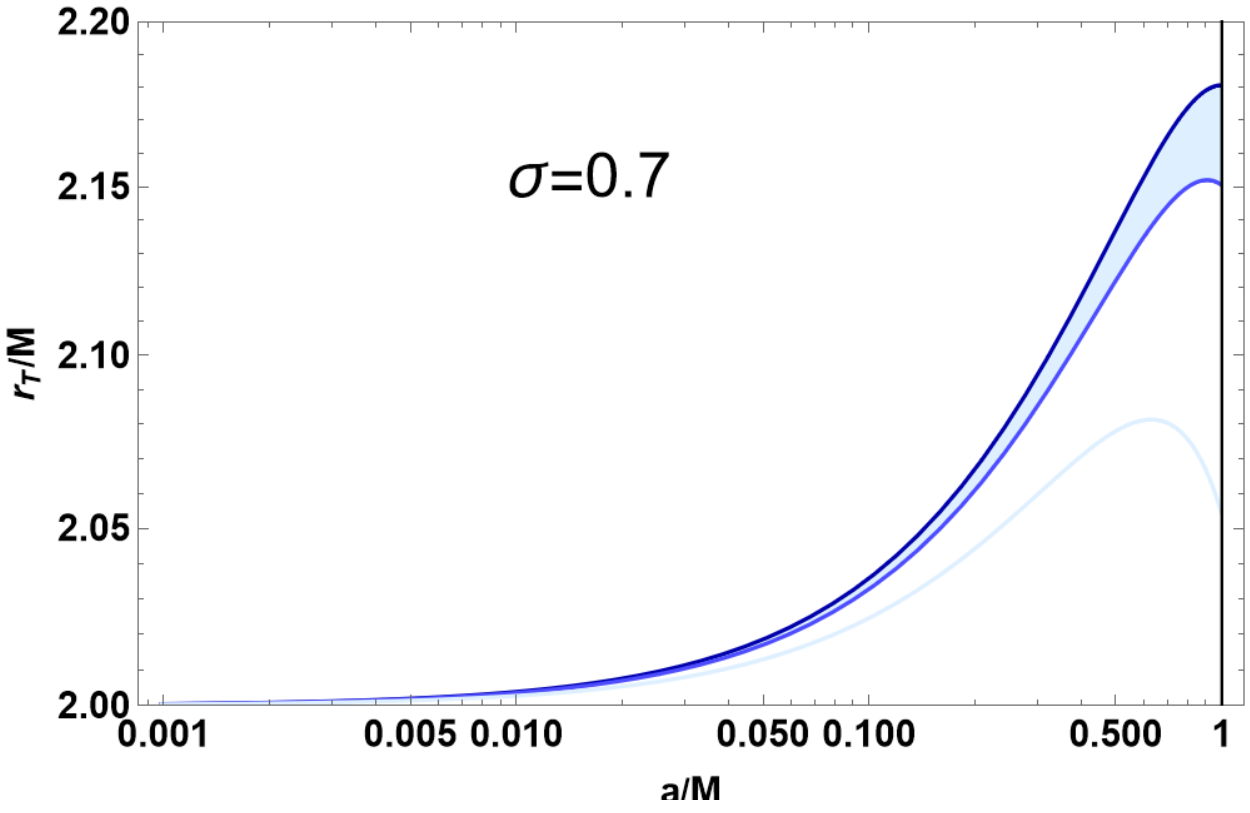}
   \includegraphics[width=6.75cm]{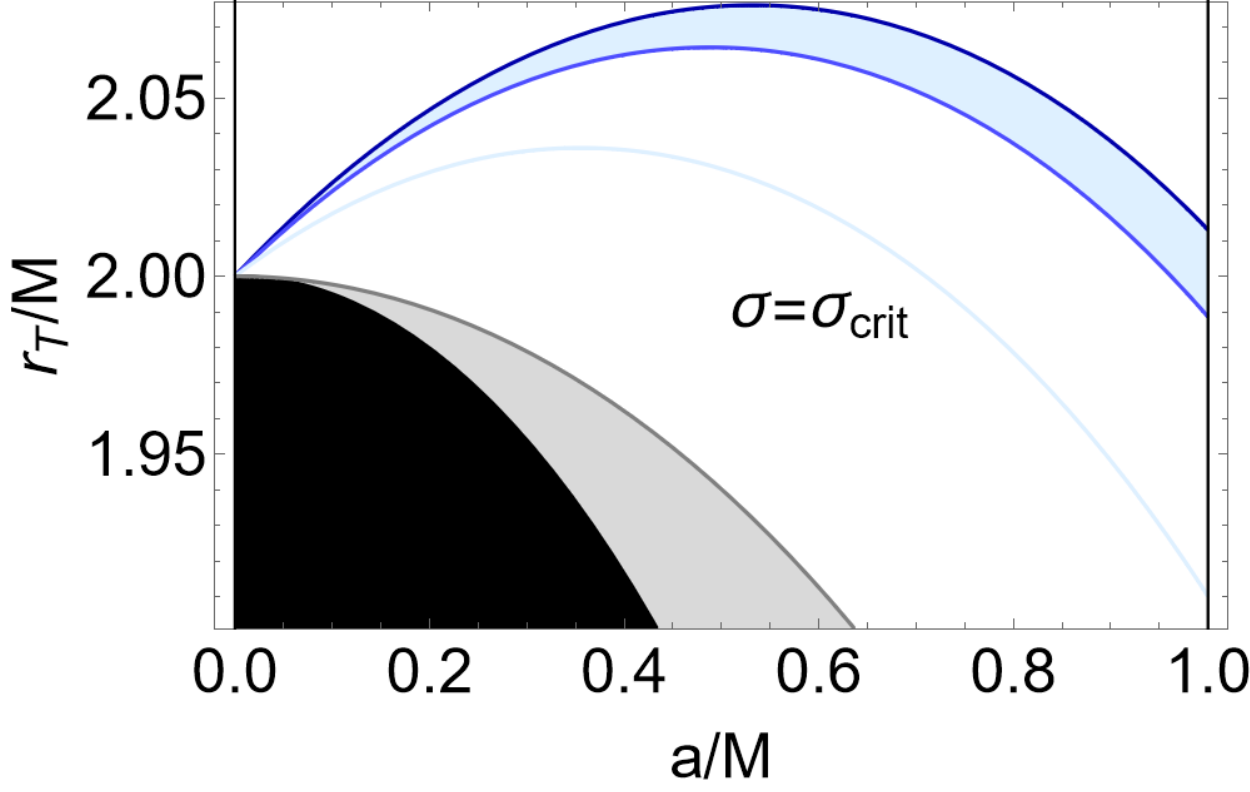}
    \includegraphics[width=6.75cm]{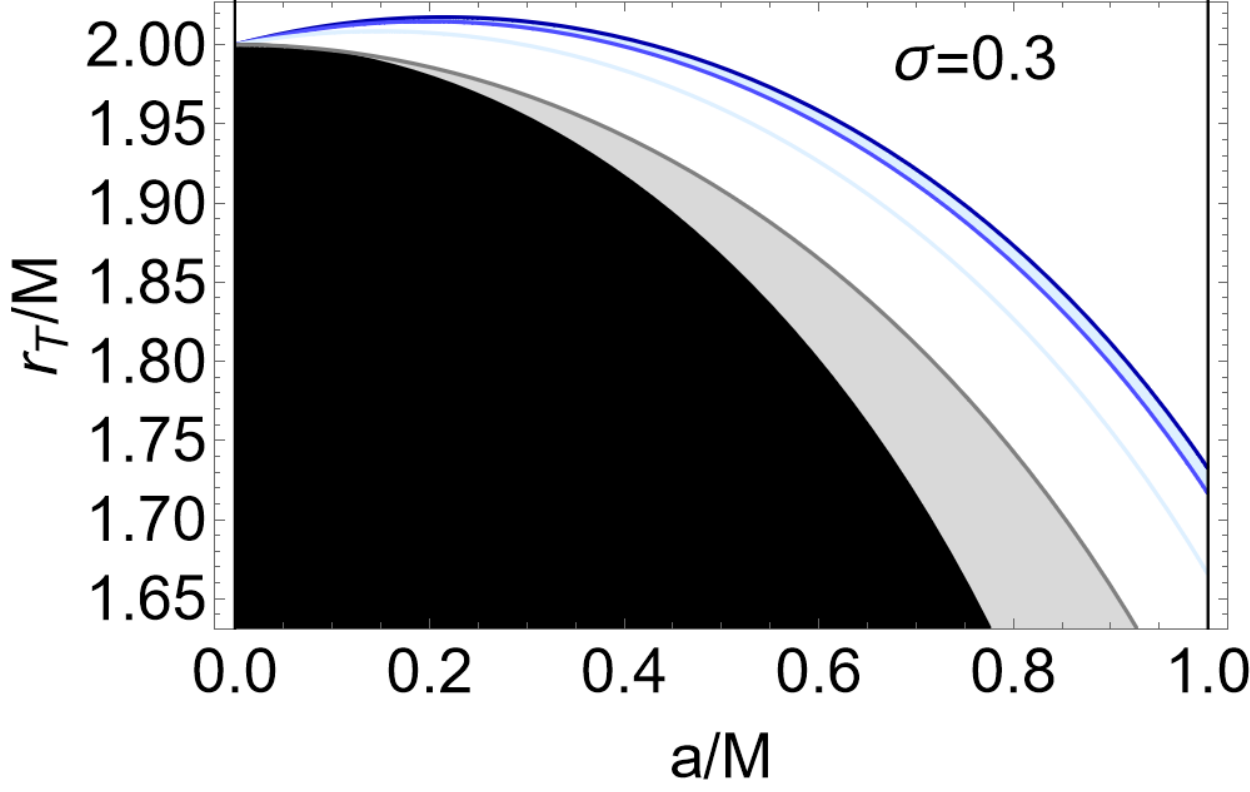}
  \caption{Extreme of the turning point radial coordinate, for fixed typical values of $\sigma$.  Black region  is the central \textbf{BH}  (region $r<r_+$ where $r_+$ is the outer \textbf{BH} horizon), the outer ergosurface is $r_{\epsilon}^+=2M$. Gray region is the outer ergoregion, $]r_+,r_{\epsilon}^+]$, for  different planes $\sigma\equiv \sin^2\theta$ signed on the panels, where $\sigma=1$ is the equatorial plane and  $\sigma_{crit}=2 \left(2-\sqrt{3}\right)\approx0.535898$. Counter-rotating flow turning points $r_\Ta$ are shown  as functions of \textbf{BH} spin-mass ratios $a/M$:  there is $r_{\Ta}(\ell_{mso}^+)\geq r_{\Ta}(\ell_{mbo}^+)\geq r_{\Ta}(\ell_{\gamma}^+) $--see Eqs\il(\ref{Eq:def-nota-ell}), plotted as dark-blue, blue and light-blue curves respectively. Note the different situations for planes smaller or larger than $\sigma_{crit}$,    the  curves  maximum as functions of the dimensionless \textbf{BH} spin, and  the spreading of the region $[r_{\Ta}(\ell_{mbo}^+),r_{\Ta}(\ell_{mso}^+)]$ (light-blue shaded)  defining the  tori driven  counter-rotating flow turning  points  corona are shown.  The  region  $[r_{\Ta}(\ell_{\gamma}^+),r_{\Ta}(\ell_{mbo}^+)]$ (white shaded) for the  turning points of counter-rotating  proto-jets driven flows is also shown. }\label{Fig:Plotrteb}
\end{figure*}

More precisely, {for a fixed value of $\ell$, function $r_\Ta(\sigma)$, defines a surface, \emph{turning sphere},  surrounding the central attractor}. We  can  identify a \emph{turning   corona}, as the  spherical shell   defined by the limiting conditions on  the radius $r_\Ta(\sigma_\Ta)$ in the range $[r_\Ta(\ell_{mbo}^+)],r_\Ta(\ell_{mso}^+)]$, for   tori driven  counter-rotating flows,  and  $[r_\Ta(\ell_{mbo}^+)],r_\Ta(\ell_{\gamma}^+)]$ for  proto-jets  driven counter-rotating flows--see Figs\il(\ref{Fig:PlotlongVie},\ref{Fig:bluecurvehereq},\ref{Fig:Plotrtea},\ref{Fig:Plotrteb},\ref{Fig:Plotrte4}).
As shown in Figs\il(\ref{Fig:Plotrte}),  the  circular region of turning points is  delimited by the radii
 $r_{\Ta}(\ell_{mso}^\pm,\sigma_\Ta)>r_{\Ta}(\ell_{mbo}^\pm,\sigma_\Ta)$.
The  turning corona radii ($r_\Ta(\ell_{\gamma}^+),r_\Ta(\ell_{mbo}^+),r_\Ta(\ell_{mso}^+)$) vary   little for  the \textbf{BH} spin and plane $\sigma$. This also implies that the flow    is located in  restricted orbital range $(r_{\Ta},\sigma_{\Ta})$, localized  in an orbital cocoon surrounding the central attractor outer ergosurface (reached at  different times $t_\Ta$ depending  on the initial data--see Figs\il(\ref{Fig:Plotrte}). Therefore the  turning flow corona  would be  easily observable  (depending on the  values of $t_\Ta$  range),  characterized possibly by an increase of   flow  temperature and  luminosity.
As the flow characteristics at the turning point have  a  little dependence on the initial data, they hold to a remarkable  extent also for  different disks models.
Radius $r_\Ta(\sigma_\Ta)$ is in fact independent explicitly from the normalization conditions, as such the corona sets the location of the turning points for the photonic as well as particle components of the flow.

\begin{figure}
\centering
  \includegraphics[width=7.5cm]{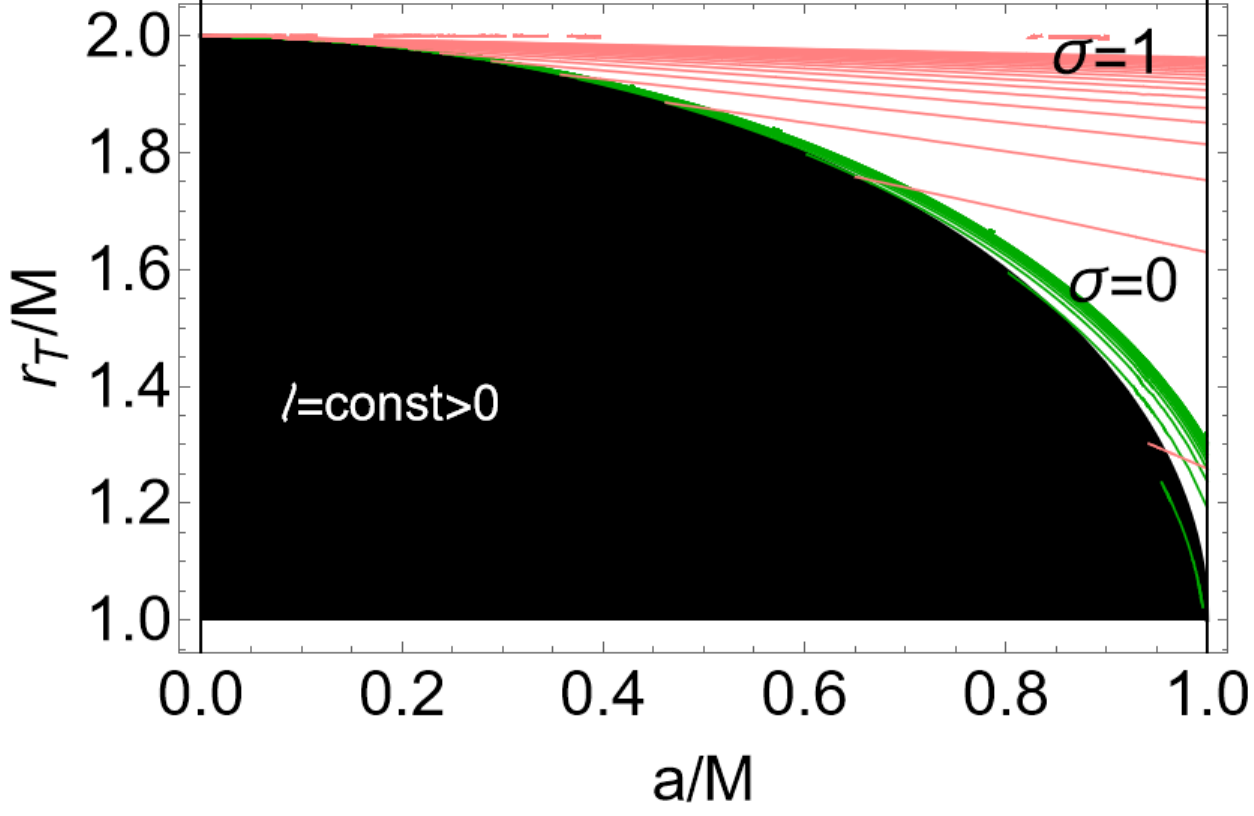}
  \caption{ Considerations on the function $r_\Ta$ for $\ell>0$. Black region  is the \textbf{BH} at $r<r_+$ with dimensionless spin $a/M$,  and $r_+$ is the outer \textbf{BH} horizon, the outer ergosurface is $r_{\epsilon}^+=2M$.  Radii $r_\Ta$  are shown  on equatorial plane $\sigma=1$ (pink) and \textbf{BH} axis ($\sigma=0$) (green).}\label{Fig:Plotrtea1a}
\end{figure}

Although  $r_\Ta$ is bounded in a  restricted  orbital range,  the  turning point radius $r_\Ta$ varies  with  $\sigma_\Ta\in [0,1]$.
The corona radii distance, $r_\Ta (\ell_
{mso}^+) -
 r_\Ta (\ell_ {mbo}^+) $, increases not monotonically  with the \textbf{BH} spin-mass ratio and with the plane $\sigma$--Figs\il(\ref{Fig:Plotrtea}).
(We also show   in Fig.\il(\ref{Fig:Plotrtea1a}) some results concerning  the case $\ell>0$.).
Decreasing  $\sigma_\Ta$, close to  the \textbf{BH} poles,  the   $r_\Ta$ range  decreases, although  the  turning points  location   variation  with   $\sigma_\Ta$ remains small--Figs\il(\ref{Fig:Plotrtea}).
The vertical and maximum vertical location (along the \textbf{BH} rotational  axis) of the turning point is  studied in Sec.\il(\ref{Sec:vertical-z}). Below we  investigate  more specifically the dependence of the turning
 point on  the plane $\sigma$ and the \textbf{BH} spin-mass ratio $a/M$,  proving the  existence  of a  $r_\Ta$  maximum for a variation of   the \textbf{BH} spin $a/M$, distinguishing therefore counter-rotating accretions for different attractors.

In Figs\il(\ref{Fig:Plotrtea}) and Figs\il(\ref{Fig:Plotrteb})  we show  the corona radii  in dependence on the  plane $\sigma_\Ta$, particularly around the limiting plane value  $\sigma=\sigma_{crit}=2 \left(2-\sqrt{3}\right)$.  For $\sigma<\sigma_{crit}$, there is  $r_\Ta<2M$  (related to the outer ergosurface location) and the  radius $r_\Ta$ decreases increasing the \textbf{BH} spin. Viceversa, at $\sigma\geq \sigma_{crit} $, turning radii   are at  $r_\Ta>2M$, decreasing  with the  spin $a/M$. The turning corona could be  therefore  a very active part of the accreting flux of matter and photons, especially  on the \textbf{BH} poles, and it is expected to be lightly more   large (and rarefied at equal flow distribution along $\sigma\in [0,1]$ )  at the equatorial plane (however the time  component $t_\Ta$, and strongly different values of the turning point, could influence significantly details on the matter distribution relevant for the observation at the turning point). The maximum $r_\Ta$ for the spin $a/M$ is in  Figs\il(\ref{Fig:Plotrteb}).

In   Figs\il(\ref{Fig:Plotrte4}) the case of slowly rotating \textbf{BHs} (small  $a/M$) and fast rotating \textbf{BHs} are shown:  for small $a/M$ the corona radii reduce to the  orbits $r_\Ta$, defining   a spherical surface  for  turning points of  particles and photons. The flow initial data however  determine $\sigma_\Ta$ and  $r_\Ta$ as independent variables, and  the time component $t_\Ta$ (related also to the  accretion process  time-scales and the details on inner disk active part where flow leaves the toroid).
The  analysis is repeated in Figs\il(\ref{Fig:Plotrteag}) for the proto-jets driven counter-rotating flows   having specific momentum $\ell^+\in \mathbf{L^+_2}=]\ell_{\gamma}^+,\ell_{mbo}^+]$.
In this analysis we  addressed the conditions for the  existence of a flow turning point and  we explored  the flow characteristics  at the turning point.  In  Sec.\il(\ref{Sec:accelerations-fluids}), we also briefly investigate   the flow  at time $t>t_\Ta$. %

\subsubsection{{Further notes on flow rotation and double turning points}}\label{Sec:turning-sign-existence}

\textbf{Flow rotation and constraints on the turning spheres}

  Turning point  definition, as locus of points where $u^\phi=0$ in terms of $\ell$, defines  a surface, turning sphere, surrounding the central \textbf{BH},
depending only on $\ell$ parameter. Here, we  also study  more in  general  function $\ell(u^\phi=0)$.
However, further conditions have to be considered for the function $\ell_\Ta$ when framed in the particles flow and tori  flows turning points, namely: \textbf{1.}  constance of  $(\Em_\Ta, \La_\Ta)$ of Eqs\il(\ref{Eq:lLE}), evaluated
at the turning point with $\dot{t}$>0, implying $\ell=$constant. The constrained turning sphere is a general property of the orbits in the Kerr  \textbf{BH} spacetime.
It is clear that function  $\ell_\Ta=$constant is a more general solution, where conditions $\Em=$constant and  $\La=$constant depend on the specific trajectory.
\textbf{2.} Second constraint is the normalization condition at the  turning point, $g_{\alpha\beta}u^\alpha u^\beta=\kappa$ (with  $u^\phi=0$).
 \textbf{3.} Third condition resolves into the description of the matter flows from the orbiting structures, translated into a constrain on the range  of values for  $\ell$, and defining the turning corona for proto-jets or accretion driven flows

The turning sphere and turning coronas are in fact a property of the background geometry, depending only on the spacetime spin.
Therefore, in particular they describe also  particles  with $\dot{r}_\Ta>0$  (outgoing particles) or particles  moving along  the central axis.

We shall study function $\ell_\Ta$ in general,       constraining it later  in the different interpretative frameworks.
An mentioned above, at the turning points the conditions
$\ell>0$    (co-rotating case) with $(\Em\gtrless0, \La\gtrless0)$ respectively, and   $g_{\alpha\beta}u^\alpha u^\beta=\kappa$ with  $\kappa=\{0,-1\} (u^t\gtrless0)$ are never satisfied  (while there is a spacelike solution ($\kappa=1$) in the ergoregion
 with  $\Em<0$ and $\La<0$ (and  $u^t>0$)).
A co-rotating $(\ell a>0)$ solution of function  $\ell_\Ta=$constant is  also studied for completeness in this work.

 For $u^t>0$, we can consider the  following four cases
 (while notation $\Ta$ as been dropped for simplicity, it is intended all the quantities be evaluated at the turning point):
\begin{description}
\item[---]For $\ell<0$ with  $(\Em>0,\La<0)$, there are no turning points in the ergoregion.

Turning points are for
 \bea\nonumber &&a\in ]0,1]\quad (\La=\La_{bh}, \Em=\Em_{bh}):\quad  [(r\in ]r_+,2] \sigma\in ]0,\sigma_{erg}[),
 (r>2, \sigma\in ]0,1])],\quad\mbox{or}
\quad
 (\sigma\in ]0,1] r>r_\epsilon^+),
\\
&&\mbox{where}\quad
 \La_{bh}\equiv -\frac{2 a r \sigma  u^t}{\Sigma},\quad
 \Em_{bh}\equiv u^t\left[1-\frac{2 r}{\Sigma}\right],\quad
 \sigma_{erg}\equiv \frac{(r-2) r}{a^2}+1,
 \eea
 (there is $\sigma_{erg}=\sigma: r_\epsilon^+(a;\sigma)=r$, for $r\in ]r_+,2]$ ).
 We consider now the second constraint, using the normalization condition at turning point.

 For   $\kappa=-1$ (flows particles) turning points are for:
 \bea\nonumber&&a\in]0,1]\quad(\dot{r}^2=\dot{r}_{bh}^2, \La=\La_{bh}, \Em=\Em_{bh}),\quad\mbox{for}:
 \\\nonumber
 && r\in ]r_+,2], \sigma\in]0,\sigma_{erg}[: (\dot{t}=\dot{t}_{bh}, \dot{\theta}^2=0), (\dot{t}>\dot{t}_{bh},  \dot{\theta}^2\in[0,\dot{\theta}_{bh}^2])];
\\\nonumber&& r>2, \sigma\in]0,1]: (\dot{t}=\dot{t}_{bh}, \dot{\theta}^2=0), (\dot{t}>\dot{t}_{bh},  \dot{\theta}^2\in[0,\dot{\theta}_{bh}^2]),\\
&&\mbox{or alternatively}
\\&&
 a\in]0,1], \sigma\in]0,1], r>r_\epsilon^+,[ (\dot{t}=\dot{t}_{bh}, \dot{\theta}^2=0); (\dot{t}>\dot{t}_{bh},  \dot{\theta}^2\in[0,\dot{\theta}_{bh}^2])],
 \eea
 where
 \bea\nonumber
 \dot{t}_{bh}\equiv\sqrt{\frac{2 r}{(r-2) r-a^2 (\sigma -1)}+1},\quad
\dot{\theta}_{bh}^2\equiv\frac{r \left[(r-2) \dot{t}^2-r\right]-a^2 (\sigma -1) \left(\dot{t}^2-1\right)}{\Sigma^2},
 \\
 \dot{r}_{bh}^2\equiv -\frac{\Delta\left[a^4 (\sigma -1)^2 \dot{\theta}^2+a^2 (\sigma -1) \left[\dot{t}^2-1-2 r^2 \dot{\theta}^2\right]+r^4 \dot{\theta}^2+r^2-(r-2) r \dot{t}^2\right]}{\Sigma^2}.
 \eea
For null--like  particles ($\kappa=0$) there is
 \bea&&\nonumber
 a\in]0,1],\quad (\dot{r}^2=\dot{r}_{bh|0}^2, \La=\La_{bh}, \Em=\Em_{bh}),\quad\mbox{for}
 \\&&\nonumber (r\in ]r_+,2], \sigma\in]0,\sigma_{erg}[, \dot{\theta}^2\in[0, \dot{\theta}_{bh|0}^2]), (r>2, \sigma\in]0,1],  \dot{\theta}^2\in[0, \dot{\theta}_{bh|0}^2]),
\quad\mbox{
or  alternatively}
\\\nonumber
&& \sigma\in]0,1], r>r_\epsilon^+,  \dot{\theta}^2\in[0, \dot{\theta}_{bh|0}^2],\quad\mbox{where}
 \\&&\nonumber
\dot{\theta}_{bh|0}^2\equiv\frac{\dot{t}^2 \left[(r-2) r-a^2 (\sigma -1)\right]}{\Sigma^2};\\&&\dot{r}_{bh|0}^2\equiv-\frac{\Delta \left[a^4 (\sigma -1)^2 \dot{\theta}^2+a^2 (\sigma -1) \left(\dot{t}^2-2 r^2 \dot{\theta}^2\right)+r^4 \dot{\theta}^2-(r-2) r \dot{t}^2\right]}{\Sigma^2}.
 \eea
 \item[---]
 For completeness we also consider  the case
$\dot{t}>0$ and $(\Em<0,\La>0)$ with $\ell<0$, where there is  no turning point.
\item[---]
We consider now the case $\ell>0$ with $(\Em>0, \La>0)$ (and   $\dot{t}>0$). There are no turning points in this case.
\item[---] The case
$\ell>0$ with $(\Em<0,\La<0)$  ($\dot{t}>0$)  is relevant in the naked singularity (\textbf{NS}) spacetime,
while in the  \textbf{BH} geometries this condition does not correspond to any orbiting structure we consider in this work \footnote{There are however turning points in the ergoregion, within these conditions
 \bea
a\in]0,M], \quad  (\La=\La_{bh}, \Em=\Em_{bh}): r\in]r_+,2[, \sigma\in]\sigma_{erg},1],
 \eea
 alternatively
 \bea
a\in]0,M],\quad  (\La=\La_{bh}, \Em=\Em_{bh}): r\in]r_+,r_\epsilon^+[,  \sigma\in]0,M].
 \eea
 However, these solutions exist only for \emph{spacelike ("tachyonic") particles}.
 Then, considering  the normalization condition with $\kappa=+1$ there is
  \bea&&
  a\in]0,M],\quad (\La=\La_{bh}, \Em=\Em_{bh},\dot{r}^2=\dot{r}_{bh|+}^2):
  \\\nonumber
  &&  r\in]r_+,2[, \sigma\in]\sigma_{erg},1]:\quad(\dot{t}\in]0,\dot{t}_{bh|+}[, \dot{\theta}^2\in[0, \dot{\theta}_{bh|+}^2]), (\dot{t}=\dot{t}_{bh|+}, \dot{\theta}^2=0),
 \\&&\nonumber
 \mbox{or}\quad
 r\in]r_+,r_\epsilon^+[,  \sigma\in]0,M]: (\dot{t}\in]0,\dot{t}_{bh|+}[, \dot{\theta}^2\in[0, \dot{\theta}_{bh|+}^2]), (\dot{t}=\dot{t}_{bh|+}, \dot{\theta}^2=0),
 \eea
 where
 \bea&&
\dot{t}_{bh|+}\equiv\sqrt{\frac{2 r}{a^2 (\sigma -1)-(r-2) r}-1},\quad\dot{\theta}_{bh|+}^2=\frac{r \left[(r-2) \dot{t}^2+r\right]-a^2 (\sigma -1) \left(\dot{t}^2+1\right)}{\Sigma^2};\\\nonumber
&&\dot{r}_{bh|+}^2\equiv-\frac{\Delta \left[a^4 (\sigma -1)^2 \dot{\theta}^2+a^2 (\sigma -1) \left(\dot{t}^2-2 r^2 \dot{\theta}^2+1\right)+r^4 \dot{\theta}^2-r \left[(r-2) \dot{t}^2+r\right]\right]}{\Sigma^2}.
 \eea}
 \end{description}
There are solutions also for  $\dot{t}<0$. In this case however energy $\Em$  should be discussed accordingly, we consider this case for Kerr  \textbf{NSs}  in \cite{new}.
 Finally we note that in this analysis we used three constants of motion,$(\Em,\La)$ and the normalization condition, while Carter constant $\Qa$ is independent on the sign of $(\Em,\La)$  and therefore from the  co-rotation or counter-rotation of the flow.

\medskip

\textbf{Double turning points}

From Figs\il(\ref{Fig:Plotrte4}) we note the existence, for large  $a$ and small  $\sigma$, of two turning points at  equal  $\ell$ and fixed  vertical axis  $z_\Ta=$constant. (There are always  two turning points $z_\Ta$ at fixed $y_\Ta$ (on the vertical direction), while on the equatorial plane there  is one turning point).  Let us focus on the \textbf{BH} geometries  turning points,  from counter-rotating fluids, evaluated at the boundary values $\{\ell_{mso}^+,\ell_{mbo}^+,\ell_\gamma^+\}$. The presence of a maximum $z_\Ta^{\max}$ of the turning point curve, solution of $\partial_{y_\Ta} z_\Ta^+=0$ is  an indication   of double turning points, where $r_\Ta^+=\sqrt{(z_\Ta^+)^2+(y_\Ta^+)^2}$ and $\sigma_\Ta=(y_\Ta^+)^2/((z_\Ta^+)^2+(y_\Ta^+)^2)$. The existence of a solution  $y_\Ta^+=y_\Ta(+)\neq 0: z_\Ta^+=r_+$ indicates the presence of double turning point at $z_\Ta^+\in[r_+,z_\Ta^{\max}[$ (and $y_\Ta\leq y_\Ta(+)$). Double turning points exist for large spins and small $\sigma$. In particular there is double turning point with $\ell_{mso}^+$ for $a>0.738315M$,  with $\ell_{mbo}^+$ for $a>0.75M$  and  with $\ell_{\gamma}^+$ for $a>0.785876M$ therefore, increasing  the magnitude of the fluid specific angular momentum, double points are for larger  \textbf{BH} spins.  There is a maximum value of  $y_\Ta^+=y_\Ta(+)>0$, increasing with the spin and, at fixed spin, decreasing with $\ell$ in magnitude. That is
solution $ z_\Ta^+=r_+$ at $y_\Ta(+)>0$ occurs for larger spin, increasing in magnitude the specific angular momentum $\ell^+$. The maximum distance from the axis of the turning point $y_\Ta(+)>0:  z_\Ta^+=r_+$, increases with the \textbf{BH} spin in the sense of  Figs\il(\ref{Fig:Plotrte4}) and decreases, increasing the angular momentum in magnitude.  More specifically there is, at $a=M$, $y_\Ta(+)=2.05274M$ ($\sigma_\Ta=0.8082,r_\Ta= 2.28336M$) for  $\ell_{mso}^+$,  $y_\Ta(+)=2.00812M$  ($\sigma_\Ta=0.801293, r_\Ta=2.24333M$)  for  $\ell_{mbo}^+$,  $y_\Ta(+)=1.8653M$ ($\sigma_\Ta=0.776753,r_\Ta=2.11645M$)   for  $\ell_{\gamma}^+$. (It is worth noting that, at $y_\Ta=2M$ there is ($\sigma_\Ta={4}/{5},r_\Ta=\sqrt{5}M$).)
\subsubsection{Analysis of the  $(r_\Ta,\sigma_\Ta)$ extreme points }\label{Sec:extreme-turning-box}
\begin{figure*}
\centering
    \includegraphics[width=5.8cm]{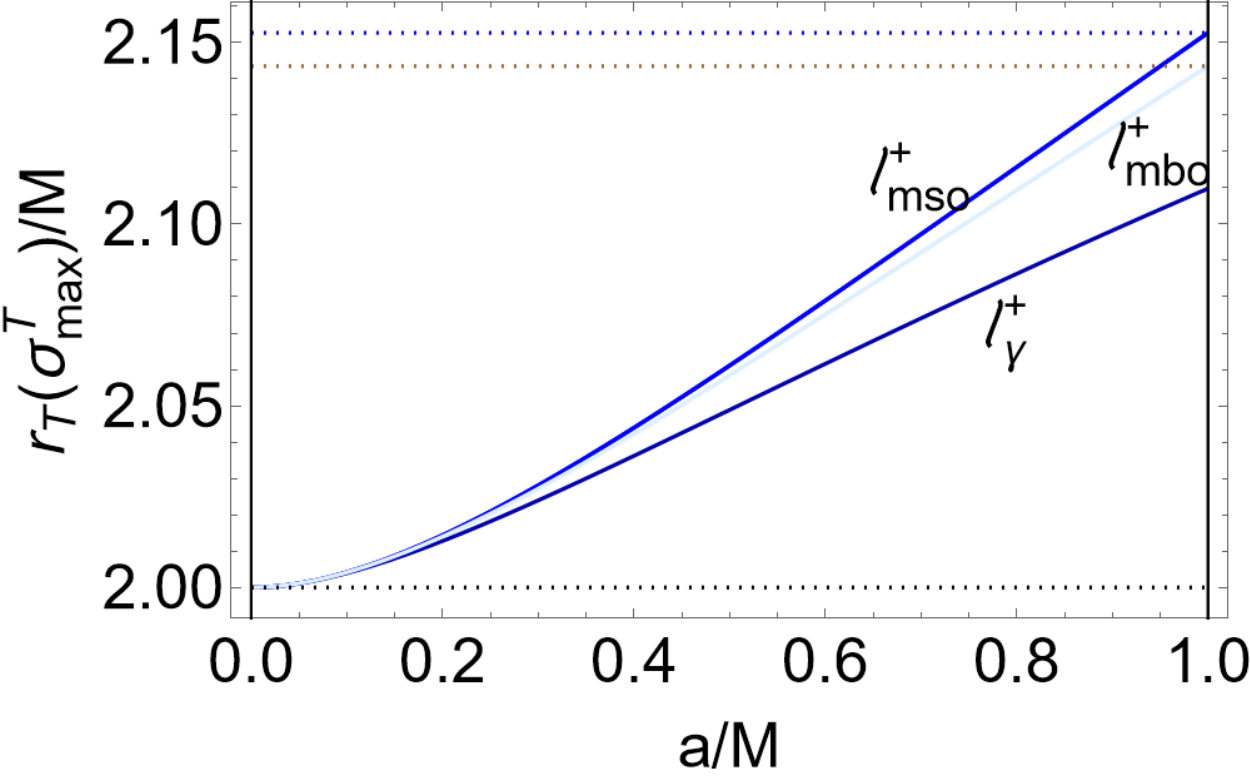}
      \includegraphics[width=5.8cm]{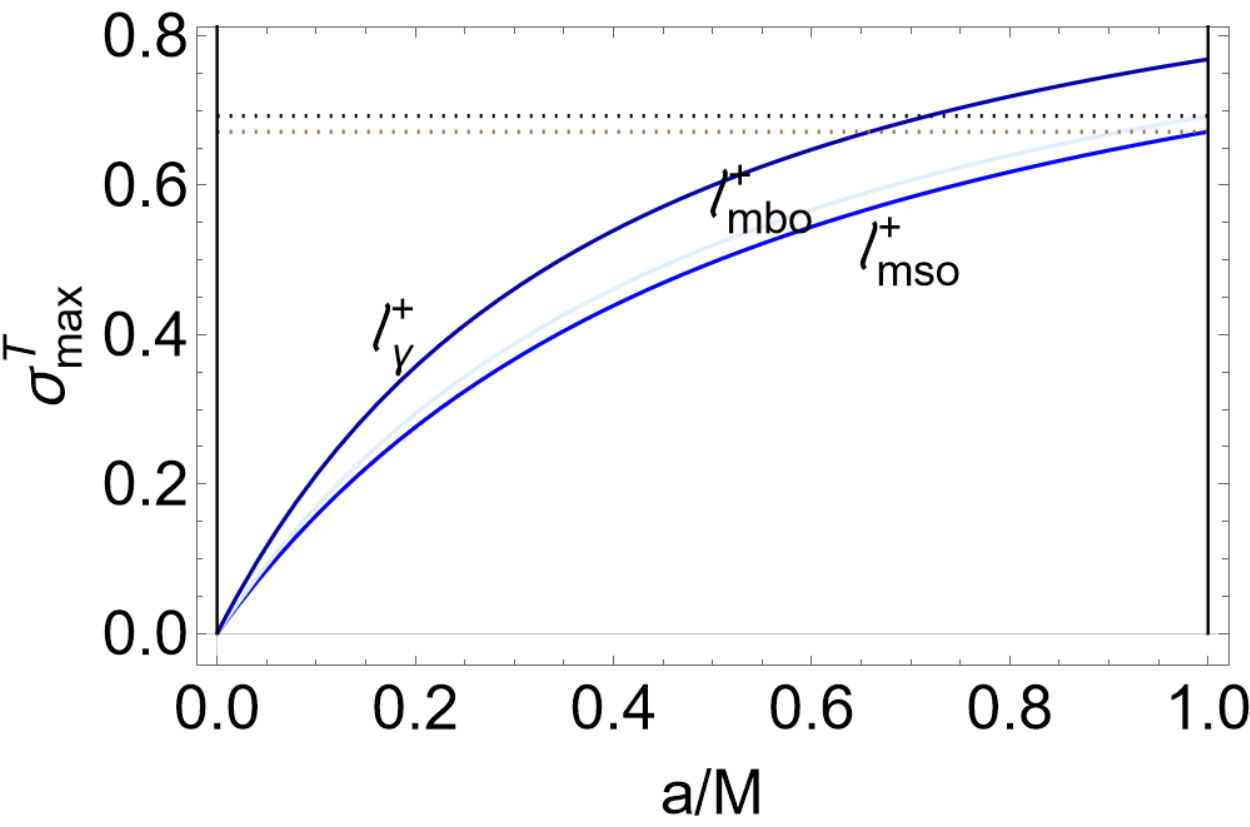}
        \includegraphics[width=5.8cm]{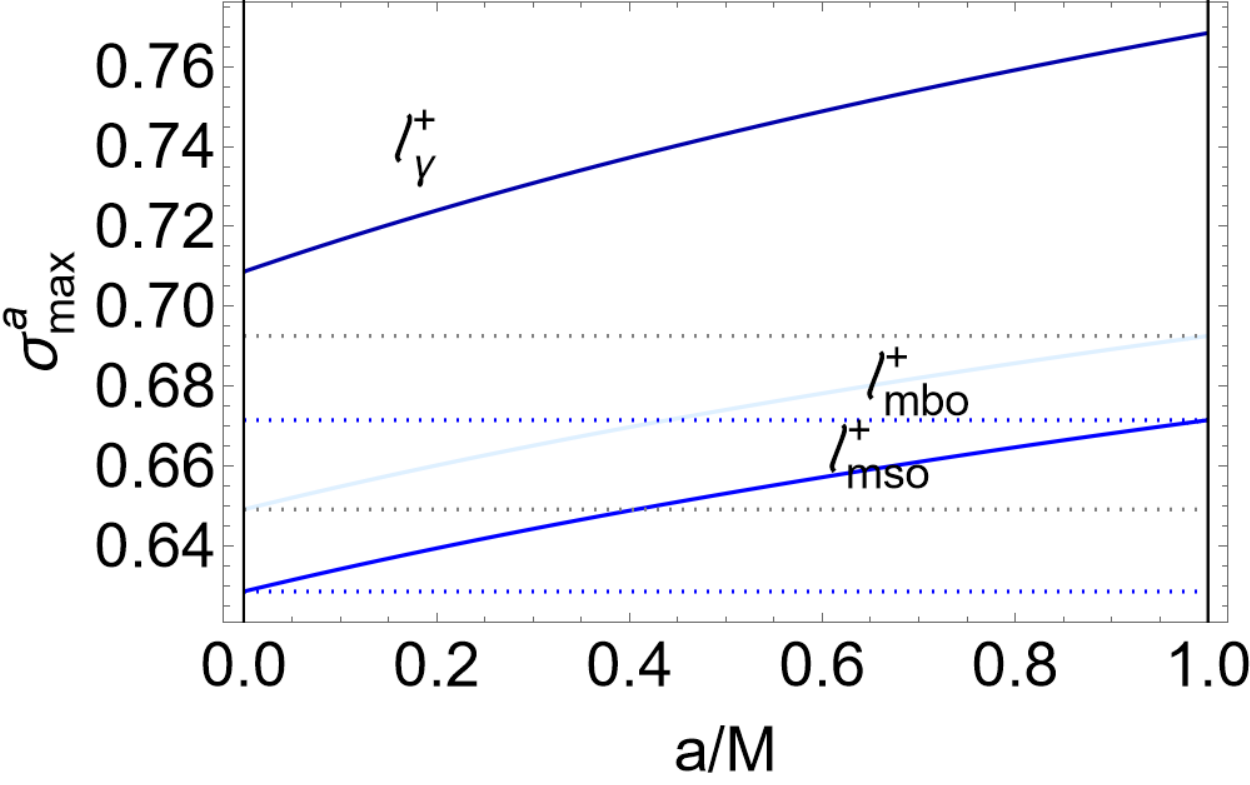}\\  \includegraphics[width=7cm]{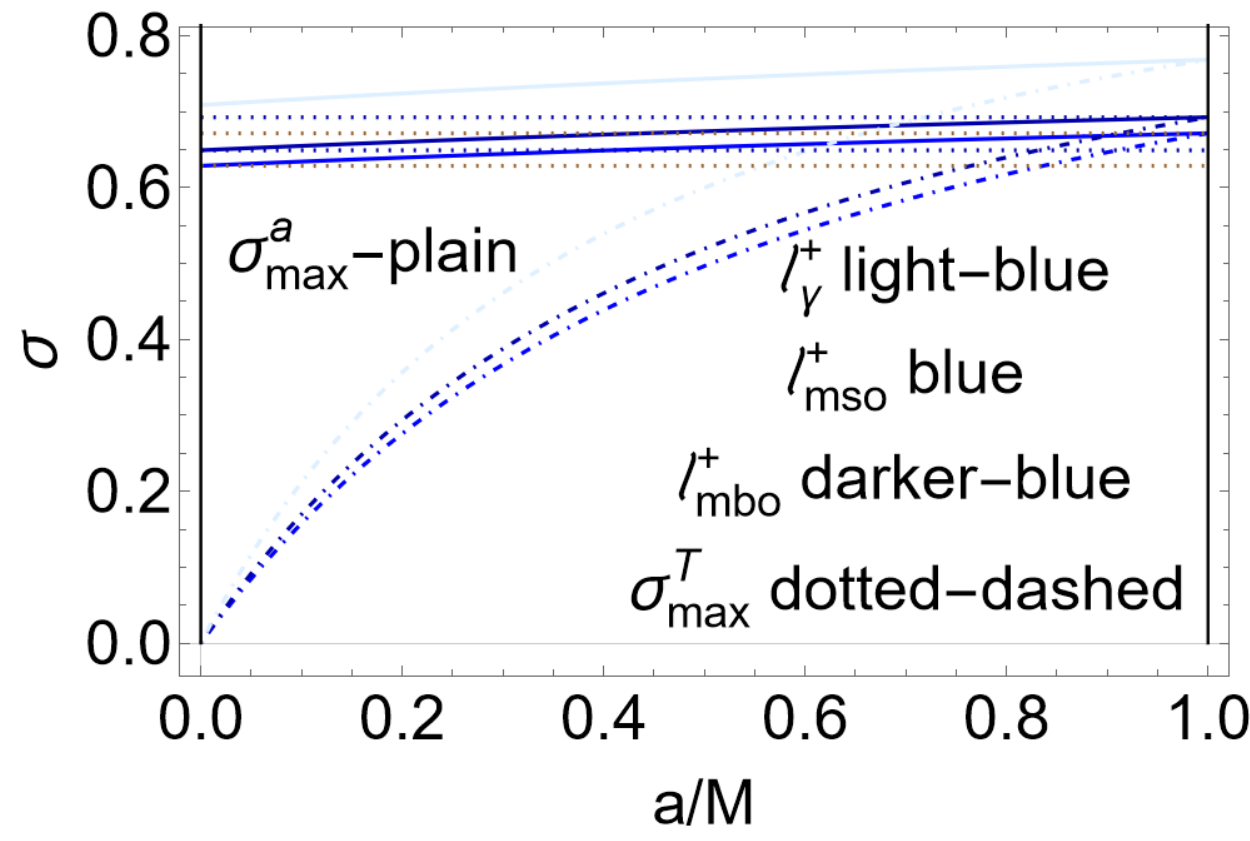}
   \includegraphics[width=7cm]{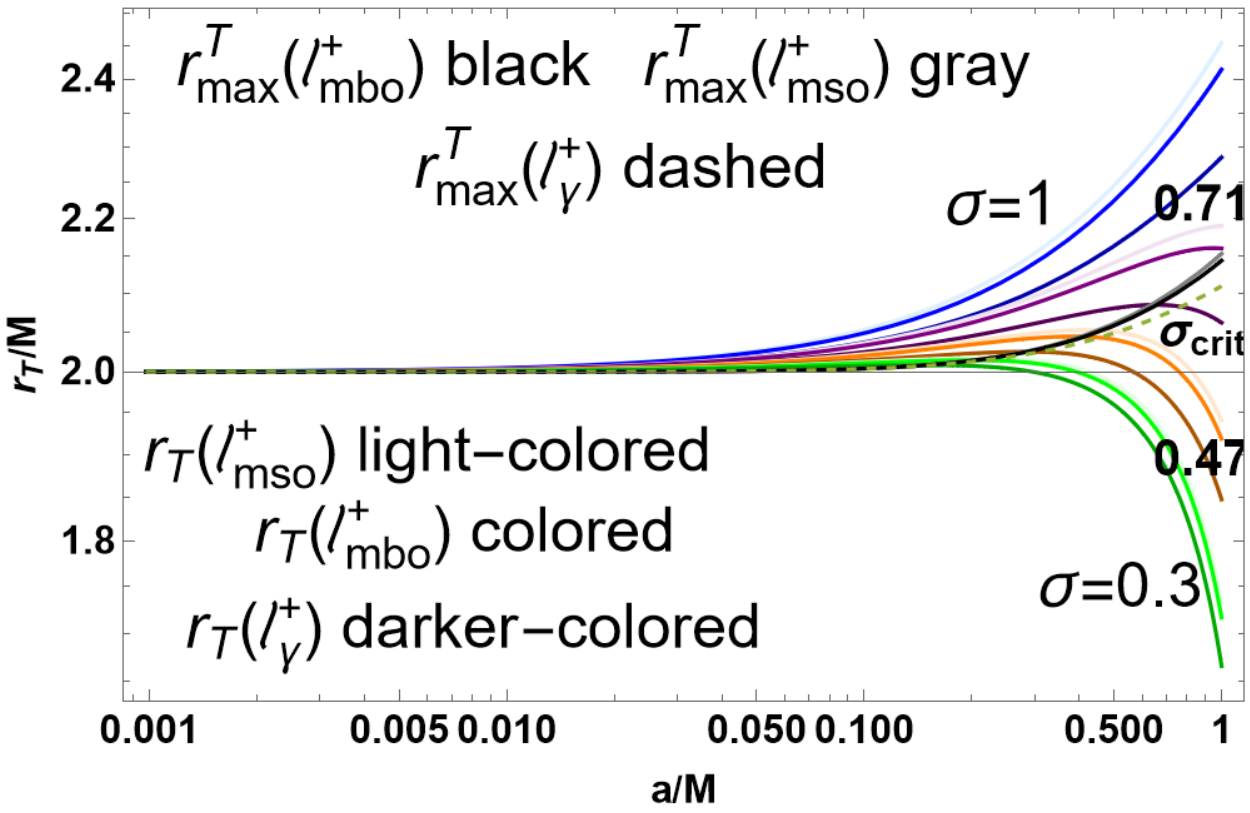}
  \caption{Counter-rotating flows. Upper line. Maximum extension of the turning radius $r_{\Ta}$ (solutions of
  $\partial_a r_{\Ta}(\sigma_\Ta)=0$ in Eq.\il(\ref{Eq:inform-amaxt})) as function of the  \textbf{BH} spin-mass ratio $a/M$.  There is $\sigma\equiv(\sin\theta)^2$.
  Blue   (light-blue) curves  are quantities evaluated on $\ell_{mso}^+$ ($\ell_{mbo}^+$),  and darker-blue curves are    quantities evaluated on $\ell_{\gamma}^+$.  Limiting momenta $\ell_{mso}^+$, $\ell_{mbo}^+$, and $\ell_{\gamma}^+$ are defined  in Eqs\il(\ref{Eq:def-nota-ell}). Dotted lines  are  maximum and minimum values of   $(r,\sigma)$, occurring for $a=0$ (the static spacetime limit) and $a=M$ (the extreme  \textbf{BH} Kerr spacetime). Upper left panel: radius
$  r_\Ta (\sigma_ {max}^\Ta) =
 r_ {max}^\sigma : \partial_ {r_\Ta} \sigma_\Ta = 0$ of Eq.\il(\ref{Eq:basi-c-pri-ncie}) as function of $a/M$. Upper center panel:  plane
 $ \sigma_ {max}^{\Ta} = \sigma_\Ta (r_ {max}^{\sigma})
 $
 of
  Eq.\il(\ref{Eq:event-zi-max-sigma}) as function of $a/M$.
 Upper right panel: Limiting  plane $\sigma_{max}^a$, regulating  the existence of the extreme spin $
  a_{{max}}^{\Ta }$ of  Eq.\il(\ref{Eq:inform-amaxt}), is shown as function of $a/M$, where $
  a_{{max}}^{\Ta }(\sigma_{max}^a)=M$. Bottom  left panel: limiting plane $\sigma_{max}^a$ and maximum plane  $\sigma_{max}^\Ta$ evaluated for $\ell_{mso}^+$, $\ell_{mbo}^+$  and  $\ell_{\gamma}^+$, plotted as   functions of $a/M$. Bottom right panel: Turning points $r_\Ta$ on $\ell_{mso}^+$ (light-coloured curves) and $\ell_{mbo}^+$ (coloured curves) on different planes $\sigma$ signed on the curves ($\sigma=1$ is the equatorial plane).  Black (gray) curve is the maximum $r_{max}^{\Ta}$ on $\ell_{mbo}^+$ ($\ell_{mso}^+$). Analysis is also performed for limiting function $r_\Ta(\ell_\gamma^+)$ for counter-rotating proto-jet driven configurations as darker-colored curves.}\label{Fig:Plotrtealg}
\end{figure*}

 There are no extremes of  the counter-rotating   flow  turning radius  $r_\Ta$
as function of $\sigma_\Ta$, and there is  $\partial_{\sigma_\Ta} r_{\Ta}>0$ as confirmed in  Figs\il(\ref{Fig:Plotrtea})\footnote{The static spacetime is a limiting case for this problem, an extreme point however exists for flows with $\ell>0$.}.
It is worth noting that there  is no extreme of
 $r_{\Ta}$ as function of   $\ell <0$.  Considering relation (\ref{Eq:add-equa-sigmata}), we see that  are no solutions of  $\partial_ {r_\Ta}\sigma_\Ta = 0 $  with  $\sigma_\Ta
    \in [0, 1]$, for  $ \ell < 0$ and $ r >
     r_ {\epsilon}^+$.
For counter-rotating flows there is  a maximum  of the turning point radius $r_\Ta$ for the variation of the attractor dimensionless spin $a/M$. This implies that the frame-dragging acts  differently for  the \textbf{BH} spin mass-ratio, distinguishing different  \textbf{BH} central attractors.
  The  conditions can be expressed  more precisely as follows:

  \medskip

There is  $ \partial_a r_{\Ta}(\sigma_\Ta)=0$  for
  \bea\label{Eq:inform-amaxt}&&
  a=a_{{max}}^{\Ta }\equiv\frac{2 \sigma  \ell }{\sigma ^2-\ell ^2(1-\sigma)} \quad \mbox{with }\quad \sigma\in[0,\sigma^a_{max}], \quad \mbox{where}
  \\\label{Eq:poi-s-dat-amax}&&
 \sigma^a_{max}\equiv \frac{1}{2} \left[\sqrt{\ell ^2 [(\ell -4) \ell +8]}-(\ell -2) \ell \right]\quad\mbox{and there is } \quad  a_{max}^{\Ta }( \sigma^a_{max})=M;  \\\
&&\nonumber \mbox{equivalently for} \quad  \sigma_{max}^\Ta\equiv \frac{ \sqrt{{\ell ^2 \left[a^2 \left(\ell ^2+4\right)+4(1- a \ell) \right]}}+ \ell(2 -a \ell)}{2 a}
    \\\label{Eq:event-zi-max-sigma}
   &&\mbox{or}\quad\ell_{max}^\Ta\equiv- \frac{\sigma }{a (1-\sigma)}\left[1+\sqrt{{ \left[1-a^2 (\sigma -1)\right]}}\right]\quad \mbox{for}\quad \sigma\in ]0,1[.
  \eea
  Note that incidentally these solutions
  coincide also with the maxima of  $\sigma_\Ta$ as function of $r_\Ta$, that is there is:
  \bea&&\label{Eq:basi-c-pri-ncie}
  \partial_ {r_\Ta} \sigma_\Ta = 0\quad\mbox{for}\quad
  r_\Ta (\sigma_ {max}^\Ta) =
 r_ {max}^\sigma\equiv\frac{1}{2} \left[\sqrt{a \left[a \left(\ell ^2+4\right)-4 \ell \right]+4}+a \ell +2\right],\quad
 \mbox{with}\quad   \sigma_ {max}^{\Ta} = \sigma_\Ta (r_ {max}^{\sigma})
  \eea
  ---Figs\il(\ref{Fig:Plotrtealg}).
 From the analysis of  $r_\Ta(\sigma _ {\max}^\Ta)$, it is clear that a maximum exists for any spin  $a/M$ (depending on  $\ell$). The maximum  increases with the BH  spin, and the maximum extension of the  corona radius is for  the extreme Kerr \textbf{BH} with $a = M$,
where  $ r_{\Ta}/M\in[2.143,
   2.153]$, while the minimum is for the case of static attractor
and coincides with the horizon $r =
 2 M$.
From the analysis of  $\sigma _ {\max}^\Ta$,  we find that the  plane  of the maximum point increases increasing the spin, reaching the  maximum for $a =
 M$ where  $\sigma_\Ta\in ]0.671,
   0.693[$.
   Similarly to $r_\Ta$,  plane $\sigma_\Ta$ varies in a    small range of values.
From the  analysis of the extremes according to the spin $a/
  M$, we find  that
the limiting value for the spin is the extreme Kerr \textbf{BH},  but for each spacetime the limiting plane is between the values for the limiting case for  extreme Kerr \textbf{BH}  and the static spacetime respectively,  that is  $\sigma\in [0.671,
   0.693]$ and $\sigma\in[0.629,
   0.649] $.
\subsection{Fluid velocities at the turning point}\label{Sec:velocity}
The  time component of the flow velocity at the turning point is:
\bea\label{Eq:itafra-regu}
&&\dot{t}_{\Ta}=\La \left(\frac{1}{\ell }-\frac{1}{a \sigma_\Ta }\right)=\Em \ell \left(\frac{1}{\ell }-\frac{1}{a \sigma_\Ta }\right)
\eea
--(see Eq.\il(\ref{Eq:tdotemLem})).
Note that for  counter-rotating flows ($\ell<0$)  there is  $\dot{t}_{\Ta}>0$ at the   turning point.
(When $\ell>0$  and $\La>0$,  there is  $\dot{t}_\Ta>0$ for  $\ell<a\sigma_{\Ta}$, which cannot occur in the tori model we consider here  where there is  $\ell>a>a\sigma_{\Ta}$--see Sec.\il(\ref{Sec:turning-sign-existence})\footnote{This implies that, at the turning point for $\ell>0$ occurring  in the ergoregion,  there is  $\dot{t}_\Ta<0$ (physically forbidden) if $\Em>0$--see also discussion in Eqs\il(\ref{Eq:lLE}) and Eqs\il(\ref{Eq:tdotemLem}).}).

Quantities   $(\sigma_\Ta, r_{\Ta}, \dot{t}_{\Ta})$  do not depend  on the Carter constant  $\Qa$ (depending however  on $\sigma$) and on the normalization condition, which represents a further constraint on the turning sphere,  therefore  they hold eventually for photons  and  matter. Notably $(\sigma_\Ta, r_{\Ta}, \dot{t}_{\Ta})$ do not depend explicitly  on the cusp  initial location  or the initial plane $\sigma_0$.
This implies that, at  the turning point,    $r_\Ta$ and $\dot{t}_\Ta$ are explicitly regulated only by the torus momentum  $\ell$, and  $\La$ or $\Em$ (in our case  $K$ parameter) for  $\dot{t}$.
Therefore,  the torus distance from the attractor  or the precise identification of the torus "emission" region is not relevant for these features of the turning point.
Nevertheless, quantities    $(t_{\Ta},\tau_{\Ta})$ depend on the initial data,  and
($\sigma_\Ta$,  $r_{\Ta}$) can be obtained separately by solving the coupled equations  for  $\dot{\sigma}$ and  $\dot{r}$,   which depend on
constant $\Qa$, and therefore  on $\sigma_0$  and  $r_0$.
These relations depend explicitly on the normalization conditions and the two constants of motion $\La$ and
 $\Em=K$ (for timelike particles). However, if the torus is cusped   then there is only one independent  parameter, being  $\ell$  sufficient to fix uniquely $\Em=K(\ell)$.

\begin{figure}
\centering
    \includegraphics[width=7cm]{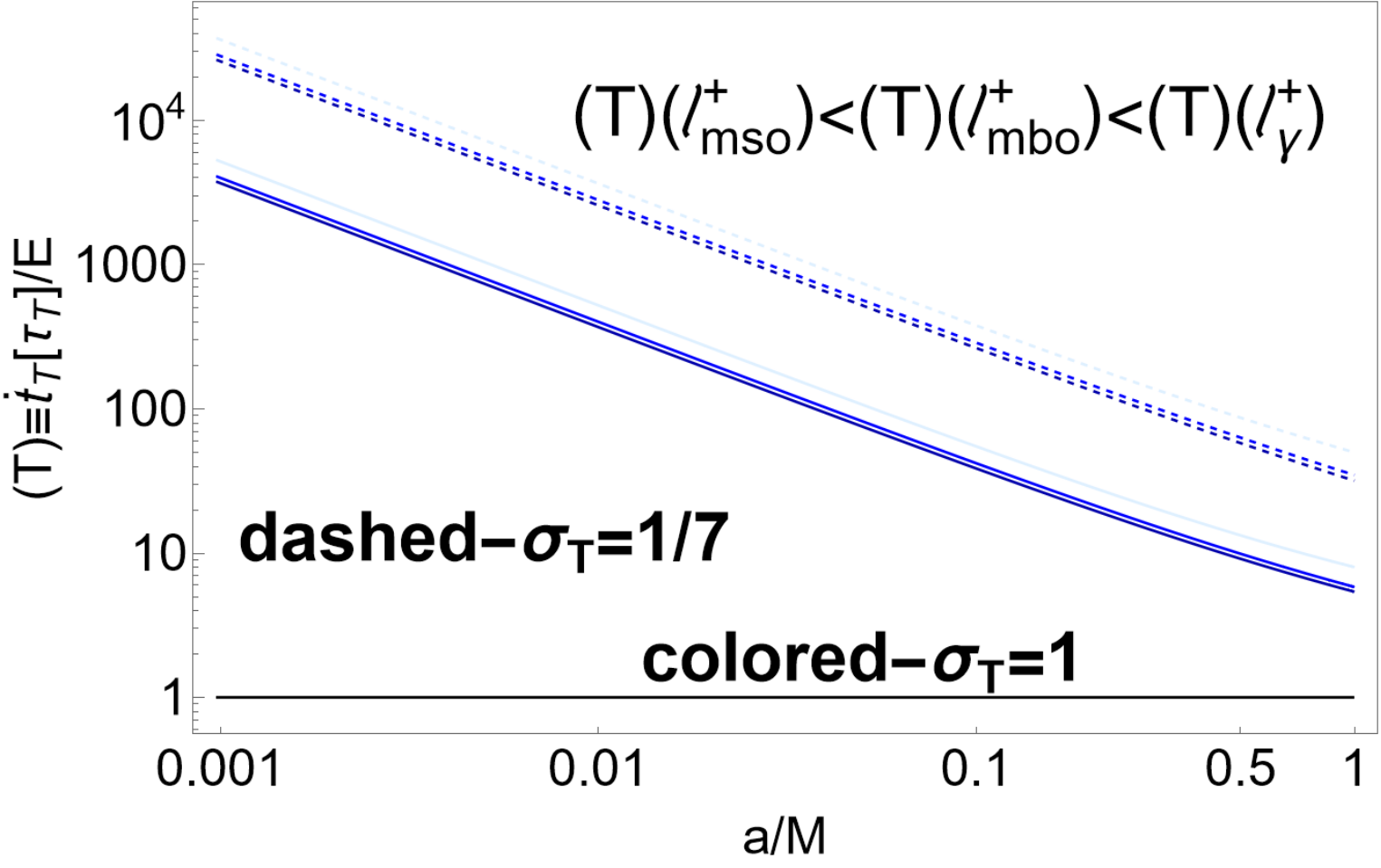}
  \caption{Quantity $\left(T\right)\equiv {\dot{t}_\Ta\left(\tau _\Ta\right)}/{\Em}$ is plotted, at the counter-rotating flow turning points,  at the extremes of the turning corona--Eq.\il(\ref{Eq:add-equa-sigmata})--- for different planes $\sigma\equiv \sin^2\theta$,   as functions of the   \textbf{BH} spin $a/M$.  Constant of motion $\Em$ is defined in Eq.\il(\ref{Eq:EmLdef}).  The \textbf{BH} equatorial plane is at $\sigma=1$. Quantity $\left(T\right)$ is evaluated at the   turning corona boundaries, for fluid specific angular momenta  $\ell=\ell_{mso}^+$, $\ell=\ell_{mbo}^+$ and $\ell=\ell_{\gamma}^+$, defined in  Eqs\il(\ref{Eq:def-nota-ell}) for  counter-rotating   tori and  proto-jets driven flows, there is  $\left(T\right)\left(\ell _{{mso}}^+\right)<\left(T\right) \left(\ell _{{mbo}}^+\right)<\left(T\right) \left(\ell _{\gamma }^+\right)$.  }\label{Fig:Plotrtewavenige}
\end{figure}

 In  {Fig.\il(\ref{Fig:Plotrtewavenige})} we can see   the evaluation of the  $\left(T\right)\equiv \dot{t}_\Ta\left(\tau _\Ta\right)/\Em$ at the turning point,    on  the turning corona  extreme in Eq.\il(\ref{Eq:add-equa-sigmata}). The analysis  points out  the small variation of these  quantities according to the fluid momenta $\ell$, being  $\left(T\right)\left(\ell _{{mso}}^+\right)<\left(T\right) \left(\ell _{{mbo}}^+\right)<\left(T\right) \left(\ell _{\gamma }^+\right)$.

Expressing $(\dot{r}_{\Ta},\dot{\sigma}_{\Ta})$ functions of
$r_{\Ta}$ we obtain
\bea
&&
\dot{r}_{\Ta}^2(r_{\Ta})=\frac{(a \ell -2 r_{\Ta}) \sqrt{\Em^2 \left[a(a- \ell) +r_{\Ta}^2\right]^2-\Delta_\Ta \left[\Em^2(a- \ell )^2+\mu^2 r_{\Ta}^2+\Qa\right]}}{2 r_{\Ta} \left[a(a- \ell) +r_{\Ta}^2\right]},
\\
&&\dot{\sigma}_{\Ta}^2(r_{\Ta})=-\frac{\ell  (a \ell -2 r_{\Ta})^2 \left[\Delta_\Ta \left[\Qa-\frac{a r_{\Ta} (\Em^2-\mu^2) [2 a+(r_{\Ta}-2) \ell]}{a \ell -2 r_{\Ta}}\right]+[2 a+(r_{\Ta}-2) \ell]\Em^2 r_{\Ta} \ell  \right]}{16 r_{\Ta}^3 \left[a(a- \ell) +r_{\Ta}^2\right]^2 [2 a+(r_{\Ta}-2) \ell]}.
\eea
Expressing $(\dot{r}_{\Ta},\dot{\sigma}_{\Ta})$ functions of
$\sigma_{\Ta}$ there is
\bea
&&\dot{\sigma}_{\Ta}^2(\sigma_{\Ta})=-\frac{\ell ^4 \left[\Em^2 (\sigma_{\Ta}-1) \left(\ell ^2-a^2 \sigma_{\Ta}\right)+\sigma_{\Ta} \left[a^2 \mu^2 (\sigma_{\Ta}-1)+\Qa\right]\right]}{16 (\sigma_{\Ta}-1) (\ell -a \sigma_{\Ta})^2 \hat{m}^2};
\\&&
\dot{r}_{\Ta}^2(\sigma_{\Ta})=-\frac{ \sqrt{ \ell ^2\Em^2\left[a  \ell -\left(\frac{\hat{m}^2}{\ell ^2}+a^2\right)\right]^2-{a \sigma_{\Ta} \left(2 \hat{m}+a \ell^2 \right) \left[\frac{\mu^2 \hat{m}^2}{\ell ^2}+ \Em^2(a-\ell )^2+\Qa\right]}}}{2 (a \sigma_{\Ta}-\ell )\hat{m}}.
\eea
Quantities  $\dot{r}$ an $\dot{\theta}$ are in Eqs\il(\ref{Eq:eqCarter-full}),   the couple   $(\dot{r},\dot{\theta})$ depends on  $\Qa$, whereas   $\dot{r}$   depends explicitly on the normalization condition,  distinguishing therefore explicitly photons from matter.
On the equatorial plane there is $\Qa=0$ only and only if $\dot{\theta}=0$ (more details on the equatorial plane case are discussed in Sec.\il(\ref{Sec:basis-eq})).
\begin{figure*}
\centering
    \includegraphics[width=7.5cm]{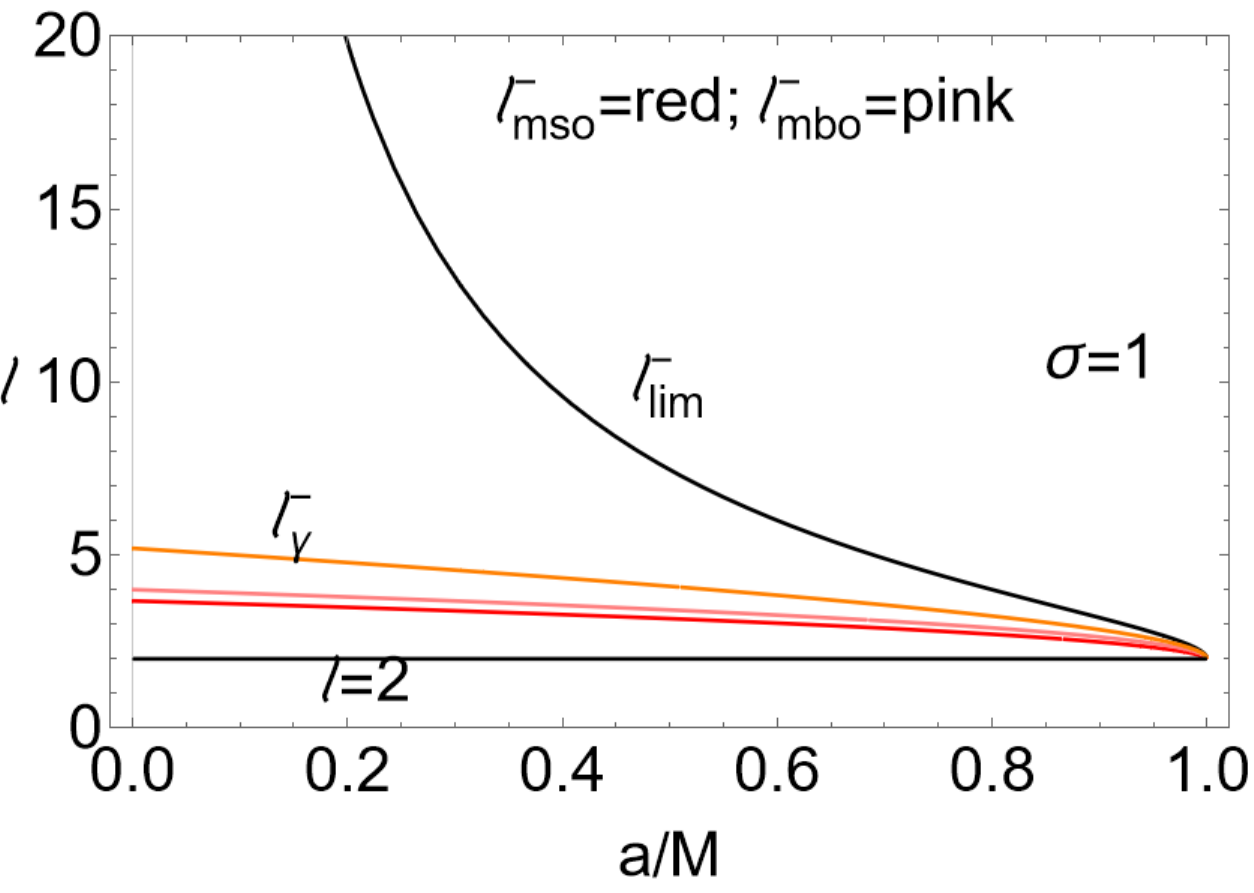}
    \includegraphics[width=8cm]{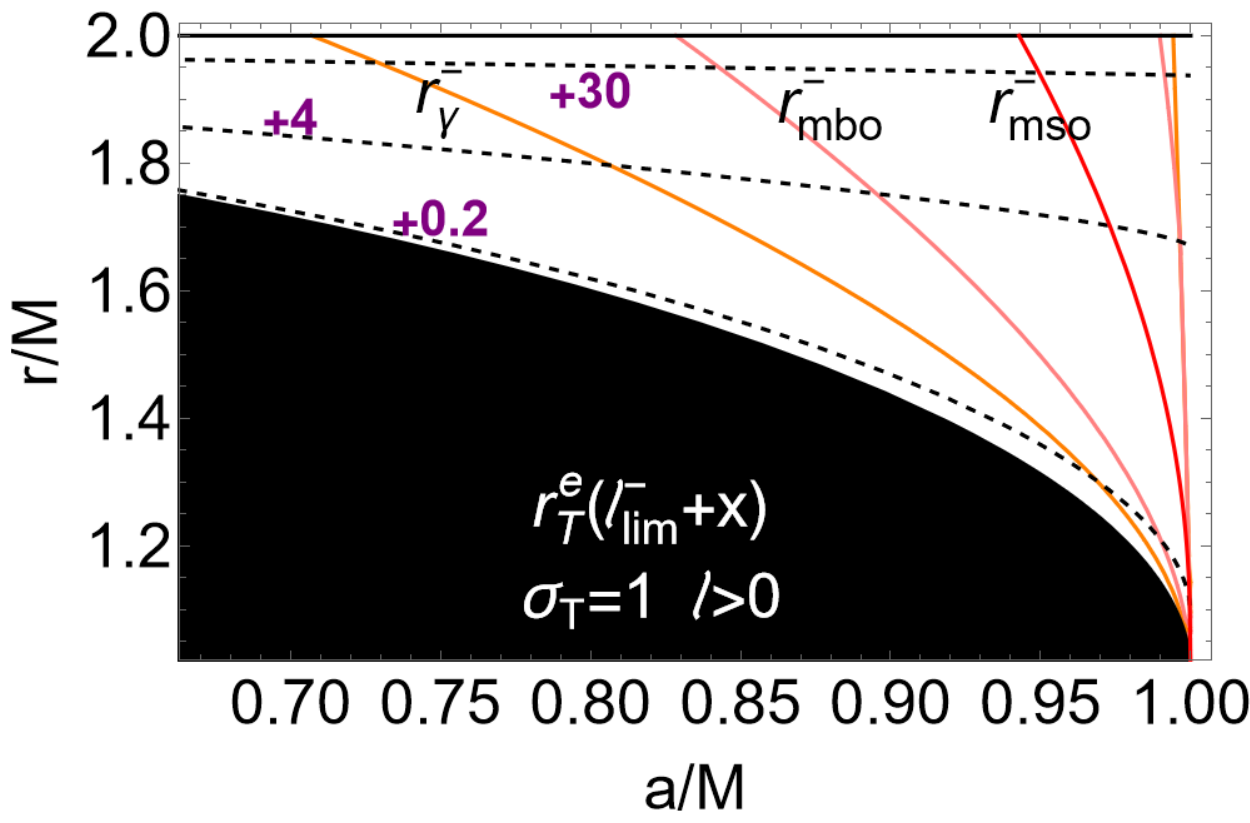}
      \includegraphics[width=7.5cm]{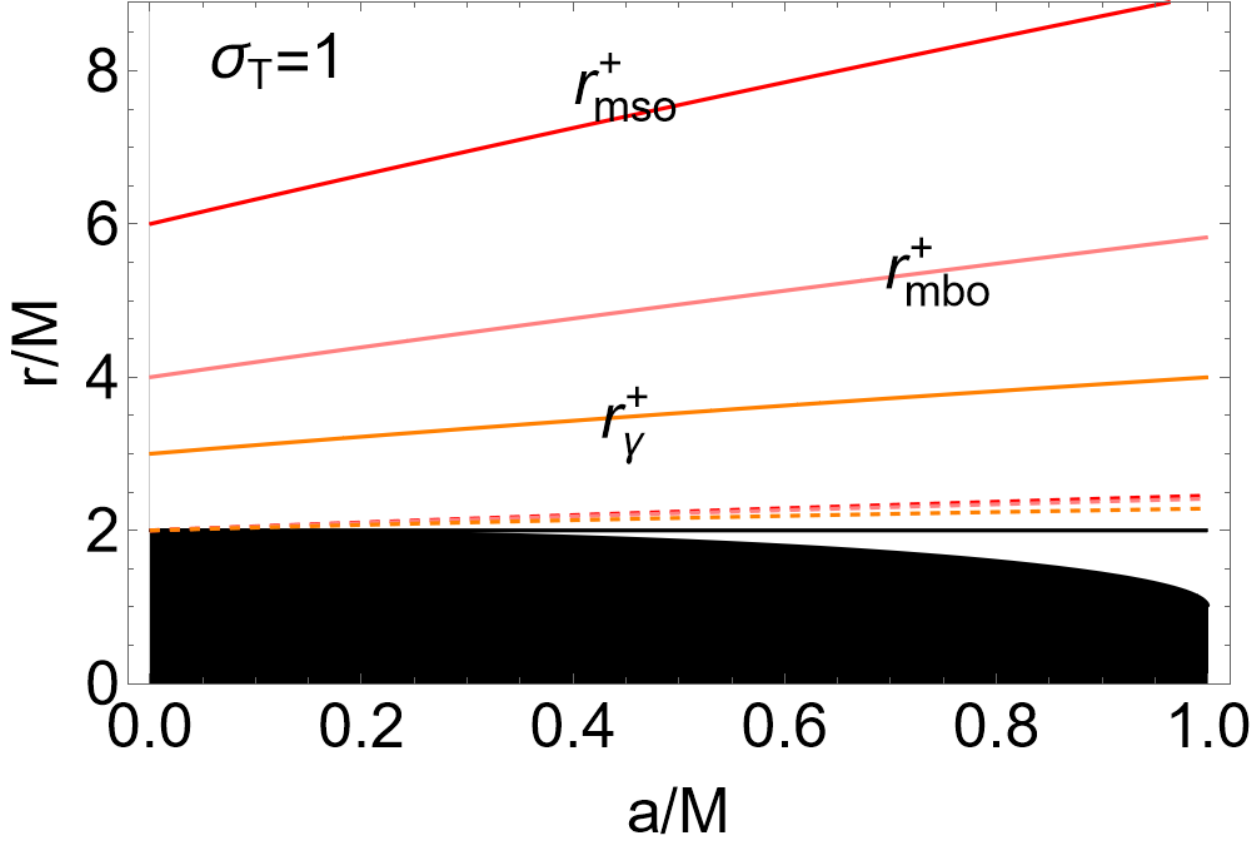}
  \includegraphics[width=8cm]{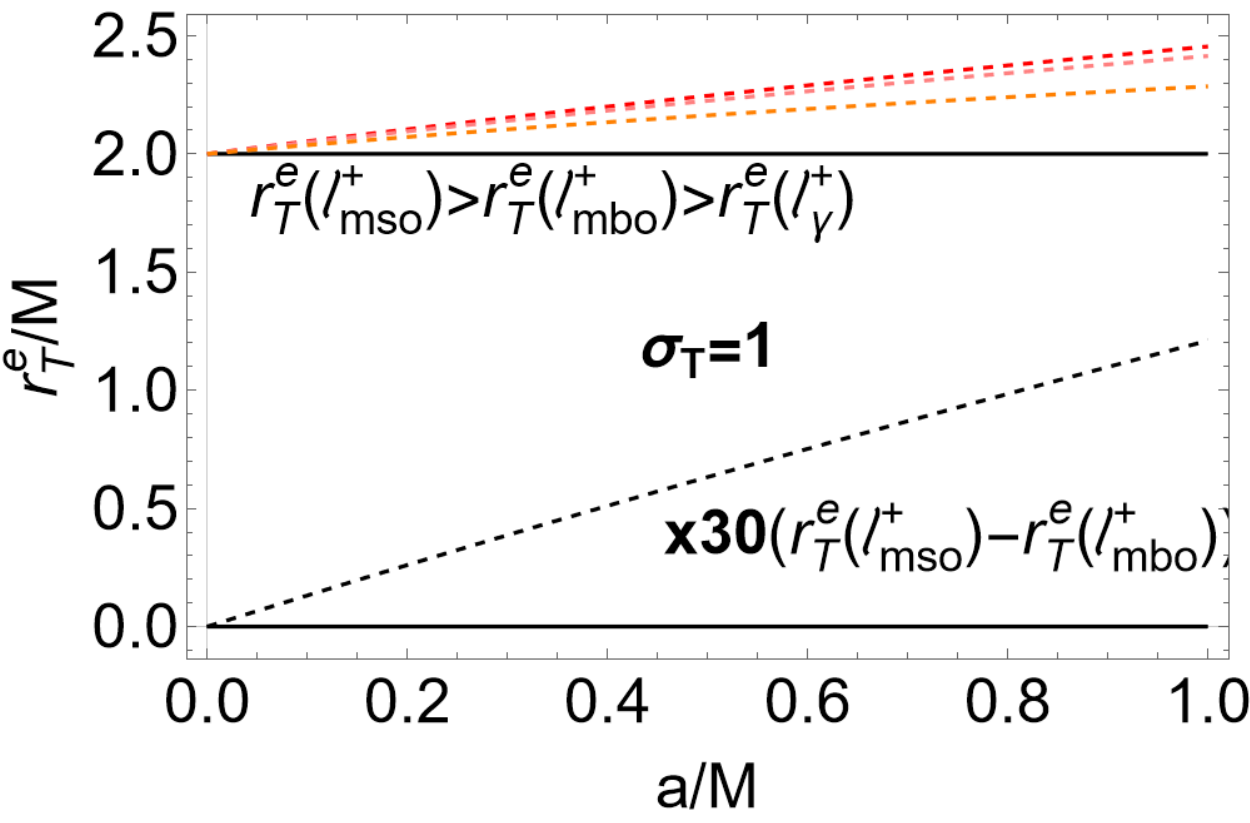}
  \caption{Allowed regions for the turning point of the azimuthal motion of the matter infalling from tori. Black region  is the central  \textbf{BH} at $r<r_+$, where $r_+$ is the outer \textbf{BH} horizon, the outer ergosurface is $r_{\epsilon}^+=2M$.
  The geodesic limiting values of the specific angular momentum $\ell^{\pm}=\{\ell_{mso}^{\pm},\ell_{mbo}^{\pm},\ell_\gamma^{\pm}\}$,   plotted as functions of the \textbf{BH} dimensionless spin, are provided in Eqs\il(\ref{Eq:def-nota-ell}). Panels show  the  function $r_\Ta(\ell)$  on the equatorial plane $\sigma_\Ta=1$ (where $\sigma\equiv \sin^2\theta$), for $\ell>0$ ($(-)$ upper line panels) and $\ell<0$ ($(+)$ bottom panels) fluids, analyzed in Sec.\il(\ref{Sec:basis-eq}.)  Upper panels show  the situation for the  $\ell>0$. Radius $r_{\Ta}^e$ is in the ergoregion $]r_+, r_{\epsilon}^+[$. Left upper panel: limiting tori specific  angular momentum $\ell_{\lim}^->\ell_{\gamma}^-$ of Eq.\il(\ref{Eq:limellcoro}) as function of the \textbf{BH} spin--mass ratio $a/M$. The function limits the existence of a  co-rotating fluid turning point at any plane $\sigma$.  The limiting value $\ell=2$ is also shown. Upper right panel:   corotating  geodesic structure,  defined  in Eqs\il(\ref{Eq:def-nota-ell}) (including the radii $r_{(mbo)}^-, r_{(\gamma)}^-$, colored correspondingly to $r_{mbo}^-, r_{\gamma}^-$), in the outer ergoregion, as functions of $a/M$. Dashed black  line  is the  radius $r_{\Ta}^e$ of Eq.\il(\ref{Eq:rte-second-sigma1}) for specific angular momentum $\ell=(\ell_{\lim}^-+x)$ for different $x$ signed on the curves.
   Bottom left panel: counter-rotating geodesic structures of Eqs\il(\ref{Eq:def-nota-ell})  and  turning radii $r^e_{\Ta}$ (dashed curves):  there is $r^e_\Ta(\ell_{mso}^+)>r^e_\Ta(\ell_{mbo}^+)>r^e_\Ta(\ell_{\gamma}^+)$ as functions of $a/M$, black line is the outer ergosurface $r_{\epsilon}^+=2M$. Right panel shows  a zoom on the   radii $r^e_\Ta(\ell_{mso}^+)>r^e_\Ta(\ell_{mbo}^+)>r^e_\Ta(\ell_{\gamma}^+)$, dashed-black line is  difference
   range $(r^e_\Ta(\ell_{mso}^+)-r^e_\Ta(\ell_{mbo}^+))$, magnified for a factor of $(\mathbf{x30})$, providing  the  maximum range for the location of tori driven counter-rotating turning points on the equatorial plane--see Eqs\il(\ref{Eq:max-cusp-eq-exte}) and (\ref{Eq:max-cusp-eq-exte1}). }\label{Fig:Plotrte}
\end{figure*}
\begin{figure*}
\centering
    \includegraphics[width=6.5cm]{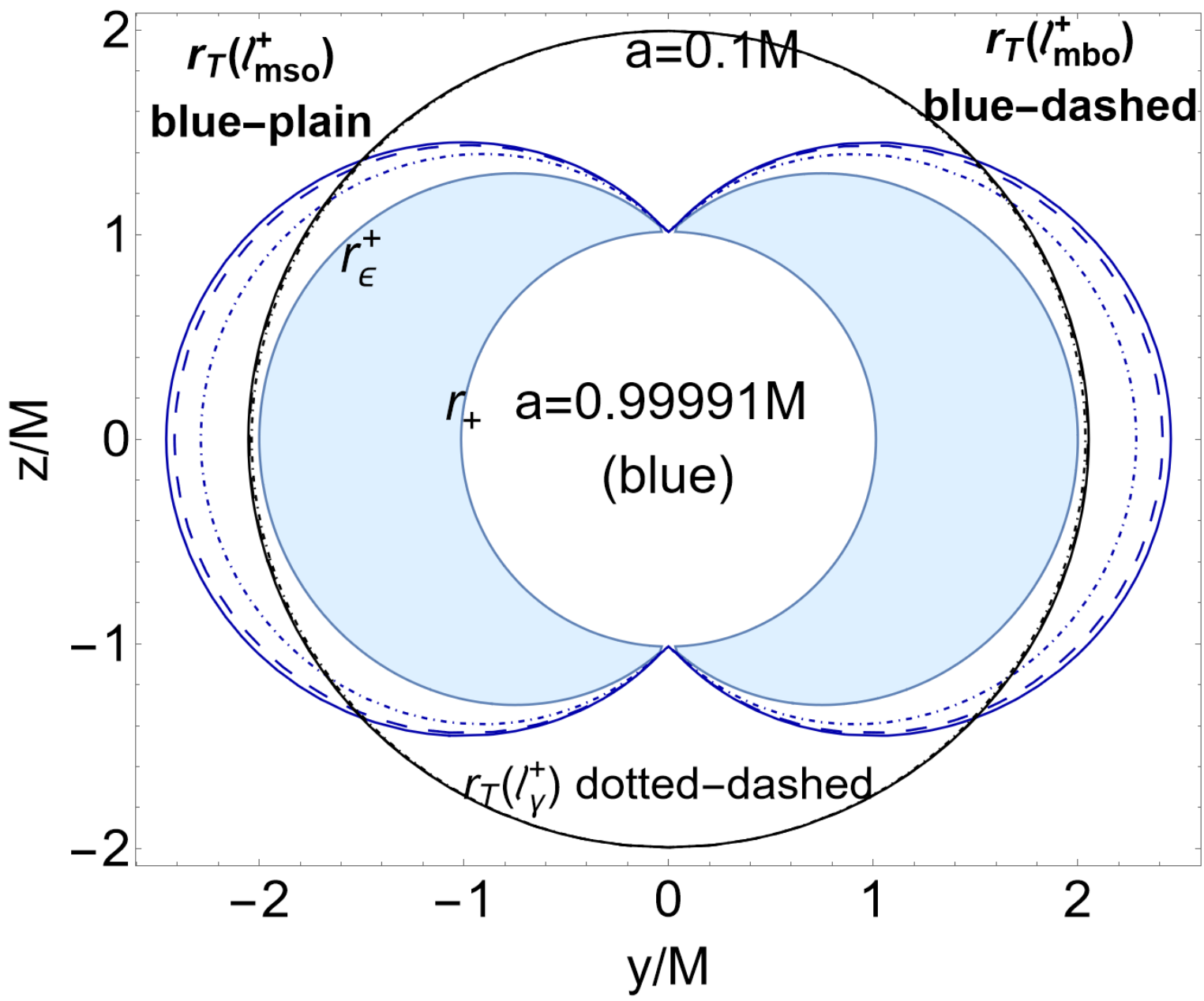}
      \includegraphics[width=6.75cm]{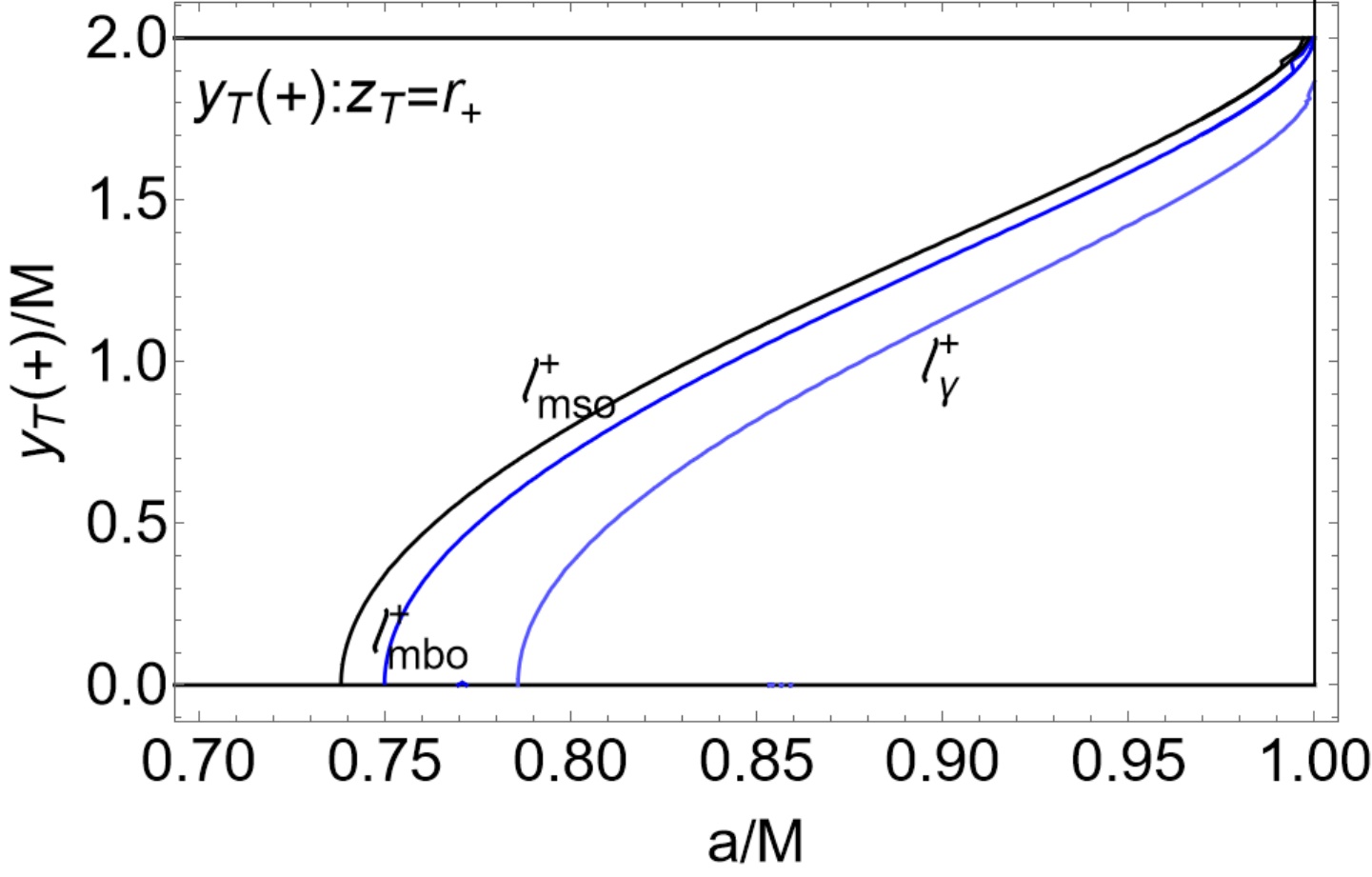}
     \\ \includegraphics[width=6.5cm]{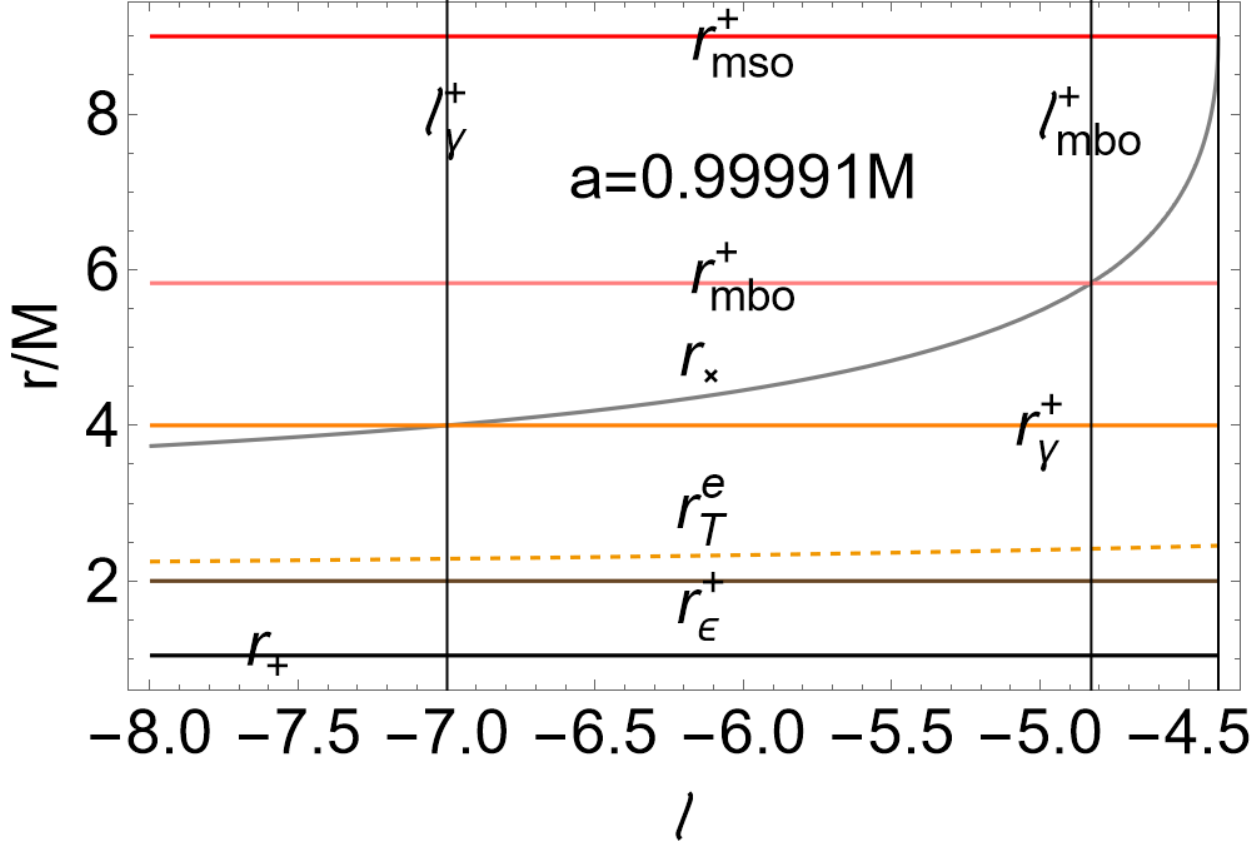}  \includegraphics[width=6.5cm]{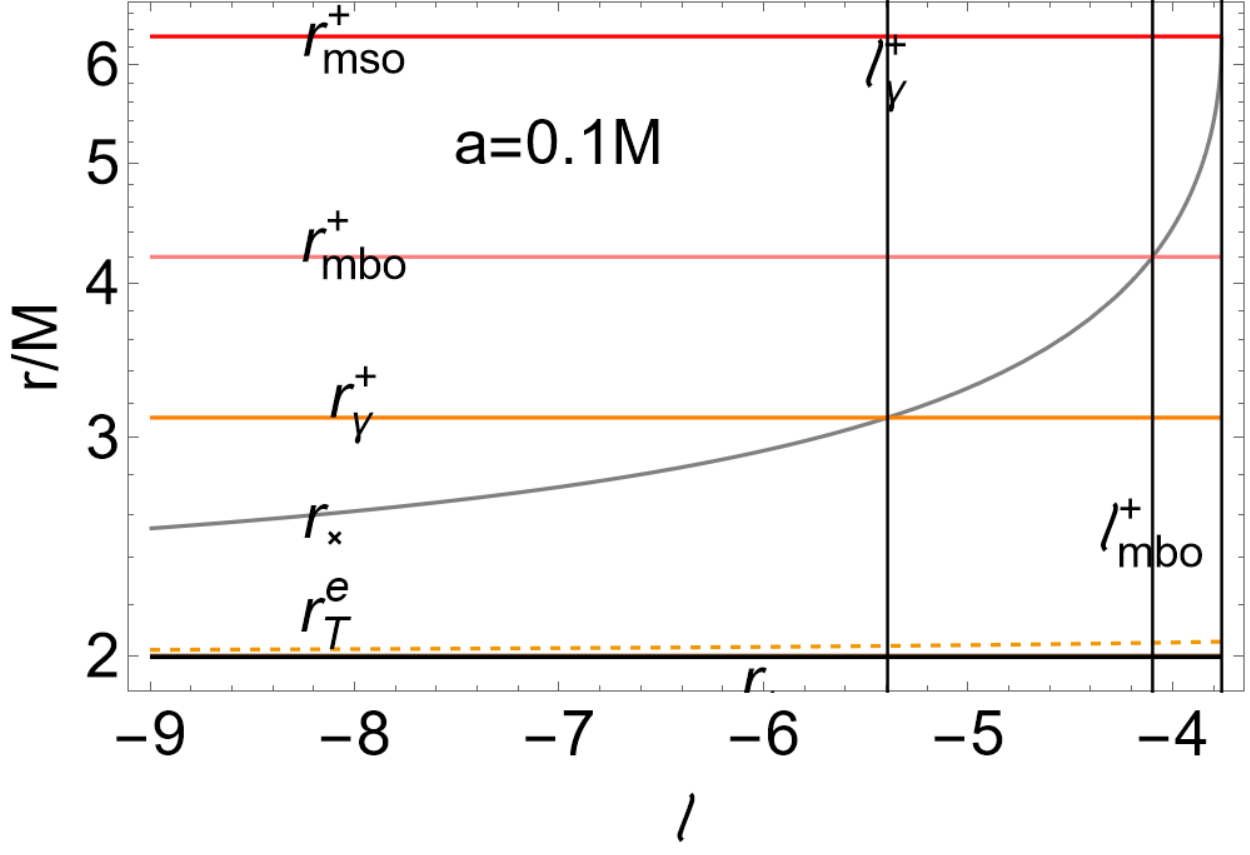}
     \\
       \includegraphics[width=6.5cm]{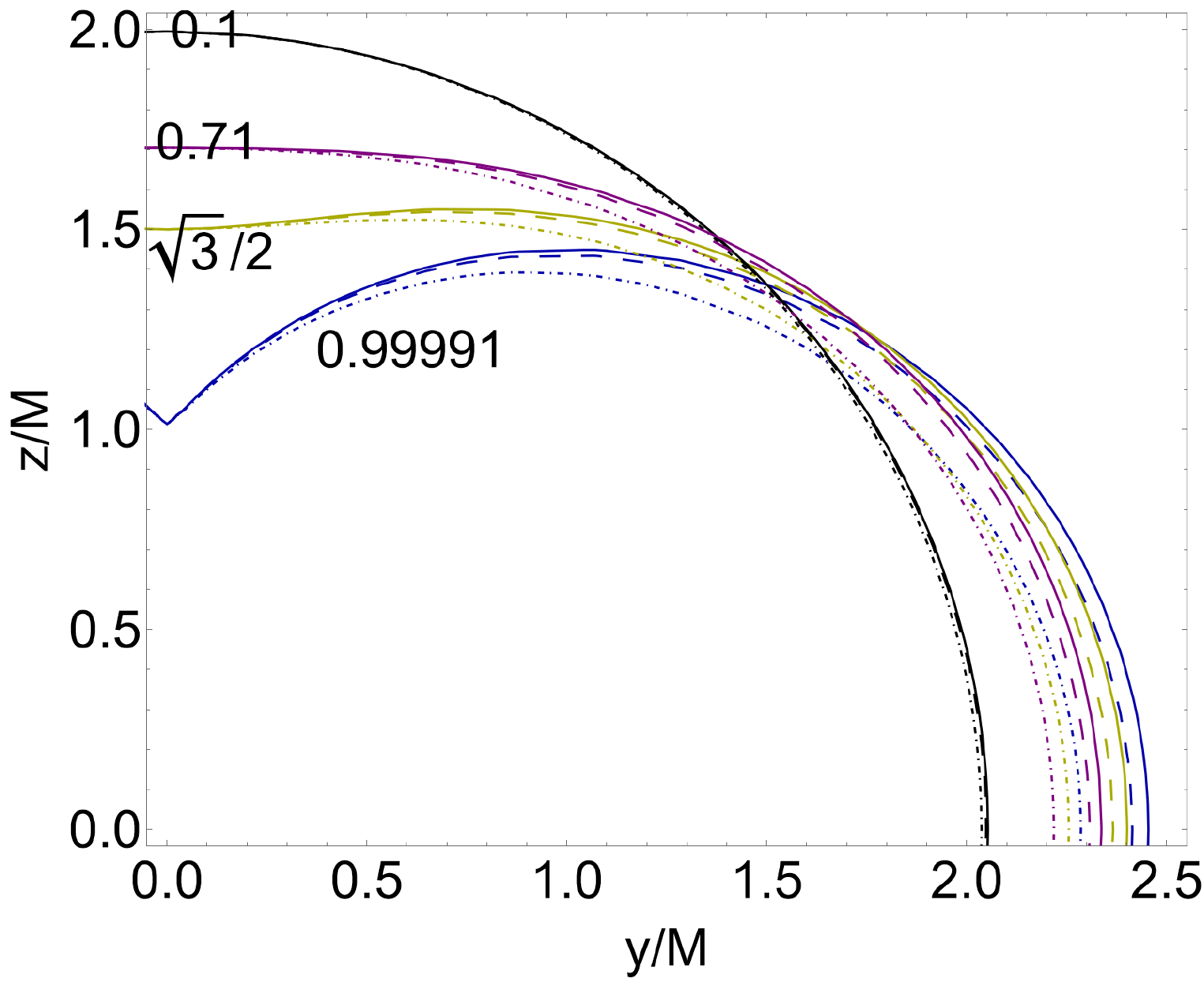}
         \includegraphics[width=6.5cm]{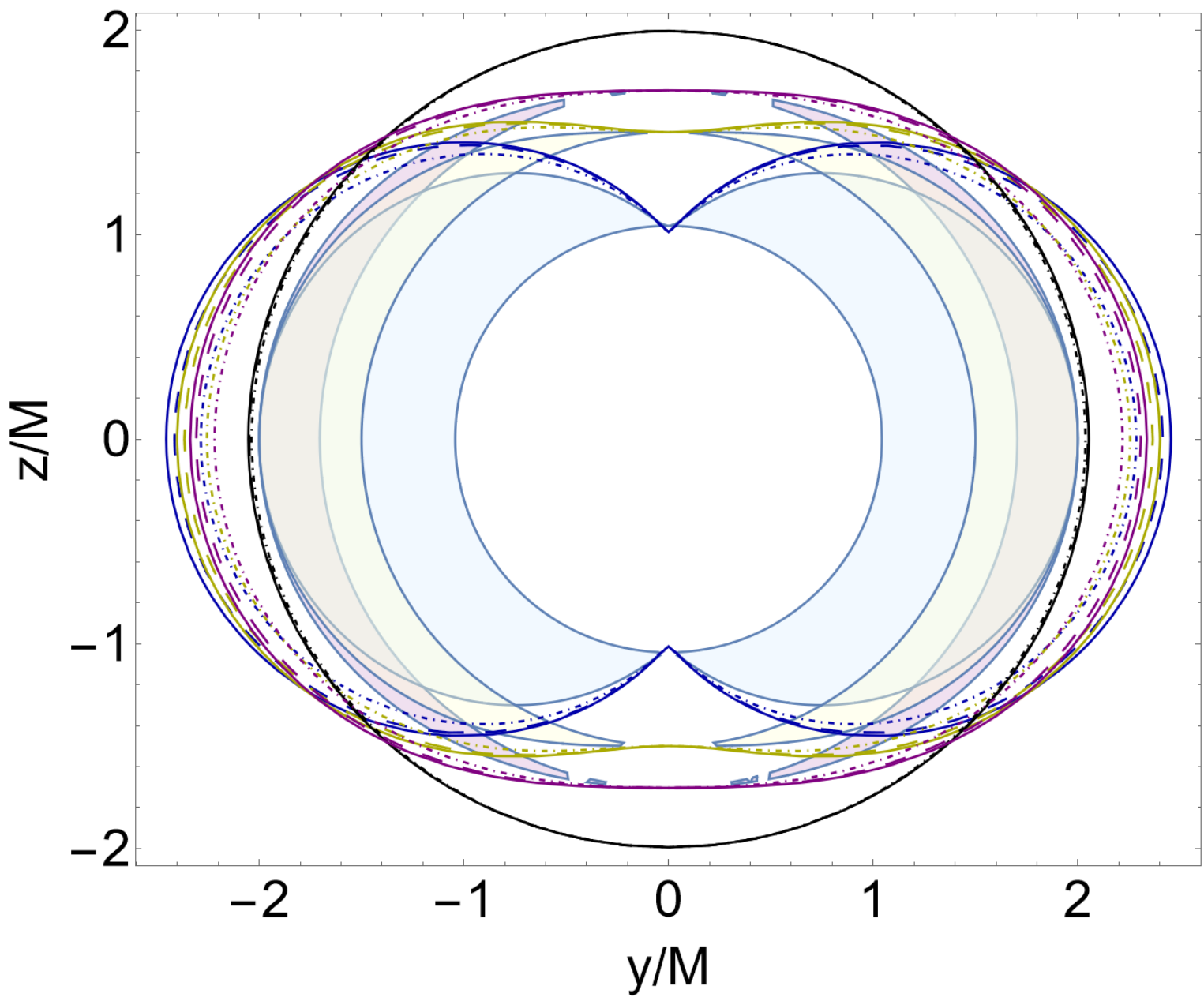}
  \caption{{Turning points of the azimuthal motion of the counter-rotating flows in the $r-\theta$ plane for both tori and proto-jets. There is $r= \sqrt{y^2+z^2}$ and $\theta=\arccos({z}/r)$. Upper left  panel: Counter-rotating flows  turning point $r_{\Ta}$ of Eq.\il(\ref{Eq:turning-point-radius}) evaluated at fluid specific angular momenta $\ell_{mso}^+$ (plain) and $\ell_{mbo}^+$  (dashed), $\ell_{\gamma}^+$  (dotted-dashed) defined in  Eqs\il(\ref{Eq:def-nota-ell}),  in the \textbf{BH} spacetime $a=0.1M$ (black curves) and $a=0.99991M$ (blue curves).  The corona defined by the
   range $r_\Ta(\ell_{mso}^+)-r_\Ta(\ell_{mbo}^+)$
   ($r_\Ta(\ell_{\gamma}^+)-r_\Ta(\ell_{mbo}^+)$) is  orbital range  of the  turning points location for counter-rotating cusped tori (proto-jets) driven flows, which reaches its maximum on the equatorial plane $z=0$--see Eqs\il(\ref{Eq:max-cusp-eq-exte}) and (\ref{Eq:max-cusp-eq-exte1}). Shaded blue region is  the outer ergoregion $]r_{+},r_\epsilon^+]$, for the \textbf{BH} with spin $a=0.99991M$. Bottom right panel:  solutions $y_\Ta=y_\Ta(+)\neq 0: z_\Ta=r_+$ ($r_+$ is the  \textbf{BH} outer horizon) indicating  the presence of double turning point at $z_\Ta>r_+$ (and $y_\Ta\leq y_\Ta(+)$).  Center panels: counter-rotating flow  turning point $r_{\Ta}^e$ on the equatorial plane  of Eq.\il(\ref{Eq:rte-second-sigma1})   and counter-rotating tori cusp $r_{\times}\in]r_{mbo}^+, r_{mso}^+]$ of Eq.\il(\ref{Eq:toricenter-inner})  (for  $\ell\in ]\ell_{mbo}^+, \ell_{mso}^+]$) as functions of the tori specific angular momentum $\ell$, for spacetime $a=0.99991M$ (left panel) and $a=0.1M$ (right panel).  Radii of  geodesic structures (horizontal lines), defined in
 Eqs\il(\ref{Eq:def-nota-ell}), and related momenta $\ell$ (vertical lines), outer horizon $r_+$ and outer ergosurface on the equatorial plane $r_{\epsilon}^+=2M$  are also plotted.  For proto-jets driven flows,  there is  $r_{\times}\in]r_{\gamma}^+,r_{mbo}^+]$  cusp of a  proto-jet for $\ell\in ]\ell_{\gamma}^+,\ell_{mbo}^+]$. Bottom panels  show    the analysis of the   upper panel for the different  \textbf{BH} spin--mass ratios $a/M$, signed  on the curves. Colored regions are  the  \textbf{BHs} outer ergoregion  $]r_{+}, r_\epsilon^+[$ (the case  $a=0.1M$ colored in black).   Bottom left panel is a close-up view  of the bottom  right panel.}}\label{Fig:Plotrte4}
\end{figure*}
In Sec.\il(\ref{Sec:accelerations-fluids})  there is a discussion on the flow  at the turning point.
\section{The  equatorial plane case}\label{Sec:basis-eq}
Motion on the equatorial plane of the Kerr central \textbf{BH}  constitutes  a relevant case for the problem of the flow turning point of  infalling matter and photons.

We can distinguish the following two cases:

\begin{description}\item[(I)] $\sigma_0=1$:  the flow trajectory starts from the \textbf{BH} and torus equatorial plane. This  situation can  be framed in the standard accretion  from a toroidal configuration  centred on the \textbf{BH} and  with symmetry and equatorial plane coincident with the \textbf{BH} equatorial plane. Accretion occurs at the torus inner edge $r_\times$ (for $\theta_0=\pi/2$). This case holds also for the proto-jets driven configurations where the cusp $r_\times$ is on the equatorial plane;
\item[(II)] $\sigma_{\Ta}=1$:  in this case the  flow turning point is on the equatorial plane.
\end{description}
Conditions \textbf{(I)} and \textbf{(II)} may hold in the same accretion  model characterized by  $\sigma_0=\sigma_\Ta=1$,  depending on the   Carter constant $\Qa$, holding for example in the special case where $\Qa=0$ and $\dot{\theta}=\ddot{\theta}=0$.

In general, on the equatorial plane, $\theta=\pi/2$, from Eqs.\il(\ref{Eq:eqCarter-full}) we find:
\bea&&\label{Eq:general-on-equatorial-plane}
\dot{\theta}^2=\frac{\Qa}{r^4},\quad \dot{t}=\frac{\La \left[a^2 (r+2)-2 a \ell +r^3\right]}{r \ell  \Delta},\quad
\dot{\phi}=\frac{\La [2 a+(r-2) \ell]}{r \ell  \Delta},\\&&\nonumber \dot{r}=\pm \frac{\sqrt{\frac{\La^2}{\ell^2}\left[a\ell -\left(a^2+r^2\right)\right]^2-\Delta \left[\frac{\La^2}{\ell^2}\left(\ell-{a}\right)^2+\mu^2 r^2+\Qa\right]}}{r^2}.
\eea
Furthermore, from the  definition of $\Em$, $\La$ and $\ell$, there is  for  $\theta=\pi/2$
\bea&&\label{Eq:que-Emealtr}
\Em=\frac{2 a \dot{\phi}+(r-2) \dot{t}}{r},\quad \La=\frac{[a^2  (r+2)+r^3]\dot{\phi}-2 a \dot{t}}{r},\quad \ell =\frac{[a^2 (r+2)+r^3] \Omega -2 a }{2 a \Omega +r-2},\quad \Qa=r^4 \dot{\theta}^2\\
&&\dot{t}=\frac{[a^2(r+2)+r^3]\Em -2 a \La}{r \Delta},\quad\dot{\phi}=\frac{2 a \Em+\La (r-2)}{r \Delta},\quad\Omega=\frac{2 a+(r-2) \ell }{a^2 (r+2)+r^3-2 a \ell},
\eea
where $\Omega$ is the relativistic angular velocity.
\subsection{Turning point on the equatorial plane: $\sigma_{\Ta}=1$}\label{Sec:argument}
From  Eqs\il(\ref{Eq:que-Emealtr}) on the turning point where  $\dot{\phi}=\Omega=0$ we find:
\bea&&\label{Eq:que-EmealtrT}
\theta_\Ta=\pi/2,\quad
\dot{\phi}_\Ta=\Omega_\Ta=0,\quad
\Em=\frac{(r_\Ta-2) \dot{t}_\Ta}{r_\Ta},\quad \La=-\frac{2 a \dot{t}_\Ta}{r_\Ta},\quad \ell =\frac{2 a}{2-r_\Ta},\quad \dot{t}=\frac{a^2  (r_\Ta+2)+ r_\Ta^3}{r_\Ta \Delta_\Ta}\Em,
\eea
see also Eqs\il(\ref{Eq:tdotemLem}), Eqs\il(\ref{Eq:itafra-regu}),
 and Eqs\il(\ref{Eq:lLE}),(\ref{Eq:add-equa-sigmata}),(\ref{Eq:turning-point-radius}). Assuming  $\dot{t}_\Ta>0$, there is $\Em\lesseqgtr0$ for $r_\Ta\lesseqgtr2M$ (located  inside and out   the ergoregion),  occuring for  $\ell\gtrless0$ respectively. (The null limiting  condition on $\Em$ in the form (\ref{Eq:que-EmealtrT}) holds for $r_\Ta=r_{\epsilon}^+=2M$ \emph{or} for $\dot{t}_\Ta=0$).  From the equation for $\dot{t}_\Ta$  we find  $\dot{t}_\Ta\Em>0$, constraining  the turning point location. (Relations in Eqs\il(\ref{Eq:que-EmealtrT}) are not independent, as on the equatorial plane $r_{\Ta}^e
\equiv 2 \left(1-{a}/{\ell }\right)$).  On the other hand, the energy $\Em(\Ta)$ does not depend explicitly on the \textbf{BH} spin $a$.
 Equally, there is $\La\lesseqgtr0$ for $\dot{t}\gtreqless0$ (where notably $\La=0$ for $\dot{t}_\Ta=0$), while  $\ell\lessgtr0$ for $r_\Ta\gtrless2M$.

Therefore,  for  flow turning point  on the attractor equatorial plane there is
\bea&&\label{Eq:rte-second-sigma1}
\ell_\Ta=\frac{2 a}{r_\Ta-2},\quad \mbox{for}\quad a\in ]0,1], \quad \ell <0\cup \ell >\ell_{lim}^-,
\\&&\label{Eq:non-dip-altru}
  r_{\Ta}=r_{\Ta}^e
\equiv 2 \left(1-\frac{a}{\ell }\right), \quad
\mbox{and}\quad \dot{t}_{\Ta}= \dot{t}_{\Ta}^e\equiv \La \left(\frac{1}{\ell }-\frac{1}{a  }\right)=\Em \ell \left(\frac{1}{\ell }-\frac{1}{a  }\right)=-\frac{\La r_{\Ta}}{2 a}
\eea
--Figs\il(\ref{Fig:Plotrte}) and Figs\il(\ref{Fig:Plotrte4}).
As discussed  in Sec.\il(\ref{Sec:fishing-particles}), the greater is the magnitude of $\ell$ (the far is the torus from the attractor) and the closer to $r_{\epsilon}^+$ the turning point is.

For counter-rotating flows,
$r_{\Ta}(\ell_{mso}^+)$ is the outer turning corona radius, therefore for  $\sigma_\Ta=1$ there is  $r_{\Ta}^e\in[2M,r_{\Ta}(\ell_{mso}^+)[$ where  $r_{\Ta}(\ell_{mso}^+)$ is maximum for the extreme \textbf{BH}.
For $a=M$ the maximum extension (for the equatorial plane) is
\bea\label{Eq:max-cusp-eq-exte}
r_{\Ta}^e/M\in [2.41421,2.45455].
\eea
Notably there is $ r_{\Ta}^e/M<r_{\gamma}^+$--see Figs\il(\ref{Fig:Plotrte})

Furthermore, as clear from Eq.\il(\ref{Eq:non-dip-altru}), for  $\sigma =
 1 $, the radius  $r_ {\Ta} $  depends only on the ratio
$\ell_a\equiv \ell/a$ (see also \cite{pugtot,ella-correlation}). There are no extreme of $r_{\Ta}^e$, on the equatorial plane,  with respect to $a/M$ and  with respect to
$\ell_a$.

From the definition of  Carter constant $\Qa$, there is from  $\sigma_\Ta=1$ (see \cite{1976BAICz..27..129B})
\bea&&\label{Eq:want-y-appl}
\dot{\theta}_\Ta^2=\frac{\Qa}{r_\Ta^4}=\frac{\Qa\ell^{4}}{16\left(\ell-a\right)^4};\quad\mbox{and}\quad  \dot{\theta}_\Ta^2=0\quad\mbox{iff}\quad \Qa=0.
\eea
This means that, within the condition $\theta_\Ta=\pi/2$, the Carter constant $\Qa$ can be different from zero and strictly positive $\Qa>0$(necessary condition for so called orbital motion  \citep{1976BAICz..27..129B})-- Eq.\il(\ref{Eq:want-y-appl}).
On the other hand there is $\dot{\theta}_\Ta^2=0$ iff $\Qa=0$. Condition $\Qa=0$  is related to condition  $\theta_0=\pi/2$ on the initial toroidal  configuration.

However  the  radial velocity component for the flow reads
\bea\label{Eq:rradialeloc}&&
\dot{r}_\Ta=\pm \frac{\sqrt{\ell^{-2}\left[a \left(\ell ^2+4\right)-4 \ell \right] \left[\La^2 (a-\ell )^2 \left[a \left(\ell ^2+4\right)-4 \ell \right]-a \ell ^4 \left[ (\La^2+4\mu^2)\left(1-\frac{a}{\ell }\right)^2+\Qa\right]\right]}}{4\left(\ell- a\right)^2},
\\\nonumber
&& \dot{r}_\Ta=\pm \frac{ \sqrt{\ell^{-2}\left[4 \ell-a \left(\ell ^2+4\right) \right] \left[\La^2(\ell[12 a(a-\ell)+4\ell ^2]-4 a^3)+a \Qa \ell ^4\right]}}{4 (a-\ell )^2},
\eea
for particles and  photons respectively. The radial velocity depends explicitly on  $(\La,\Qa,\mu,\ell)$
(sign $\pm$, for the ingoing flow  $(-)$ or  outgoing  flow $(+)$, is not fixed).

\begin{figure}
\centering
    \includegraphics[width=7.5cm]{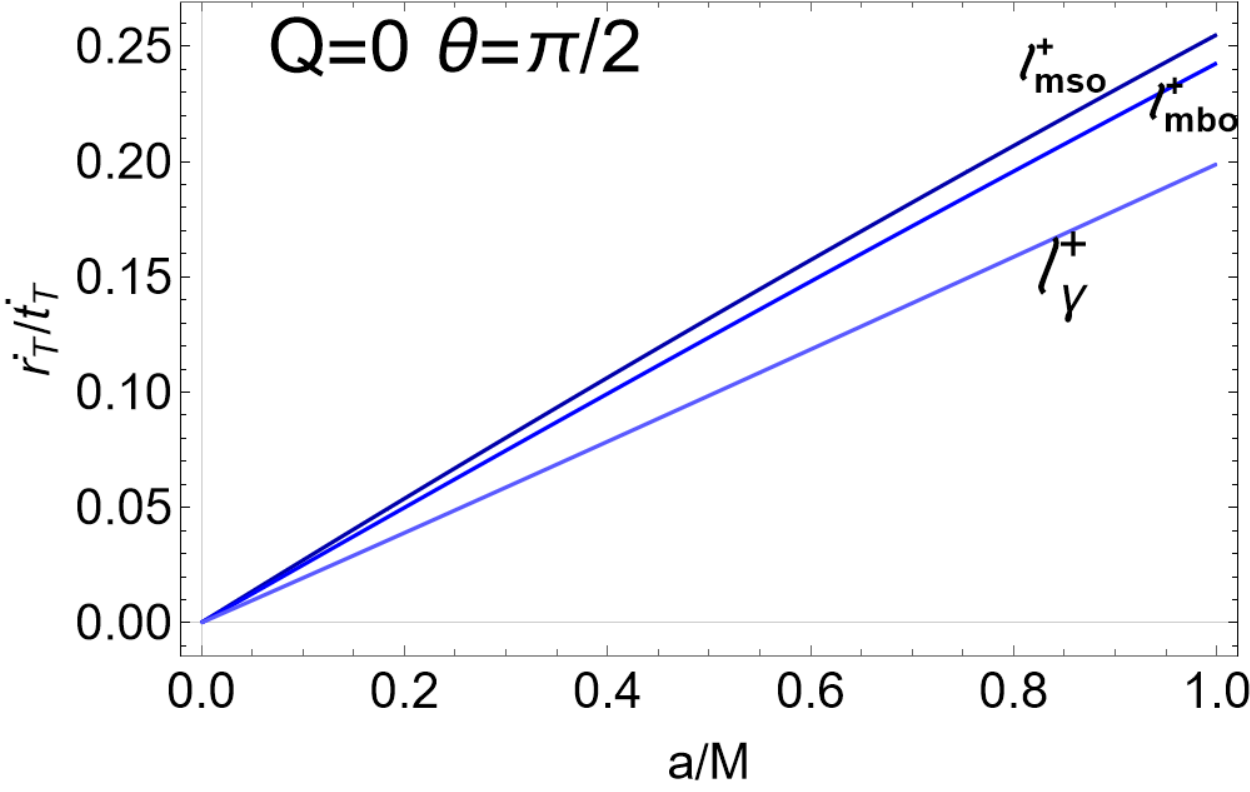}
  \caption{Relativistic photon radial velocity $\dot{r}_\Ta/\dot{t}_\Ta$  of Eq.\il(\ref{Eq:pick-twopiIsquare})  is plotted as functions of the   \textbf{BH} spin $a/M$. The radial velocity is evaluated  at the turning point  for  photons,   on the equatorial plane,  with Carter constant $\Qa=0$ and  for the specific angular momenta   $\ell_{mso}^+$, $\ell_{mbo}^+$ and $\ell_{\gamma}^+$ defined in   Eqs\il(\ref{Eq:def-nota-ell}) for  tori and  proto-jets driven  counter-rotating   flows. }\label{Fig:PlotlongVieanormali}
\end{figure}

From Eqs\il(\ref{Eq:rradialeloc}), for   $\Qa=0$ there is:
\bea\label{Eq:pick-twopiIsquare}
\dot{r}_{\Ta}=\pm \frac{\sqrt{-\frac{(a-\ell )^2 \left[a \left(\ell ^2+4\right)-4 \ell \right] \left[\La^2(\ell -a)+a \mu^2 \ell ^2\right]}{\ell ^2}}}{2 (a-\ell )^2},\quad  \dot{r}_{\Ta}=\pm \frac{ \sqrt{\frac{\La^2 (a-\ell )^3 \left[a \left(\ell ^2+4\right)-4 \ell \right]}{\ell ^2}}}{2 (a-\ell )^2}.
\eea
for particles and  photons respectively.
Note that the  radial velocity   does not depend on the impact parameter only, but depends explicitly also on $\La$. The photonic $(\mu=0)$ relativistic velocity $\dot{r}_\Ta/\dot{t}_\Ta$ at the turning point, for $\Qa=0$,  depends on $\ell$ only (there is $\ell \La>0$), and  this case is shown in {Fig.\il(\ref{Fig:PlotlongVieanormali})} for $\ell$ in the range bounded by the liming momenta  $\ell^+_{mso}>\ell_{mbo}^+>\ell_{\gamma}^+$ for counter-rotating tori and proto-jets driven photons --Figs\il(\ref{Fig:Plotexperthousandp}).
The  photons relativistic radial velocity depends on  the impact parameter $\ell$ inherited from the toroidal initial configurations,   increasing in magnitude  with the \textbf{BH} spin  and decreasing  with the increase of $\ell$ in magnitude, being therefore  greater (in magnitude) for the tori-driven flows with respect to the proto-jets driven flows. On the other hand, the range of values for the relativistic radial velocity  is larger for proto-jet driven flows, and increases with the \textbf{BH} spin $a/M$, distinguishing  photons  from proto-jets and tori driven flows,  and narrowing the   photon  component radial velocities at the turning point in the  tori driven  counter-rotating flows.
\subsubsection{Conditions on the counter-rotating flows with Carter constant $\Qa=0$}\label{Sec:basisCQu0}
From
Eq.\il(\ref{Eq:eich}),
it is clear that values of
$\Qa$ are limited by the  constants of motion $(\mu, \Em, \La)$, differing  explicitly for photons and matter, when  $\theta\neq\pi/2$ \emph{or}  $(\Em^2 - \mu^2) \neq \left(\frac{\La}{ a\sqrt{\sigma}}\right)^2$. The Carter constant is not restricted by
the \textbf{BH} spin $a/M$  on the equatorial plane $\theta=\pi/2$, \emph{or for} $\mu^2=\Em^2$, \emph{or for}
$(\Em^2 - \mu^2) = \left(\frac{\La}{ a\sqrt{\sigma}}\right)^2$. (The second condition on the particle energy is related to the limiting conditions distinguishing proto--jets and tori driven flows. This condition and the third relation is briefly discussed below.)

 According to   Eq.\il(\ref{Eq:want-y-appl}) a zero Carter constant  implies
 \bea
\Qa=0:\quad (\dot{\theta})^2= (\dot{\theta}_\zeta)^2 \equiv\frac{(\sigma -1) [a^2 \sigma  (\mu^2-\Em^2)+\La^2]}{\sigma  \Sigma^2}.
 \eea
At the  turning point, where  $r=r_\Ta$, on a general plane  $\sigma_\Ta\in [0,1]$, there is
\bea
\Qa=0:\quad(\dot{\theta}_\Ta)^2= \frac{(\sigma_{\Ta} -1) \left[a^2 \sigma_{\Ta}  \left(\mu^2-\frac{\La}{\ell }\right)+\La^2\right)}{\sigma_{\Ta}  \left[\frac{\left(\sqrt{a^2 \sigma_{\Ta} ^2+\ell ^2 \left[a^2 (\sigma_{\Ta} -1)+1\right]-2 a \sigma_{\Ta}  \ell }+a \sigma_{\Ta} -\ell \right)^2}{\ell ^2}-a^2 \sigma_{\Ta} +a^2\right]^2}.
\eea
If the turning point is on the equatorial plane (and $\Qa=0$) then there is, according to Eq.\il(\ref{Eq:want-y-appl}), $\dot{\sigma}_\Ta=0$.
(Only the equation for the radial velocity $u^r$ depends explicitly  on $\Qa$).

Let us consider explicitly   the condition $\dot{\theta}=0$:
 \bea&&\label{Eq:generali}
\mbox{it holds for}\quad  \Qa = (1 - \sigma)\left [a^2 (\mu^2 - \Em^2) + \frac{\La^2}{\sigma} \right],
\eea
and there is
\bea
\label{Eq:positiv-negativesecond-condit}
 &&
 \Qa=0,\quad \mbox{for}\quad
  \sigma =
   1  \quad \mbox{\emph{or}}\quad  (\Em^2 - \mu^2) =\left(\frac{\La}{ a\sqrt{\sigma}}\right)^2,
   \eea
  this condition distinguishes photons ($\mu=0$) and matter ($\mu>0)$, and accretion driven ($\Em<\mu$) from proto-jets driven ($\Em>\mu$) flows.
The  condition of  Eq.\il(\ref{Eq:generali}),   implies
\bea&&\label{Eq:post-stor}
\dot{\theta}=0,\quad \dot{\phi}=\frac{\La [2 a+(r-2) \ell]}{r \ell  \Delta}
\quad \dot{r}=\pm \frac{\sqrt{r \left(\frac{\La^2 \left[a^2 (r+2)-4 a \ell +r^3-(r-2) \ell ^2\right]}{\ell ^2}-\mu^2 r \Delta\right)}}{r^2},\quad\dot{t}=\frac{\La \left[a^2 (r+2)-2 a \ell +r^3\right]}{r \ell  \Delta};
\eea
where, at the turning point, there is in particular
\bea
&&
\dot{\theta}_\Ta=0,\quad \dot{\phi}_\Ta=0,\quad\dot{r}_\Ta=\pm \frac{ \sqrt{\frac{(a-\ell )^2 \left[4\ell-a \left(\ell ^2+4\right)\right] \left[a(\mu^2 \ell ^2-\La^2)+   \La^2 \ell \right]}{\ell ^2}}}{2 (a-\ell )^2},\quad \dot{t}_\Ta=\La \left(\frac{1}{\ell }-\frac{1}{a}\right).
\eea
Nevertheless  the second condition on Eq.\il(\ref{Eq:positiv-negativesecond-condit}) constrains the  tori    with  the conditions
   \bea
      \left(\frac{\Em}{a \sqrt{\sigma}}\right)^2= \frac{ \mu^2 }{a^2 \sigma -\ell^2},\quad\mbox{and}\quad \left(\frac{\La}{a \sqrt{\sigma}}\right)^2= \frac{\mu^2 \ell^2}{a^2 \sigma -\ell^2},
   \eea
but condition  $a^2 \sigma -\ell ^2\geq0$  does not hold for the tori considered in this model (where $\ell<\ell_{mso}^+$)--see also Eqs\il(\ref{Eq:itafra-regu},\ref{Eq:tdotemLem}).
 On the other hand, the first condition of Eq.\il(\ref{Eq:positiv-negativesecond-condit})  implies
\bea
 \Qa=0,\quad \mbox{for}\quad
  \sigma = 1  \quad\mbox{and}\quad \dot{\theta} = 0
\eea
and therefore reduces to  Eq.\il(\ref{Eq:post-stor}). If instead there is  $\sigma =
       1$ but  $\dot{\theta}\neq 0$  then $\Qa>0$.
        If viceversa  there is $\sigma =
       1$ then there is $\Qa=0$  only if $\dot{\theta} = 0$.

We summarize  as follows: for matter ($\mu>0$) there is
 $\Qa=0$ if  $\theta_0=\pi/2$ \textit{and} $\dot{\theta}_0=0$ which can be the initial condition on the flow  or   at  the turning point $r_\Ta$.

If the initial data on the flow trajectory  are on the equatorial plane,   the flow  has initial non--zero poloidal velocity only if  $\Qa>0$ (see Sec.\il(\ref{Sec:Carter-sign-discussion}) for a discussion on the Carter constant sign).
If the turning point is on the equatorial plane  then  the  poloidal velocity can also be non--zero, meaning that   the flow can cross (vertically) the equatorial plane\footnote{It is worth noting   that the initial conditions on the flow are substantially dependent only on the  conditions on the specific angular momentum, constrained by the limits provided through  the  background geodesic structures and the data on the inner edge location of Eq.\il(\ref{Eq:toricenter-inner}), providing eventually an  upper and lower bound to the turning point. Therefore results discussed here are partly applicable to the case of different initial conditions on the fluids,  and  may be relevant also   for the case of tori misalignment. }.

\textbf{{General conditions on the  Carter constant for the counter-rotating flow }}\label{Sec:Carter-sign-discussion}
Using Eq.\il(\ref{Eq:eich})
there is,  for photons and  particles ($\mu^2\geq 0$)  with $a\neq0$
\bea&&\label{Eq:positiveQ-negaitvecont}
\Qa<0 \quad\mbox{for}\quad \sigma\in ]0,1[,\quad \Em^2>\mu^2,\quad  \left(\frac{\La}{a\sqrt{\sigma}}\right)^2< (\Em^2-\mu^2),\quad\mbox{and}\quad (\dot{\theta})^2\in [0, (\dot{\theta}_{\zeta})^2[
\eea
(excluding the poles $\sigma=0$ and the equatorial plane $\sigma=1$). The condition (\ref{Eq:positiveQ-negaitvecont}) on the energy  describes  proto-jets driven flows (where $K_\times>1$). Notably these conditions are independent from the corotation or counter-rotation of the flow.
It should be noted  that  $\Qa\nless0$, where $ (\dot{\theta}_\zeta)^2 =0$,  and    $
\dot{\theta}^2>0$
is  always verified on the equatorial plane for  $r\geq r_{\epsilon}^+$.
 \subsection{Flow from the equatorial plane ($\sigma_0=1$) and general considerations  on initial configurations}\label{Sec:point-actua}
  Here we consider counter--rotating flows  emitted from the equatorial plane,  assuming therefore  $\sigma_0=1$,  with Eqs\il(\ref{Eq:general-on-equatorial-plane}) as initial data for  the  accreting flows\footnote{For a perturbed initial condition on the tori driven  flows we could  consider a non--zero   (small) component of the  initial poloidal velocity.  Then  $\dot{\theta}_0\neq1$  implies that in no following point of the trajectories,  and particularly  at the turning point,  there is $\theta=\pi/2$ and $\dot{\theta}=0$ (as   $\Qa>0$).}.

Condition  $\dot{r}=0$ defines  the fluid   effective potential.
{As proved in Eq.\il(\ref{Eq:positiveQ-negaitvecont}),  the Carter constant must be positive or zero  on  the equatorial plane.}
We  discuss below  the  conditions where $\Qa\geq0$ with  $\ell<0$ and $\dot{r}=0$ (according to effective potential definition), introducing  the following energy function  and   limiting momenta
\bea&&
\Em_g\equiv \sqrt{\frac{\Delta \left(\mu^2 r^2+\Qa\right)}{r \left[a^2 (r+2)-4 a \ell +r^3+(2-r) \ell ^2\right]}},\quad
\ell_g^\pm\equiv\frac{2 a \pm r	 \sqrt{\Delta}}{2-r}.
\eea
For counter-rotating fluids ($\ell<0$) considering $\Em>0$ with $r>2M$ (corresponding to tori or proto-jets on the equatorial plane), there is  $\dot{r}_0=0$ in the following cases:
\bea&&\mbox{for}\quad \Qa>0, \quad \mu\geq 0,\;a\in[0,1],\; r>2,\;\ell\in] \ell_g^+,0[,\; \Em=\Em_g
\\&&
\mbox{for }\quad \Qa=0:\quad
(\mu=0, a\in [0,1], r>2,\; \ell =\ell_g^+,\; \Em>0)\quad\mbox{and}
\\
&& \qquad \qquad \qquad (\mu>0,\; a\in [0,1],\; r>2,\;\ell\in]\ell_g^+,0[,\; \Em=\Em_g),
\eea
where we distinguished   photons and matter  in   the counter-rotating flows. (Note these conditions  have been found from  the conditions on  $\dot{r}^2=0$ for $\theta=\pi/2$, therefore, although framed in the set of tori initial data considered here,  they can hold also in other points of the flow trajectories, therefore  notation $(0)$,  referring to the initial point $r=r_0$, has not been emphasized\footnote{For completeness we report  also the   case $\ell>0$ and $\Em>0$ where we consider   $r>r_+$ (as the corotating torus can also be located in the ergoregion).
For $ \ell>0$ and  $ \Qa>0$, for matter and photons ($\mu\geq0$) with $\dot{r}_0=0$,  solution is for $\Em=\Em_g$ in the following cases
 \bea
  &&
\mbox{for}\quad a=0:\; r>2,\; \ell\in]0,\ell_g^-[; \\
&&\mbox{for}\quad  a\in ]0,1]:  \left(r\in]r_+,2[,\; \ell\in ]0,\ell_g^+[\cup\ell> \ell_g^-\right);
(r\geq 2,\; \ell\in]0,\ell_g^-[),
 \eea
 distinguishing  the Schwarzschild  $(a=0)$ and the  Kerr background $(a>0)$.
For zero Carter constant, and  $(\ell>0, \Qa=0)$, there is    $\dot{r}=0$ with  $\Em>0$  and  $\ell =\ell_g^\pm$.
More specifically, for photons ($\mu=0$)  $\ell =\ell_g^\pm$
 \bea
 &&\mbox{for}\quad a=0:\; r>2,\ell=\ell_g^-;\quad\mbox{for}\quad
 a\in ]0,1]:\;\left(r\in]r_+,2[,\; \ell =\ell_g^\pm\right);\quad (r=2,\ell=\ell_g^-);\quad (r>2,\ell=\ell_g^+).
 \eea
 For matter ($\mu>0$) in the static spacetime there is
\bea && a=0:\; r>2,\; \ell\in]0,\ell_g^-[,\; \Em=\Em_g.
\eea
In the Kerr spacetime, there is the  solution $\Em=\Em_g$ in the following cases
\bea
&& a\in ]0,1]:\; \left(r\in]r_+,2[,\;
 (\ell\in]0,\ell_g^+[\cup\ell >\ell_g^-)\right);\;
\left(r=2,\; \ell\in\left]0,\frac{a^2+2}{a}\right[,\;
\right);\quad  (r>2,\; \ell\in]0,\ell_g^-[),
 \eea
 distinguishing the ergoregion $]r_+,r_{\epsilon}^+[$ and the region $r>r_\epsilon^+$.}). Clearly, the limiting conditions on the fluid momenta have  to be  combined with the conditions on the  tori momenta  $\ell=\ell^{\pm}$  discussed in Sec.\il(\ref{Sec:tori-models}).

Finally, note that if  initially
 there is   $\dot{\theta}_0=0$  then   Eq.\il(\ref{Eq:generali}) holds, and   the motion  can also be on planes different from the   equatorial plane, with non--zero  Carter constant.
 \section{Turning points of the  counter-rotating proto-jet  driven flows }\label{Sec:proto-jets-driven}
Proto-jet driven flows  are characterized by a high centrifugal component of the fluids force balance with   $\ell\in ]\ell_{\gamma}^+,\ell_{mbo}^+]$ for counter-rotating flows and  $\ell\in ]\ell_{mbo}^-,\ell_{\gamma}^-]$ for co-rotating flows--see Sec.\il(\ref{Sec:tori-models}). The high centrifugal component can lead to a destabilization of the fluid equilibrium (according  to the P-W instability  mechanism) leading to the formation of a matter cusp with parameter value   $K=K_\times>1$,  corresponding to open boundary conditions at infinity (i.e. at the corresponding  outer edge of the toroidal  configurations) with  matter funnels along the \textbf{BH} rotational axis--see   Figs\il(\ref{Fig:PlotlongVie})\footnote{As shown in  Figs\il(\ref{Fig:PlotlongVie}),  open toroidal  configurations  are also  obtained within  different conditions on the fluid specific angular momentum $\ell$  and  energy $K$, with  matter funnels  from the inner Roche lobe of  quiescent tori,  or with momenta lower then minimum $\mp\ell_{mso}^\pm$,  and even for  very low $K$ --\citep{pugtot,proto-jets,open,long,ella-jet,sadowtree}.}.

Here we consider counter-rotating  proto-jets  from the  cusp ("launch" point associated to a minimum of the hydrostatic pressure)  on the \textbf{BH} equatorial plane. According to Sec.\il(\ref{Sec:tori-models}) there is  $r_\times\in [r_{\gamma}^+, r_{mbo}^+]$ and  $r_{center}^+\in[r_{(mbo)}^+, r_{(\gamma)}^+]$ respectively--Figs\il(\ref{Fig:Plotbalchocasep}).

\begin{figure*}
\centering
    \includegraphics[width=7.5cm]{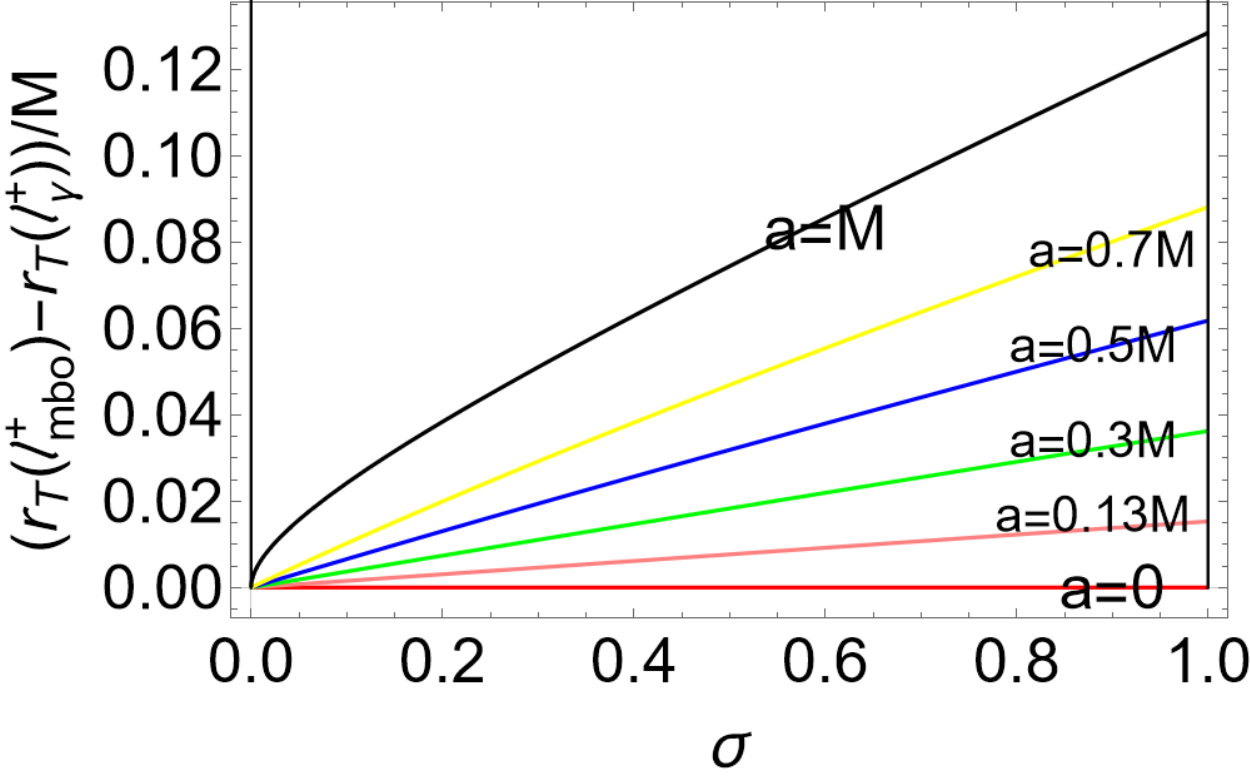}
  \includegraphics[width=7.5cm]{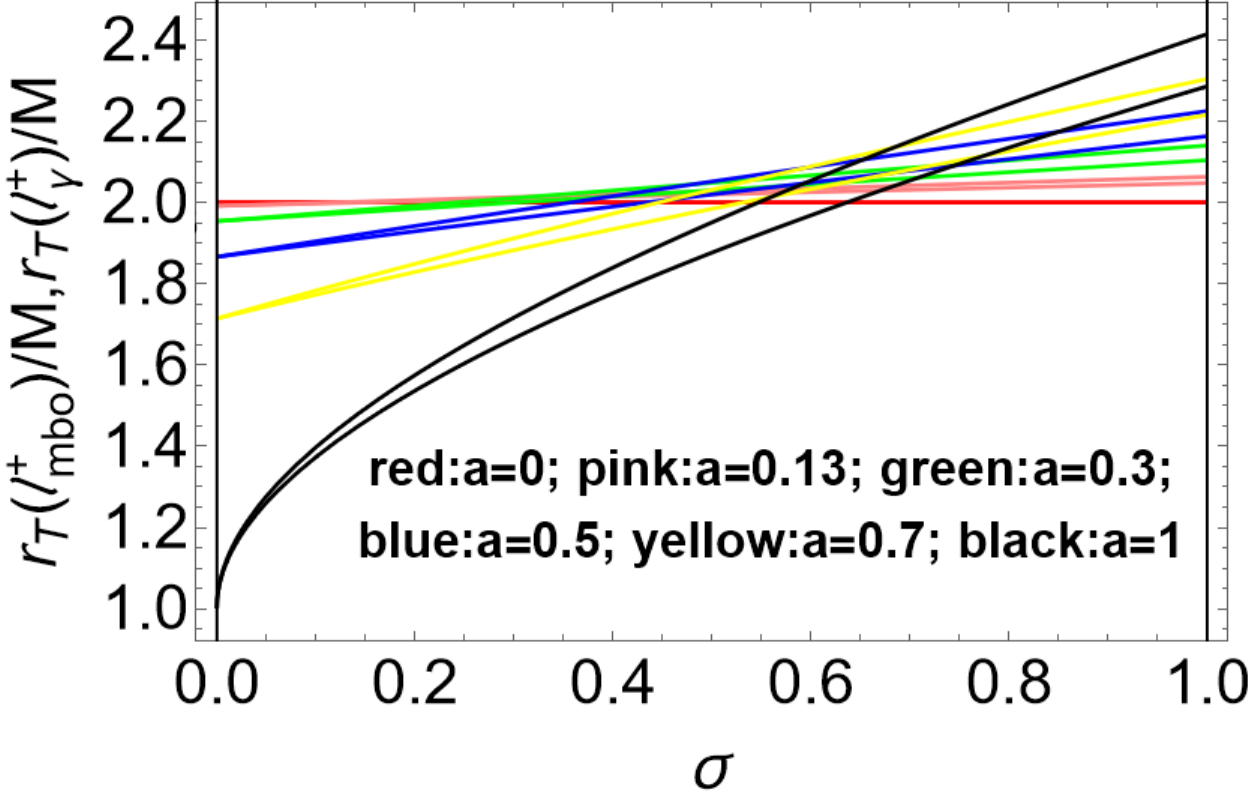}
   \includegraphics[width=7.5cm]{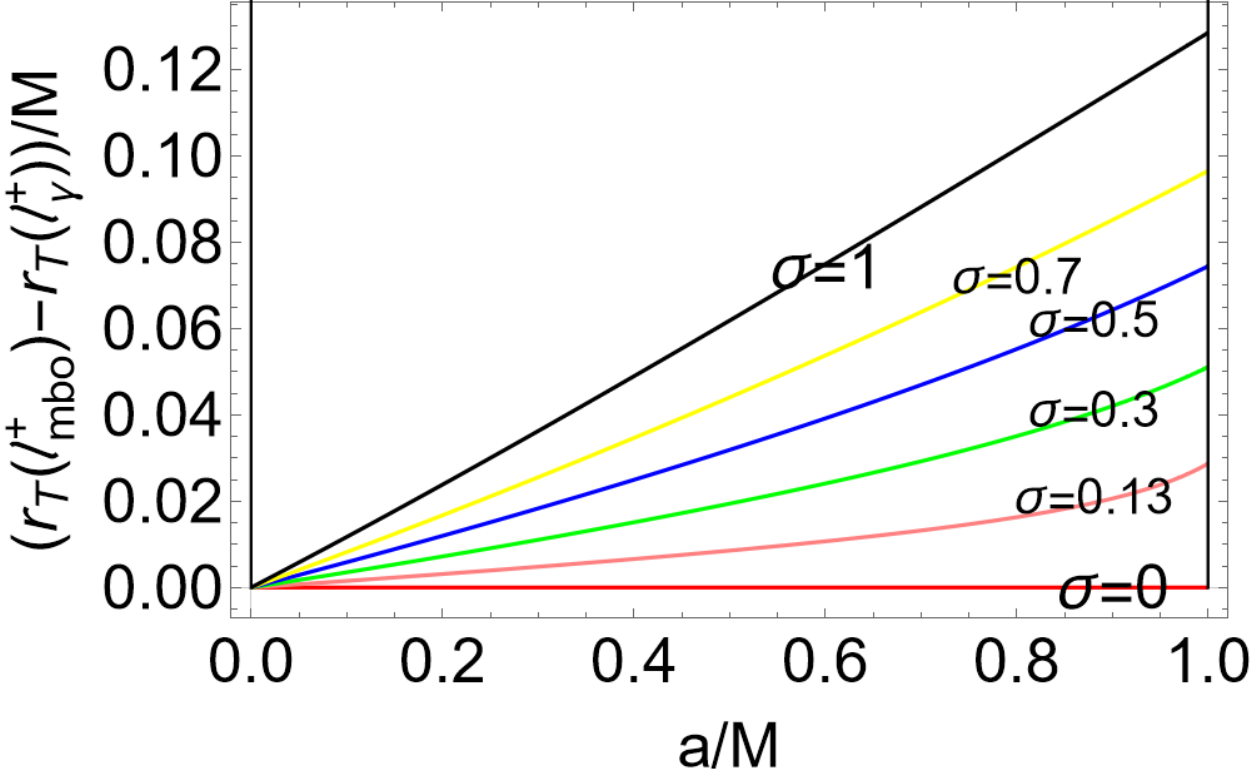}
    \includegraphics[width=7.5cm]{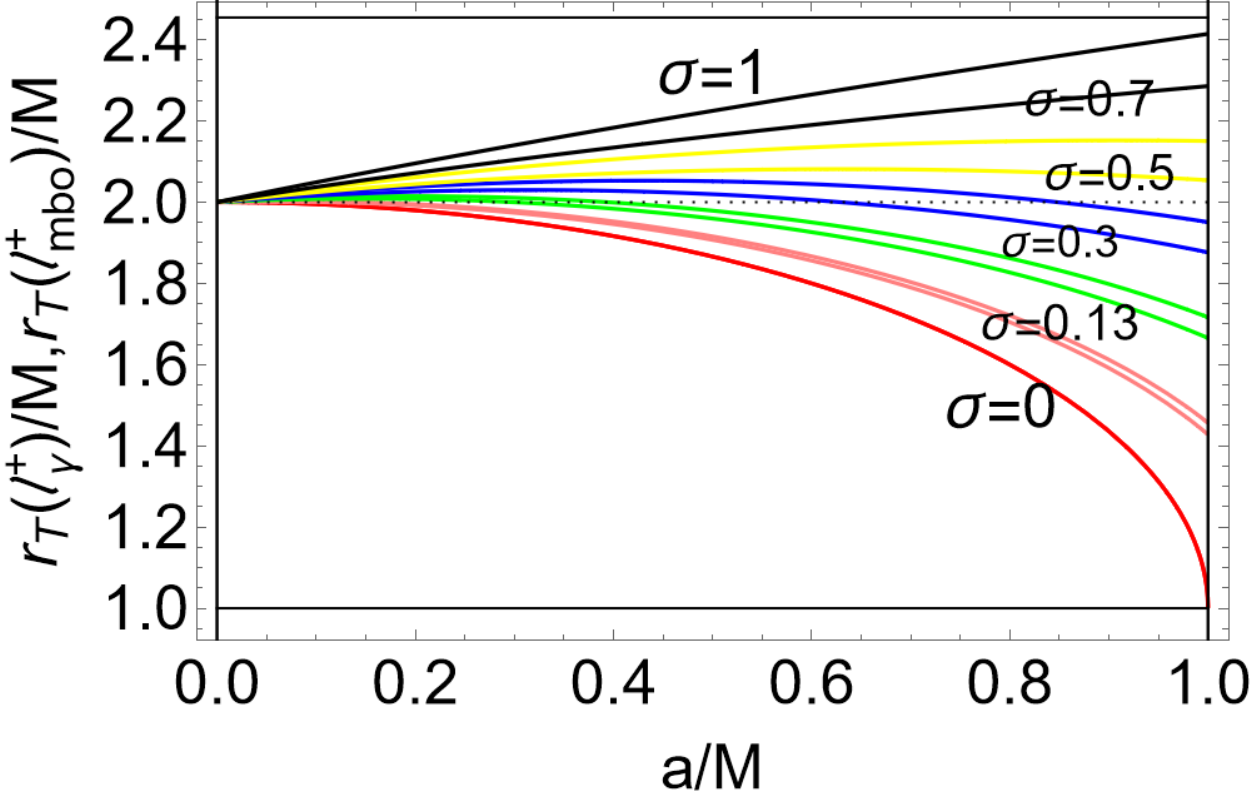}
  \caption{Proto-jets counter-rotating flow turning radius $r_{\Ta}$ evaluated for fluid specific angular momentum $\ell_{mbo}^+$ and $\ell_{\gamma}^+$ of  Eqs\il(\ref{Eq:def-nota-ell})). The turning point  corona  radius  $r_{\Ta}(\ell_{mbo}^+)-r_{\Ta}(\ell_{\gamma}^+)$ is shown as function of the plane $\sigma\equiv \sin^2\theta$  for different \textbf{BH} spin $a/M$ signed on the curve  in the upper--left panel  and as function of the \textbf{BH} spin  $a/M$ for different planes $\sigma$ signed on the curves in the bottom--left panel.  Radii  $r_{\Ta}(\ell_{mbo}^+)>r_{\Ta}(\ell_{\gamma}^+)$ are shown   in the upper--right  panel  as functions of the planes $\sigma$,  for different \textbf{BH} spin $a/M$ and as functions of the \textbf{BH} spin  $a/M$ for different planes $\sigma$  in the bottom--right  panel. The corresponding  analysis for the cusped tori counter-rotating driven flows  is realized  in Figs\il(\ref{Fig:Plotrtea}).}\label{Fig:Plotrteag}
\end{figure*}

We explore the possibility that the counter-rotation fluid feeds a jet with  initial  flow direction   $\ell<0$, investigating  the existence of a  counter-rotating flow turning point particularly at  a plane $\sigma_\Ta<1$, representing a more articulated vertical  structure of the  proto-jets flow (along the axis of the central \textbf{BH}). {(For $\Em>0$ and $\La<0$, condition $u^\phi<0$ holds, independently on the normalization condition, only for $r>r_\epsilon^+$.)} We should distinguish, at $\sigma_\Ta\neq1$, the ingoing flows, defined by $\dot{r}_\Ta<0$, from outgoing flows, defined by $\dot{r}_\Ta>0$.
Figs\il(\ref{Fig:Plotrteag})  show how the situation is  similar to the accretion driven flows (see Figs\il(\ref{Fig:Plotrtea})) and therefore  the turning  point  is bounded according to  a turning circular  corona, defined by the momenta $-\ell^+\in [-\ell_{mbo}^+,-\ell_\gamma^+]$, whose extension  for  the proto-jets driven flows is smaller then of the tori driven corona, closer to the ergosurface and  contained in the   turning corona for tori driven flows. Similarly to the tori driven turning  corona, the corona for proto-jet driven flows  is rather  small, expecting therefore a   fluid turning  point with a centralization of matter and photons  in a very narrow orbital region in planes $ \sigma_\Ta \in [0,1] $ and   regulated by the time components $ {t}_\Ta$ and $\tau_\Ta$ evaluated for  the two limiting momenta $(\ell_{mbo}^+,\ell_\gamma^+)$--see Figs\il(\ref{Fig:Plotrte4}).  For  $\ell=\ell_{\gamma}^+$,   the turning radius is  very close  to the ergoregion.  From  Figs\il(\ref{Fig:Plotrte}),   it can be seen that on the equatorial plane the turning radius  is smaller then the turning radius for  $\ell_{mbo}^+$ but larger then $r_\epsilon^+$ and the distance increases with the spin. We note also that, despite the  proto-jets and tori driven flow coronas are close, the proto-jets configurations are not related  to the cusped tori as  there is $\ell\in\mathbf{ L_1}$ for cusped tori and $\ell\in\mathbf{L_2}$ for proto-jets. Furthermore these limits hold for particles and photons (note that  these  results do not depend explicitly on $K$ or on the normalization condition on the particles flow).
Results of this analysis are shown in Figs\il(\ref{Fig:Plotrteb},\ref{Fig:Plotrtealg},\ref{Fig:Plotrte}).
 In  Figs\il(\ref{Fig:Plotrteb}) there is the analysis in dependence on  the plane $\sigma$. Decreasing  $\sigma$ (with respect to the reference  critical value  $\sigma_{crit}$) the situation for proto-jets driven flows  is  different from the accretion driven flows.  For smaller \textbf{BH} spins  $a/M$,
 the turning point is more depended on the \textbf{BH} spin, distinguishing  turning points located closer  to the \textbf{BH} axis,  $\sigma<\sigma_{crit}$, or on  the equatorial plane ($\sigma=1$).
In Figs\il(\ref{Fig:Plotrteag}) we can compare  the counter-rotating  proto-jets driven  flows corona  with  Figs\il(\ref{Fig:Plotrtea}) for the counter-rotating  cusped tori driven  flows, completing  the analysis  of  Figs\il(\ref{Fig:Plotrte4}).
Radius $r_{\Ta}(\ell_{\gamma}^+)$ is the closest to the ergosurface, making the proto-jets driven corona more internal, i.e. closer to the ergosurface, than the accretion disks driven corona, with a larger   spacing between the corona radii and a stronger variation with   plane $\sigma_\Ta$ (Figs\il(\ref{Fig:Plotrteag})) and the  \textbf{BH}   spin (Figs\il(\ref{Fig:Plotrteb})).
The proto-jet driven turning point corona is  easily  distinguishable from  the  tori driven turning point corona, being located in two separated  orbital regions.
(It should be noted that Eq.\il(\ref{Eq:protre-spess-comunci}) is sufficient to assure that for  very small $\sigma$ (i.e. close to the  \textbf{BHs} poles) $r_\Ta$  closes on the horizon (in the adopted coordinate frame), as evident  also from Figs\il(\ref{Fig:Plotrte4}).
It is clear that if $\sigma_\Ta\approx0$, then quantity   $\dot{t}_\Ta\to +\infty$, as in Eqs\il(\ref{Eq:tdotemLem}), but $\sigma_\Ta$  and  $r_\Ta$  are bounded  as  $r_\Ta>r_\epsilon^+$.).
 \section{Verticality of the  counter-rotating flow turning point}\label{Sec:vertical-z}
Consider the vertical flat coordinate $z\equiv r\sqrt{(1-\sigma)}$. Using the coordinate  $z_\Ta=r_\Ta\sqrt{(1-\sigma_\Ta)}$ for the turning point,  there is
\bea
z_\Ta=r_\Ta \sqrt{\frac{r_\Ta [2 a+(r_\Ta-2) \ell]}{a (2 r_\Ta-a \ell )}}= \frac{\sqrt{1-\sigma_\Ta } \left[\ell -a \sigma_\Ta-\sqrt{a^2 \sigma_\Ta ^2+\ell ^2 \left[a^2 (\sigma_\Ta -1)+1\right]-2 a \sigma_\Ta  \ell } \right]}{\ell }
\eea
and, for very large $\ell$  in magnitude, $z_\Ta$ tends   to the ergosurface $z_\epsilon^+$ in agreement with Eq.\il(\ref{Eq:protre-spess-comunci}).

In agreement with the analysis of  Sec.\il(\ref{Sec:extreme-turning-box}),  $z_\Ta$ has no extreme  as function of     $\ell$, but the vertical coordinate $z_\Ta$ decreases with magnitude of $\ell$ for  proto-jet and tori  driven counter-rotating  flows.
Tori corresponding to very large $(-\ell^+)$ are  located far from the attractor (the far  the faster spinning is the central \textbf{BH}), and tend to be large and stabilized against the P-W instability, with a consequent regular topology (absence of a torus cusp).

\begin{figure*}
\centering
  \includegraphics[width=6.5cm]{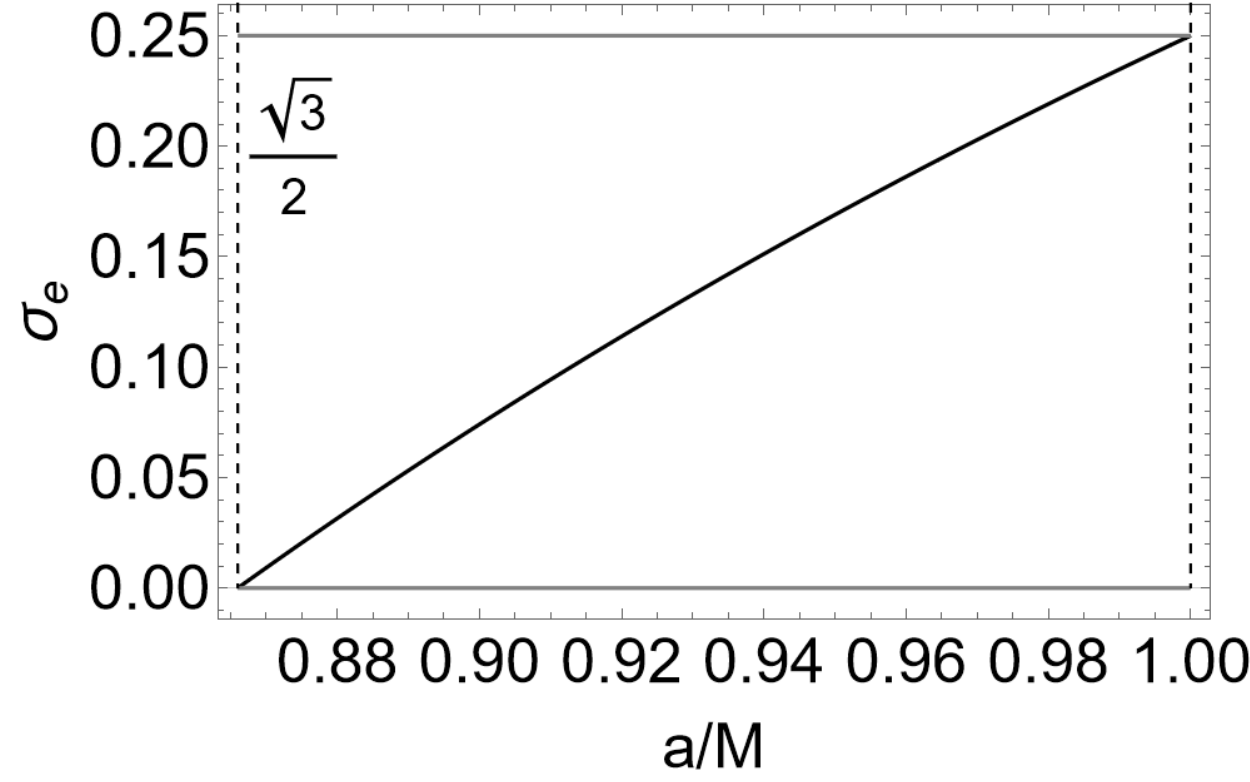}
 \includegraphics[width=6.5cm]{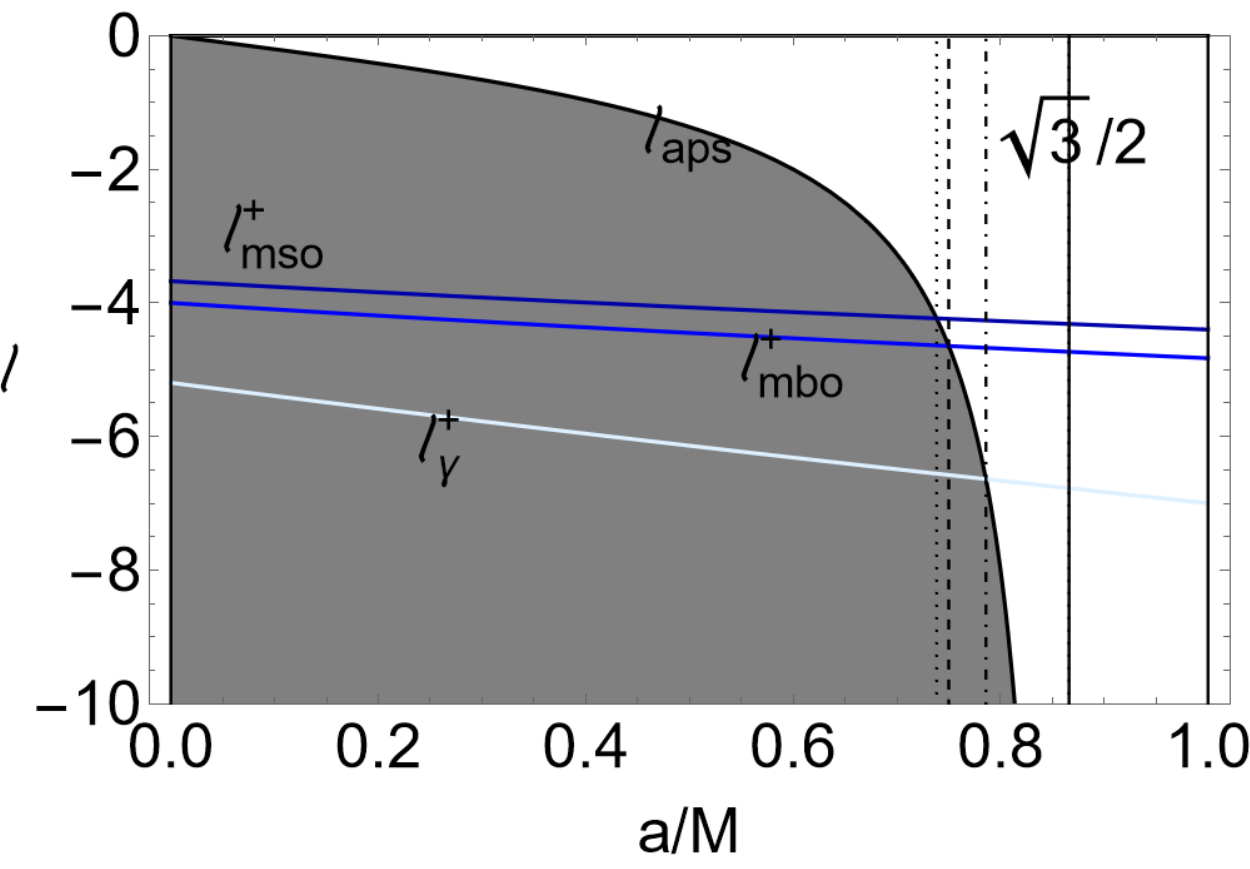}
   \includegraphics[width=6.5cm]{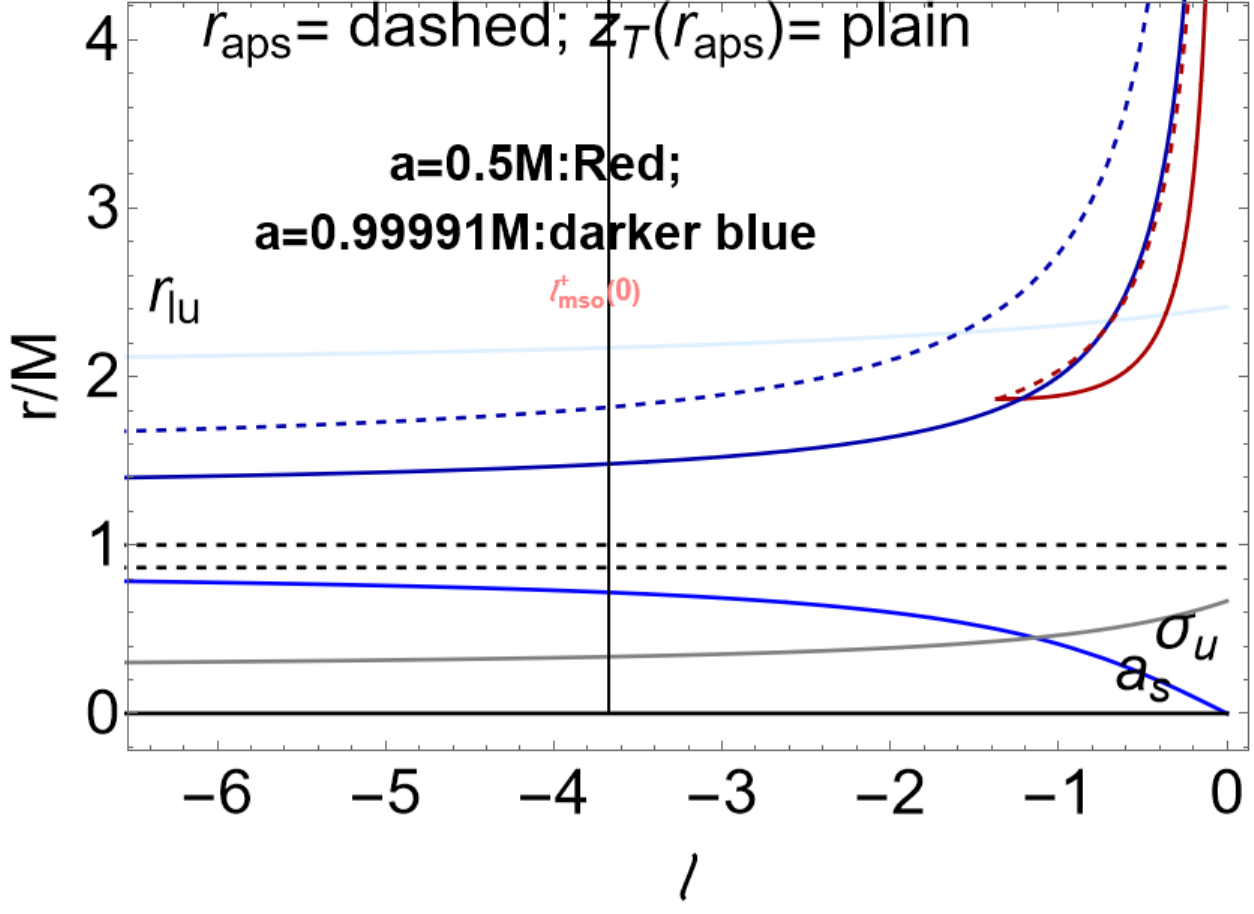}
      \includegraphics[width=6.5cm]{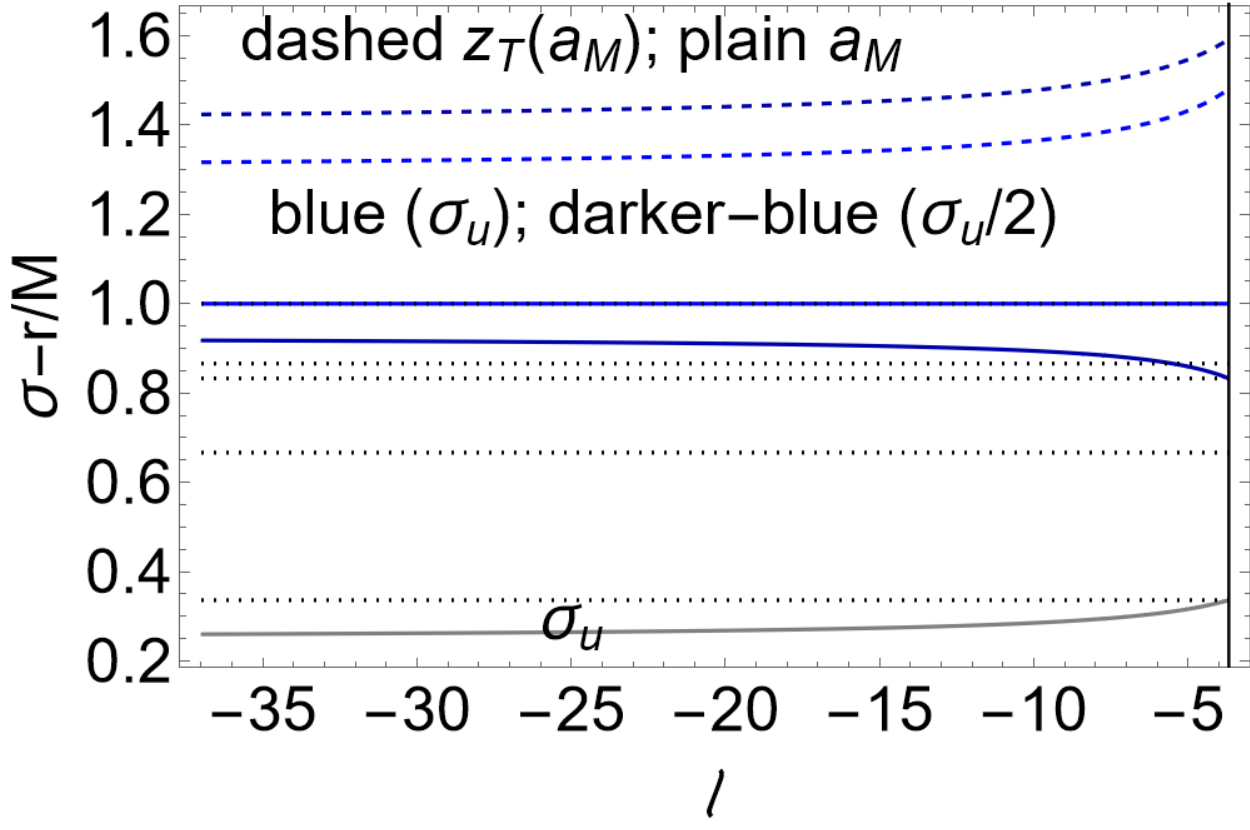}
       \caption{Analysis of the  maximum vertical position $z_\Ta$ of the counter-rotating $(\ell<0)$ flow turning point $r_\Ta$ of Sec.\il(\ref{Sec:vertical-z}). There is $z_\Ta=r_\Ta\sqrt{(1-\sigma_\Ta)}$, where $\sigma\equiv \sin^2\theta$ and $\ell$ is the fluid specific angular momentum.  Upper left panel: Limiting plane $\sigma_e$ of Eq.\il(\ref{Eq:sigmausigmae}) as function of the central \textbf{BH} spin-mass ratio  $a/M$.
       Limiting spin $a/M=\sqrt{3}/2$ is shown as dotted line. According to Eq.\il(\ref{Eq:particla-zn}),  plane $\sigma_e$  constrains the solutions $\partial_{\sigma_\Ta}z_\Ta=0$.  Upper right panel: limiting momentum $\ell_{aps}$ of Eq.\il(\ref{Eq:def-lupsetal}) is plotted as function of the \textbf{BH} spin-mass ratio $a/M$.
        Momentum  $\ell_{aps}$ regulates the  analysis of Eqs\il(\ref{Eqs:sof-A6po8}) for the solutions of $ \partial_r z_\Ta=0$.  Limiting momenta  $\ell_{mso}^+$, $\ell_{mbo}^+$ and $\ell_\gamma^+$, defined in  Eqs\il(\ref{Eq:def-nota-ell}), are also shown. Dotted lines are the limiting conditions for flow, with  the specific angular momentum     $\ell_{aps}=\{\ell_{mso}^+,\ell_{mbo}^+,\ell_\gamma^+\}$ respectively. Bottom left panel: Limiting plane $\sigma_u$ of Eq.\il(\ref{Eq:sigmausigmae}), radius $r_{aps}$ of Eq.\il(\ref{Eq:raps}) with vertical coordinate $z_{\Ta}(r_{aps})$, limiting spin $a_s$ of
Eq.\il(\ref{Eq:as}) and limiting radius $r_{lu}$ of Eqs\il(\ref{Eq:rplusplusulp}) as functions of fluid specific momenta $\ell$. The situation for the \textbf{BH} spin $a=0.5$ (red curve) and $a=0.99991M$ (darker blue curve) are shown. Radius $r_{lu}$ is a limiting function in the analysis of Eq.\il(\ref{Eq:anali-zza}) for the solutions of $
 \partial_a z_\Ta=0$; spin $
 a_s$  is a limiting function in the analysis of  Eqs\il(\ref{Eqs:sof-A6po8a}) for the solutions $
 \partial_r z_\Ta=0$, while  $
 \sigma_u$ is a limiting plane in the analysis of Eq.\il(\ref{Eq:particla-za}) for the solutions of $
 \partial_{\sigma_\Ta}z_\Ta=0$.
Bottom right panel shows the   limiting plane  $\sigma_u$ and
 solution $a=a_M$ of Eqs\il(\ref{Eq:aMdef}) (plain colored curves) in Eq.\il(\ref{Eq:particla-za}) for  solutions of $
 \partial_{\sigma_\Ta}z_\Ta=0$. Vertical coordinate $z_\Ta(a_M)$ (dashed-curves) for $\sigma=\sigma_u$ (blue curves) and $\sigma=\sigma_u/2$ (darker blue curve) are  also shown.}\label{Fig:Plotrtesam}
\end{figure*}

\begin{figure*}
\centering
\includegraphics[width=6.5cm]{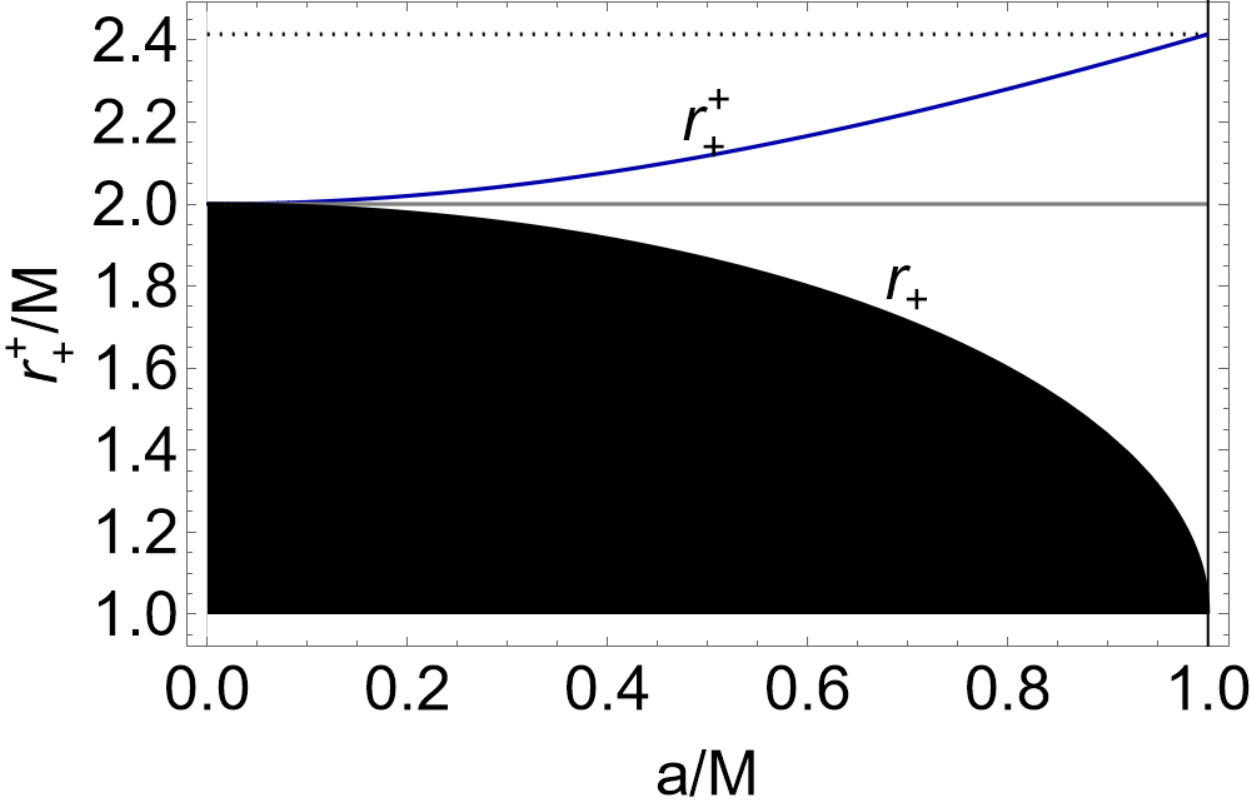}
 \includegraphics[width=6.5cm]{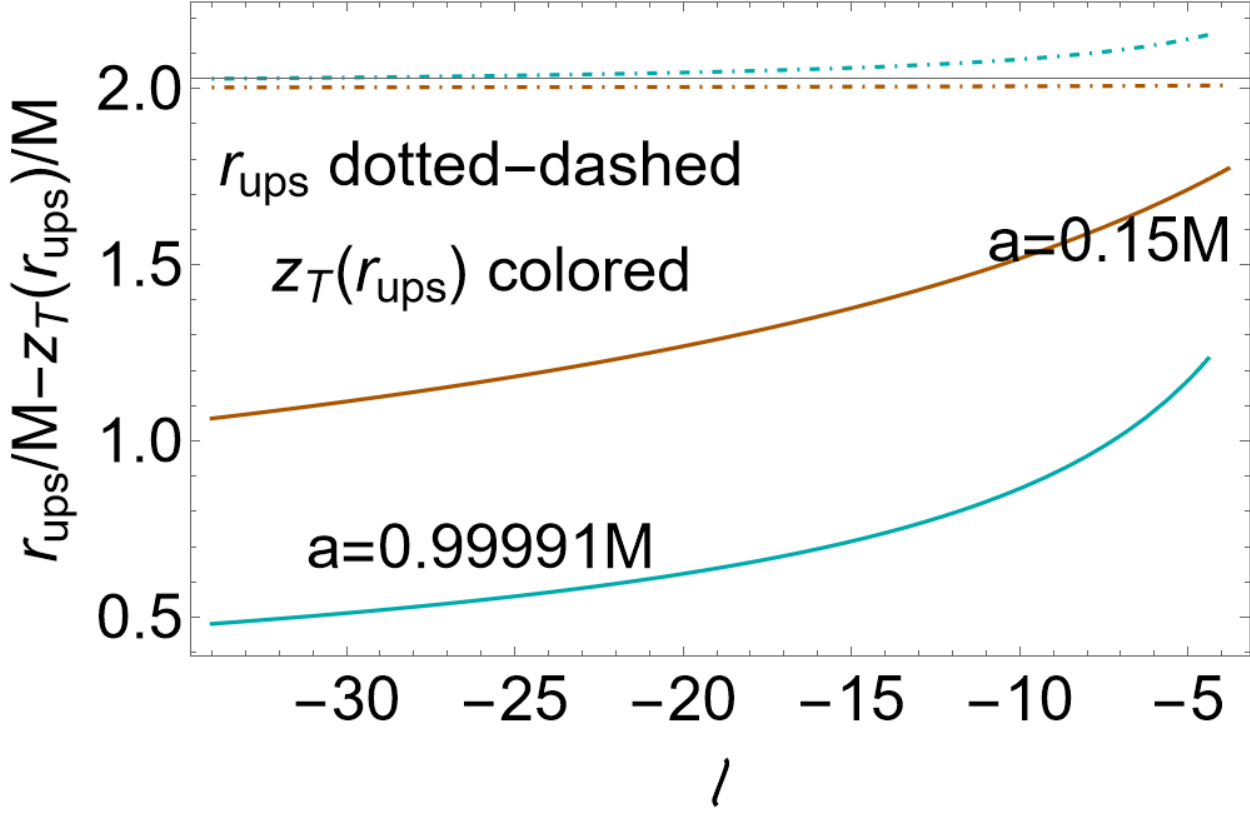} \\
   \includegraphics[width=6.5cm]{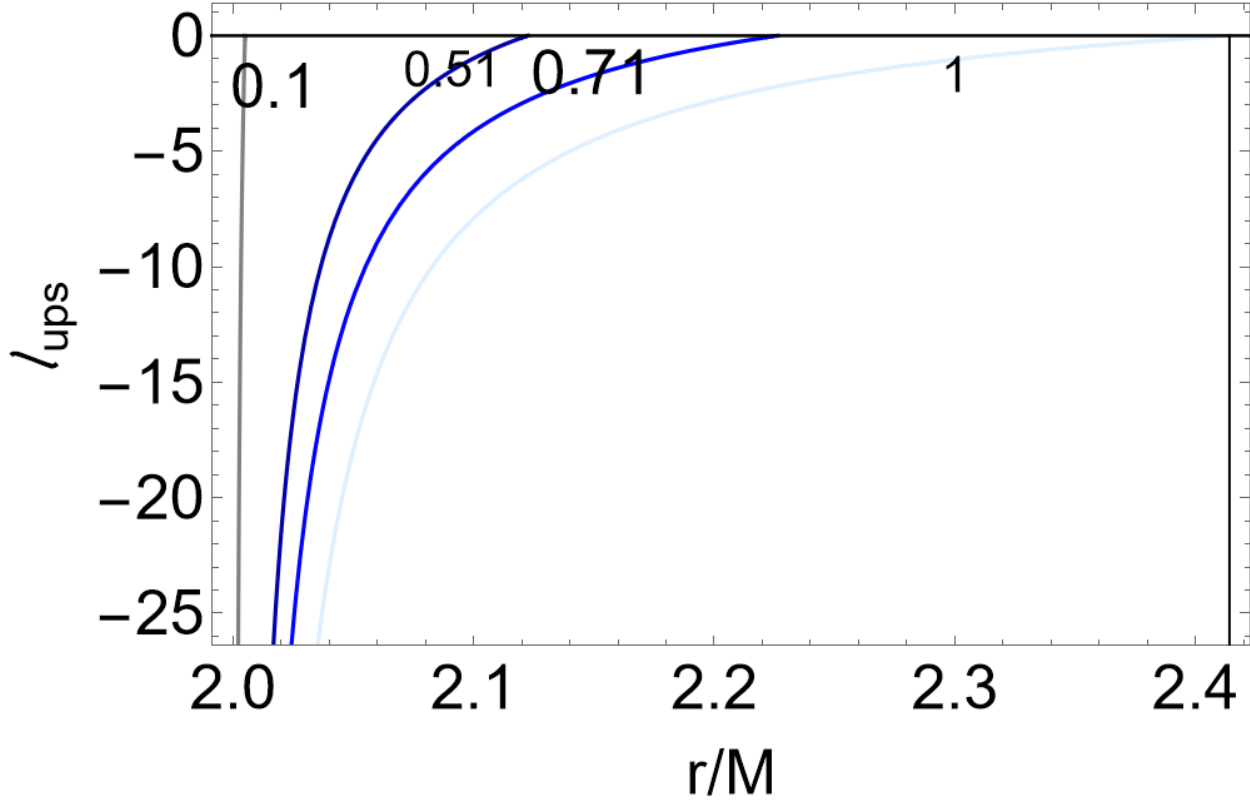}
  \includegraphics[width=6.5cm]{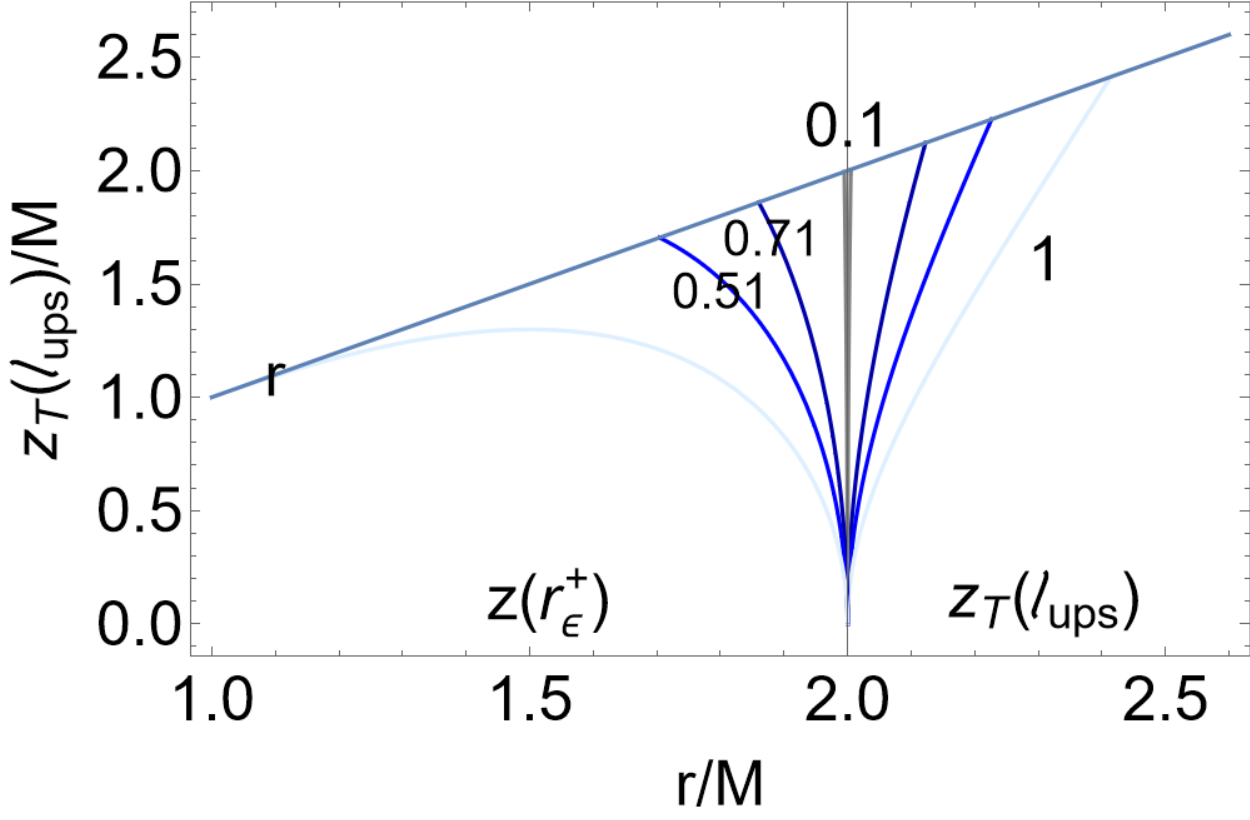}
  \caption{Analysis of the  maximum vertical position of the counter-rotating flow turning point of Sec.\il(\ref{Sec:vertical-z}). There is $z_\Ta=r_\Ta\sqrt{(1-\sigma_\Ta)}$. Upper left panel: black region is the central \textbf{BH} with $r<r_+$, where $r_+$ is the \textbf{BH} horizon.  Radius $r_+^+$ of Eq.\il(\ref{Eq:rplusplusulp}), is plotted as function of the \textbf{BH} spin--mass ratio $a/M$, considered in the analysis of Eq.\il(\ref{Eq:tal-16-gm-rr}), governing the solutions of  $\partial_a z_{\sigma_\Ta}=0$. Right upper panel:  radii $r_{ups}$ of Eq.\il(\ref{Eq:rplusplusulp}), (dotted-dashed curves), maximum vertical coordinate of the counter-rotating flow turning point,  solution of $\partial_a z_{\sigma_\Ta}=0$ and the vertical coordinate $z_\Ta(r_{ups})$ (colored curves) as functions of the fluid specific angular momentum $\ell$, for  spin $a=0.15 M$ (red curves) $a=0.99991M$ (darker-cyan curve)--see analysis of Eq.\il(\ref{Eqs:rups}). Below panels: specific momenta $\ell_{ups}$ of Eqs\il(\ref{Eq:def-lupsetal}) (left panel) as function of the radial distance from the attractor $r/M$ and the vertical coordinate of the turning point $z_\Ta(\ell_{ups})$ as function of $r/M$ (right panel) for different \textbf{BH} spin-mass ratios $a/M$ signed on the curves. In the right panel the ergosurface vertical coordinate $z(r_{\epsilon}^+)$ is also shown, see analysis of Eqs\il(\ref{Eq:tal-16-gm-rr}) and (\ref{Eq:tal-16-gm-rra}).}\label{Fig:Plotrtesamb}
  \end{figure*}

\begin{figure*}
\centering
  \includegraphics[width=5.8cm]{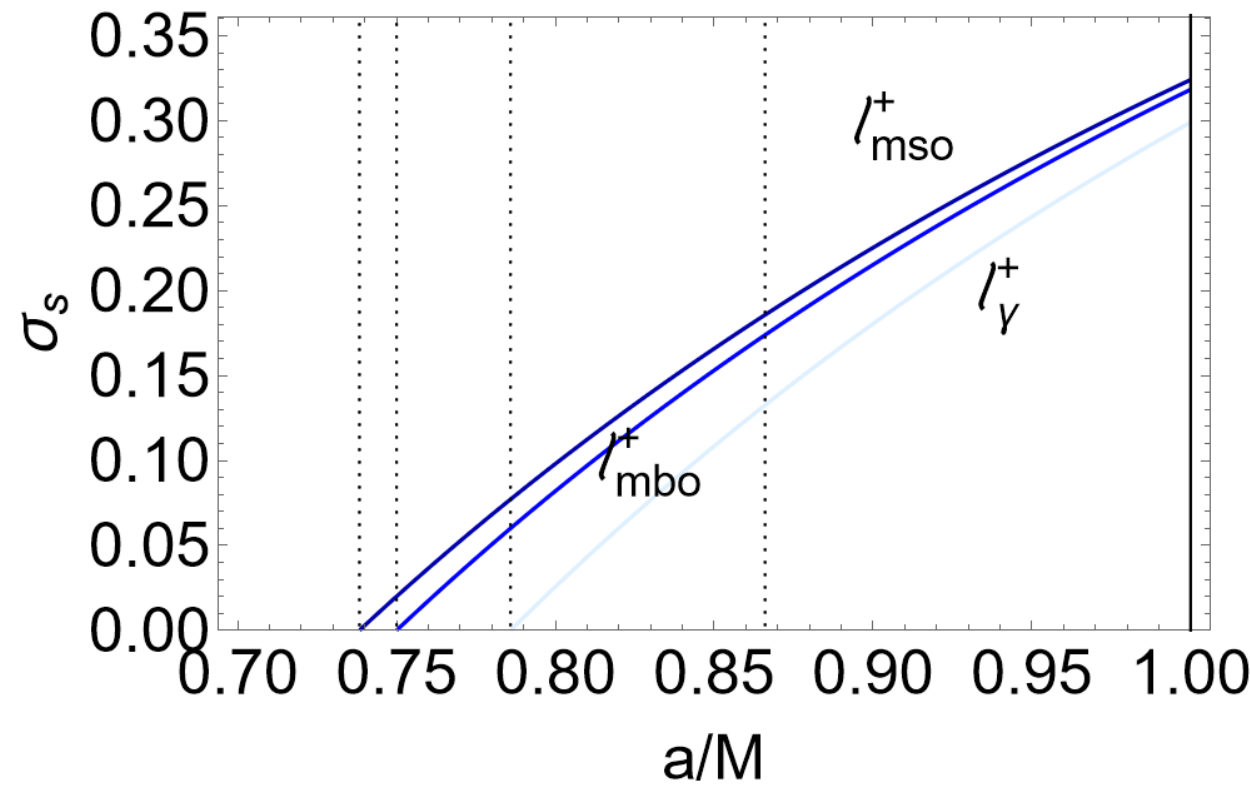}
    \includegraphics[width=5.8cm]{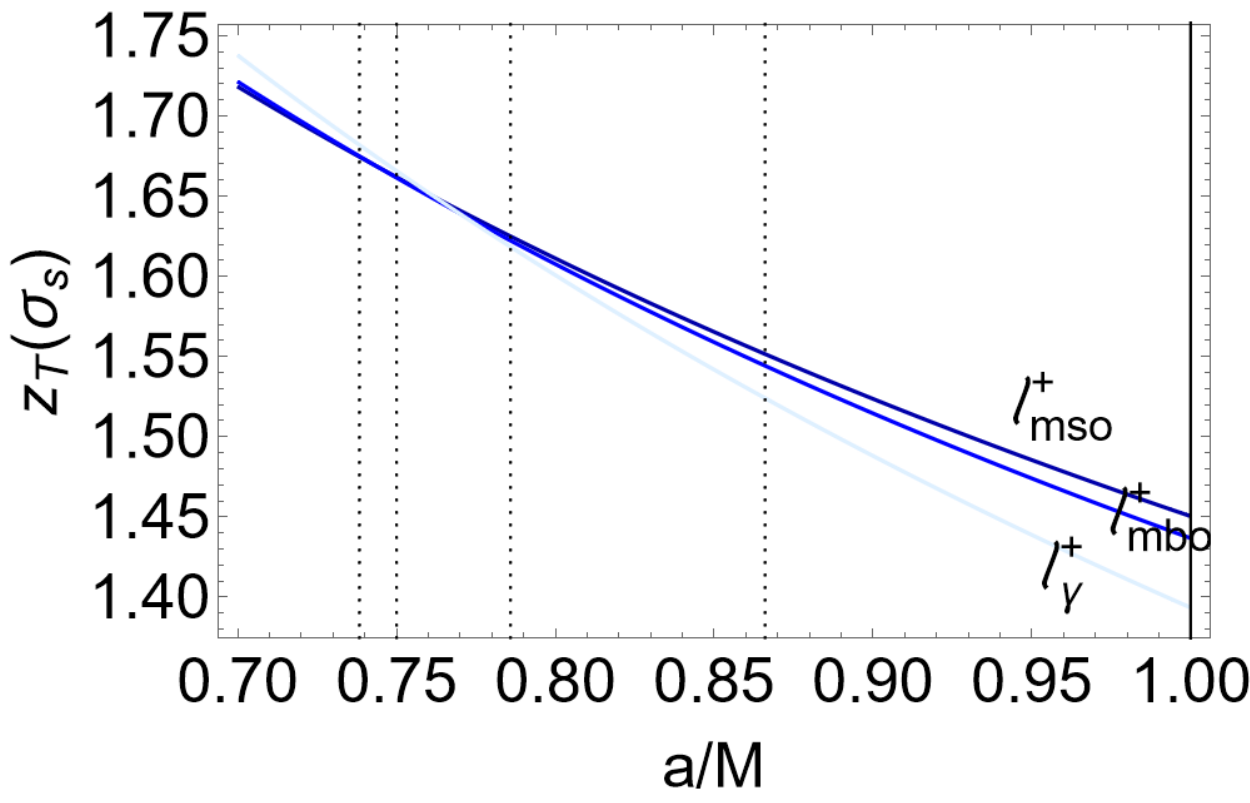}
   \includegraphics[width=5.8cm]{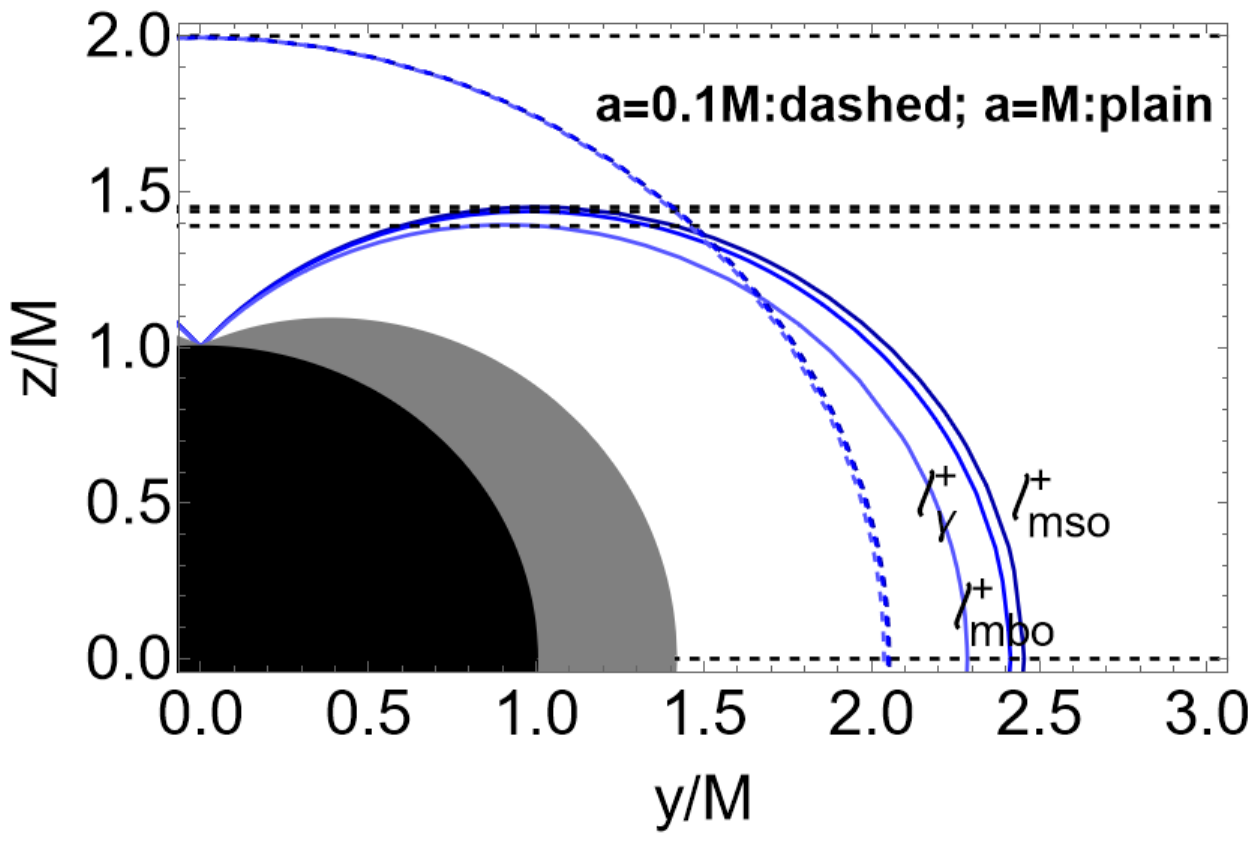}
       \caption{Left panel: maximum plane  $\sigma_s$ of Eq.\il(\ref{Eq:sigmas-manu-sens}), for fluid specific  momenta $\ell_{mso}^+$, $\ell_{mbo}^+$ and $\ell_\gamma^+$, defined in  Eqs\il(\ref{Eq:def-nota-ell}), signed on the curves. Center panel: the maximum  vertical coordinate  of the turning point of the counter-rotating flow as function of  the spin-mass ratio of the central \textbf{BH} (solutions of
 $\partial_{\sigma_\Ta}z_\Ta=0$) of Eq.\il(\ref{Eq:particla-z}). Dotted lines are the limiting condition for  $\sigma=0$ (\textbf{BH} poles) on $\ell_{mso}^+$, $\ell_{mbo}^+$ and $\ell_\gamma^+$. There is $z_\Ta=r_\Ta\sqrt{(1-\sigma_\Ta)}$, where $\sigma\equiv \sin^2\theta$--see Sec.\il(\ref{Sec:vertical-z}).{Right panel: analysis of the turning sphere  maximum vertical point $z_\Ta$. Black region is the \textbf{BH} $r<r_+$ (with outer horizon $r_+$), gray region is $r\in ]r_+,r_\epsilon^+]$, where $r_\epsilon^+$ is the outer ergosurface. Dashed  (plain) curves are the turning spheres for $a=0.1M$ ($a=M$).}}\label{Fig:Plotrtesamd}
\end{figure*}

\begin{figure*}
\centering
        \includegraphics[width=5.65cm]{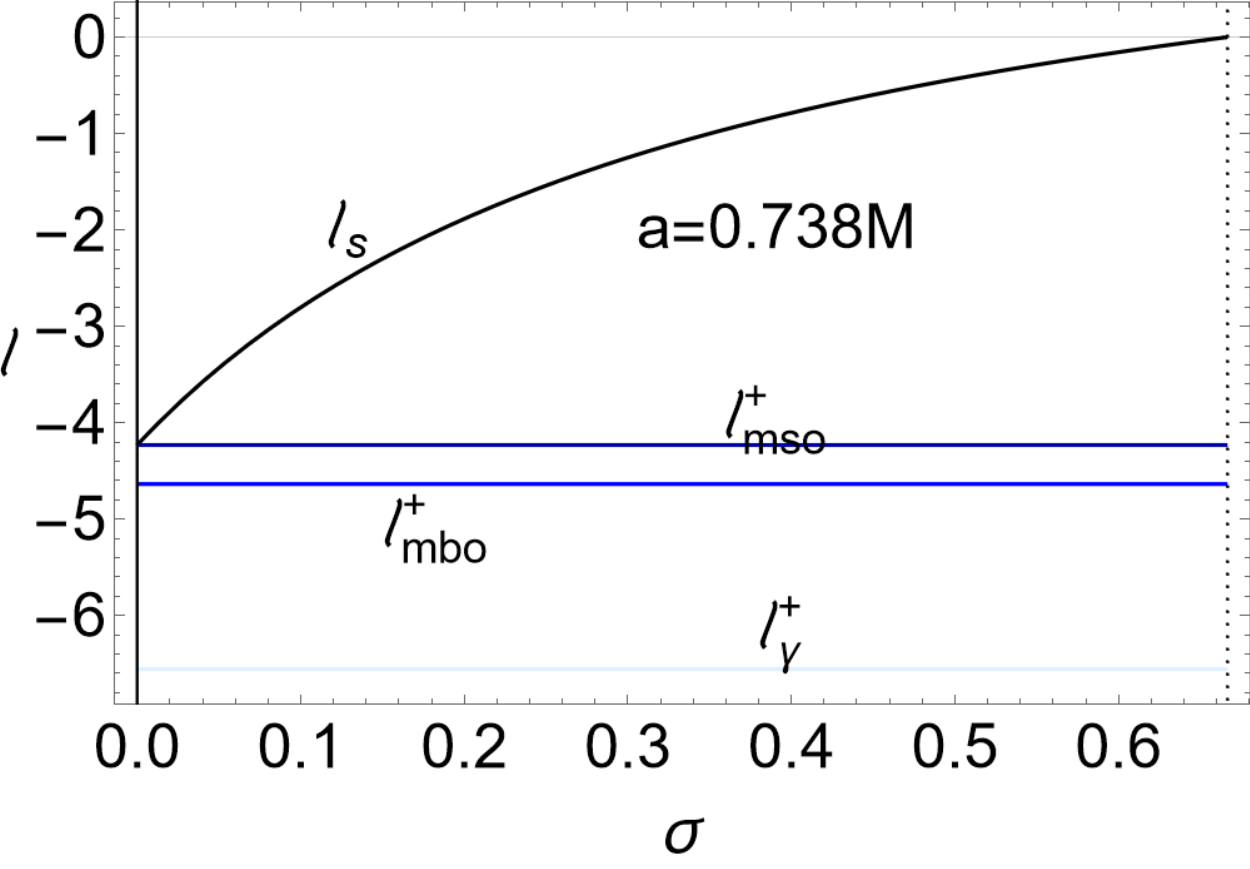}
  \includegraphics[width=5.65cm]{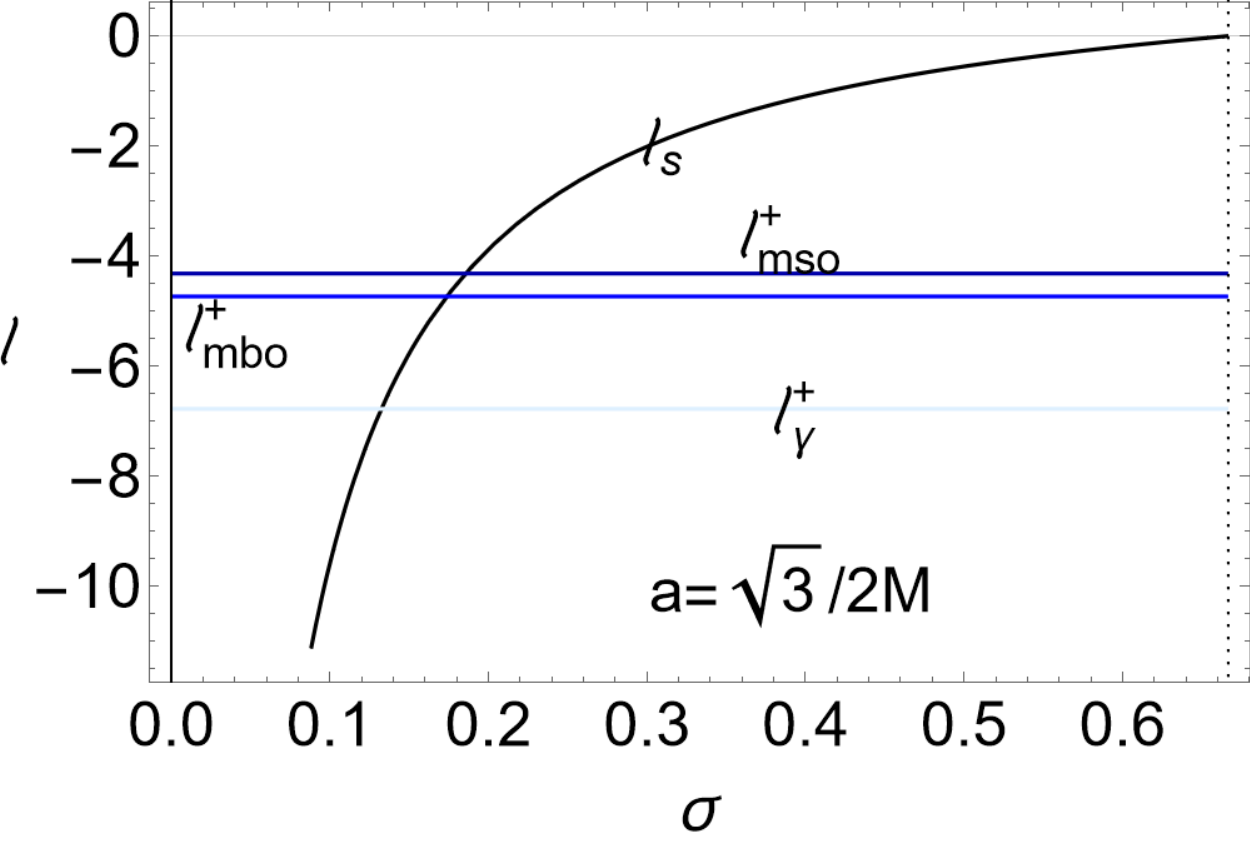}
  \includegraphics[width=5.65cm]{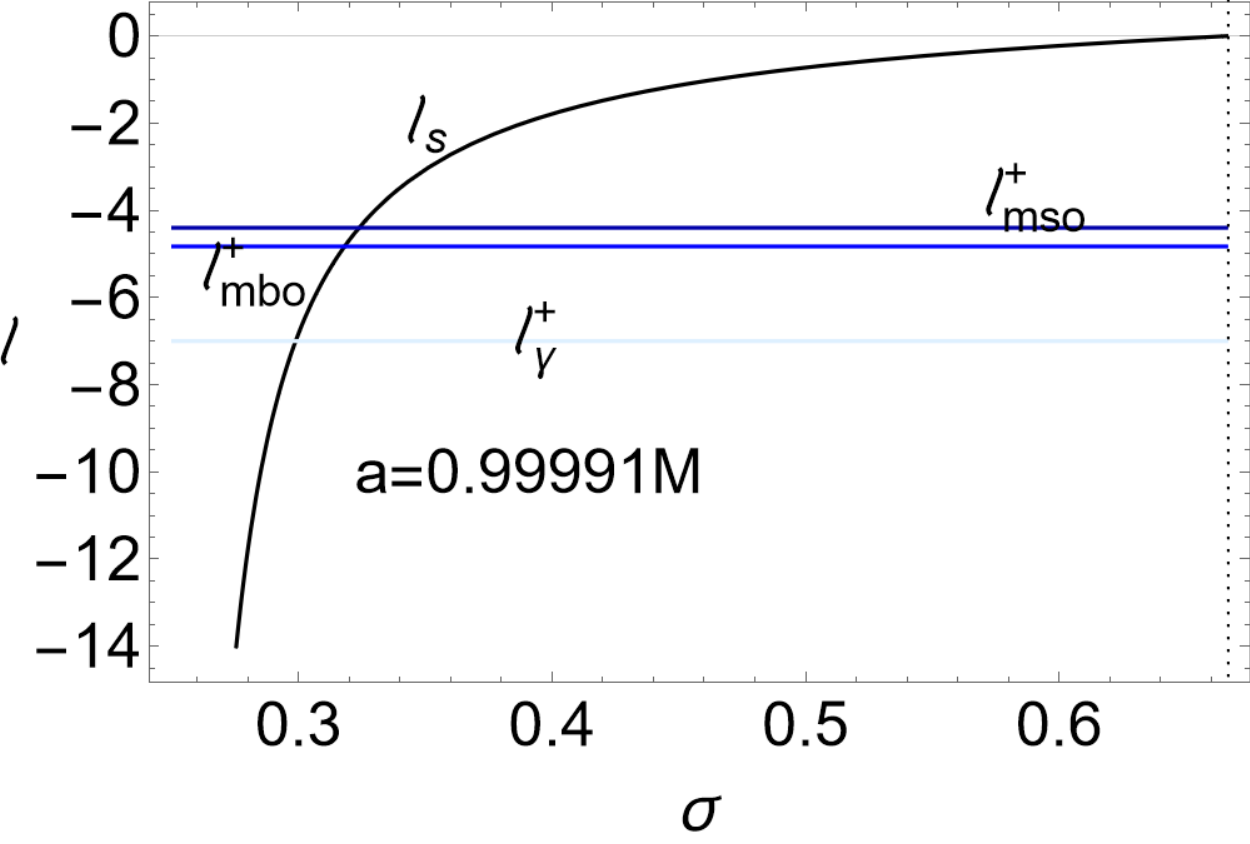}\\
  \includegraphics[width=5.65cm]{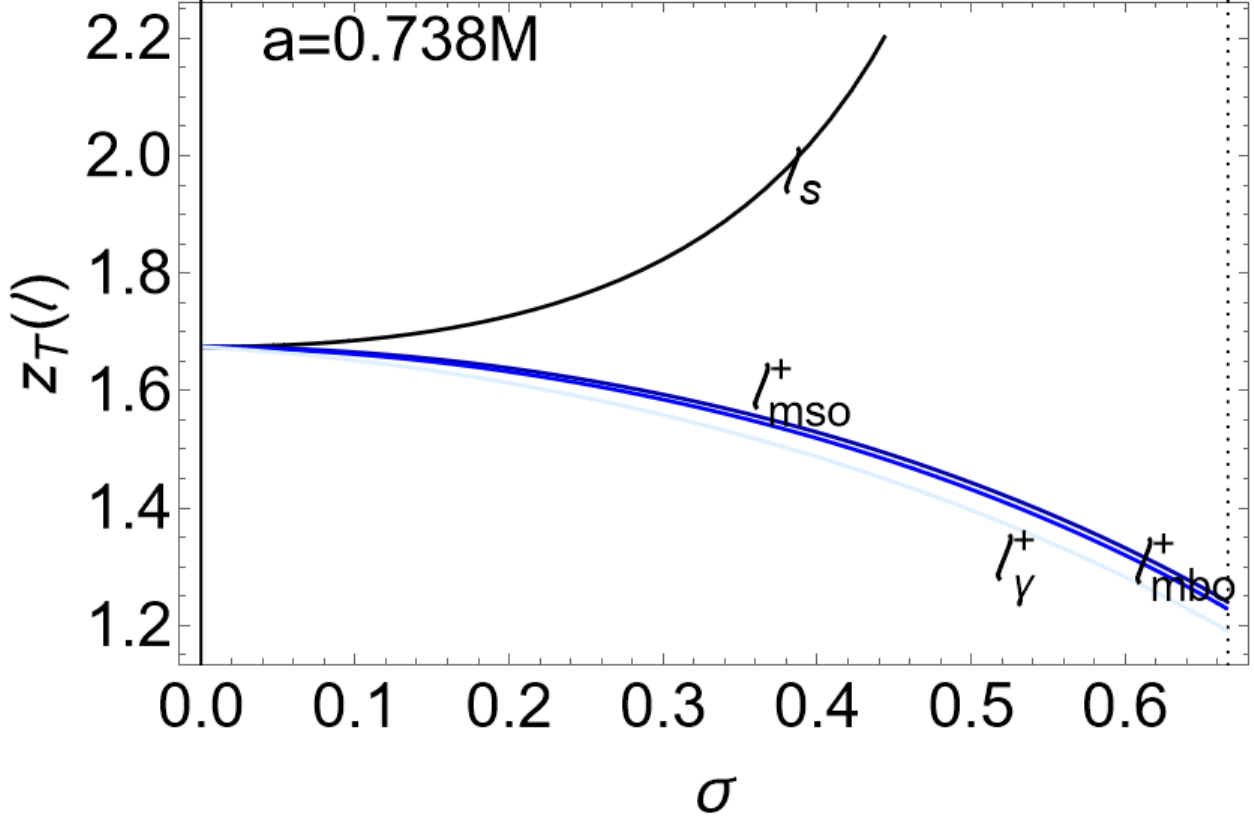}
  \includegraphics[width=5.65cm]{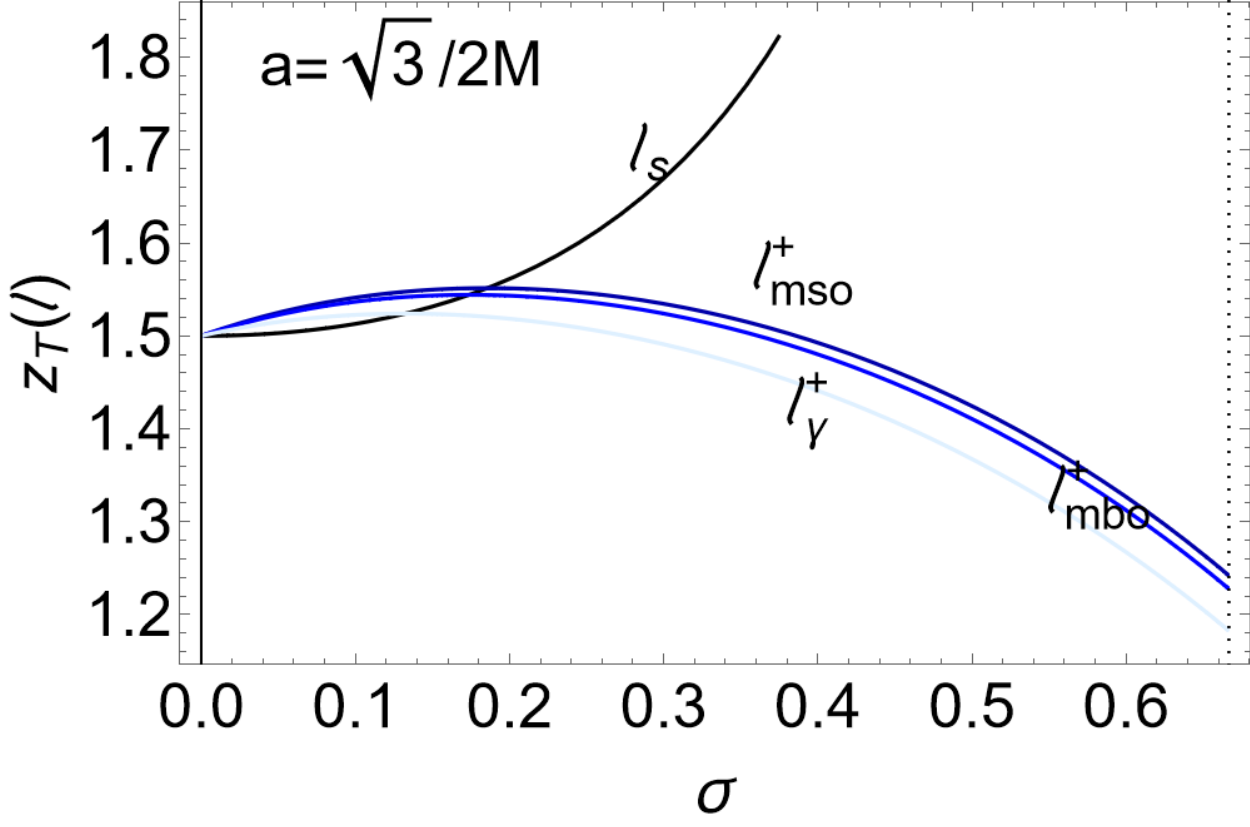}
  \includegraphics[width=5.65cm]{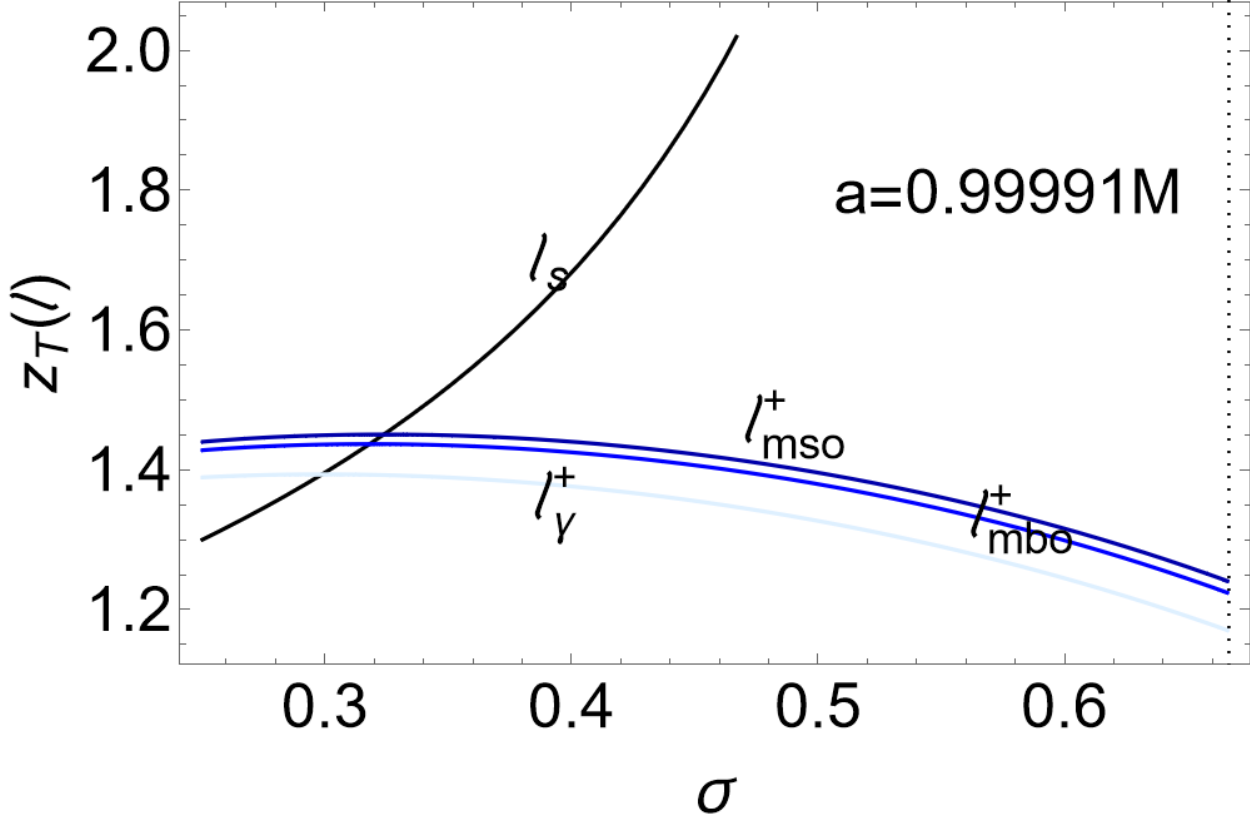}
  \caption{Analysis of the  maximum vertical position $z_\Ta=r_\Ta\sqrt{(1-\sigma_\Ta)}$  of the counter-rotating flow turning point of Sec.\il(\ref{Sec:vertical-z}). Fluid specific angular momentum, $\ell_s$ of Eqs\il(\ref{Eq:effective-ells-mg16}) (upper panels) and vertical coordinate of the turning point $z_{\Ta}(\ell)$ (bottom panels) evaluated on $\ell_s$ and  for momenta $\ell_{mso}^+$, $\ell_{mbo}^+$ and $\ell_\gamma^+$ defined in  Eqs\il(\ref{Eq:def-nota-ell}) signed on the curves are plotted as functions of the plane $\sigma\equiv \sin^2\theta$. Different \textbf{BH} spins $a/M$, signed on the panels are considered. Momentum $\ell_s$ is  solution of
$\partial_{\sigma_\Ta}z_\Ta=0$--see  Eq\il(\ref{Eq:particla-zn}).}\label{Fig:Plotrtesamc}
\end{figure*}

Below we consider     $z_\Ta$  as function  of the \textbf{BH}   $a/M$,  the radius  $r_\Ta$ and the plane $\sigma_\Ta$,
 introducing  the following spin functions:
\bea&&\label{Eq:as}
a_s\equiv\frac{\ell +\sqrt{\ell ^2 \left(3 \ell ^2+4\right)}}{2(\ell ^2+1)},\quad
a_{ps}\equiv\frac{\sqrt{(r-2) \left[(r-2) \ell ^2+4 r\right]}-(r-2) \ell }{2},\\\label{Eq:aMdef}
&&a_M\equiv\frac{(5 \sigma -2) \ell }{\chi}+2 \sqrt{\frac{(1-\sigma) \ell ^2 \left[4(1- \sigma) +3 \ell ^2\right]}{\chi^2}}\quad\mbox{where}\quad \chi\equiv(\sigma +2) (3 \sigma -2)+4 (\sigma -1) \ell ^2
\eea
where spins $a_s$ and $a_M$ are shown in Figs\il(\ref{Fig:Plotrtesam}).
We also introduce the radii
\bea\label{Eq:rplusplusulp}
&&r_{lu}\equiv \frac{{\ell +2}+\sqrt{\ell ^2-4 \ell +8}}{2},\quad r_+^+\equiv 1+\sqrt{1+a^2},\quad r_{ups}\equiv \frac{a \ell +2+\sqrt{a^2 \left(\ell ^2+4\right)+4(1- a \ell)}}{2},\\
&&\label{Eq:raps}
r_{aps}\equiv \frac{a \left(\ell ^2-2\right)+\ell  \left[\sqrt{\frac{\left[\ell(a \ell-1)+a \right] \left[a \left(\ell ^2+4\right)-4 \ell \right]}{\ell ^2}}+2\right]}{3 \ell },
\eea
where
 $r_+^+$ and $r_{ups}$ are plotted in Figs\il(\ref{Fig:Plotrtesamb}), while radius  $r_{aps}$ and $r_{lu}$, are shown in Figs\il(\ref{Fig:Plotrtesam}).
Furthermore, we define the
momenta
\bea\label{Eq:def-lupsetal}
&&\ell_{ups}\equiv\frac{r}{a}-\frac{a}{r-2},\quad\ell_{aps}\equiv 2a\left[\frac{1}{4 a^2-3}-2 \sqrt{\frac{1-a^2}{\left(4 a^2-3\right)^2}}\right],\\
&&\label{Eq:effective-ells-mg16}\ell_s\equiv a\left[\frac{(5 \sigma -2)}{3-4 a^2 (1-\sigma)}-2 \sqrt{\frac{ (1-\sigma) \left[a^2 (\sigma +2) (3 \sigma -2)+4(1- \sigma)\right]}{\left[3-4 a^2 (1-\sigma)\right]^2}}\right].
\eea
Momentum $\ell_{ups}$ is shown in Figs\il(\ref{Fig:Plotrtesamb}),   $\ell_{aps}$ is shown in Figs\il(\ref{Fig:Plotrtesam}) while  $\ell_s$ is in Figs\il(\ref{Fig:Plotrtesamc}). Finally  we consider the
 planes:
\bea\label{Eq:sigmas-manu-sens}
&&\sigma_s\equiv\frac{2}{3} \left[\sqrt{\frac{\left[\ell(a \ell-1)+a\right]\left[a \left(\ell ^2+4\right)-4 \ell \right]}{a^2}}-\ell ^2-1\right]+\frac{5 \ell }{3 a},\\\label{Eq:sigmausigmae}
&&\sigma_u\equiv\frac{\ell (5 -2 \ell)-2+2 \sqrt{(\ell -2)^2 [(\ell -1) \ell +1]}}{3},\quad\sigma_e\equiv\frac{4 a^2-3}{4 a^2},
\eea
where
plane $\sigma_s$ is shown in  Figs\il(\ref{Fig:Plotrtesamd}) and planes $(\sigma_e,\sigma_u)$ are shown in Figs\il(\ref{Fig:Plotrtesam}).

For $\ell <0$ and $\sigma\in[0,1]$ there are the following  extremes for the turning point vertical coordinate  $ z_\Ta$ as functions of $r$:
 \bea&&\label{Eqs:sof-A6po8a}
 \partial_r z_\Ta=0:\quad \mbox{for}\quad  r=r_{aps},\quad\mbox{for}\quad a\in ]a_s,1],
\\&&\label{Eqs:sof-A6po8}
\mbox{or alternatively for}\quad a\in \left[\frac{\sqrt{3}}{2},1\right]\quad\mbox{and}\quad  \left(a\in \left]0,\frac{\sqrt{3}}{2}\right],\quad \ell \in ]\ell_{aps},0[\right).
 \eea
These results also point out the limiting spin  $a/M={\sqrt{3}}/{2}\approx0.866025$, distinguishing  fast rotating from slowly rotating \textbf{BHs}. Radius  $r_{aps}$, limiting momentum $\ell_{aps}$ and limiting spin $a_s$ are shown in Figs\il(\ref{Fig:Plotrtesam}).
 Therefore, for $\ell <0$ and $\sigma\in[0,1]$ there are the following extremes  of the vertical coordinate $ z_\Ta$ with the \textbf{BH} spin--mass ratio:
 \bea&&\label{Eq:anali-zza}
 \partial_a z_\Ta=0:\quad\mbox{for}\quad a=a_{ps}\quad\mbox{and }\quad r\in ]2,r_{lu}],
 \eea
or equivalently
 \bea
 &&\label{Eqs:rups}
\quad\quad\quad\quad\quad\quad\mbox{for}\quad  r=r_{ups} \quad\mbox{and}\quad a\in ]0,1],\;
\eea
alternatively
\bea
 \label{Eq:tal-16-gm-rr}
 &&\quad\quad\quad\quad\quad\quad\mbox{for}\quad \ell =\ell_{ups}, \quad\mbox{and}\quad a\in ]0,1],\quad r\in ]2, r_{+}^+[,\quad \mbox{equivalently}
 \\&&\label{Eq:tal-16-gm-rra}\quad\quad\quad\quad\quad\quad\mbox{for}\quad \ell =\ell_{ups}, \quad\mbox{and }\quad   a\in ]a_{\pm},1], \quad r/M\in ]2,\sqrt{2}+1[,
 \eea
 where $a_{\pm}\equiv \sqrt{r(2M-r)}$
 is the horizons curve in the plane $a-r$. It is immediate to see that the radius $r=r_\Ta$ is upper bounded by the limiting value $r_\Ta=(\sqrt{2}+1)M=2.41421M$, according to the analysis Sec.\il(\ref{Sec:fishing-particles}) and Eq.\il(\ref{Eq:max-cusp-eq-exte}).  (There is $z_\Ta(\ell_{ups})=r \sqrt{{(r-2M) r}/{a^2}}$, Figs\il(\ref{Fig:Plotrtesamb}), while the ergoregion is $z_{\epsilon}^+=r \sqrt{{(2M-r) r}/{a^2}}=r(a_{\pm}/a)$).
 Limiting radius $r_{lu}$ is shown in Figs\il(\ref{Fig:Plotrtesam}), momentum $\ell_{ups}$, radius $r_{ups}$, vertical momentum $z_\Ta(r_{ups})$ and limiting radius $r_+^+$
are shown in  Figs\il(\ref{Fig:Plotrtesamb}). There is $z_{\Ta}(r_{ups})<1.77M$,  decreasing  with the increase of the dimensionless spin $a/M$.
 For $\ell<0$,  the extremes of  $z_\Ta$ according to the plane $\sigma$ are
 \bea&&\label{Eq:particla-za}
 \partial_{\sigma_\Ta}z_\Ta=0:\quad\mbox{for}\quad \left(\sigma\in [0 ,\sigma_u],\; a=a_M\right),\quad\mbox{or equivalently}
 \\\label{Eq:particla-z}
 && \quad\quad\quad\mbox{for}\quad \left(\sigma =\sigma_s,   a\in ]a_s,1]\right); \quad\left(a=a_s,\; \sigma =0\right),
 \eea
 or alternately
 \bea
 \label{Eq:particla-zn}
 &&\mbox{for}\quad\ell =\ell_s:\quad\mbox{and} \quad\left(a\in \left]0,\frac{\sqrt{3}}{2}\right[,\;  \sigma\in\left[0,\frac{2}{3}\right[\right);\; \left(a=\frac{\sqrt{3}}{2},\;\sigma\in\left]0,\frac{2}{3}\right[\right); \left(a\in\left]\frac{\sqrt{3}}{2},1\right],\; \sigma\in\left]\sigma_e,\frac{2}{3}\right[\right).
 \eea
 Solution $\ell_s$  is shown in Figs\il(\ref{Fig:Plotrtesamc}).
Limiting  plane $\sigma_e$  and solution $a_M$ are shown in Figs\il(\ref{Fig:Plotrtesam}).
 Plane $\sigma_s$  is in  Figs\il(\ref{Fig:Plotrtesamd}). In this analysis we single out the limiting critical plane $\sigma=2/3\approx0.666667$ and  spin $a/M=\sqrt{3}/2$,  showing  the different situation for slowly spinning attractors and fast attractors, and turning points closer or farer from the \textbf{BH} poles. We can  note the different situations for   the  counter-rotating flows from the cusped tori and proto-jets driven flows. Considering   Figs\il(\ref{Fig:Plotrtesamd}), there is $a\geq 0.74 M$ and   $\sigma_s<0.35$, with $\sigma_s$ increasing  with the spin and $z_\Ta<1.75 M$ (generally) decreasing with the spin $a/M$.  There exists a discriminant spin $a\approx 0.77M$. For slower spin  $a/M$, the vertical coordinate turning point is higher for the proto-jet driven flow than for cusped tori flow. For larger \textbf{BH} spins, the situation is inverted and the regions for turning points in proto-jets driven flows spread, according to the different planes and decrease  with increasing \textbf{BH} spin.
The  turning point therefore is not a characteristic of the matter funnels or photons jets  structures, collimated along  the \textbf{BH} axis, but  remains a defined vertical structure in a cocoon surrounding the ergosurface  and closing on the outer ergosurface $r_\epsilon^+$, more  internal with respect to the   tori driven flows turning corona.
{This aspect was also partially dealt with in Sec.\il(\ref{Sec:turning-sign-existence}),
in relation to the analysis of Figs\il(\ref{Fig:Plotrte4}) for the double turning  points at fixed $(\ell,z)$. The presence of a vertical maximum is  an indication of the  double turning point on the vertical axi. Constrained  by the condition
$ z_\Ta=r_+$ at $y_\Ta(+)>0$, double turning points are possible for fast spinning \textbf{BHs}, e.g.,  with $a>0.74$ for  fluids with $\ell=\ell_{mso}^+$, and  $a>0.75$  for fluids with  $\ell=\ell_{mbo}^+$.
 Here we specify these results regarding the presence of the maximum. In Figs\il(\ref{Fig:Plotrtesamd}) is the analysis of the vertical maximum ($\partial_y{_\Ta} z_\Ta=0$) for different \textbf{BH} spins $a/M$ and momenta $\ell<0$. The maximum increases, decreasing the \textbf{BH} spin with the limiting situation of $a\geq0$ and $z_\Ta\leq 2M$. Then it is maximum at $\ell_{mso}^+$ (decreases with the magnitude of $\ell$), and the  bottom boundary of the maximum $z_\Ta$  occurs for the extreme \textbf{BH} with $a=M$.  Then, at $a=M$ there is the maximum $z^{\max}_\Ta=1.451M$ for $\ell=\ell_{mso}^+$, $z^{\max}_\Ta=1.437M$ for $\ell=\ell_{mbo}^+$, and $z^{\max}_\Ta=1.39M$ for $\ell=\ell_{mso}^+$.}
\section{Flow thickness and counter-rotating tori energetics}\label{Sec:flow}
For the  tori and  proto-jets   counter-rotating driven flows turning point,  is located  at $r>r_\epsilon^+$ . %

However we can  study the   frame--dragging  influence on the  accretion flow from the counter-rotating  tori  considering   the flow thickness   of the super-critical  tori  (with $K=K_s\in ]K_\times,1[$) throats.
We start by analyzing   the thickness of the accretion flow in the counter-rotating configurations.  (A comparative analysis with the corotating flows can be found for example in \cite{Multy,letter,long}).

\begin{figure*}
\centering
      \includegraphics[width=5.5cm]{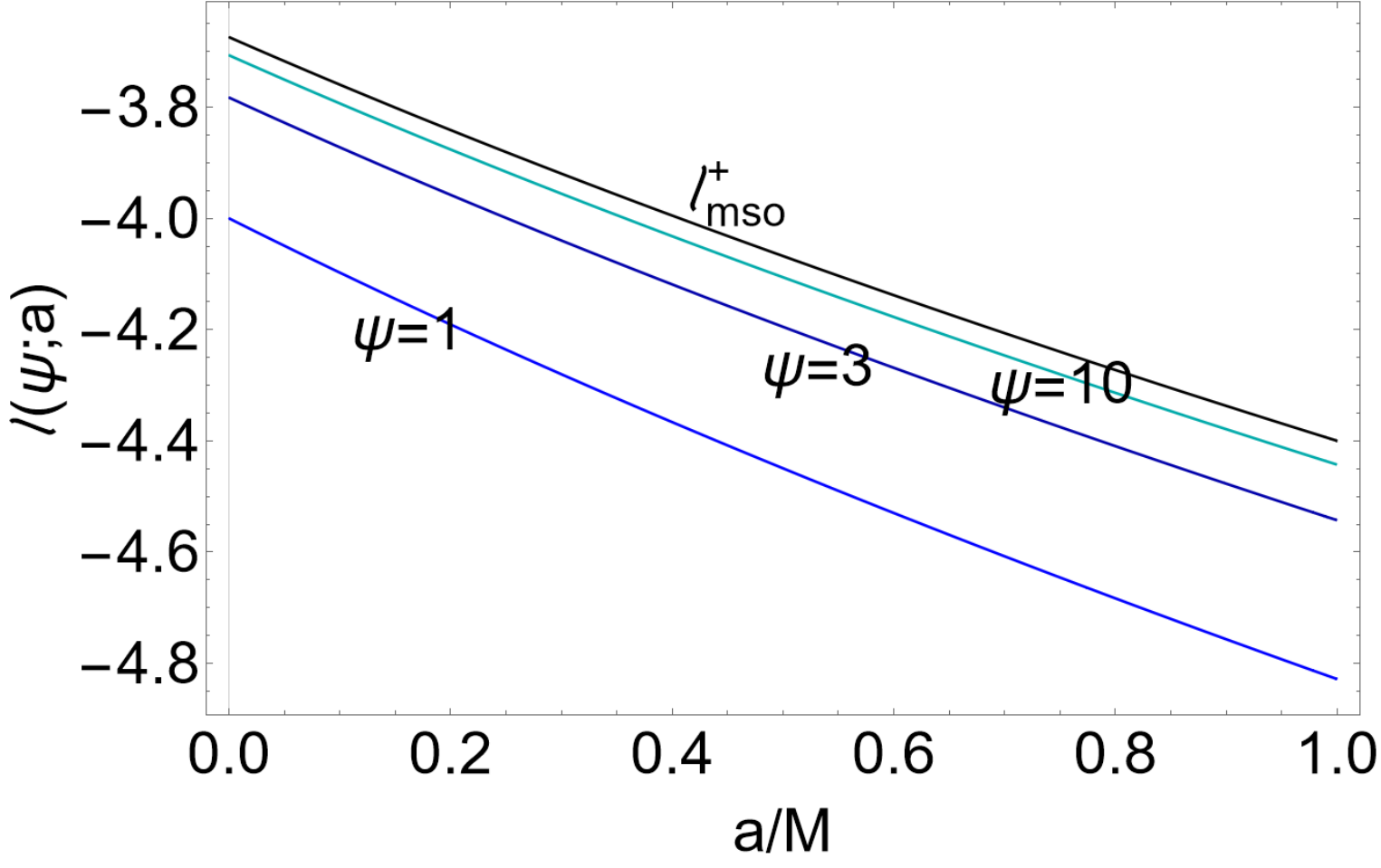}
    \includegraphics[width=5.5cm]{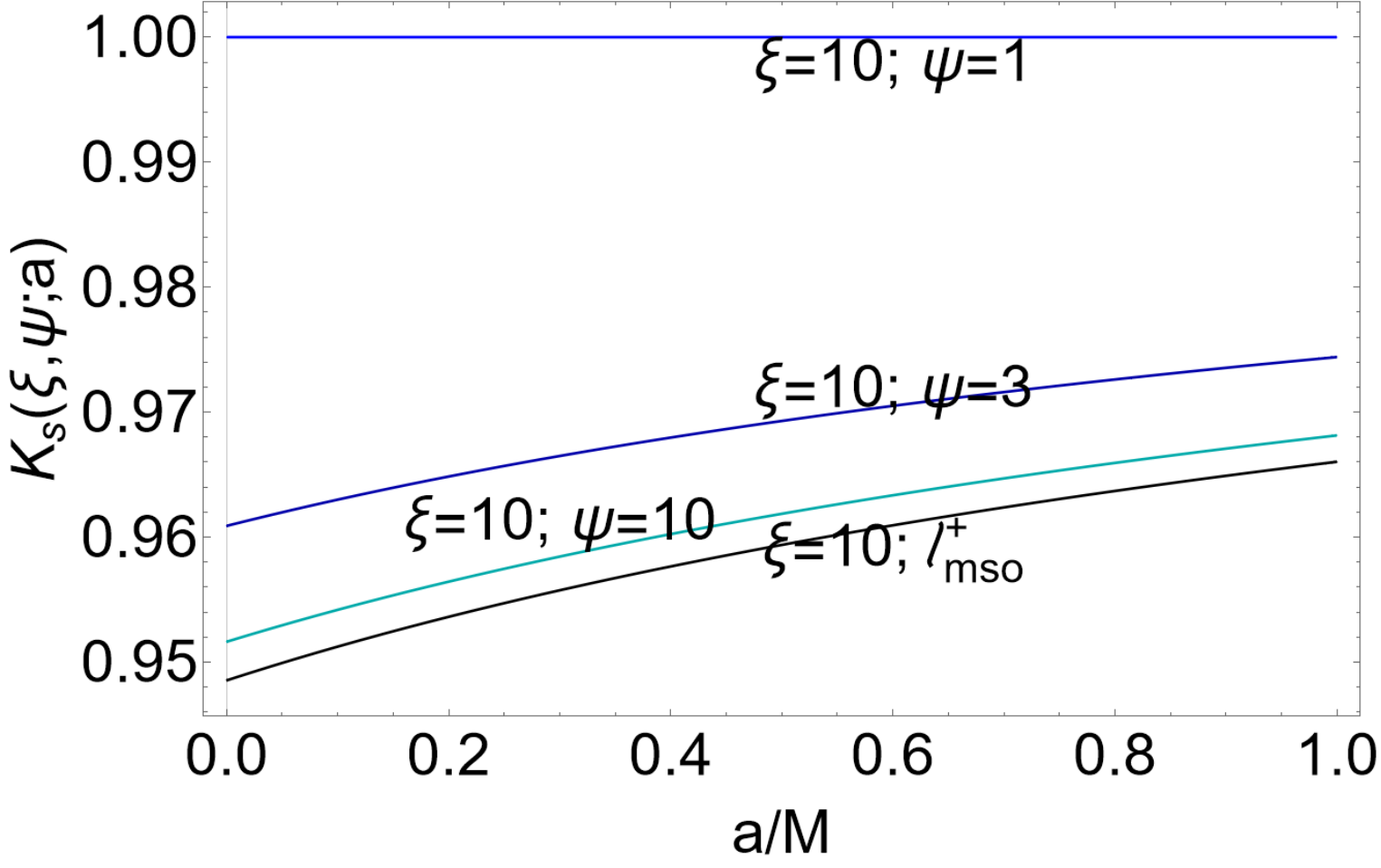}
      \includegraphics[width=5.5cm]{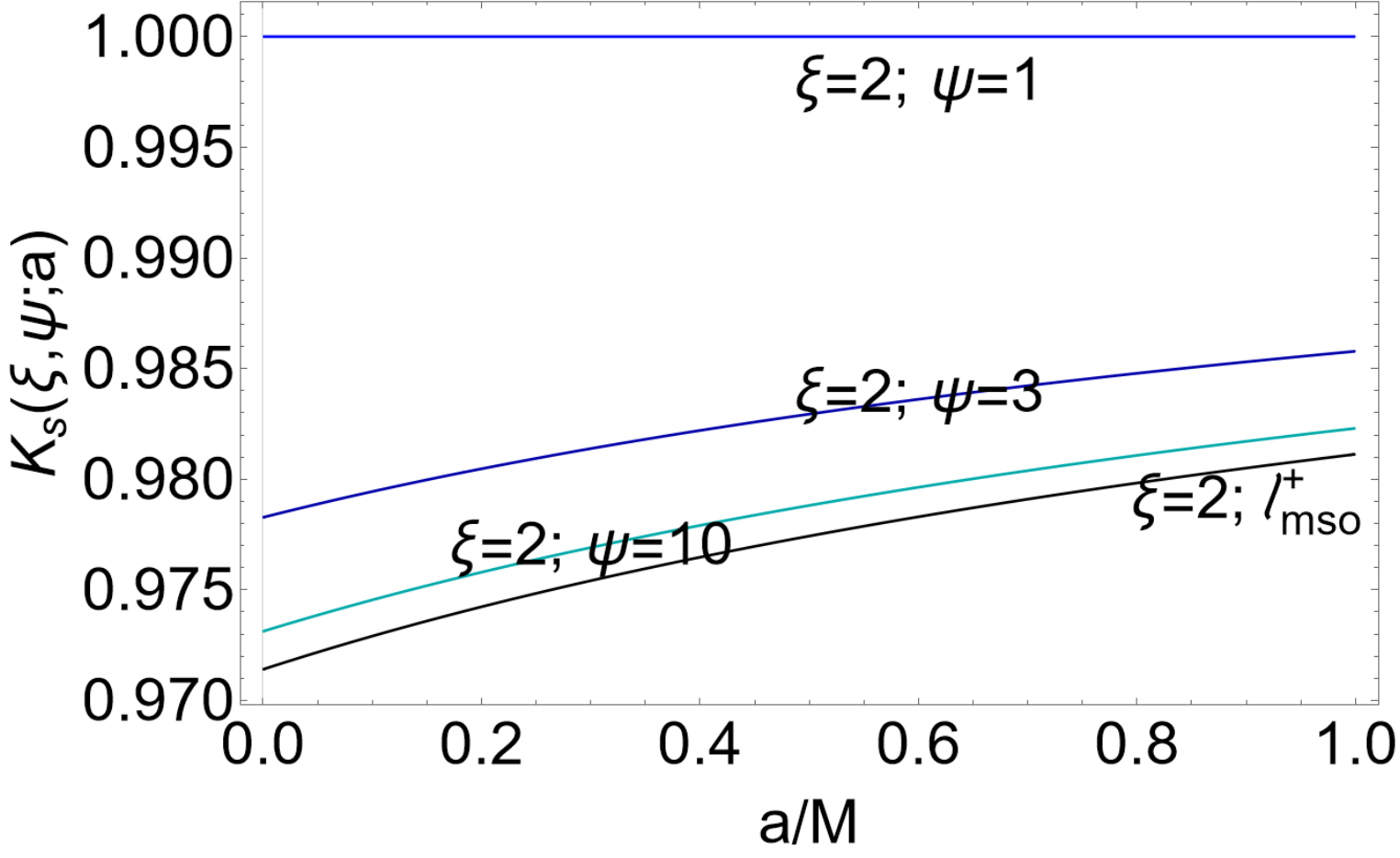}
          \includegraphics[width=5.5cm]{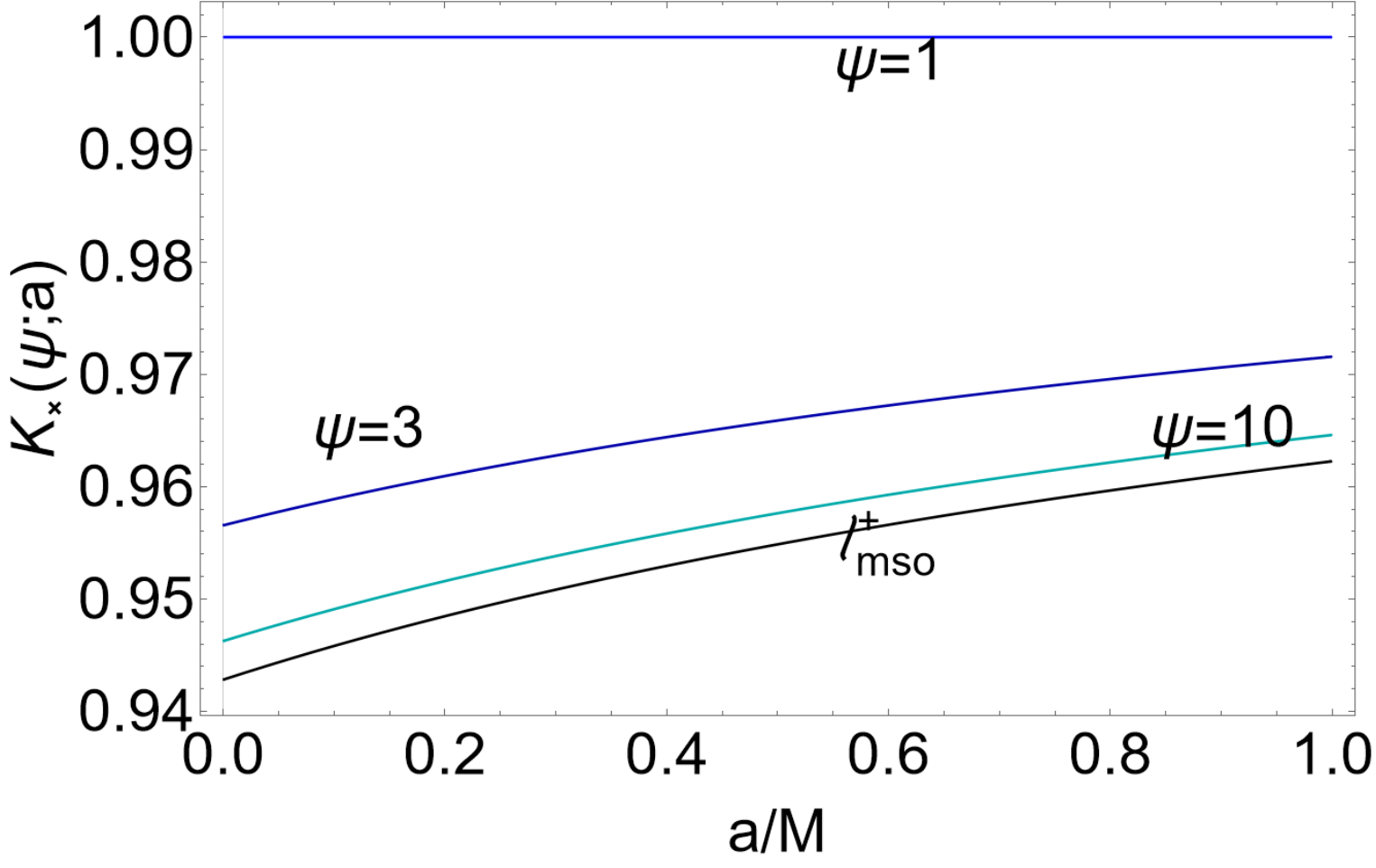}
            \includegraphics[width=5.5cm]{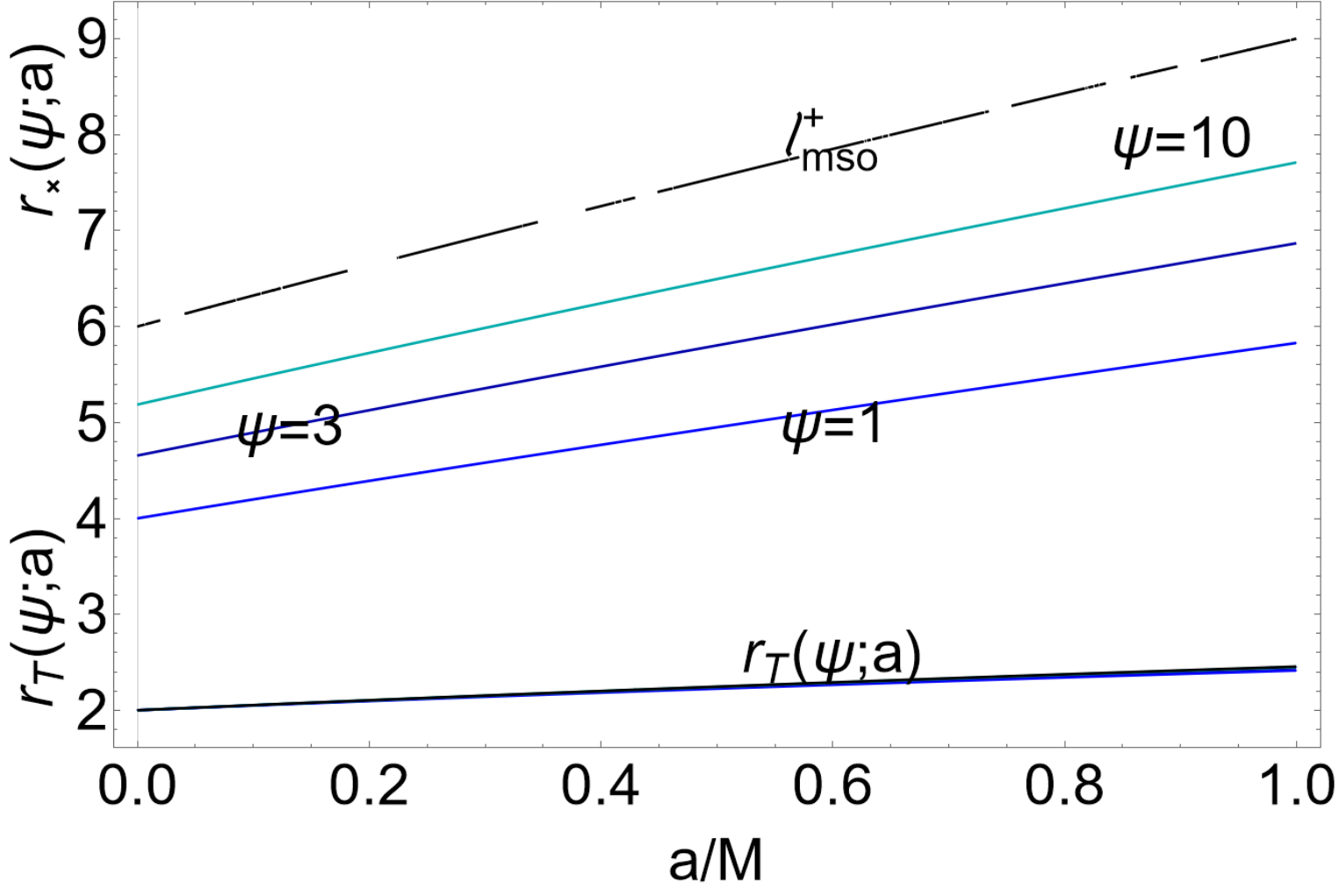}
              \includegraphics[width=5.5cm]{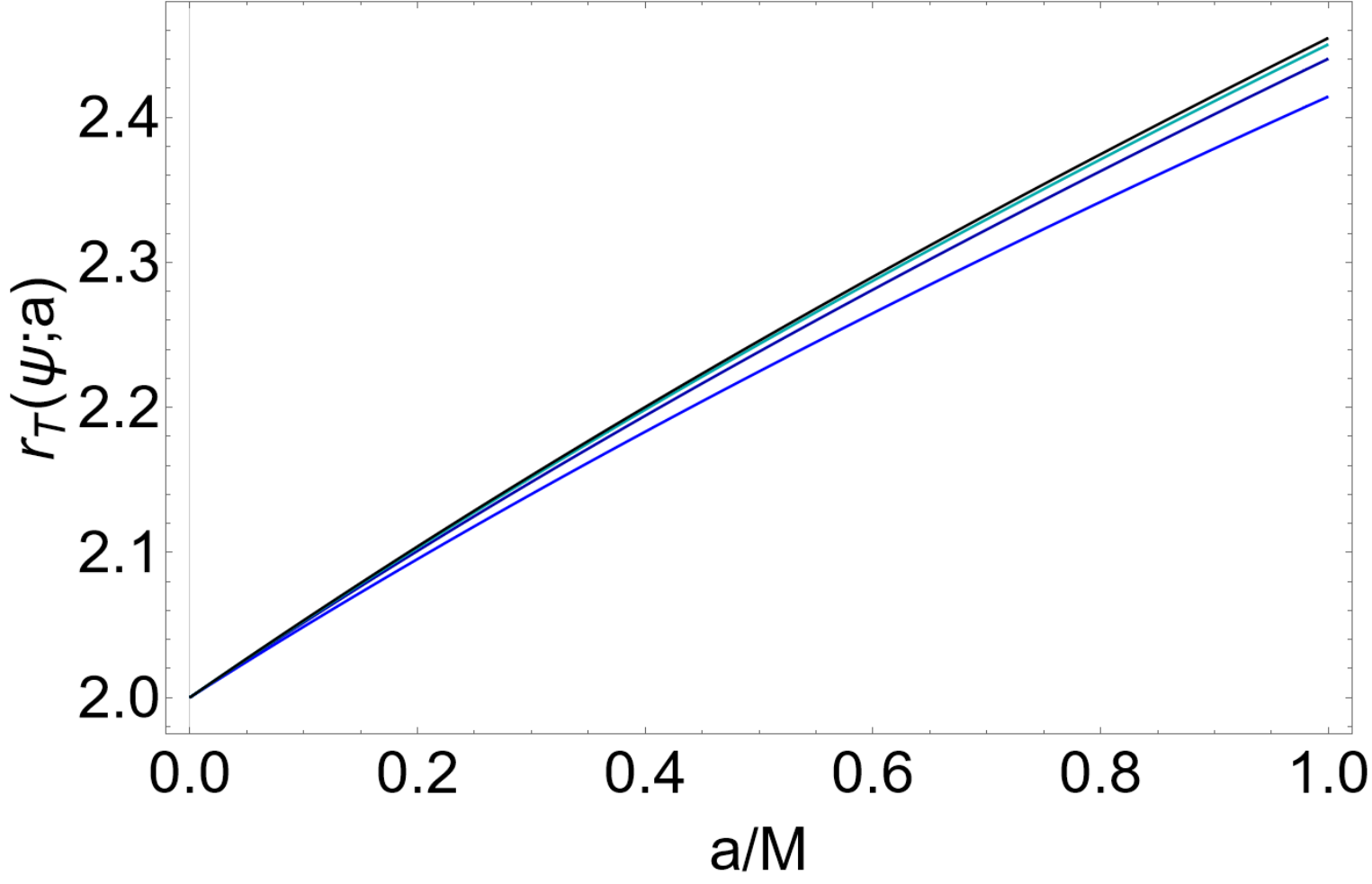}
           \caption{Upper left panel counter-rotating fluid specific angular momentum $\ell_{ecc}(\psi; a)$ of Eq.\il(\ref{Eq:psixi}) for different values of $\psi:\ell\in[\ell_{mbo}^+,\ell_{mso}^+]$ for the maximum extension of the turning corona.
Limiting fluid specific  momenta $\ell_{mso}^+$ and  $\ell_{mbo}^+$  are defined in   Eqs\il(\ref{Eq:def-nota-ell}).  Central and right upper panels:  quantities $K_s\in[K_\times,1]$, regulating the flux thickness of  Eq.\il(\ref{Eq:psixi}) for different values of $\psi$ signed on the curves for $\xi=10$ (central panel) and $\xi=2$ (right panel). Bottom left panel: parameter $K$ evaluated at the torus cusp for different values of the momenta parameter $\psi$ signed on the curves. Center bottom panel: cusp location $r_\times$ for different values of $\psi$ and correspondent  turning point location $r_\Ta$. Bottom right  panel shows a zoom  of the $r_\Ta$ for different values of $\psi$ according to the color choice of the central panel. }\label{Fig:PPlotrangqwals}
\end{figure*}

\begin{figure*}
\centering
\includegraphics[width=5cm]{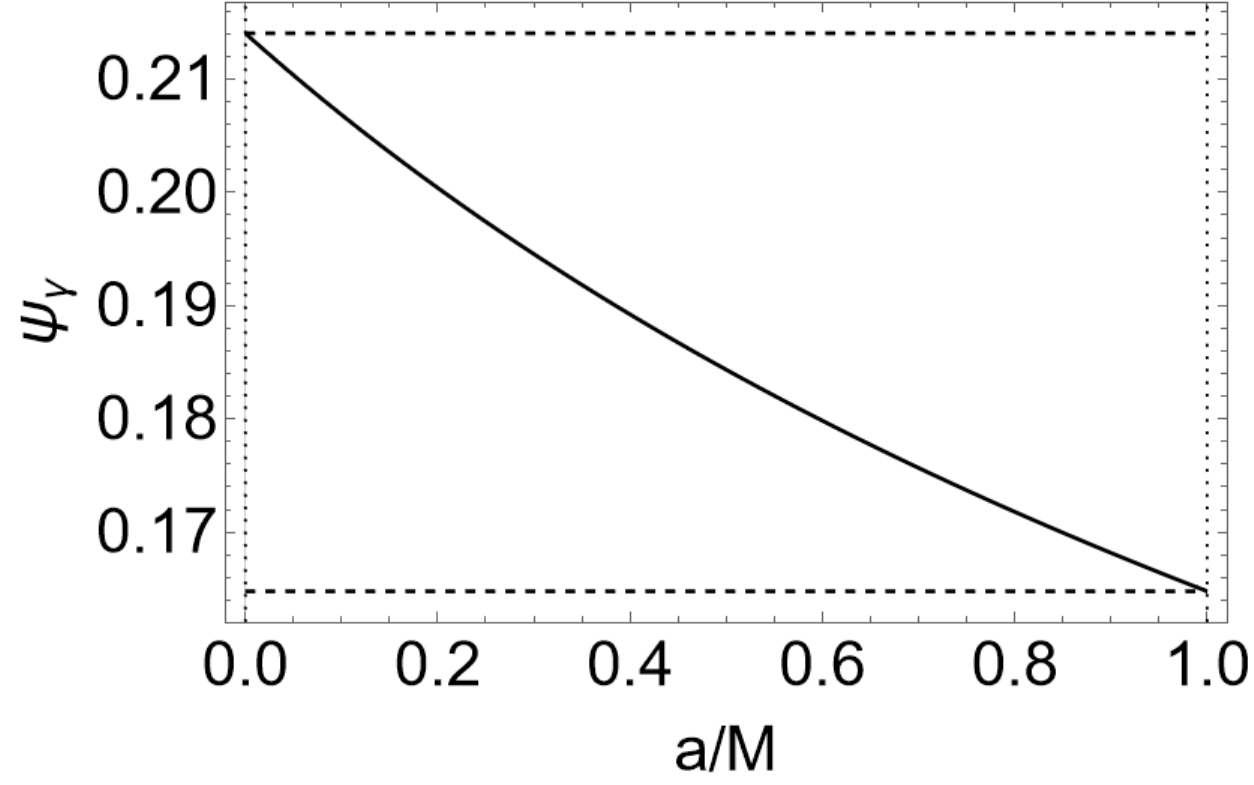}
\caption{Parameter $\psi_\gamma: \ell_{ecc}=\ell_\gamma^+$ as function of the  \textbf{BH} spin-mass ratio $a/M$. Counter-rotating fluid specific angular momentum $\ell_{ecc}(\psi;a)$ is  defined  in Eq.\il(\ref{Eq:psixi}), there is $\ell_\gamma^+=\ell(r_\gamma^+)$ where $r^+_\gamma$ is the counter-rotating photon last circular orbit.}\label{Fig:Plotpsigamma}
\end{figure*}

According to the analysis in  \cite{Japan}--see also \cite{long,letter,Multy}--we can relate  some energetic characteristics of the orbiting disks to the thickness of the super-critical tori flows (with   $K=K_s\in ]K_\times, 1[$ for accretion driven  flows). Considering  polytropic  fluids, at the inner edge (cusp)  the flow is essentially pressure-free.
To consider all possible cases,
we  fix   the cusp location, the throat thickness and  location,  fluid momenta $\ell=\ell_{ecc}$ and  $K=K_s$ parameter  according to  the following definitions:
\bea&&\label{Eq:psixi}
K_s(\xi,\ell;a)\equiv K_{\times}+\frac{1-K_\times}{\xi},
\quad \ell_{ecc}(\psi;a)\equiv \ell_{mso}^++\frac{\ell_{mbo}^+-\ell_{mso}^+}{\psi},\quad\mbox{where}\quad  K_\times(\ell;a)= K(r_\times),
\eea
where $ r_\times$ is in Eqs\il(\ref{Eq:toricenter-inner}) and  $(\xi,\psi)$ are two positive constants regulating  the
momentum $\ell$ and the $K$ parameter in  the accretion driven range of values--Figs\il(\ref{Fig:PPlotrangqwals}), with
\bea\nonumber
\xi\in [1,+\infty],\quad \psi\in [1,+\infty],\quad
\lim\limits_{\xi \rightarrow 1} K_s=1,\quad
\lim\limits_{\xi \rightarrow +\infty} K_s=K_\times,\quad\lim\limits_{\psi \rightarrow 1}\ell_{ecc}=\ell_{mbo}^+,\quad
\lim\limits_{\psi \rightarrow +\infty}\ell_{ecc}=\ell_{mso}^+.
\eea
(However    in the range   $\psi\in[0,1[$,  momenta $\ell_{ecc}(\psi)$ describe proto-jet driven flows where $\ell\in \mathbf{L_2}$, or tori  with $\ell\in\mathbf{L_3}$).  The limiting value of $\psi$   in this case  is  $\psi_\gamma\equiv ({\ell_{mbo}^+-\ell_{mso}^+})/({\ell_\gamma^+-\ell_{mso}^+})$, which is a function of $a/M$--Fig.\il(\ref{Fig:Plotpsigamma}).

  While  $r_\times$ is the cusp location fixed by $\ell=\ell_{ecc}$, radius    $r_s<r_\times$ is related to  the accreting matter flow thickness and determined by the parameter $K_s$--see Figs\il(\ref{Fig:PPlotrangqwals}).
Consider  counter-rotating  tori
with pressure
 $p=\kappa \varrho^{1+1/n}$, where $\gamma\equiv 1+1/n$ is the polytropic index and $\kappa $ is a polytropic constant. The
mass-flux, the  enthalpy-flux (related to the  temperature parameter),
and  the flux thickness can be  estimated
as \emph{$\Gamma$--quantities}, having  general form  $\Gamma (r_\times,r_s,n)=\beta_1(n,\kappa)(W_s-W_{\times})^{\beta_2(n)}$, where $\{\beta_1(n,\kappa),\beta_2(n)\}$ are  functions of the polytropic index and constant  and
  $W=\ln K$. (More specifically the $\Gamma$--quantities are:
 the $\mathrm{{{Enthalpy-flux}}}=\mathcal{D}(n,\kappa) (W_s-W)^{n+3/2}$;
the  $\mathrm{{{Mass-Flux}}}= \mathcal{C}(n,,\kappa) (W_s-W)^{n+1/2}$ and
the  $\mathcal{\bar{L}}_{\times}/\mathcal{\bar{L}}= \mathcal{B}/\mathcal{A} (W_s-W_{\times})/(\eta c^2)$,
  which is  the  fraction of energy produced inside the flow and not radiated through the surface but swallowed by central \textbf{BH}.    While $\mathcal{\bar{L}}$ is  the total luminosity, and  $(\mathcal{D}(n,\kappa),\mathcal{C}(n,\kappa),\mathcal{A}(n,\kappa),\mathcal{B}(n,\kappa))$ are functions of the polytropic index and polytropic constant, $\dot{M}$ is the total accretion rate where, for a stationary flows, $\dot{M}=\dot{M}_{\times}$ and
$\eta\equiv \mathcal{L}/\dot{M}c^2$  is the efficiency.)

\begin{figure*}
\centering
     \includegraphics[width=5.6cm]{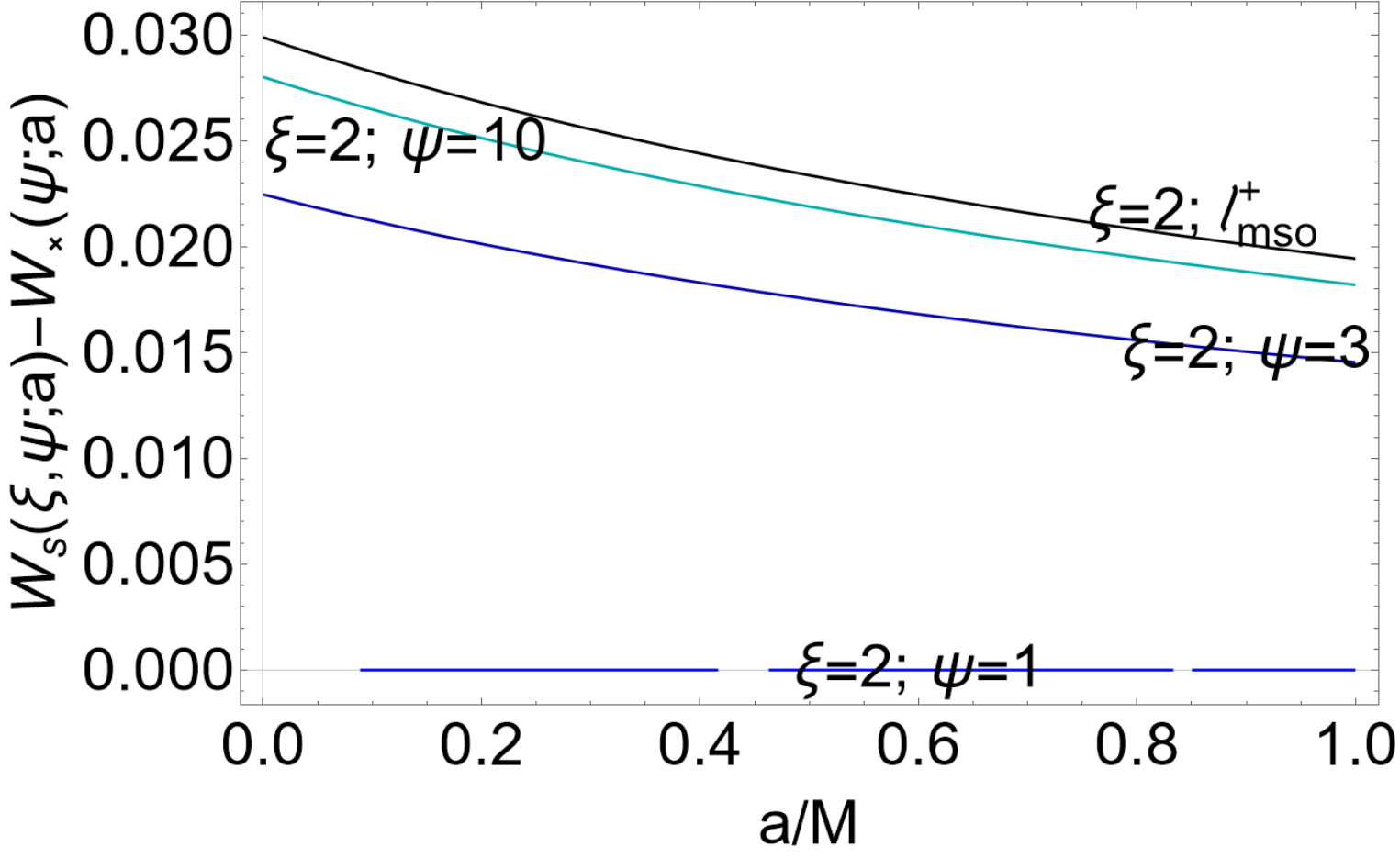}
          \includegraphics[width=5.6cm]{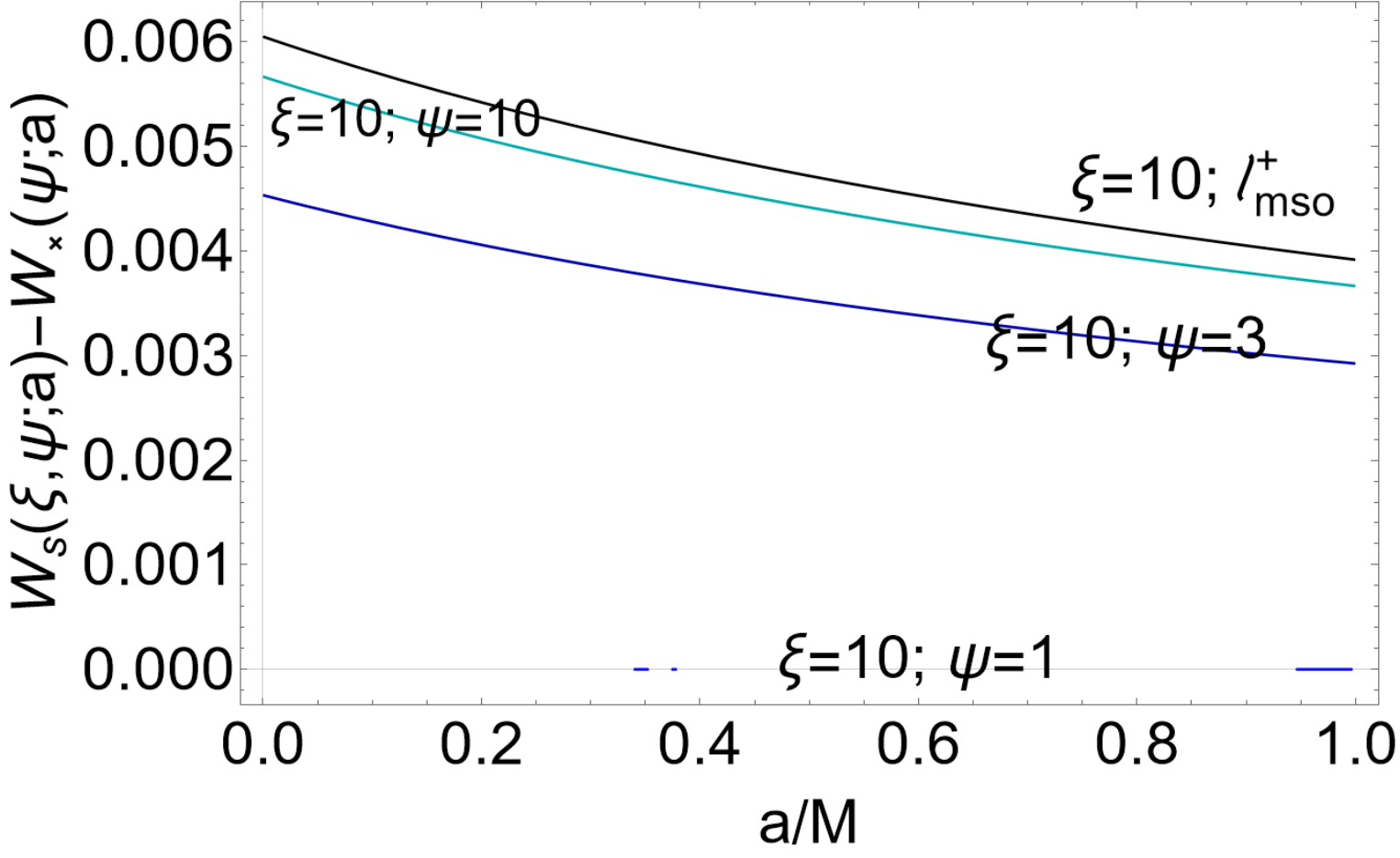}
        \includegraphics[width=5.6cm]{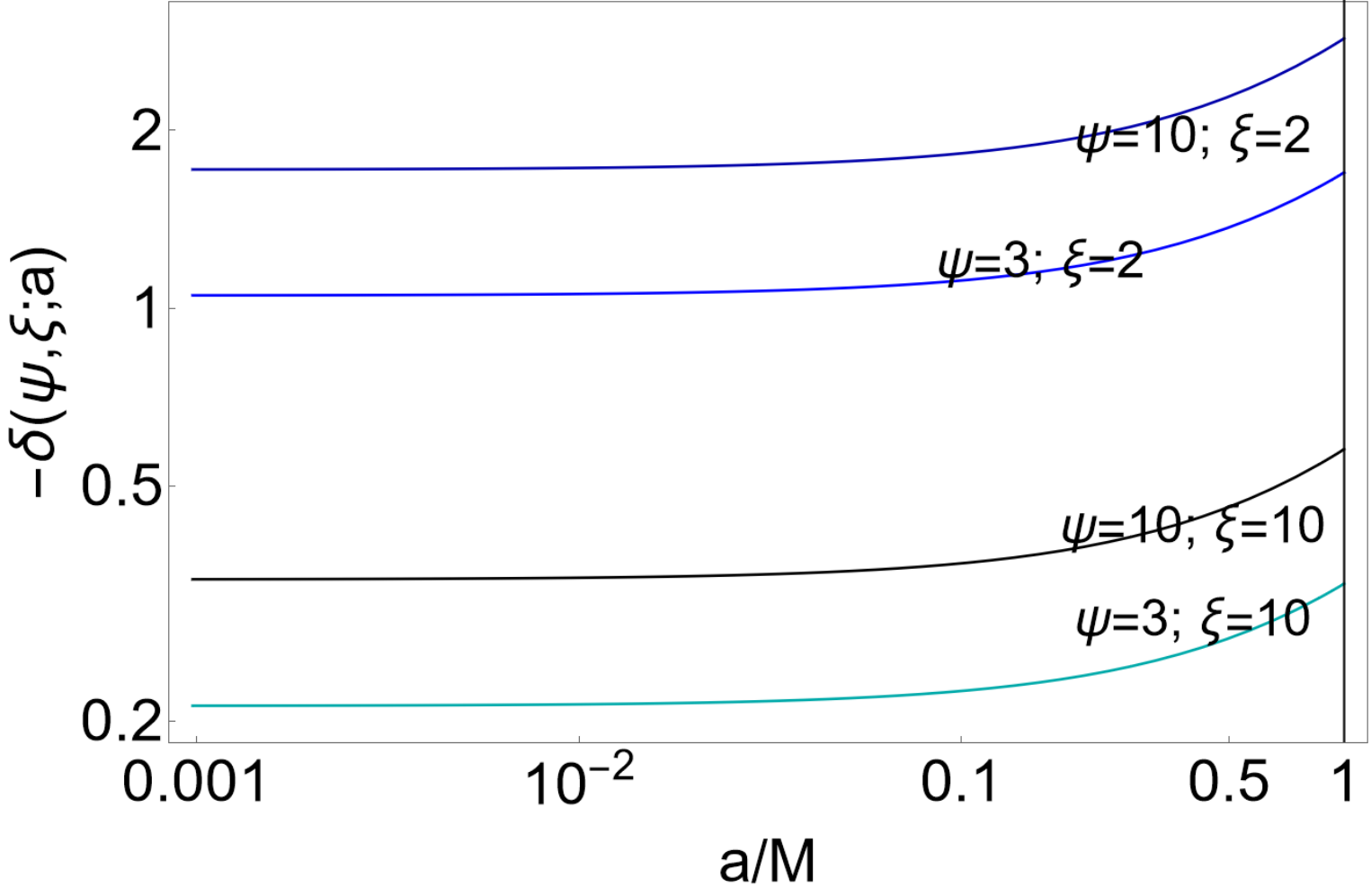}
\caption{Difference $ W_s-W_\times$,  defining  the $\Gamma$--quantities introduced  in Sec.\il(\ref{Sec:flow}),  where $W\equiv \ln K$ for $\xi=2$ (left panel) and $\xi=10$ (center panel) are shown as functions of the \textbf{BH}  spin-mass ratio $a/M$, at different fluid momenta $\ell$ regulated by the $\psi$ parameters signed on the curves--see  Eq.\il(\ref{Eq:psixi}). The $\Gamma$--quantities regulate the  mass--flux, the  enthalpy--flux (related to the  temperature parameter),
and  the flux thickness. Right panel:  $\delta$--quantities (regulating    the cusp luminosity, measuring the
rate of the thermal--energy    carried at the  cusp,   the disk accretion rate,   and  the  mass flow rate through the cusp i.e., mass loss accretion rate) defined in Sec.\il(\ref{Sec:flow}) are shown for different values of $\psi$ (regulating the momenta) and $\xi$ (regulating   $K_s\in]K_\times,1[$) signed on the curves.}\label{Fig:Plotrangqwalsldopo}
\end{figure*}

We  examine also the \emph{$\delta$-quantities},  having  general form  $\delta={\Gamma}(r_\times,r_s,n) r_\times/\Omega(r_\times)$; where  $\Omega(r_\times)$ is the relativistic angular  frequency  at  the   tori cusp  $r_\times$ where the pressure vanishes.
 The  $\delta$-quantities  regulate   the cusp luminosity, measuring the
rate of the thermal-energy    carried at the  cusp,   the disk accretion rate,   and  the  mass flow rate through the cusp (i.e., mass loss accretion rate). (More specifically, the $\delta$--quantities are:
 the cusp luminosity
$\mathcal{L}_{\times}={\mathcal{B}(n,K) r_{\times} (W_s-W_{\times})^{n+2}}/{\Omega(r_{\times})}$,
 the disk accretion rate     $\dot{m}= \dot{M}/\dot{M}_{Edd}$ (compared to the characteristic Eddington accretion rate);
the  mass flow rate through the cusp
 $\dot{M}_{\times}={\mathcal{A}(n,K) r_{\times} (W_s-W_{\times})^{n+1}}/{\Omega(r_{\times})}$,  where $(\mathcal{A}(n,\kappa),\mathcal{B}(n,\kappa))$ are functions of the polytropic index and polytropic constant.).
In the analysis of  Figs\il(\ref{Fig:Plotrangqwalsldopo})  we  assumed $\beta_1=\beta_2=1$.
It is clear that these  quantities, depending on the details of the toroidal models, provide  a wide estimation for more refined  tori models. Nevertheless  with this analysis    these quantities  can be estimated in relation with fluid thickness  (regulated by $W_s-W_\times$), the tori  distance from the central attractor  (cusp location  $r_\times\in [r_{mbo}^+, r_{mso}^+]$  fixing  also the center of maximum density and pressure in the disk)  and the \textbf{BH} spin.

The toroidal models used in this analysis  are shown in Figs\il(\ref{Fig:PPlotrangqwals}) in terms of   $\ell_{ecc}$, cusp location $r_\times$, the  flow turning radius $r_\Ta$,  and parameter values $K_s$ and $K_\times$.
As clear from   Figs\il(\ref{Fig:PPlotrangqwals}) and Figs\il(\ref{Fig:Plotrangqwalsldopo}),  the  $\Gamma$--quantities decrease with the \textbf{BH} spin $a/M$, with  decreasing  $K_s\in]K_{\times},1[$, and decrease with  increasing $\ell\in]\ell_{mbo}^+,\ell_{mso}^+]$ in magnitude.  The $\delta$--quantities  increase with the \textbf{BH} spin mass ratio,  with  increasing  $K$ and with the decrease of $\ell$ in magnitude. For greater values of $\Gamma$-- and $\delta$--quantities,  the flow  thickness, regulated by  the  $K$  parameter (larger values of  $(K_s-K_\times)$ correspond to  larger flow thickness) is   mostly dependent on the background  properties, especially for fast spinning attractors. (For fast spinning attractors  the flow thickness  is  in fact largely  independent on the tori details and properties).
Furthermore, for the counter-rotating flows,   the closer to the \textbf{BH} the tori are  (smaller magnitude of the $\ell$) the greater  $\Gamma$-- and $\delta$--quantities  are\footnote{The $\delta$-- and $\Gamma$--quantities depend on the polytropic index and constant through  the functions  $(\beta_1, \beta_2)$. The dependence on the polytropic  affects the dependence on the  \textbf{BH} spin mass ratio. We show this different behaviour in dependence according to $\beta_2\neq1$ in Figs\il(\ref{Fig:Plotrangqwalsldopoprova2}) and Figs\il(\ref{Fig:Plotmetripreda1}),  where we note that, according to different values of $\beta_2$, $\delta$--quantities decrease or increase with the \textbf{BH} spin-mass ratio according also with the analysis of  \cite{Multy,letter}.}.
 The counter-rotating  tori energetics  are mostly dependent from the characteristics of the accreting tori   for slowly  spinning attractors  (where   similarities are with the  corotating  tori), which are closer to the attractor  and with smaller magnitude of momenta (note that the  flow reaches  the central  attractor with negative $\ell$ {and, on each trajectory, there  could also be  two turning points}).
\begin{figure*}
\centering
    \includegraphics[width=5.5cm]{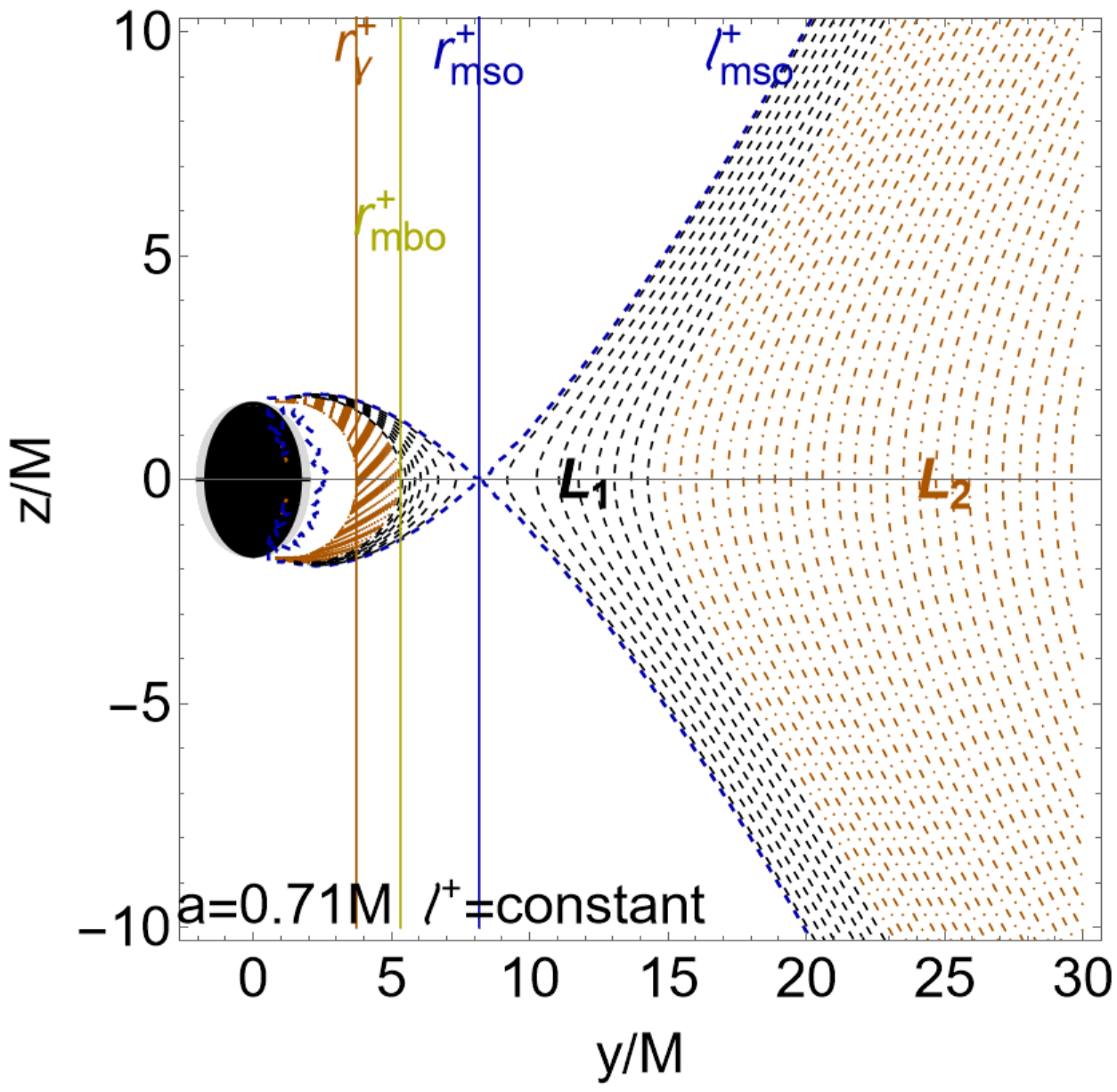}
              \includegraphics[width=5.5cm]{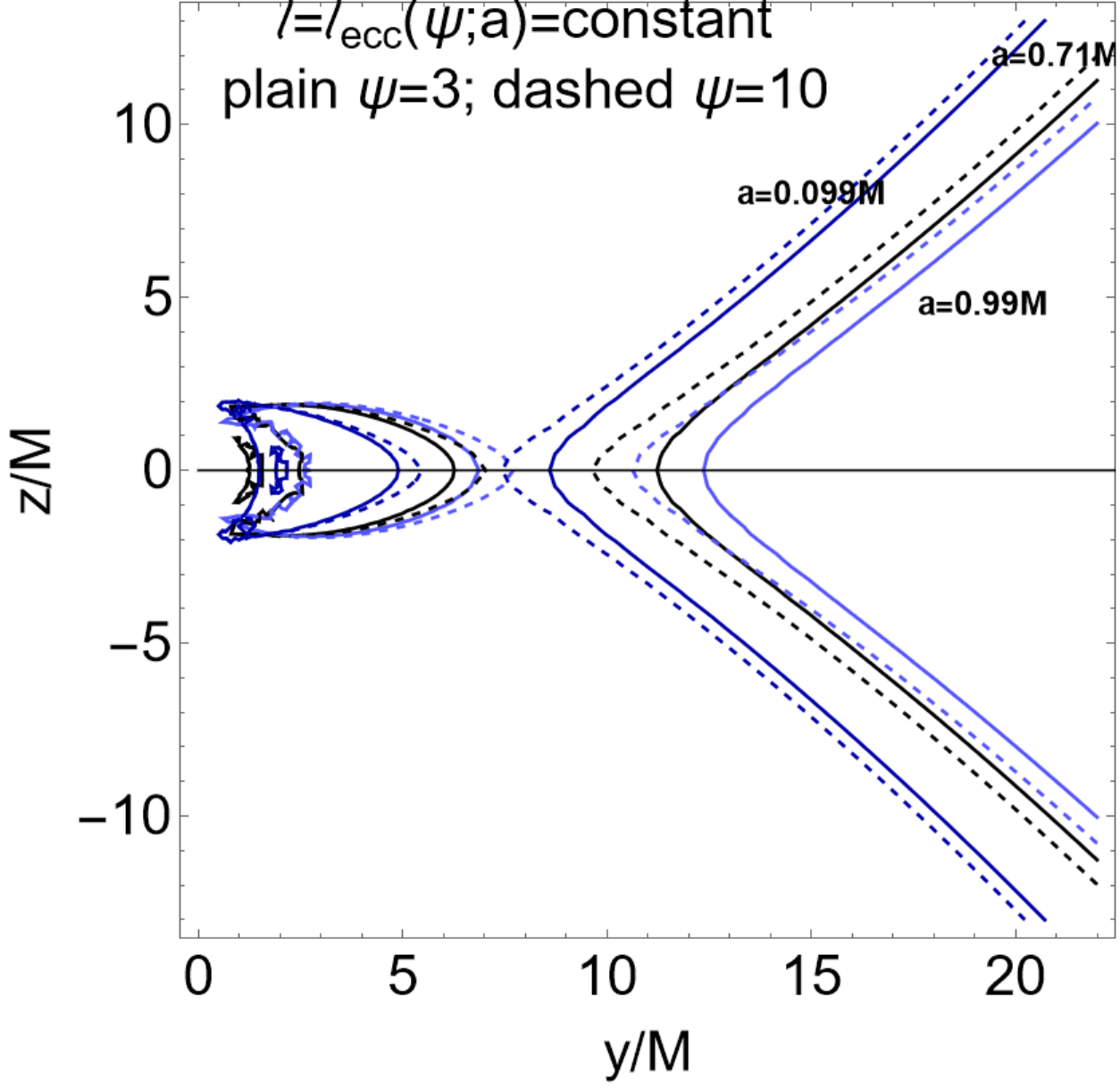}
                \includegraphics[width=6.5cm]{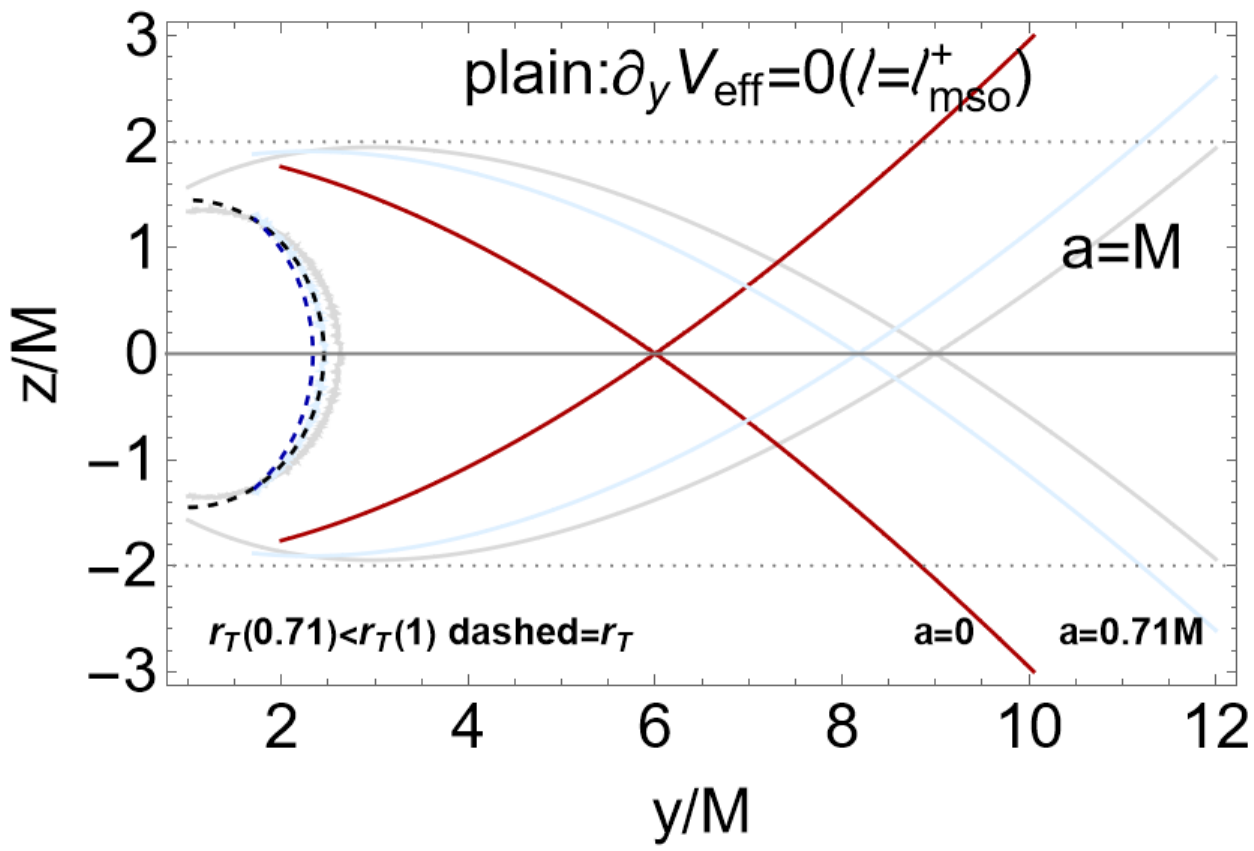}
          	\includegraphics[width=5.5cm]{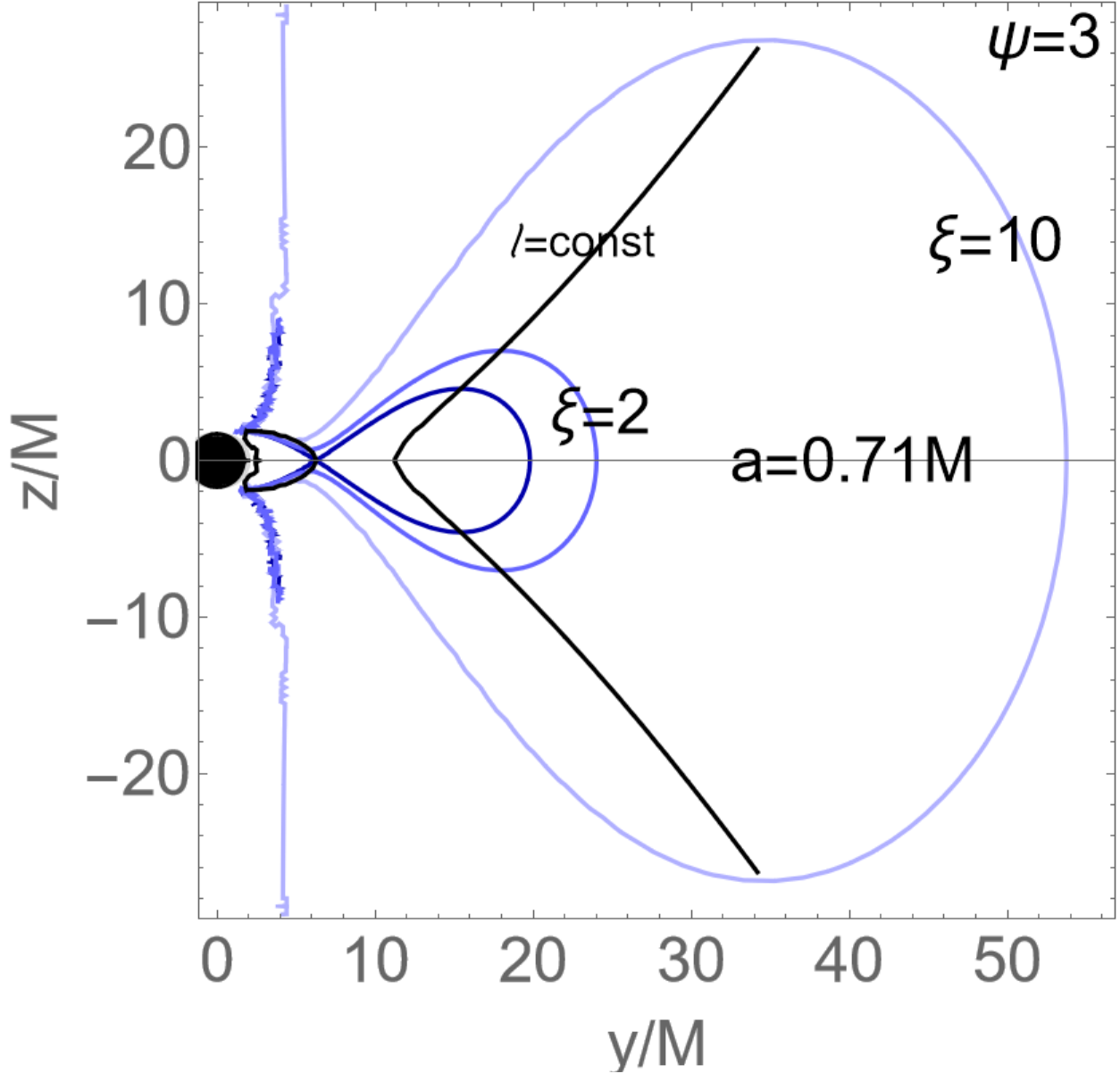}
      \includegraphics[width=5.5cm]{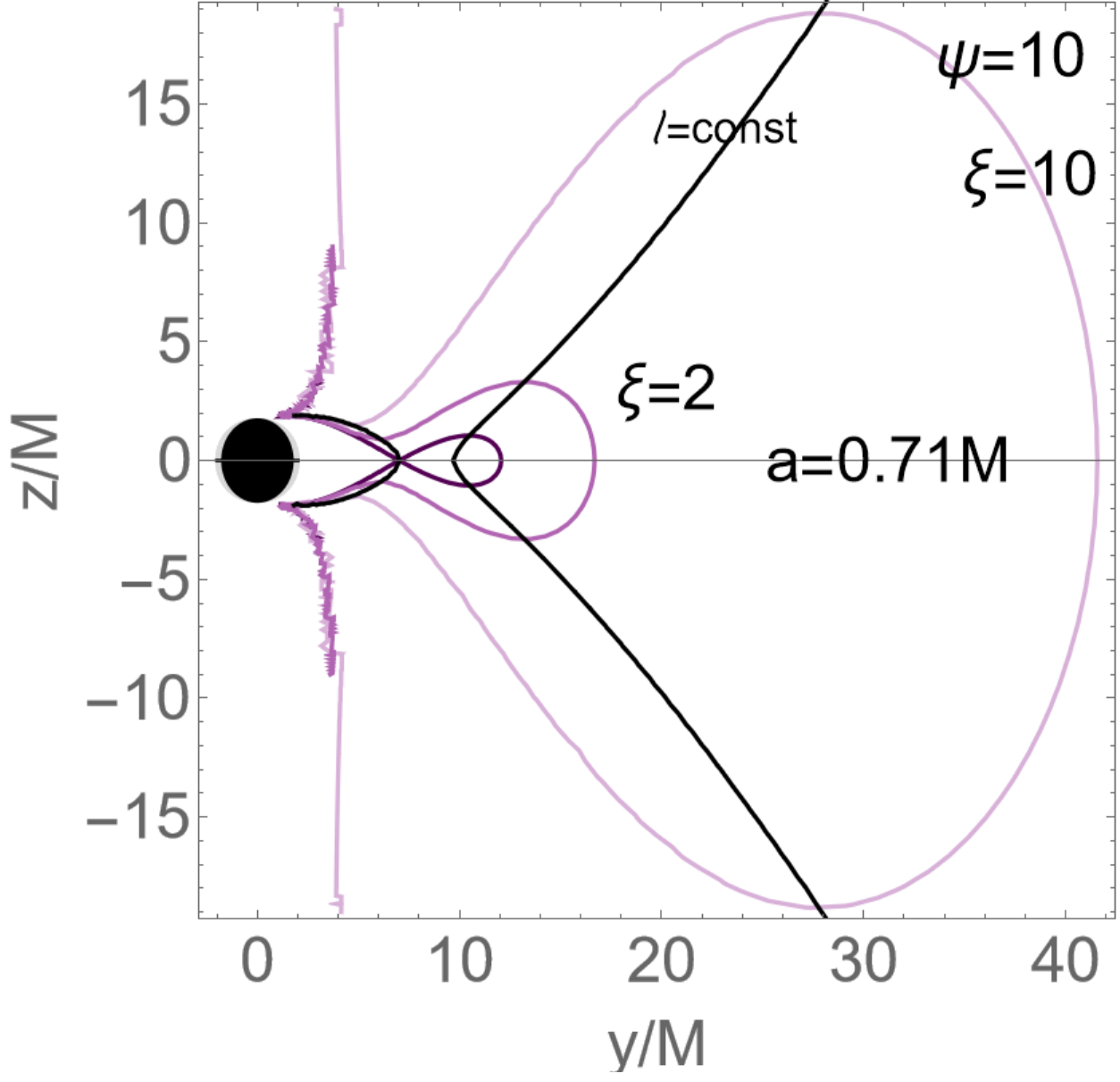}
       \includegraphics[width=6.5cm]{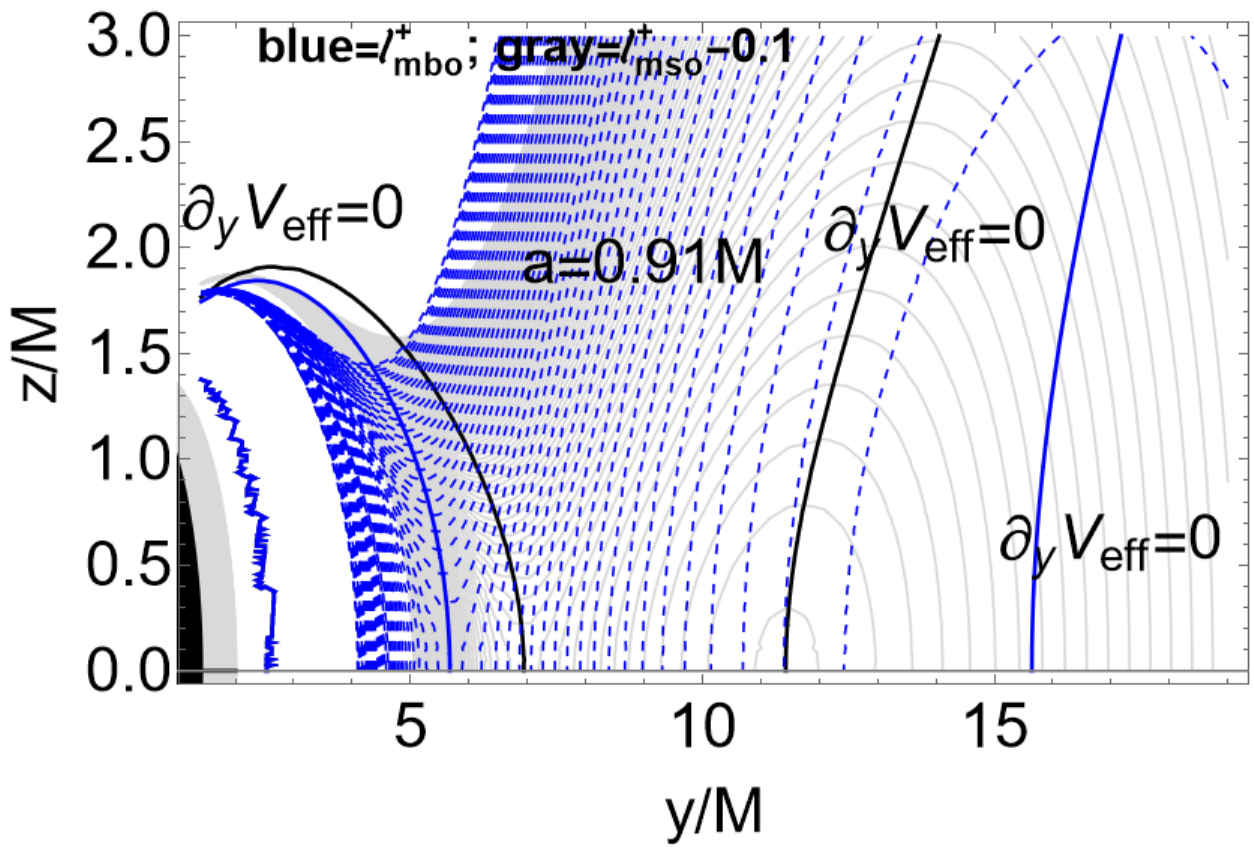}
      \caption{Tori models: equidensity surfaces and surfaces of constant $\ell$,  related to the analysis of Sec.\il(\ref{Sec:flow}). Black region is the central \textbf{BH}, gray region is the outer ergoregion, there is $r=\sqrt{z^2+y^2}$ and $\sigma =(\sin\theta)^2={y^2}/({z^2+y^2})$, and $\ell$ is the fluid specific angular momentum. Upper left panel: solutions of equation $\ell^+=\partial_yV_{eff}(a;y,z,\ell)=0$, coincident with $\ell^+(x,y,a)=$constant are shown  in the  \textbf{BH} spacetime with spin $a=0.71M$. Momenta $\ell$ are in the range $\mathbf{L_2^+}\equiv[\ell_{\gamma}^+,\ell_{mbo}^+]$, for counter-rotating proto-jets driven flows (dotted--dashed dark-- yellow curves), and range  $\mathbf{L_1^+}\equiv[\ell_{mbo}^+,\ell_{mso}^+]$, for counter-rotating cusped tori driven flows (dashed  black curves). Momenta and geodesic radii shown in the panels are defined in Eqs\il(\ref{Eq:def-nota-ell}). Curves connect  the center of maximum density and pressure  with the tori geometrical maxima in the range $y=r>r_{mso}^+$,  and the cusp $r_\times$ (minimum of pressure), fixed by  momentum $\ell^+=$constant with the minima of the matter throat $(K>K_\times)$, providing thus indication of tori (and matter  throat) thickness. Central upper panel shows the  curves  $\ell=$constant  for different \textbf{BH} spins $a/M$ signed on the curves and for momenta $\ell_{ecc}$ with $\psi=3$ and $\psi=10$ of Eq.\il(\ref{Eq:psixi}). Right  upper panel:  limiting counter-rotating  solutions for $\ell=\ell_{mso}^+$, in the \textbf{BH} spacetimes   $a=0$ (Schwarzschild), $a=0.71M$ and $a=M$ (extreme Kerr \textbf{BH}), and the related turning radius $r_\Ta(\sigma)$, plotted as inner dashed curves. Below panels: cusped and super--critical   tori at $\psi=3$ (left) and $\psi=10$ (center), where $\psi$ fixes the fluid momenta $\ell_{ecc}$, and different $\xi$ signed on the curves ($\xi$ fixes the $K_s$ parameter) of Eq.\il(\ref{Eq:psixi}). Black curves are solutions of $\ell^+(y,z,a)=$constant. Right panel: toroidal surfaces in the \textbf{BH} spacetimes $a=0.91M$ at  $\ell=(\ell_{mso}^+-0.1)$(gray line) and $\ell_{mbo}^+$ (blue-dashed lines). Blue  and black plain lines are curves  $\ell(x,y,z):\partial_y V_{eff}=0$ constant.}\label{Fig:PlotsuppoB}
\end{figure*}

\medskip

\textbf{Maximum density and pressure points and tori thickness}

\medskip

Introducing the  coordinates $(y,z): r=\sqrt{y^2+z^2}$ and $\sigma =(\sin\theta)^2={y^2}/({z^2+y^2})$,  solutions of $\ell^+=\partial_yV_{eff}(a;z,y,\ell)=0$, coincident with solutions of  $\ell^+(x,y,a)=$constant, shown in Figs\il(\ref{Fig:PlotsuppoB})\footnote{This result is framed in the set of results known as  von Zeipel theorem, holding  for barotropc tori in axisymmetric spacetimes \citep{zanotti,Koz-Jar-Abr:1978:ASTRA:,M.A.Abramowicz,Chakrabarti,Chakrabarti0}. The surfaces known as   von Zeipel's cylinders, are defined by the conditions:
$\ell=\mbox{constant}$ and $\Omega=\mbox{constant}$. More precisely, the von Zeipel condition states that  the surfaces of constant
pressure coincide with the surfaces of  constant density if and only if the surfaces with the  angular momentum $\ell=
\mbox{constant}$ coincide with the surfaces with constant angular  velocity. In the stationary  and axisymmetric  spacetimes, the family of von
Zeipel's surfaces  does not depend on the particular rotation law of the fluid,
$\Omega=\Omega(\ell)$, but on the background
spacetime only. In the case of a barotropic fluid,   von Zeipel's theorem guarantees that
the surfaces $\Omega=\mbox{constant}$ coincide with the surfaces  $\ell=\mbox{constant}$. },  are the curves  connecting  the center of maximum density and pressure for barotropic fluids (uniquely fixed by  the $\ell$ parameter) with the tori  geometrical maxima (regulating the tori thickness and fixed by $K$) in the range $y=r>r_{mso}^+$ for cusped tori, and  the curve connecting the cusp $r_\times$ (minimum of pressure, uniquely fixed by  the $\ell$ parameter), with the extremes of the flow  throat  (regulating the flow thickness and fixed by $K$), and therefore provides the throat thickness for super-critical tori--Figs\il(\ref{Fig:PlotsuppoB}).
Therefore, the throat thickness, similarly to the turning point function $r_\Ta(\sigma_\Ta)$, is governed by the  fluid specific angular momentum.

\medskip

\textbf{The maximum thickness of the  accretion throat}

\medskip

{ More specifically,
solution $\ell(x,y,z):\partial_y V_{eff}=0$ provides, for fixed $a/M$, the curve $\mathcal{C}^+_{\ell^+}$, representing the  torus center, i.e. maximum of fluid pressure  (on $z=0$), and  torus geometrical maximum (on $z>0$), at $r>r_{center}$,  for any  $K\geq K_{center}$ and, at $y=r<r_{inner}$ (torus inner edge) the curve $\mathcal{C}^-_{\ell^+}$, containing  the  minimum of fluid pressure (on $z=0$), i.e. torus cusp $r_\times$ or proto-jets cusp $r_j$, and the  geometrical minimum (on $z>0$)  of the throw boundary  (region $y<r_{\times}$) for the over-critical tori ($K>K_\times$). Therefore curve $\mathcal{C}^-_{\ell^+}$  provides in  this sense a \emph{definition of  throat thickness}--see  Figs\il(\ref{Fig:PlotsuppoB})-\emph{below panels}, while it is clear that the Roche lobe generally increases with $\ell$.
The analysis  of Fig.\il(\ref{Fig:PlotsuppoB})-\emph{upper left  panel} shows, for $y<r_{inner}$ that the higher  $\mathcal{C}^-_\ell$ curve (larger throat thickness)  is, at fixed $a/M$, for the curve $\mathcal{C}^-_{\ell_{mso}^+}$ defined by  momentum $\ell=\ell_{mso}^+$. For $|\ell^+|>|\ell_{mso}^+|$,  curves $\mathcal{C}^-_\ell$ are upper bounded by the curve  $\mathcal{C}^+_{\ell_{mso}^+}$, at $\ell=\ell_{mso}^+$, which is also the more extended on the equatorial plane (cusp located  far on the equatorial plane).  Fig.\il(\ref{Fig:PlotsuppoB})-\emph{upper center  and right panel}  show  that in this sense   the throat thickness  increases with the \textbf{BH} spin (increasing the \textbf{BH} spin $a/M$, the curves $\mathcal{C}^-_{\ell^+}$ (at $\ell(x,y,z)=$constant) stretch far from the central attractor with $y=r_{mso}^+$ (cusp of the curve at $\ell_{mso}^+$).). }

As  shown in  Figs\il(\ref{Fig:PlotsuppoB}),  the maximum thickness  of the flow  throat  for super-critical tori  is provided by the limiting solution with  $\ell^+=\ell_{mso}^+$, and therefore   determined \emph{only} by the background properties through  \textbf{BH} dimensionless spin.  The  maximum accretion throat thickness increases with the \textbf{BH}  spin $a/M$, reaching  its maximum at $a=M$. As the  cusp moves outwardly on the equatorial plane $(z=0)$  with increasing \textbf{BH} spin (and tori angular momentum magnitude),  the counter-rotating flow throat extends on the equatorial plane. The turning points $r_\Ta$ is a bottom limit of the throat as shown in  Figs\il(\ref{Fig:PlotsuppoB})--(upper-right-panel),  with  $\ell_{mso}^+$ being also  the outer boundary of the turning corona.

Although  the counter-rotating orbiting  tori may be also very large, especially at large \textbf{BH} spins,  the momentum  is limited in  $\mathbf{L_1}$ and
the width of the  throat, dependent only on  $K_\times-1$, seen  in Figs\il(\ref{Fig:PPlotrangqwals}) remains  very small  and included in a region whose vertical coordinate    $z_{throat}\in[-2M,2M]$  and generally  $z_{throat}>z_\Ta$. Therefore the \textbf{BH} energetics would  depend   on its spin  $a/M$ (as the  tori energetics  essentially depends on the \textbf{BH} spin) rather than on   the properties of the counter-rotating fluids or  the tori masses--Figs\il(\ref{Fig:Plotrangqwalsldopo}).
The fluid contributes to the \textbf{BH} characteristic parameters  (spin $J$, total mass  $M$ and then rotational mass $M_{rot}$\citep{ella-correlation}),  with matter of momentum  $\ell<0: \ell\in ]\ell_{mbo}^+,\ell_{mso}^+[$, for the matter swallowed by the attractor, having a  turning point far from the ergosurface and the accretion throat--Figs\il(\ref{Fig:Plotrangqwalsldopo}).
This implies also that  the maximum amount of matter swallowed by the \textbf{BH} from the counter-rotating tori considered here  is constrained   by the limiting configurations with $\ell=\ell_{mso}^+$.
\section{On the fluids  at the turning point }\label{Sec:accelerations-fluids}
\begin{figure*}
\centering
    \includegraphics[width=5.5cm]{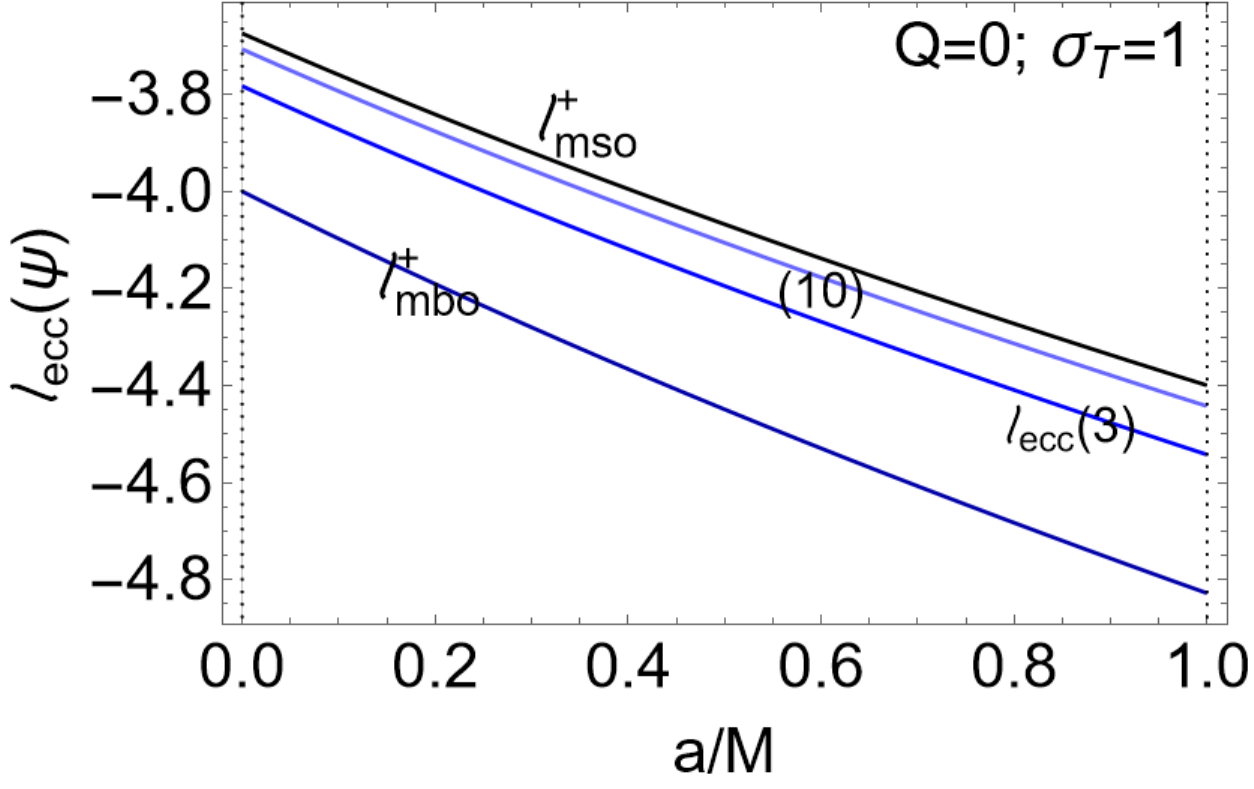}
        \includegraphics[width=5.5cm]{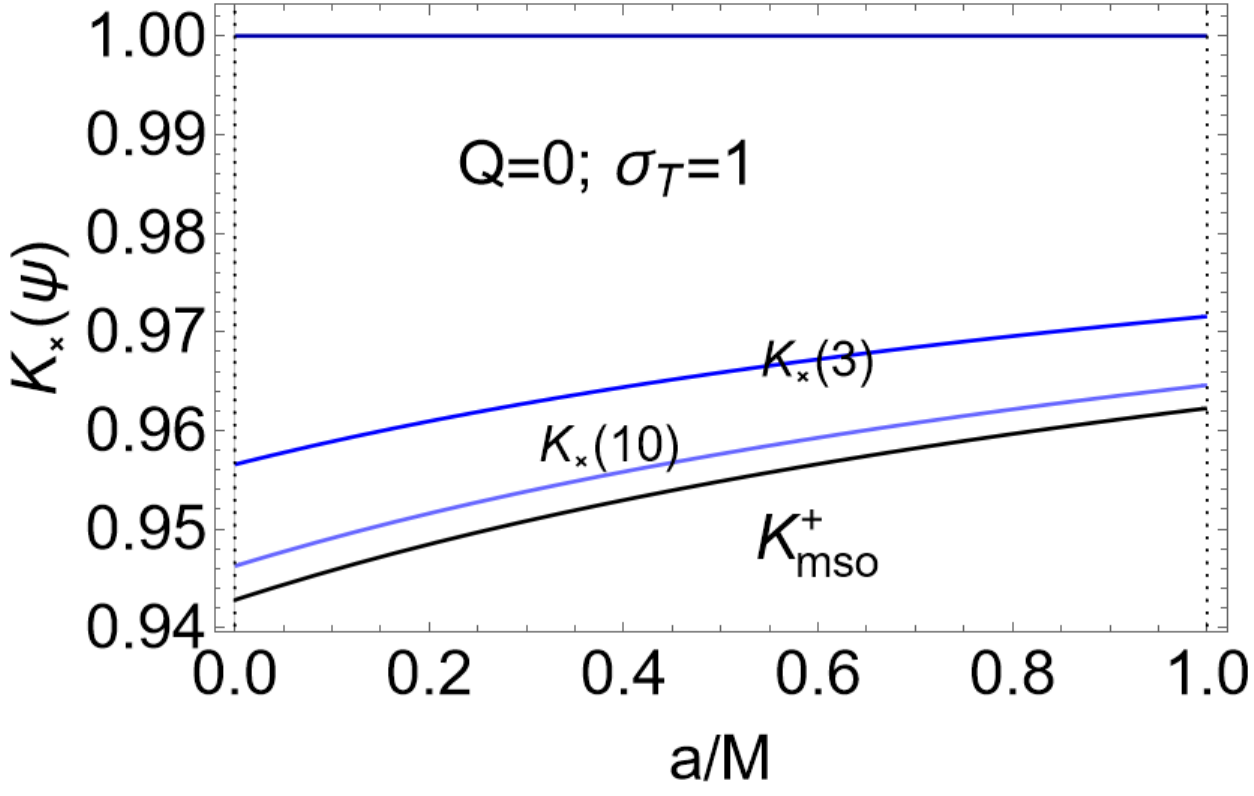}
              \includegraphics[width=5.5cm]{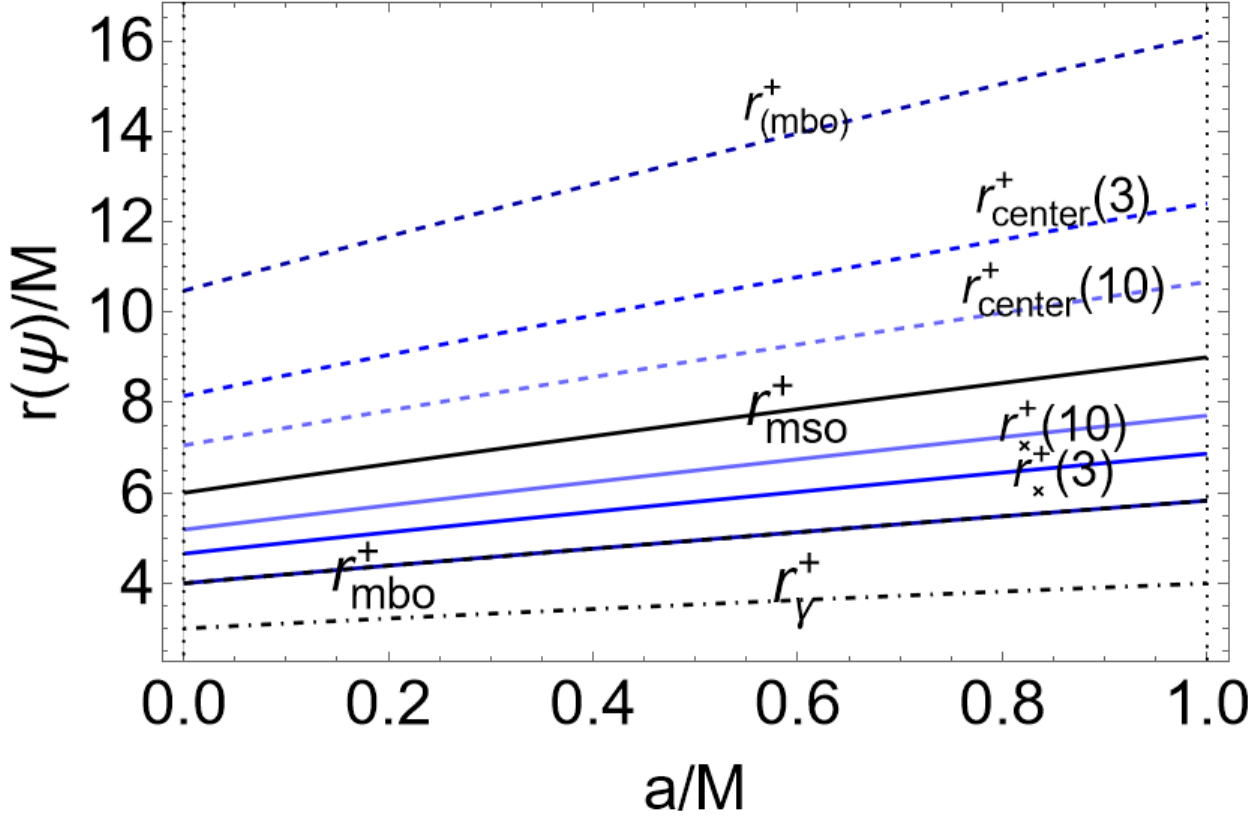}
    \includegraphics[width=6.5cm]{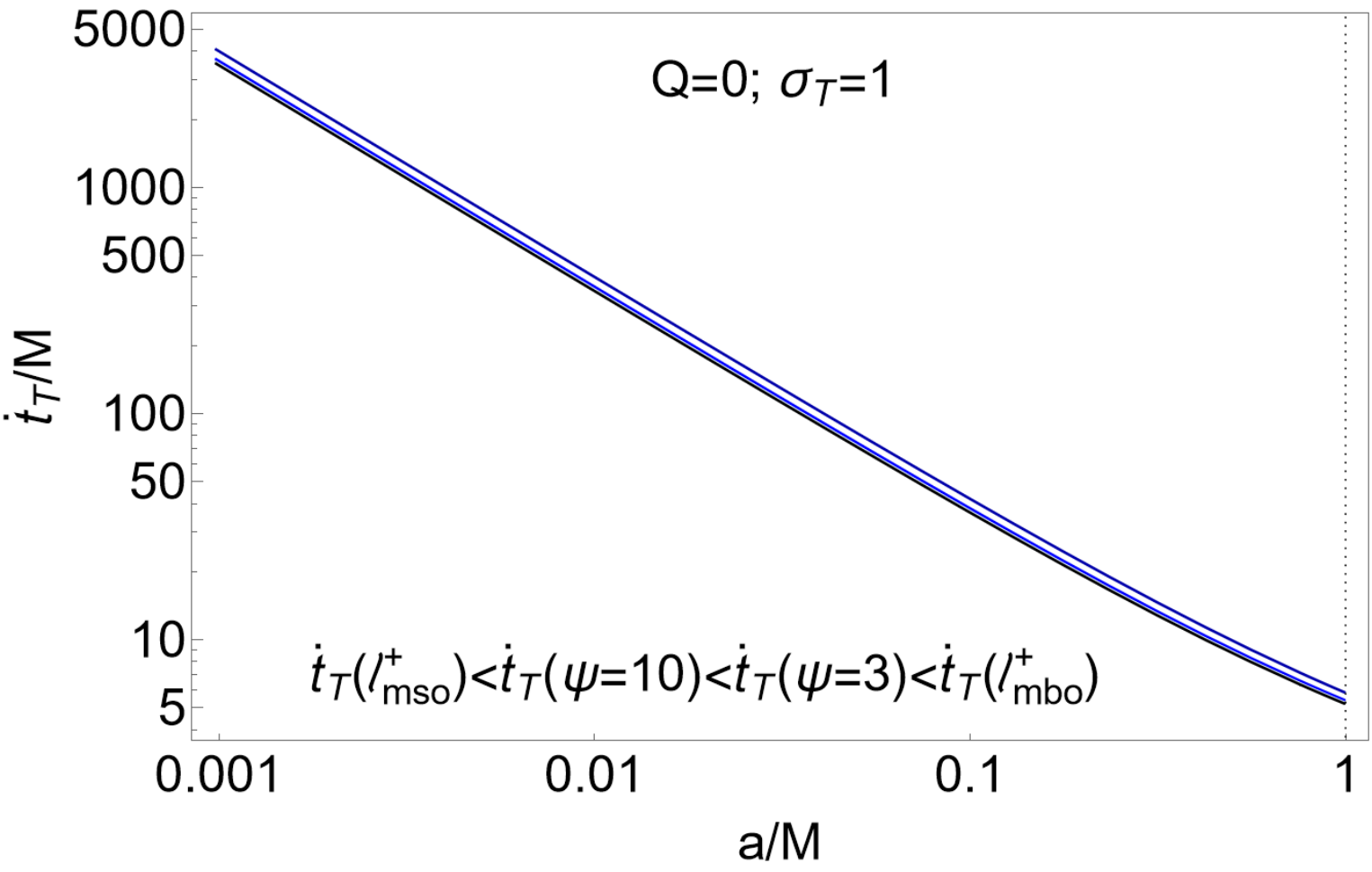}
     \includegraphics[width=6.5cm]{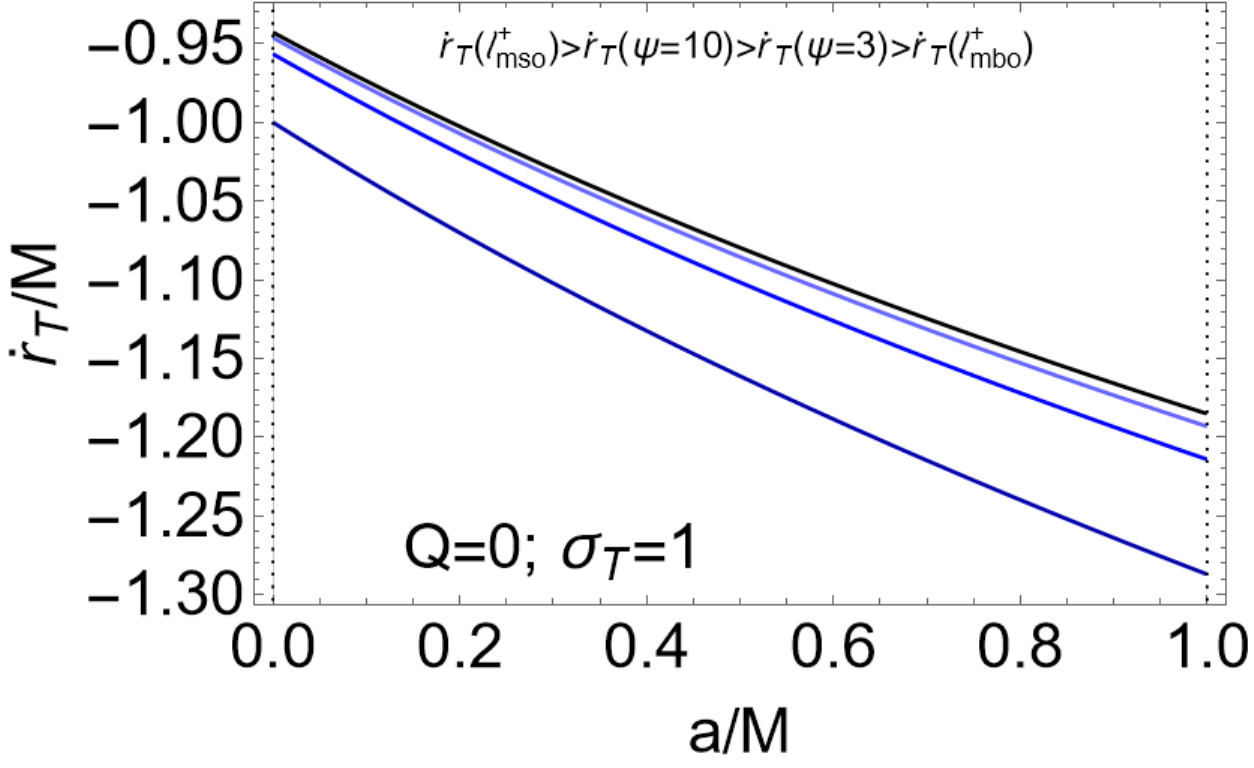}\\
          \includegraphics[width=6.5cm]{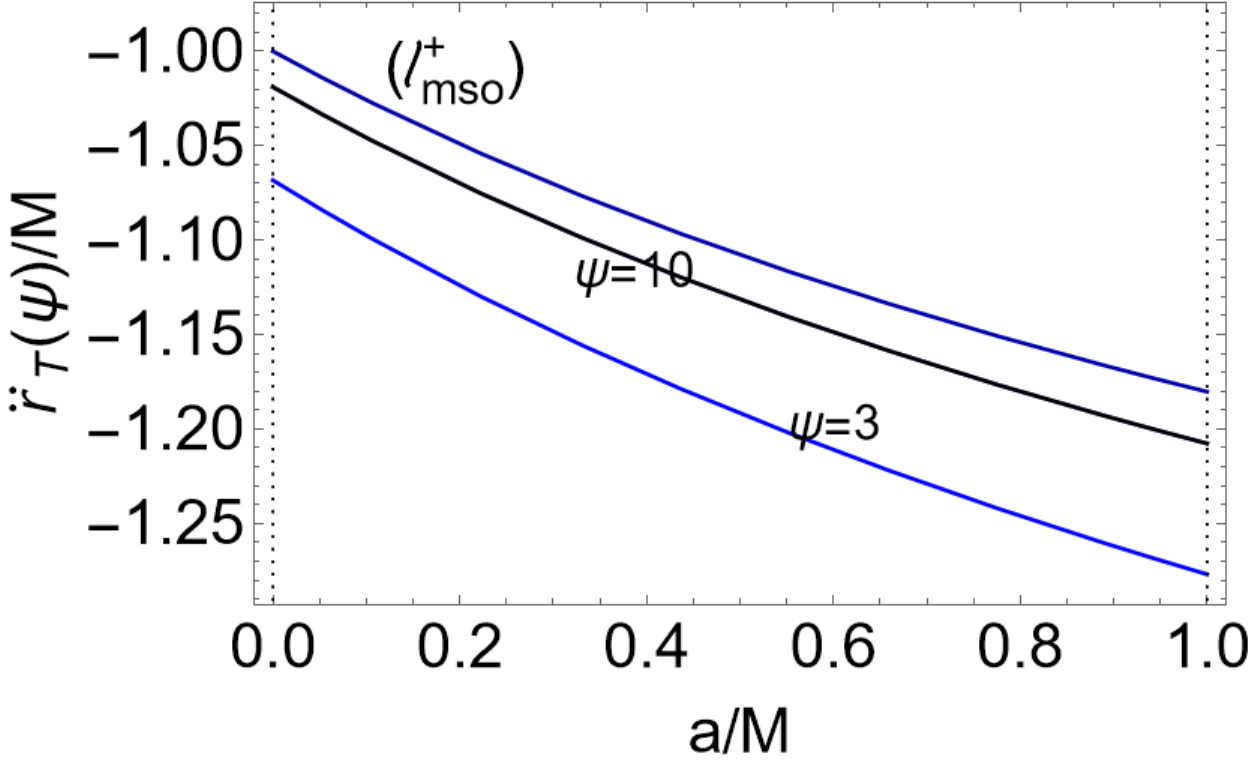}
        \includegraphics[width=6.5cm]{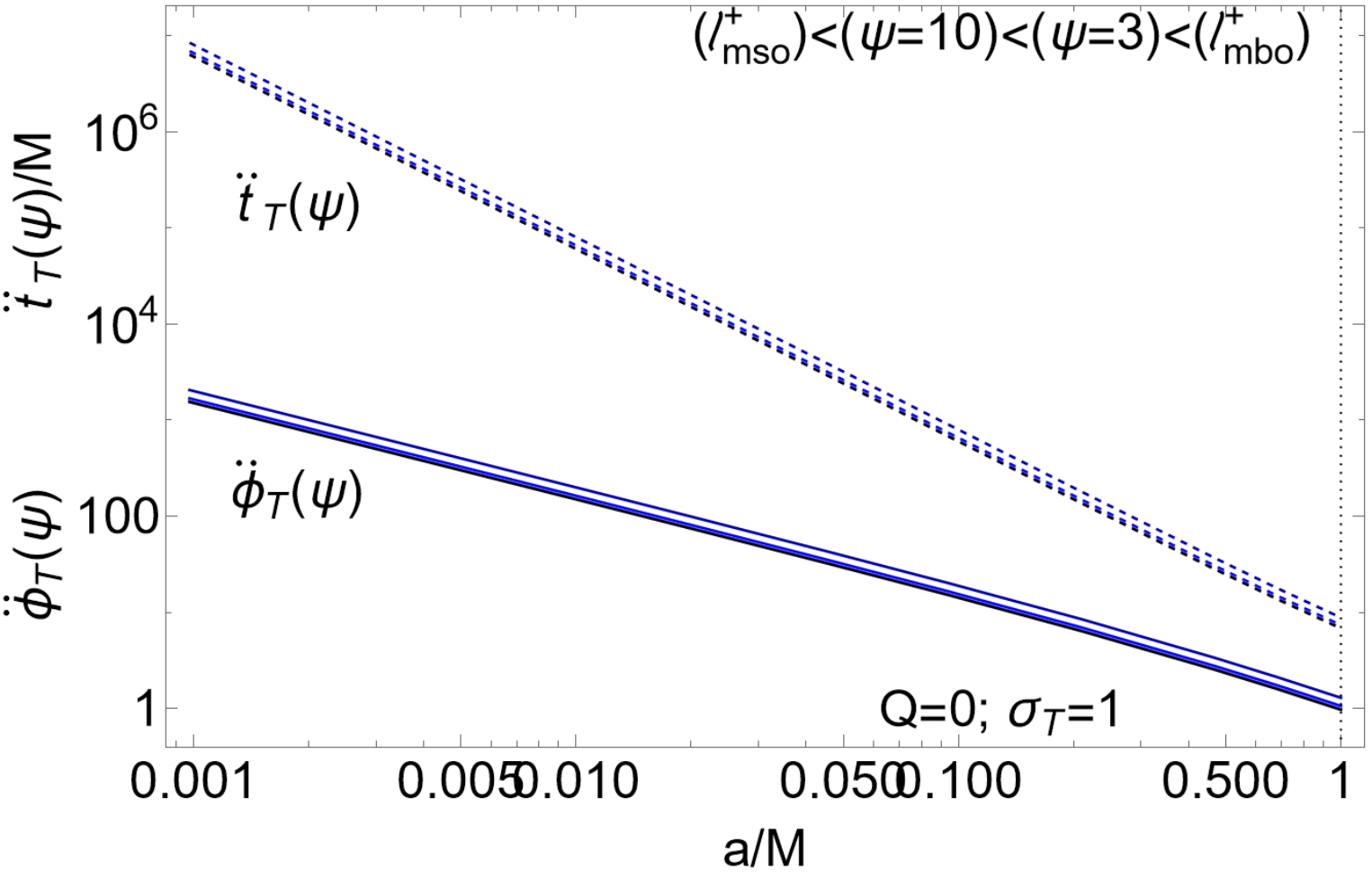}
  \caption{Analysis of the ingoing particles ($\dot{r}_\Ta<0$) at the tori driven counter-rotating  fluid turning point on the equatorial plane $\sigma_\Ta=1$ ($\sigma\equiv \sin^2\theta$), with Carter constant $\Qa=0$  and $\dot{\theta}=\ddot{\theta}=0$. Upper panels: tori models fixing the initial data for the infalling counter-rotating flows. Upper--left panel: fluid angular momentum Eq.\il(\ref{Eq:psixi}), functions of the \textbf{BH} spin-mass ratio $a/M$, for different values of the parameter $\psi$. Momenta      $\ell_{mso}^+$, $\ell_{mbo}^+$ and $\ell_{\gamma}^+$  and radii   $r_{mso}^+$, $r_{mbo}^+$, $r_{(mbo)}^+$ and $r_{\gamma}^+$ defining the   turning corona boundaries are defined in  Eqs\il(\ref{Eq:def-nota-ell}). Upper center panel:  cusped tori energy parameter $K$ evaluated at the cusp $r_\times$ of the counter-rotating torus, for different values of the  fluid specific angular momentum parameter $\psi$ signed on the curves.Upper-right panel: tori cusp location $r_\times$ and center $r_{center}$ as functions of the \textbf{BH} spin $a/M$, for different fluid angular momenta $\ell_{ecc}$. Center line panels: time velocity component $\dot{t}_\Ta$  (left panel) and radial component   $\dot{r}_\Ta$  (right panel) at the turning point of Eq.\il(\ref{Eq:general-on-equatorial-plane}), functions of the \textbf{BH} spin $a/M$ for different momenta  $\ell_{ecc}$.
  Below line  panels: fluid acceleration $\ddot{r}_\Ta$  (left panel) and $\ddot{t}_\Ta, \ddot{\phi}_\Ta$  (right panel) at the turning point, functions of the \textbf{BH} spin $a/M$ for different $\ell_{ecc}$. }\label{Fig:Plotexpermeant}
\end{figure*}

\begin{figure*}
\centering
          \includegraphics[width=6.5cm]{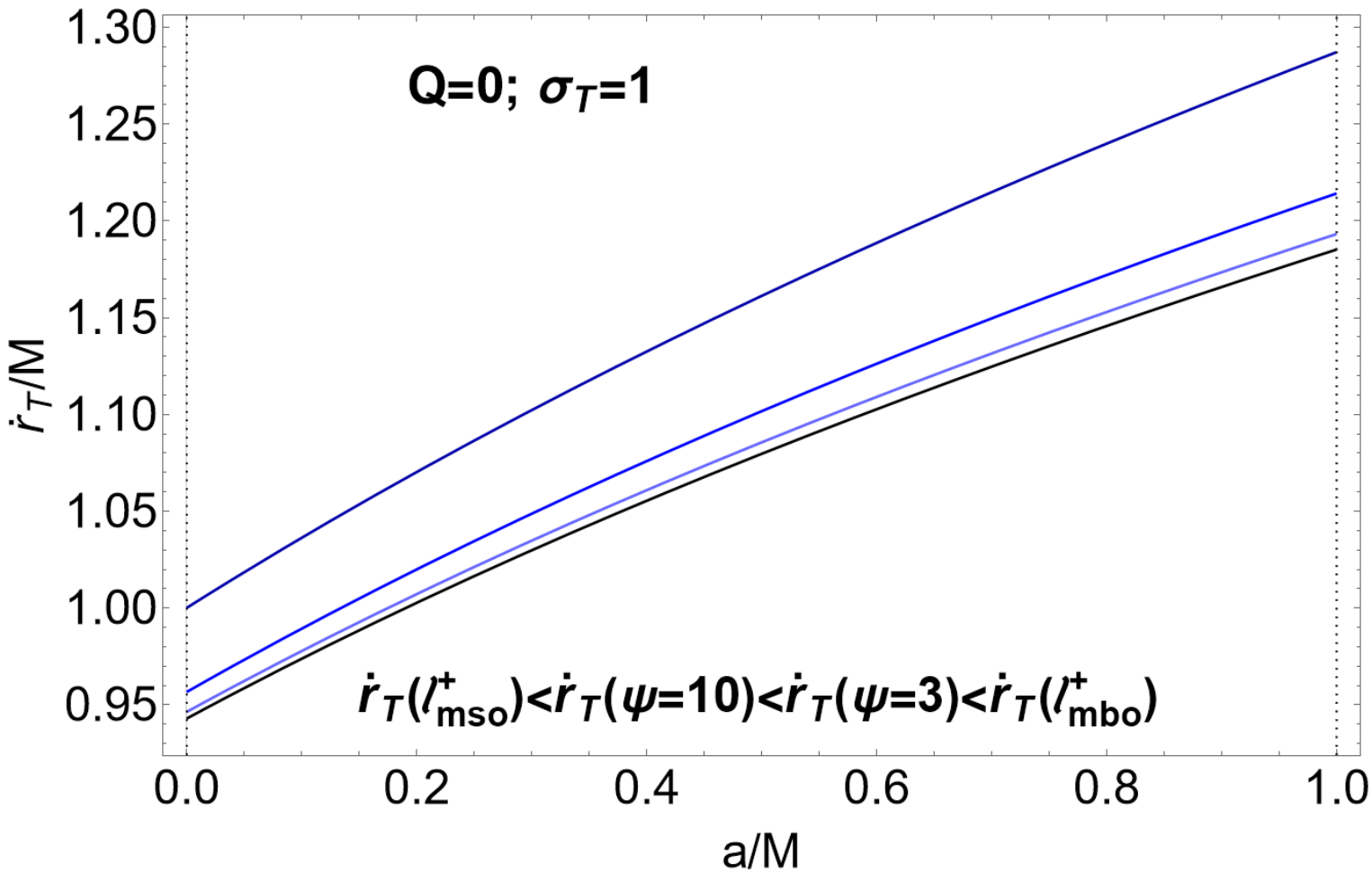}
        \includegraphics[width=6.5cm]{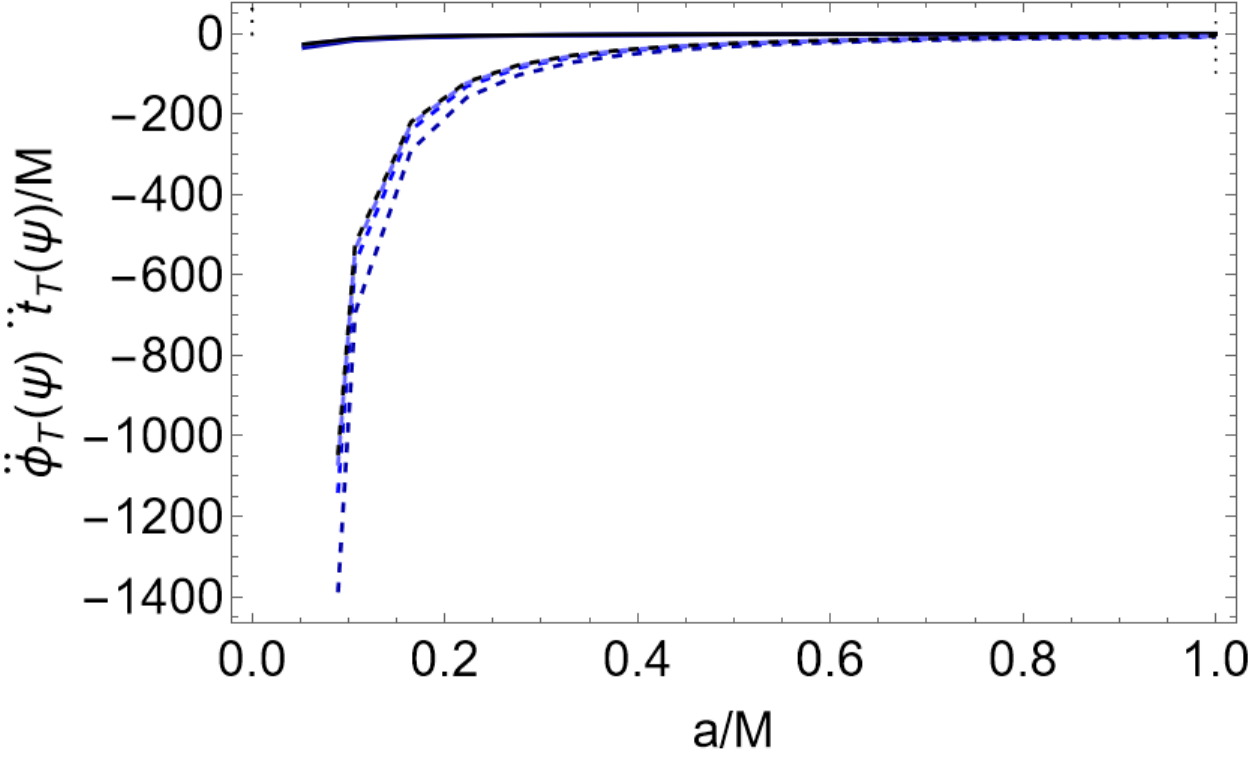}
  \caption{Tori driven counter-rotating flow particles accelerations  $(\ddot{\phi}_\Ta, \ddot{t}_\Ta)$ (right panel)  and  radial velocity  $\dot{r}_\Ta$ (left panel) at the fluid turning point on the equatorial plane $\sigma_\Ta=1$  (where $\sigma\equiv \sin^2\theta$), with Carter constant $\Qa=0$ and $\dot{\theta}=\ddot{\theta}=0$,  for outgoing particles $\dot{r}_\Ta>0$ from  the toroidal models of Figs\il(\ref{Fig:Plotexpermeant}). Velocity  $\dot{r}_\Ta$ and accelerations $(\ddot{\phi}_\Ta, \ddot{t}_\Ta)$ differentiate  outgoing particles from the ingoing particles cases of Figs\il(\ref{Fig:Plotexpermeant}). }\label{Fig:Plotexperthousandp}
\end{figure*}

\begin{figure*}
\centering
          \includegraphics[width=5.5cm]{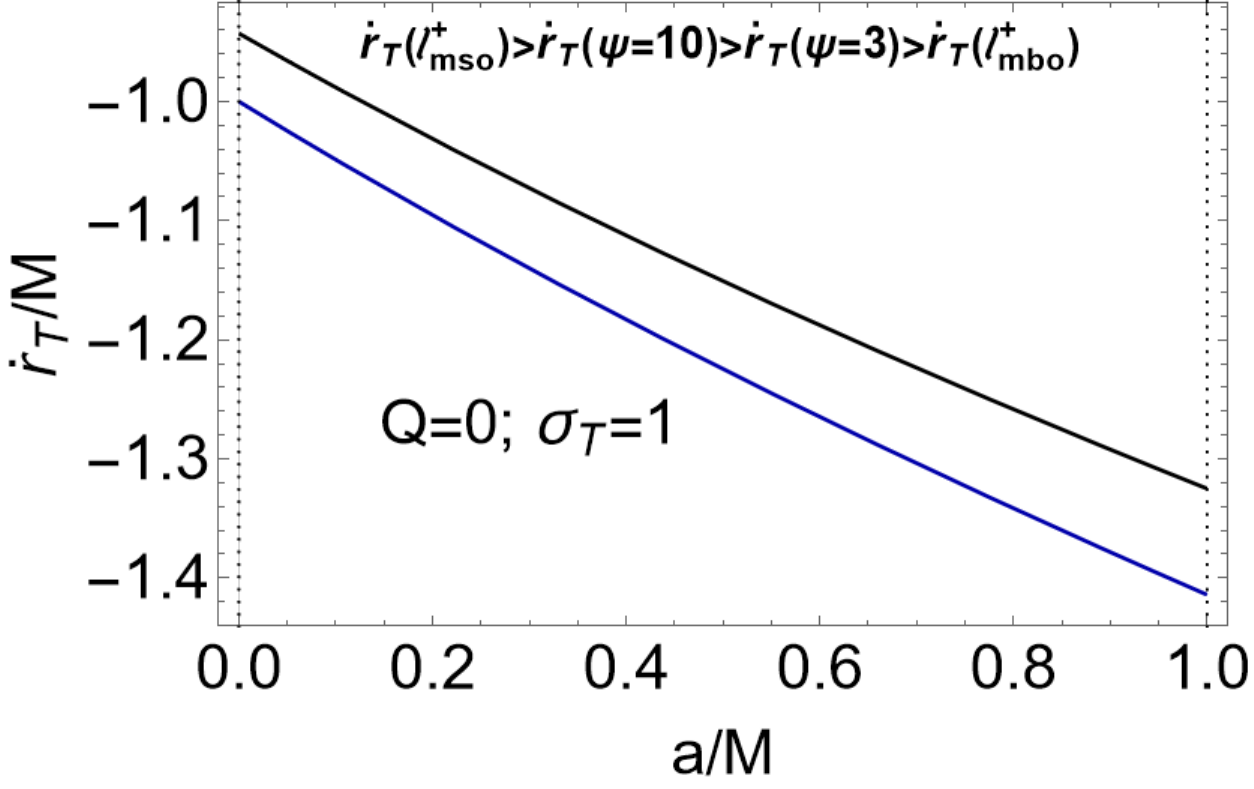}
        \includegraphics[width=5.5cm]{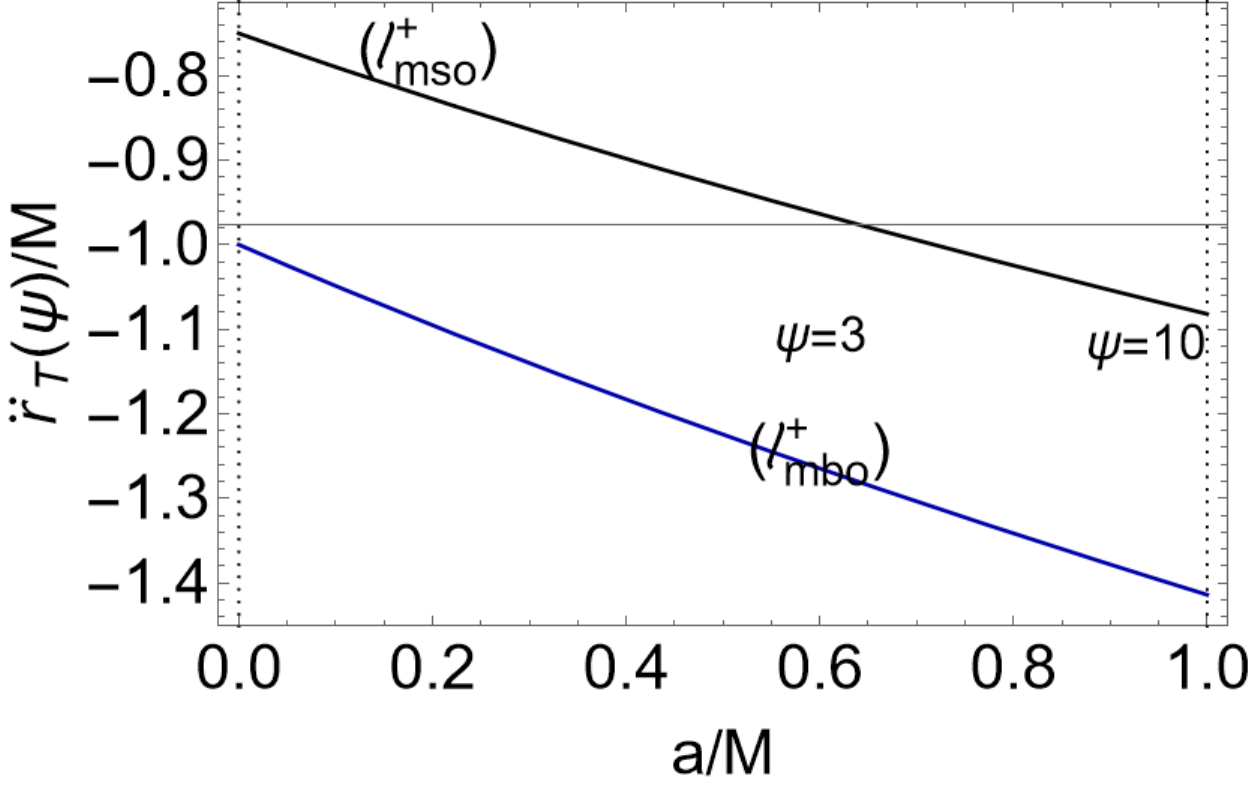}
        \includegraphics[width=5.5cm]{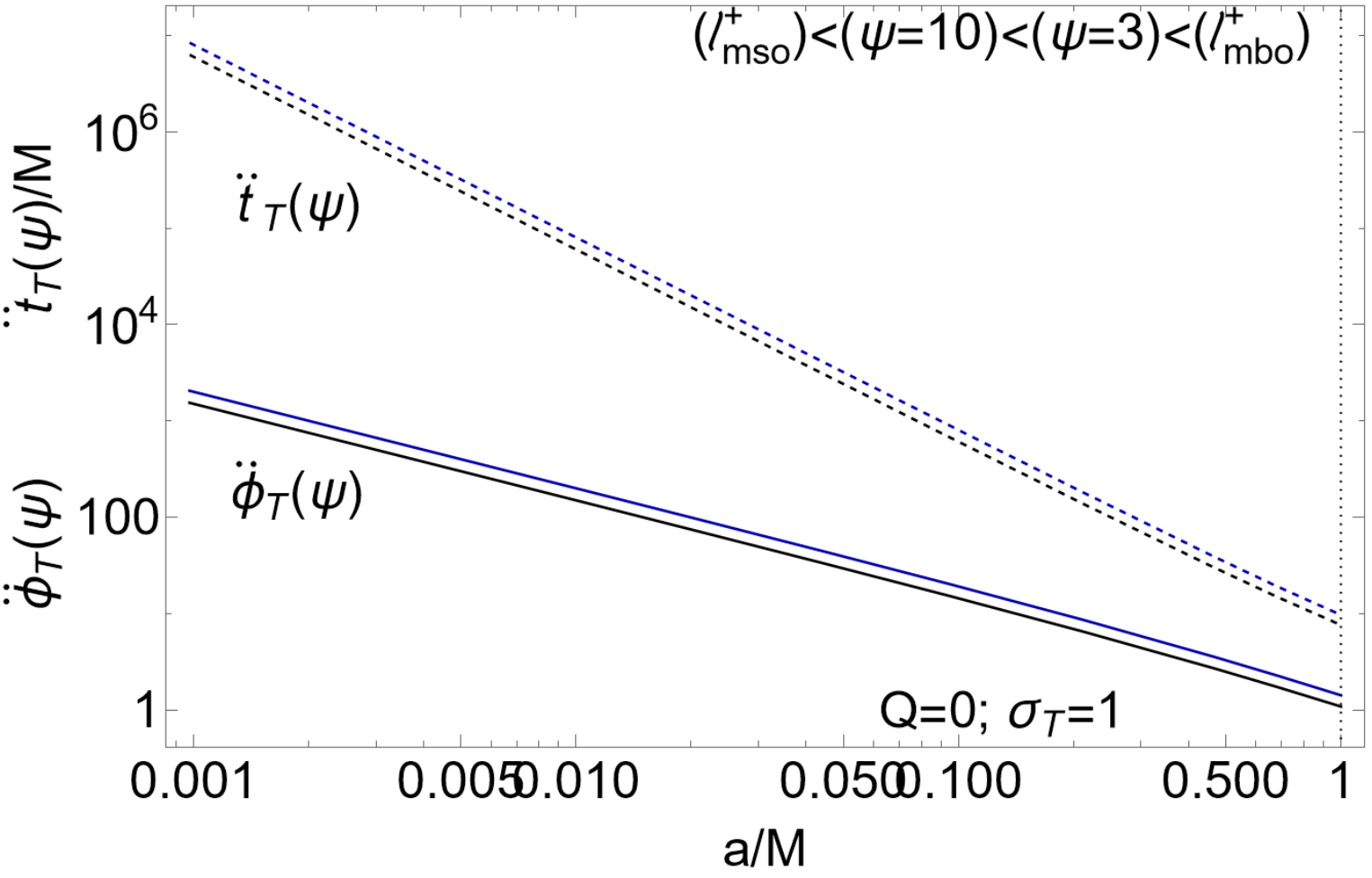}
        \\
     \includegraphics[width=5.5cm]{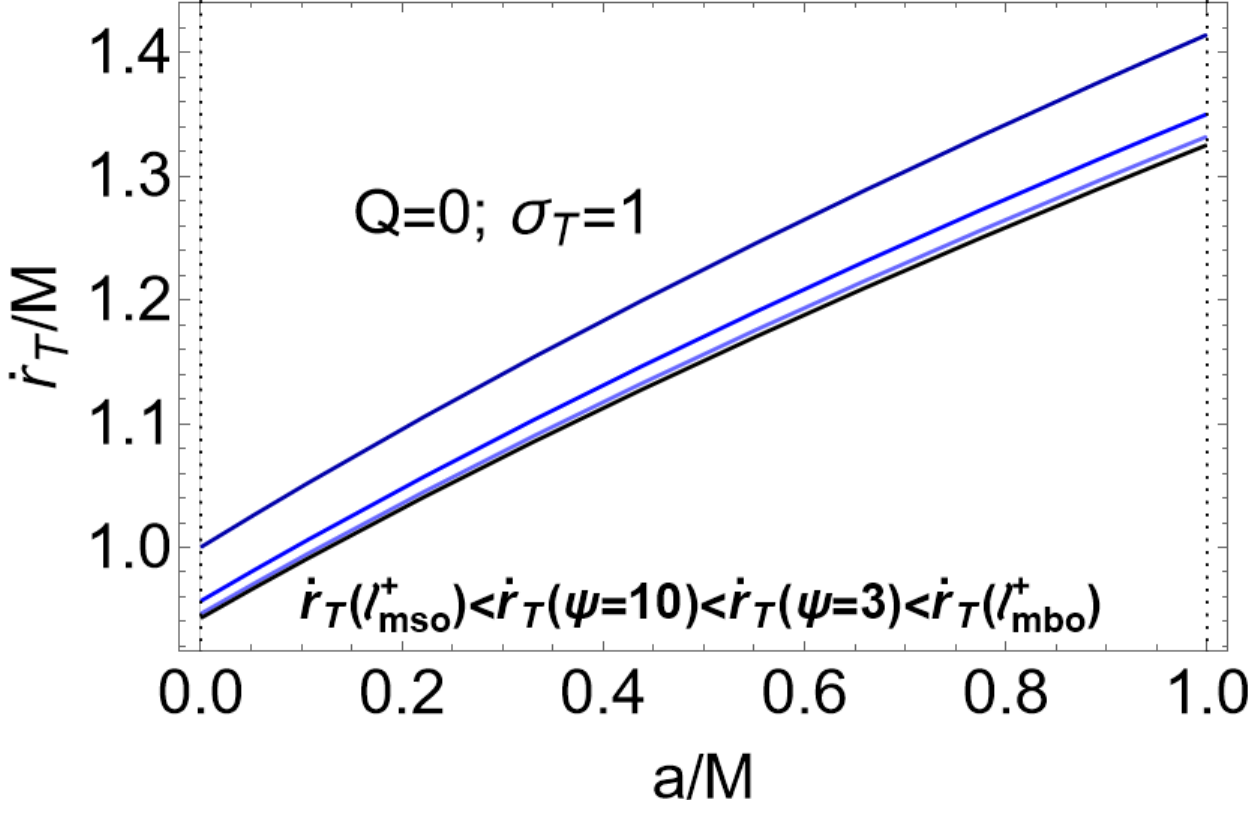}
        \includegraphics[width=5.5cm]{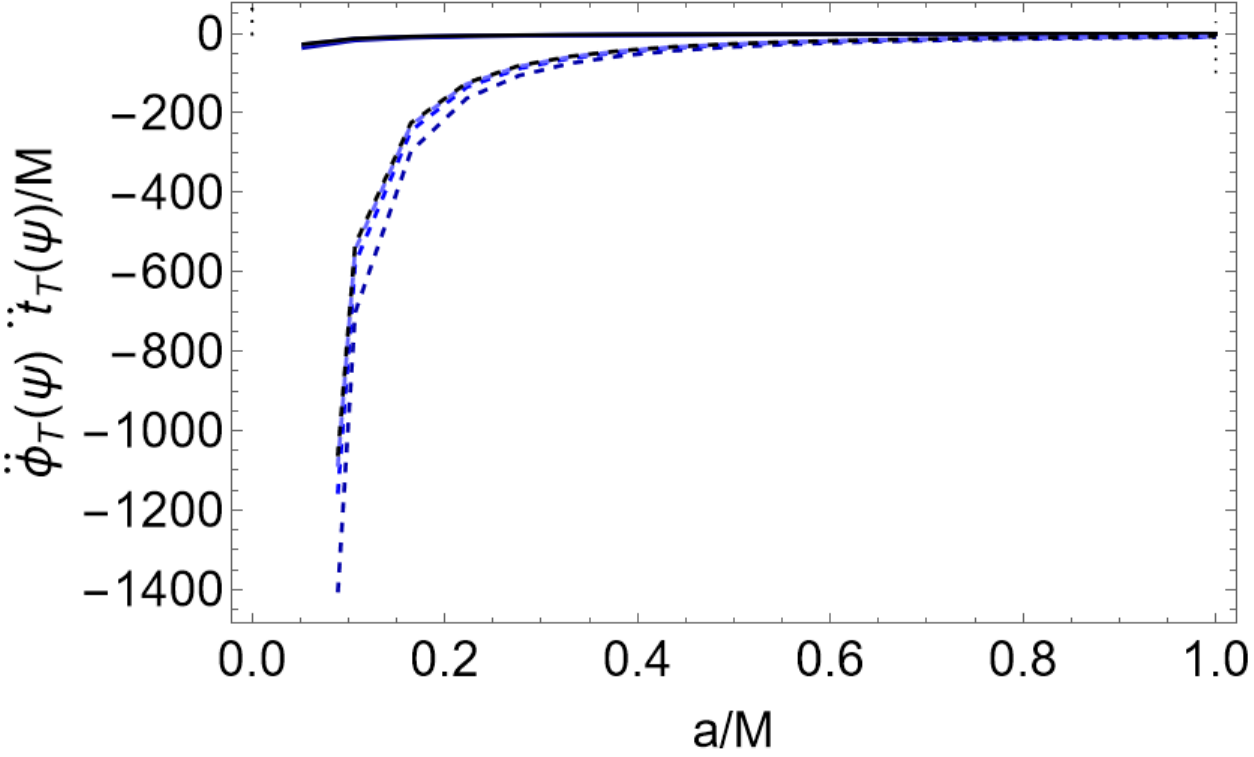}
  \caption{Accelerations components  and radial velocities  at the tori-driven  counter-rotating fluid turning point on the equatorial plane $\sigma_\Ta=1$, where $\sigma\equiv \sin^2\theta$, with Carter constant  $\Qa=0$ and $\dot{\theta}=\ddot{\theta}=0$,   for ingoing (upper panels) and outgoing (below panels)  photons  for the toroidal models of Figs\il(\ref{Fig:Plotexpermeant}). }\label{Fig:Plotexperthousanphton}
\end{figure*}

\begin{figure*}
\centering
    \includegraphics[width=5.5cm]{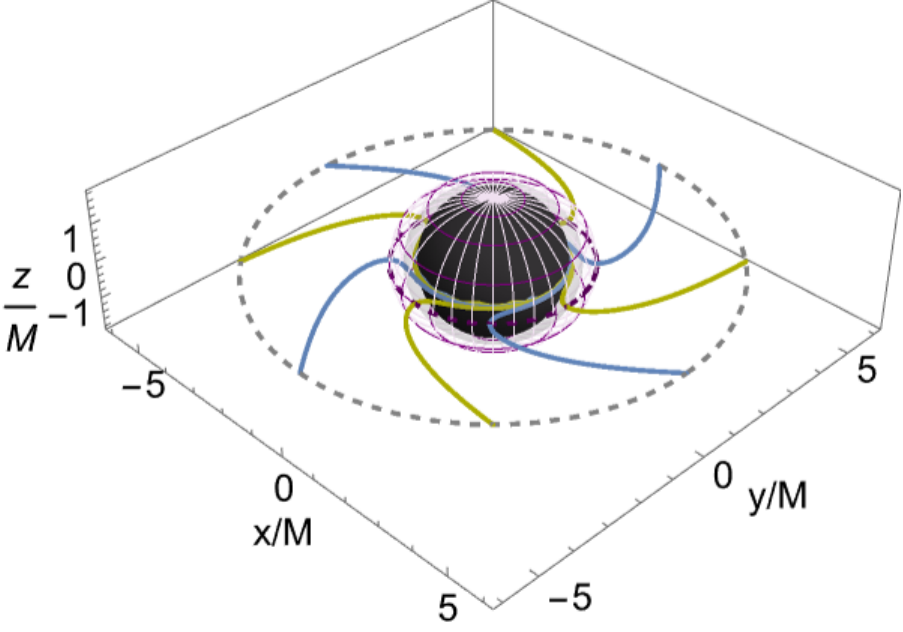}
    \includegraphics[width=5cm]{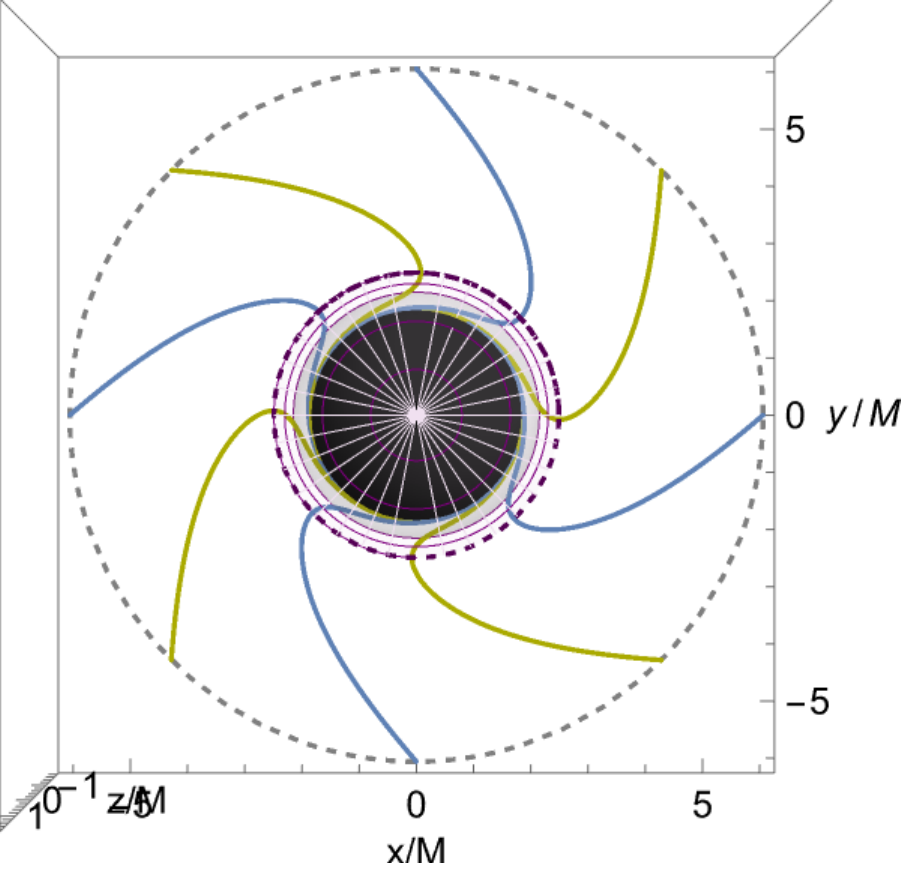}
   \caption{Tori driven counter-rotating flows photons  turning points in the \textbf{BH} spacetime with spin-mass ratio  $a/M=0.71$ of the  cusped  tori of  Figs\il(\ref{Fig:bluecurvehereq}). The torus  has  fluid specific angular momentum   $\ell=-4.5$,  where is $\{z=r \cos\theta,y=r \sin \theta \sin\phi,x=r \sin\theta \cos\phi\}$ in dimensionless units. The limiting fluid specific angular momenta, defined  in Eqs\il(\ref{Eq:def-nota-ell}), are  $ \{\ell_{mso}^+=-4.21319,\ell_{mbo}^+=-4.61534,\ell_{\gamma}^+=-6.50767\}$. The cusped  torus  is    orbiting the equatorial plane of the central \textbf{BH}. Left  and right panels show a  front and above view  of the  counter-rotating  flow stream from the torus inner edge (cusp-dashed gray curve)  to the central \textbf{BH}. Black region is the  central  \textbf{BH}  (region $r<r_+$, radius $r_+$ is the outer horizon). Flow turning point $r_{\Ta}=2.31556M$ of Eqs\il(\ref{Eq:turning-point-radius}),(\ref{Eq:max-cusp-eq-exte1}), (\ref{Eq:rte-second-sigma1})  is plotted as the deep-purple curve.
   The initial radial velocity normalization conditions for photons have been used to find  $\dot{r}_0$. Radius $r_{\Ta}$
   lies  in the  turning corona defined by the
   range $(r_\Ta(\ell_{mso}^+)-r_\Ta(\ell_{mbo}^+))$. Gray region is the outer ergosurface, light-purple shaded region is the region $r< r_\Ta(\sigma_\Ta)$ (where $\sigma\equiv\sin^2\theta$)--
see Figs\il(\ref{Fig:Plotrte4}).In Figs\il(\ref{Fig:bluecurvehereq}) is the analysis of matter particles.}\label{Fig:bluecurvehereqb}
\end{figure*}

 In   Figs\il(\ref{Fig:bluecurvehereq}) we show
the flow  from a counter-rotating  torus with a turning point on  the equatorial plane,  however the  fluid particles trajectories  at $\tau>\tau_\Ta$ ($t>t_\Ta$)  depend on the  fluid initial data.  In this section we analyze   the  flow particles accelerations   $\dot{u}_{\Ta}$    at the  turning point.
 To simplify the discussion  we limit the analysis to matter particles ($\mu=1$) at the turning point on the equatorial plane, where   $\Qa=0$, $\theta_\Ta=\pi/2$ and $\dot{\theta}_\Ta=\ddot{\theta}_\Ta=0$, and using  the  second-order differential equations  of the geodesic motion within the constrain  provided by the  normalization condition.
Tori models for the counter-rotating flows  are defined by the function $\ell=\ell_{ecc}(\psi,a)$ of Eq.\il(\ref{Eq:psixi}), using the condition $\ell_{ecc}=\La/\Em$ where $\Em\equiv K(r_{\times})$ and  $r_{\times}$  is  given by Eq.\il(\ref{Eq:toricenter-inner}), and by using the  turning point radius $r=r_\Ta^e$ for  outgoing/ingoing  particles defined by   $\dot{r}_\Ta\gtrless0$  respectively-- Figs\il(\ref{Fig:Plotexpermeant}).
(The radial acceleration is independent from the radial velocity sign as  the dependence from the even power of $\dot{r}$ depends on  $\dot{\theta}_\Ta$ which,  in the  case  considered here, is zero).
{In fact, as already mentioned, the turning coronas are a background property, depended only on the \textbf{BH} dimensionless  spin $a/M$. Each turning sphere depends on the specific angular momentum $\ell$, indifferently for ingoing or outgoing particles or  with other velocity components  at the turning point, for example with motion along the vertical axis, therefore here   we complete this analysis considering the   general frame  where   there can be  particles  and photons with an outgoing radial component of the velocity.

The central \textbf{BH} is not  isolated and  there will be matter, and  photons radiated in any direction, the eventual turning points of the general trajectories will be located  on the turning spheres, equal  for photons and particles. The analysis of Figs\il(\ref{Fig:Plotexpermeant}),(\ref{Fig:Plotexperthousandp}), (\ref{Fig:Plotexperthousanphton}), (\ref{Fig:bluecurvehereqb})  gives us an indication on the trajectories crossing the turning sphere, connected with the tori parameters  fixed  in the  study of the tori  energetics. We also considered counter-rotating photons for comparison with the infalling matter. }
We fix  the velocities and accelerations at the flow   turning point for fixed toroidal cusped  models  of Eq.\il(\ref{Eq:psixi})   ($\psi=$constant)  to describe the situation  with  variation of  the \textbf{BH} spin and the fluid specific angular momentum $\ell\in \mathbf{L_1}$.
We also show the tori  inner parts in   different tori models.  We can note how $\dot{t}_\Ta$ decreases with the spin $a/M$  and the momenta $\ell$ in magnitude, while the radial (infalling) velocity increases in magnitude with the spin (distinguishing  fast spinning from  slowly spinning attractors)  and increases with the decreasing momenta in magnitude.
The analysis of the toroidal acceleration  defines   $\phi_\Ta$ as extreme of the $\phi(\tau)$,  as for assumption there is
$\dot{\phi}_\Ta\lesseqgtr0$ for $\tau\lesseqqgtr\tau_\Ta$ respectively.  As there is $\ddot{\phi}_\Ta(\tau)>0$, the derivative $\dot{\phi}(\tau)$ increases.
To  complete the analysis we also consider the  outgoing condition  in  Figs\il(\ref{Fig:Plotexperthousandp}) and the case of photons in  Figs\il(\ref{Fig:Plotexperthousanphton})---see also Figs\il(\ref{Fig:bluecurvehereqb})--
confirming   that the tuning point function, $r_\Ta(\sigma_\Ta)$, is substantially the same for photons and particles\footnote{Consider $\Qa\neq0$, for plane different from the equatorial plane, assuming  $\dot{\theta}_0=0$,
 there is   $\Qa=\cos^2\theta_0 \left[a^2 \left(1-{\La^2}/{\ell ^2}\right)+\La^2 \csc ^2\theta_0\right]$, with
 $r_0(\theta_0)$ fixed according to the fixed tori models and   $r_\Ta(\theta_\Ta)$  in Eq.\il(\ref{Eq:turning-point-radius}). However conditions on
$\ell=\ell_{ecc}$ and $K=K_\times(\ell_{ecc})$ are constant on all the   torus surface, and we can  recover  $\Em$  and $\La$,  from $\ell$ and $K$.}.
\section{Discussion and conclusions}\label{Sec:discussion}
  Kerr background frame-dragging  expresses in the formation of a  turning point  of  accreting matter flow from the counter-rotating orbiting tori (and proto-jets), defined by condition $u^{\phi}=0$ on the torodial velocity of the flow. In this article  we discuss the (necessary) existence  of the turning point for the counter-rotating flow, characterizing  the flow  properties  at the turning point.
The turning point function, $r_\Ta(\sigma_\Ta)$ {or equivalently $\sigma_\Ta(r_\Ta)$ has been studied in the more general case and then specialized for the case of flow from orbiting tori}. Fluid velocity components  at the turning point are  studied in Sec.\il(\ref{Sec:velocity}).
In Sec.\il(\ref{Sec:accelerations-fluids})  there are some considerations on the
    fluid  accelerations  at the turning point.
Counter-rotating flows turning points  are  located (under special conditions on the particles energy parameter $\Em$) out of the ergoregion  (turning points with $\ell>0$ are located in the ergoregion for timelike particles).
{Turning points
are largely independent from the details of the tori models  and  the normalization condition, depending  on the fluid specific angular momentum $\ell$ only, describing therefore    photon and matter components.
At fixed $\ell$,  turning points are located on a spherical surface (turning sphere) surrounding the  central attractor. The connection with the flows associated with the orbiting torus leads to the   individuation   of  a spherical shell, turning corona, surrounding the central \textbf{BH},  whose outer and inner boundary  surfaces are  defined by  the fluid angular momentum     $(\ell_{mso}^+,\ell_{mbo}^+)$ respectively  and  $(\ell_{mbo}^+,\ell_{\gamma}^+)$ for accretion driven turning points and  proto-jets driven turning points respectively. Turning coronas depend only on the \textbf{BH} spin  $a/M$, describe turning points for particles  regardless from the tori models, and are to be considered a  background property. }

The torus and proto-jets driven  turning coronas are  a narrow  annular region close and external to the  \textbf{BH} ergosurface. The torus driven corona is separated from and more external    to   the proto-jets driven corona.

  The independence of the  turning point radius on  the details of initial data on the flow (details of tori modes, accretion mechanisms) and the small extension of  of the annular region, narrow the flows turning points identification.
 The  coronas are larger at the \textbf{BH} equatorial plane  (where  they are also the farthest from the central attractor) and smaller on the \textbf{BH} poles.
 The separation between the tori driven and proto-jets driven coronas,  distinguishes the two flows, while each annular region sets the turning points for  matter and photons as well.
 We singled out also properties of the flow at the turning points distinguishing photon from matter components in the flow, and proto-jets driven and tori driven accreting flows.
The turning corona can be   a very active part of the accreting flux of matter and photons, especially  on the \textbf{BH} poles, and lightly more   rarefied  at the equatorial plane,   and it can  be characterized  by an increase of the flow luminosity and temperature.
However,  observational properties  of this region can  depend strongly on the  processes timescales,  in this investigation considered  in terms of the times $(t_\Ta, \tau_\Ta)$ the flow reaches the turning points.

Main properties of the turning points and the flow in the corona depend  on the  background properties mainly, on the flow initial constant momentum $\ell$, which is limited by the  $(\ell_{mso}, \ell_{mbo},\ell_\gamma)$ functions of the  \textbf{BH} spin-mass ratio only.
Function $\ell_{mso}^+(a)$ sets the maximum extension of the torus turning corona, and define the maximum throat thickness (which is also related to several energetic properties of the tori).
As  shown in  Figs\il(\ref{Fig:PlotsuppoB}),  the maximum throat thickness     for super-critical tori  is provided by the limiting solution with  $\ell^+=\ell_{mso}^+$, and therefore   determined {only}  by  \textbf{BH} dimensionless spin. This implies also that  the maximum amount of matter swallowed by the \textbf{BH} from the counter-rotating tori considered here  is constreined  by the limiting configurations with $\ell=\ell_{mso}^+$, and   flow reaches  the attractor with  $\ell<0$.
Therefore the fast spinning  \textbf{BH} energetics would  depend essentially  on its spin  $a/M$ rather than on   the properties of the counter-rotating fluids or  the tori masses--Figs\il(\ref{Fig:Plotrangqwalsldopo}).

The turning radius however can have an articulated dependence on the spin and plane, with the occurrence of some maxima.
In Figs\il(\ref{Fig:Plotrtea}) and Figs\il(\ref{Fig:Plotrteb})  we show  the corona radii  in dependence from different planes $\sigma_\Ta$, particularly around the limiting plane value  $\sigma=\sigma_{crit}=2 \left(2-\sqrt{3}\right)$.  For $\sigma<\sigma_{crit}$, there is  $r_\Ta<2M$  (being related to the outer ergosurface location and therefore the plane $\sigma$) and radius $r_\Ta$ decreases increasing the \textbf{BH} spin. Viceversa, at $\sigma\geq \sigma_{crit} $, turning radii   are at  $r_\Ta>2M$, decreasing  with the  spin $a/M$.
Functions  $r_\Ta$ and  $\sigma_\Ta$ have
 extreme values  which we considered in Sec.\il(\ref{Sec:extreme-turning-box}).

The analysis of the  turning sphere vertical maximum  is  particularly relevant for the jet and proto-jet particles, having a vertical component of the velocity. The proto-jets particle turning points are  closer to the ergosurface on the vertical axis  (and  on the equatorial plane). We have shown that  the  turning  sphere  vertical  maximum  is greater for  accretion driven flows and  and minimum for proto-jets driven flows (decreasing  with the magnitude of $\ell$), and decreases with the \textbf{BH} spin $a/M$  (in the limit  $a=0$ and $z_\Ta=2M$). It is maximum at $\ell_{mso}^+$ (the proto-jets turning point verticality  is lower then the  accretion  driven turning points). The  bottom boundary of the maximum $z_\Ta^{\max}$  occurs for the extreme \textbf{BH} with $a=M$, where  there is $z_\Ta=1.451M$ for $\ell=\ell_{mso}^+$, $z_\Ta=1.437M$ for $\ell=\ell_{mbo}^+$, and $z_\Ta=1.39M$ for $\ell=\ell_{mso}^+$ Double turning points (at fixed $\ell$ and $z_\Ta$), studied in  Sec.\il(\ref{Sec:turning-sign-existence}) in Figs\il(\ref{Fig:Plotrte4}) and Sec.\il(\ref{Sec:vertical-z}), are related  to the presence of a maximum. There is a   double point at $z_\Ta\in[r_+,z_\Ta^{\max}r[$  and $y_\Ta\leq y_\Ta(+)$ where $y_\Ta(+): z_\Ta= r_+$, for  $a>0.738$ for   flows with $\ell=\ell_{mso}^+$ and for  $a>0.75$ per flows with  $\ell=\ell_{mbo}^+$.
We thus conclude  that the azimuthal turning points of both the flows from
accretion counter-rotating tori and jets can have interesting  astrophysical consequences.

\appendix
\section{Polytropics and  tori energetics}
\begin{figure*}
\centering
 \includegraphics[width=5.5cm]{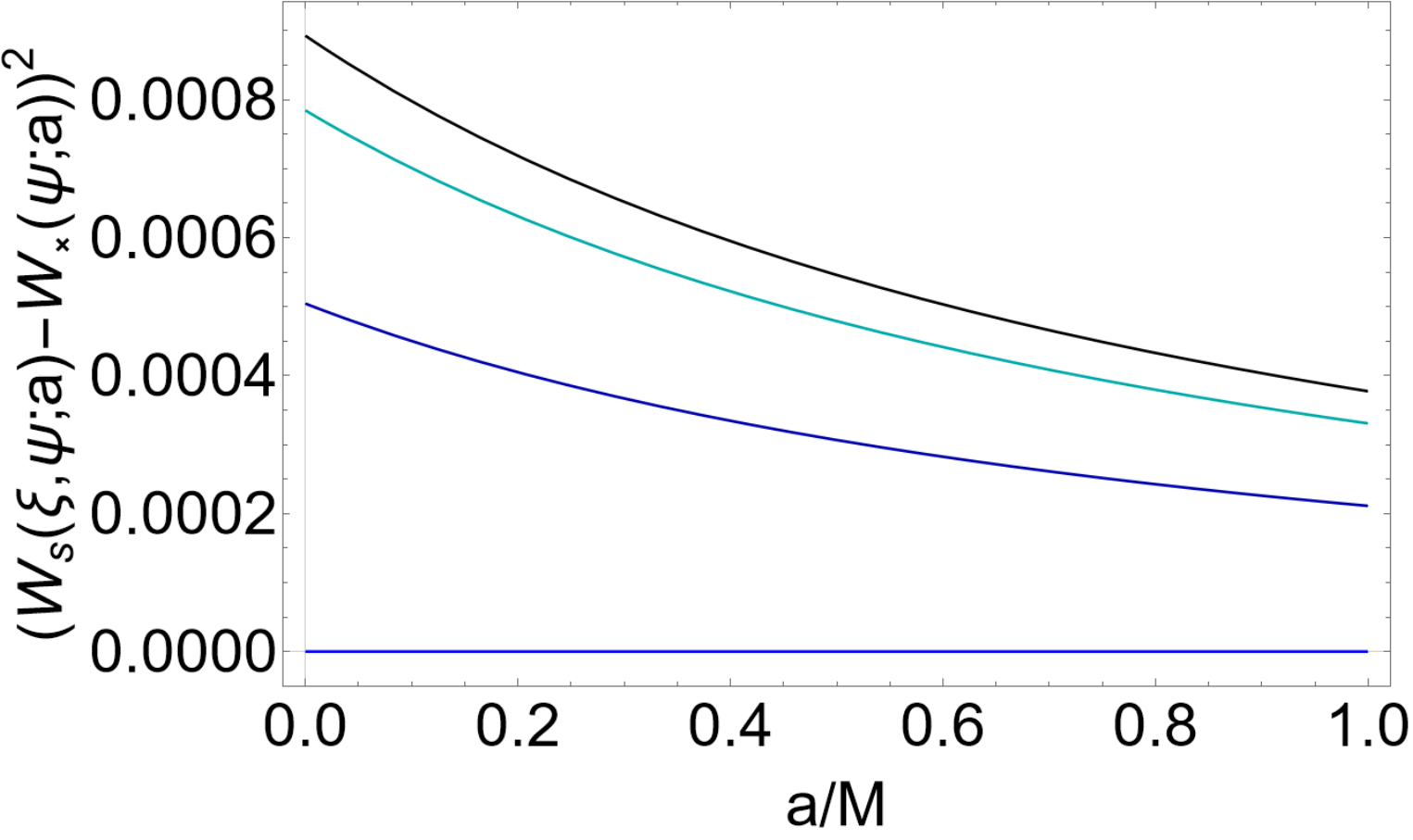}
 \includegraphics[width=5.5cm]{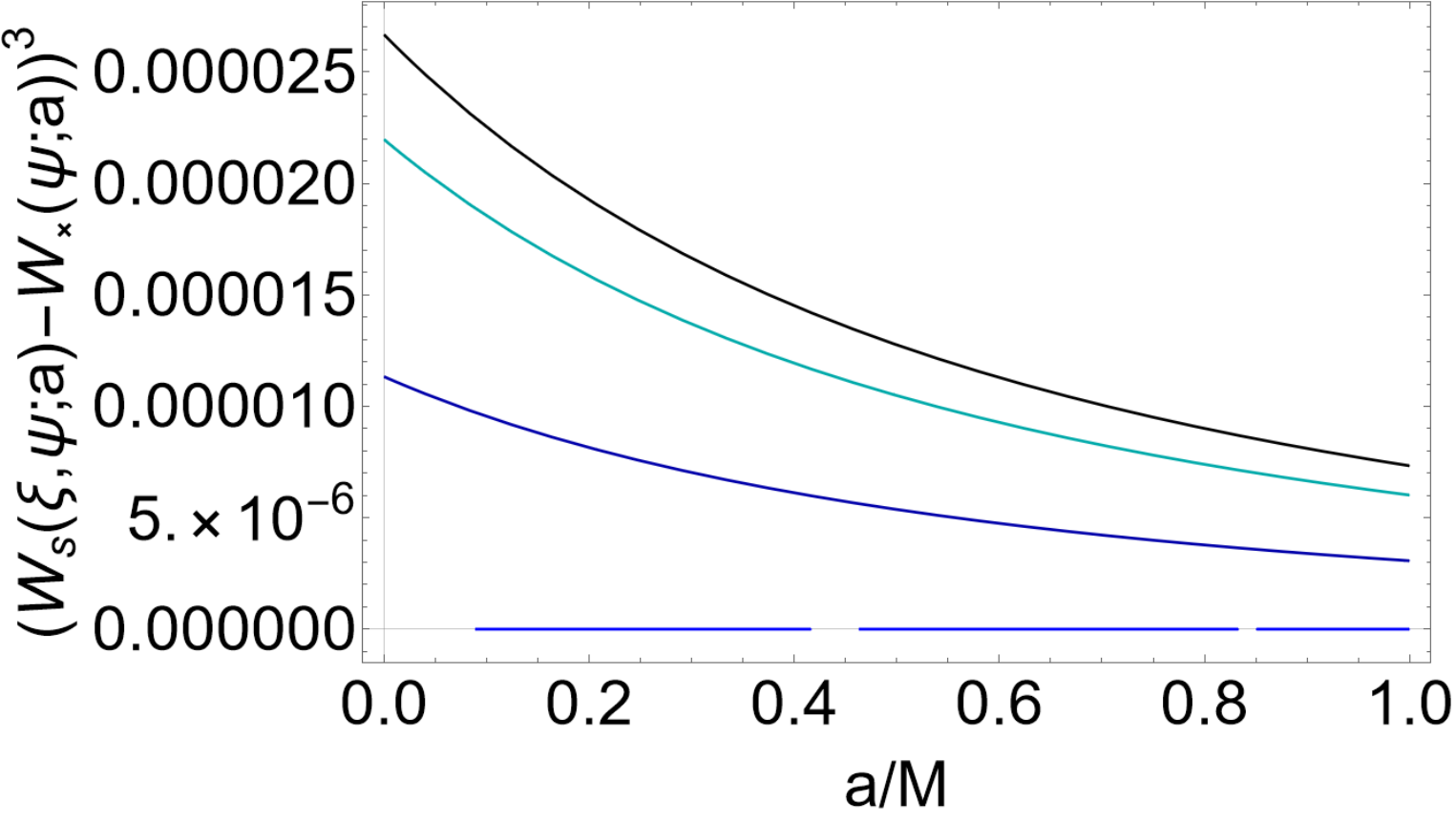}
 \includegraphics[width=5.5cm]{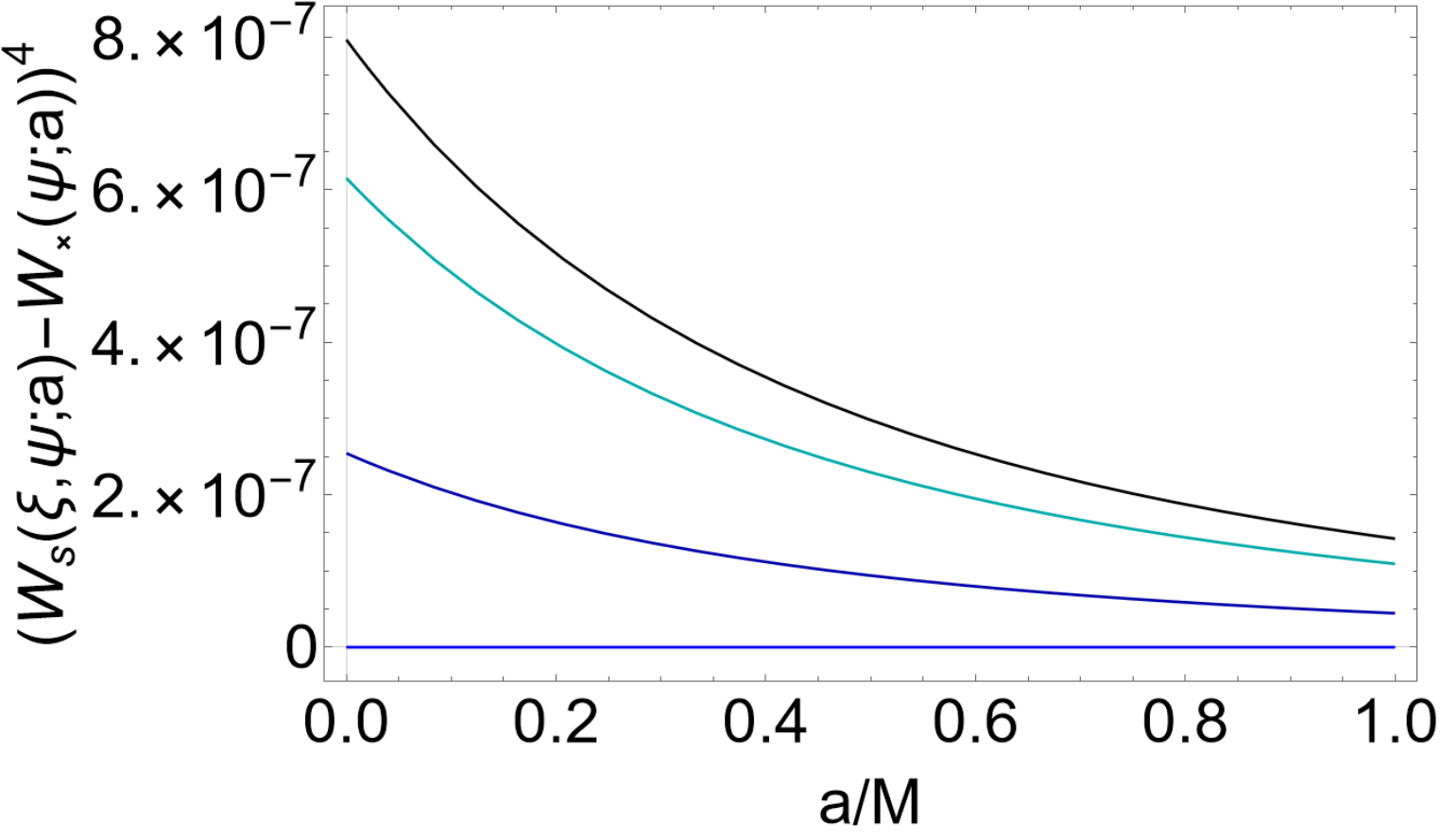}
   \\
 \includegraphics[width=5.5cm]{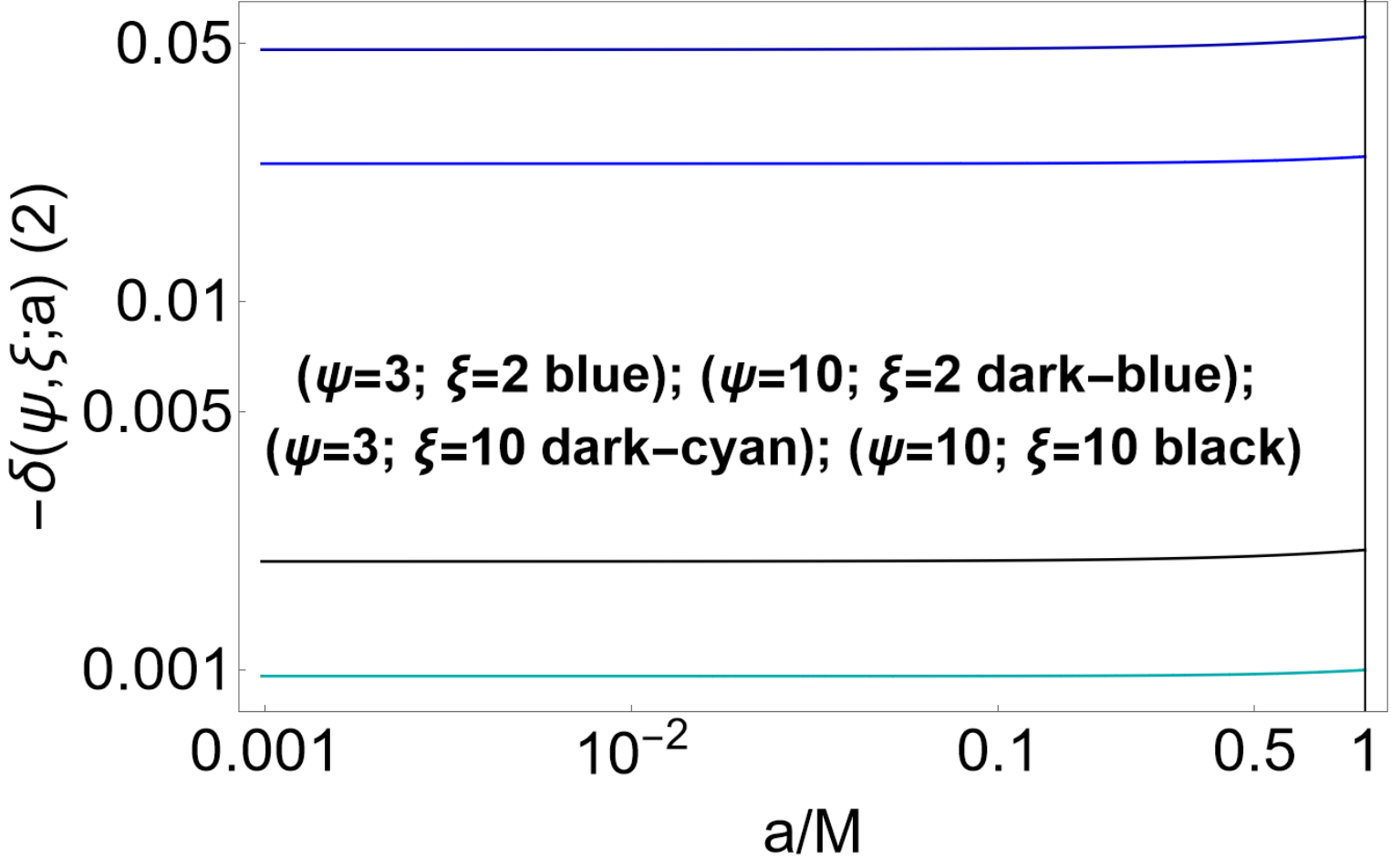}
 \includegraphics[width=5.5cm]{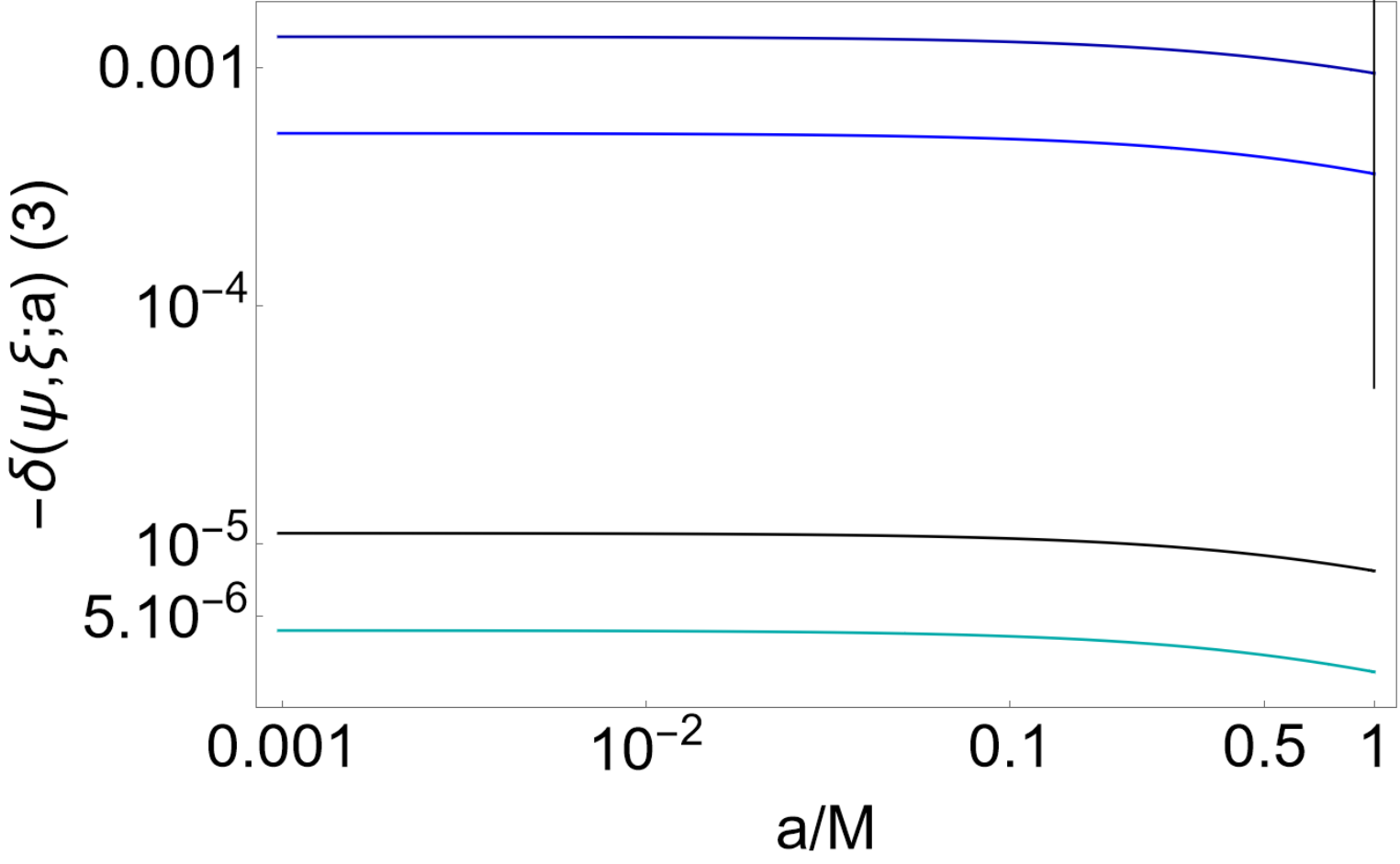}
 \includegraphics[width=5.5cm]{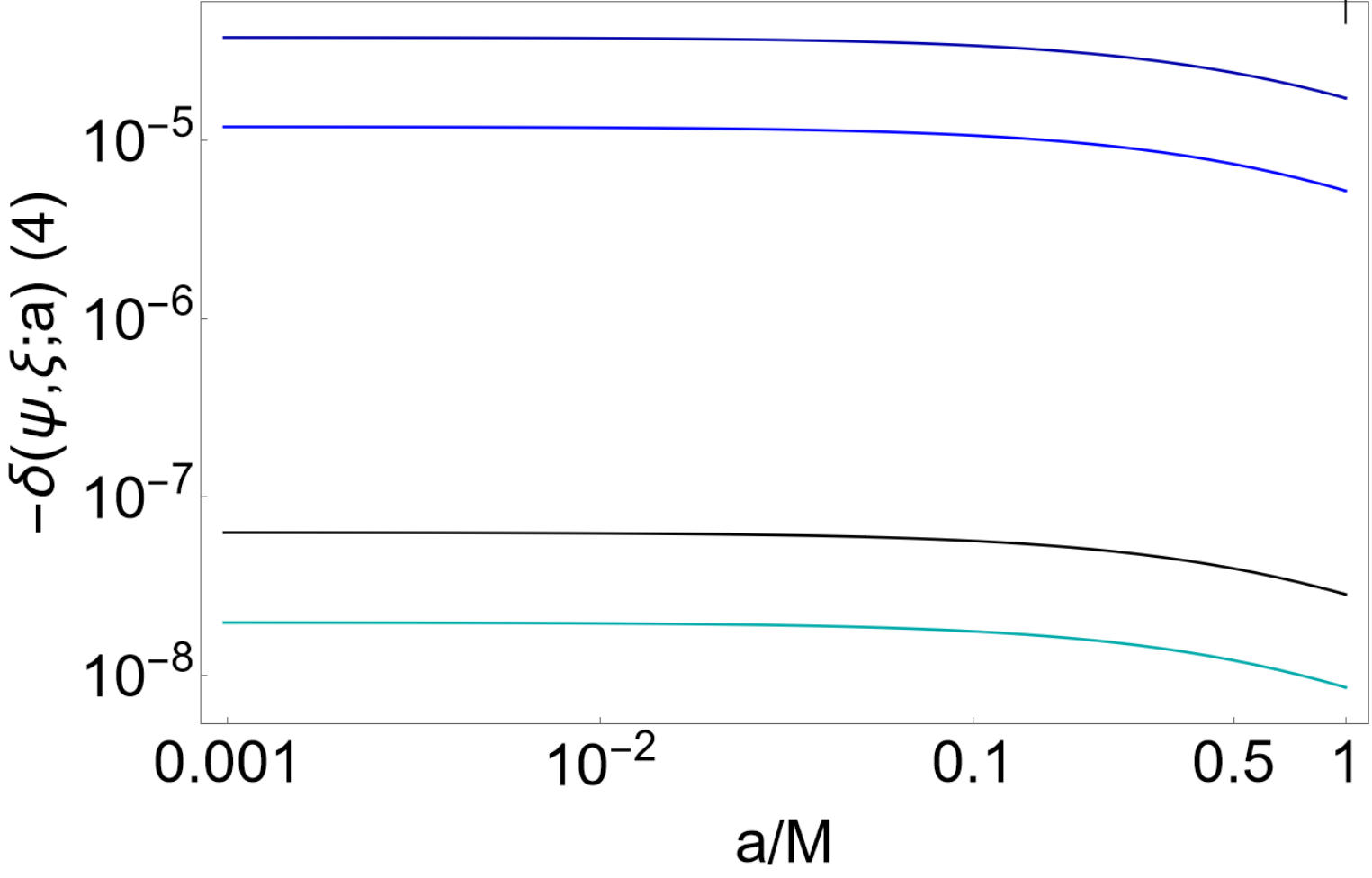}
  \caption{Difference $(W_s-W_\times)^{\nu}$ (upper panels),  defining  the $\Gamma$--quantities introduced  in Sec.\il(\ref{Sec:flow}), and the $\delta$-quantities (below panels)  $-\delta(\nu)$ are shown as functions of the \textbf{BH}  spin-mass ratio $a/M$, at different fluid momenta $\ell$ regulated by the $\psi$ parameters signed on the curves and different $K_s\equiv e^{W_s}$ parameters regulated by $\xi$, and different ${\nu}$--see  Eq.\il(\ref{Eq:psixi}). The $\Gamma$--quantities regulate the  mass--flux, the  enthalpy--flux (related to the  temperature parameter),
and  the flux thickness.  $\delta$--quantities regulate    the cusp luminosity, the disk accretion rate,   and  the  mass flow rate through the cusp i.e., mass loss accretion rat--see also Figs\il(\ref{Fig:Plotrangqwalsldopo}). \emph{$\Gamma$--quantities}, have general form  $\Gamma (r_\times,r_s,n)=\beta_1(n,\kappa)(W_s-W_{\times})^{\beta_2(n)}$, where  here $\beta_1(n,\kappa)=1$ and $\beta_2(n))\equiv\nu$ are  functions of the polytropic index and constant.
 \emph{$\delta$-quantities}, while $\delta={\Gamma}(r_\times,r_s,n) r_\times/\Omega(r_\times)$; where  $\Omega(r_\times)$ is the relativistic angular  frequency  at  the   tori cusp  $r_\times$.
   Blue curves correspond to   $(\psi =3;\xi =2)$; dark-blue curves to $(\psi =10;\xi =2)$,  dark-cyan curves to  $(\psi =3;\xi =10)$; and black curves to $(\psi =10;\xi =10)$.}\label{Fig:Plotrangqwalsldopoprova2}
\end{figure*}

In Figs\il(\ref{Fig:Plotrangqwalsldopoprova2})  we  focus on the relation  between  different polytropics and  tori energetics,  following the analysis of Sec.\il(\ref{Sec:flow}).

\begin{figure*}
\centering
 \includegraphics[width=6.5cm]{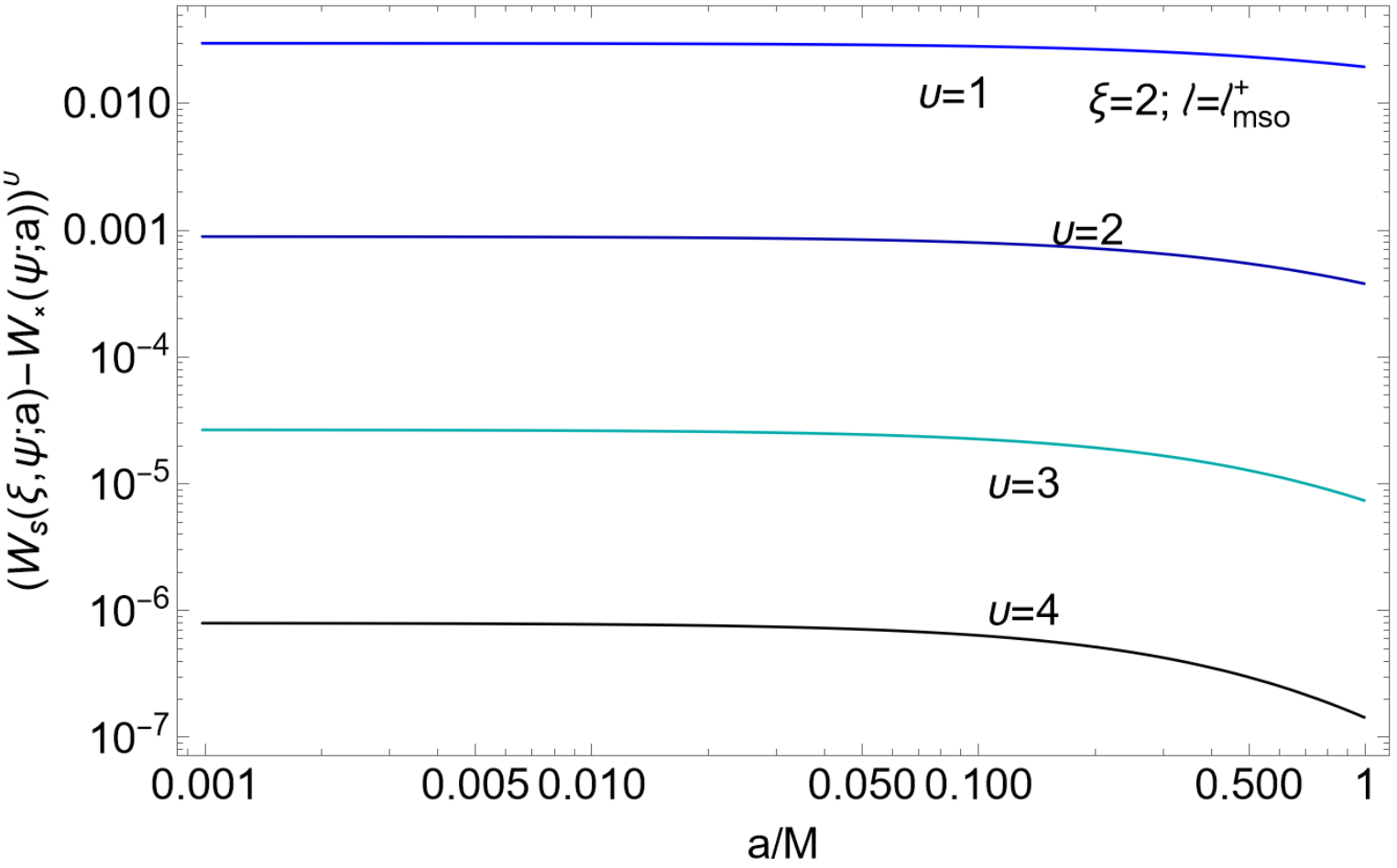}
  \includegraphics[width=6.5cm]{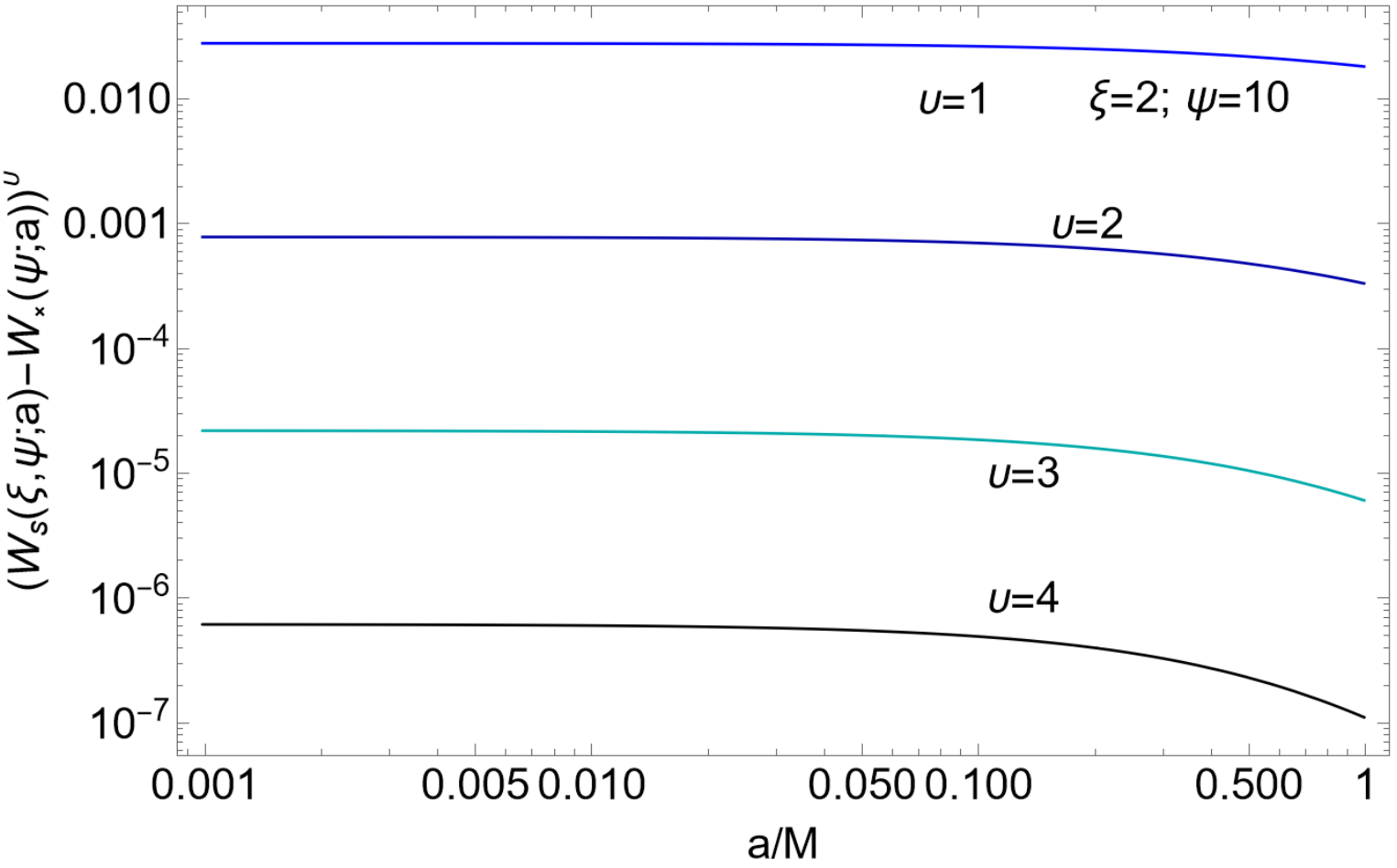}\\
   \includegraphics[width=6.5cm]{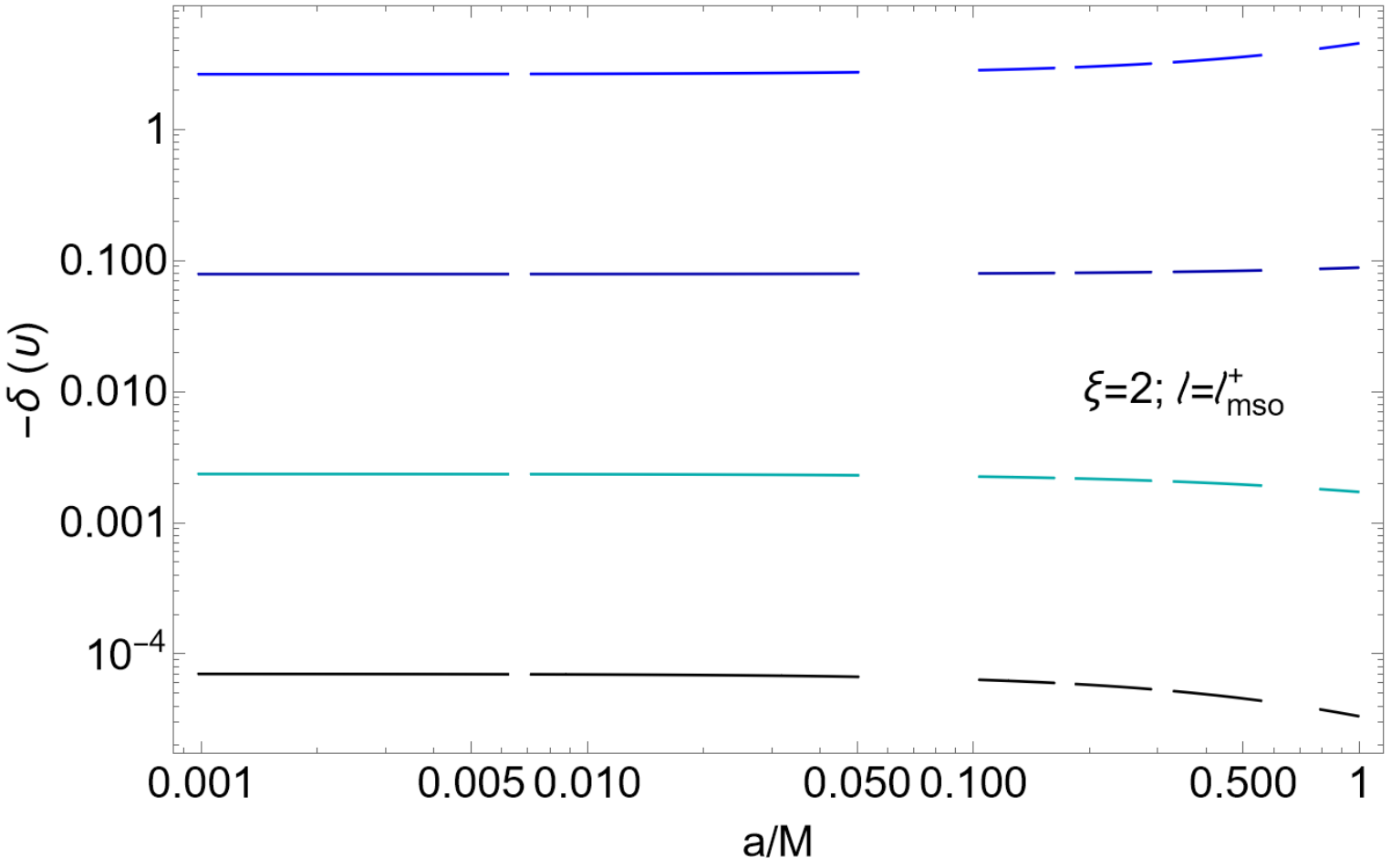}
 \includegraphics[width=6.5cm]{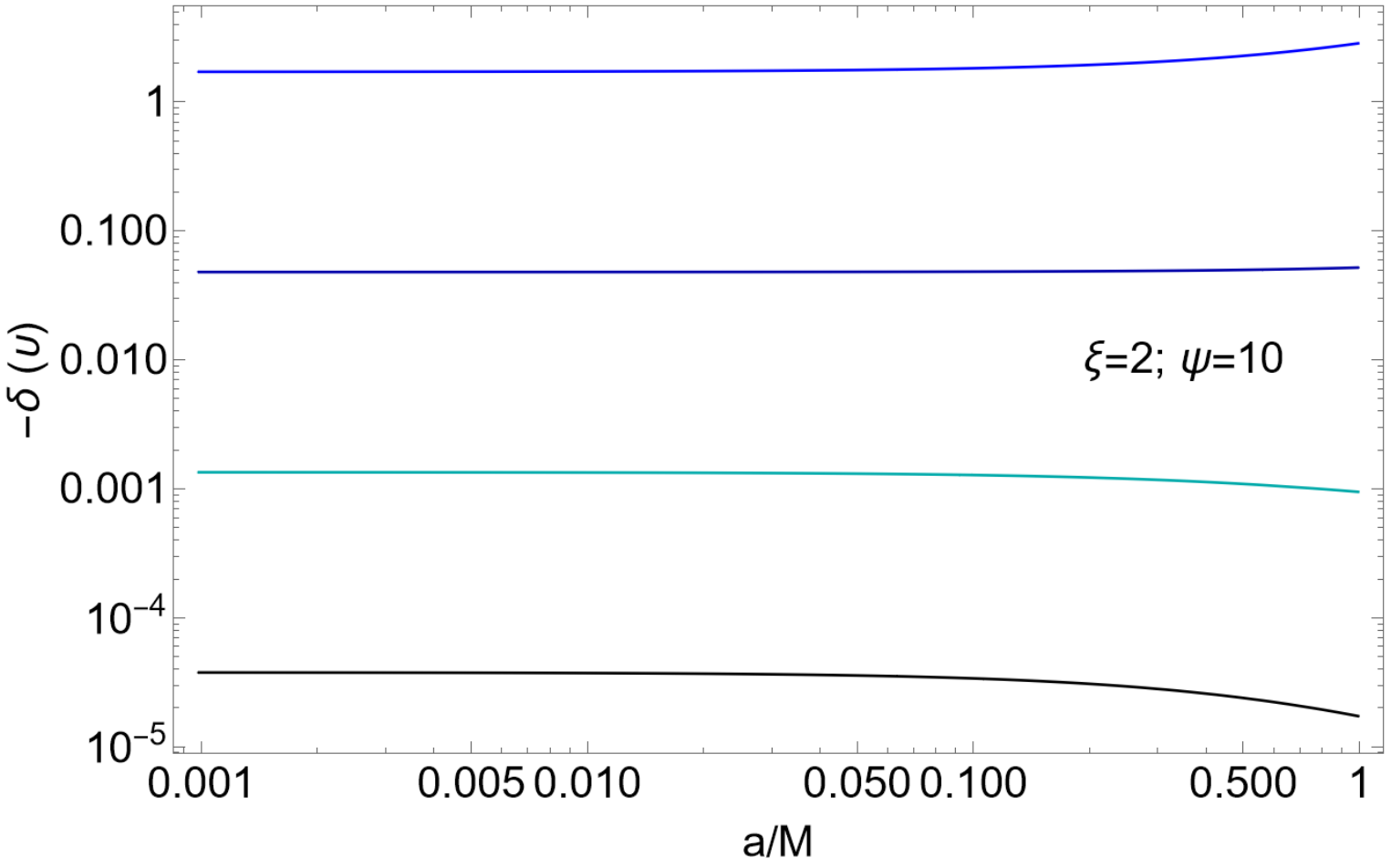}
    \caption{Upper panels show difference $(W_s-W_\times)^{\nu}$  defining  the $\Gamma$--quantities, and below panels show $\delta$-quantities,   as functions of the \textbf{BH}  spin-mass ratio $a/M$, at different fluid momenta $\ell$ regulated by the $\psi$ parameters and different $K_s\equiv e^{W_s}$ parameters regulated by $\xi$, and different ${\nu}$, signed on the curves--see  Eq.\il(\ref{Eq:psixi}) and   Sec.\il(\ref{Sec:flow}). Panels show the situations for different $(\xi,\psi)$ or $(\xi, \ell)$,  where  blue curves correspond to   $\nu=1$; dark-blue curves to  $\nu=2$,  dark-cyan curves to  $\nu=3$; and black curves to $\nu=4$. The $\Gamma$--quantities regulate the  mass--flux, the  enthalpy--flux (related to the  temperature parameter),
and  the flux thickness.  $\delta$--quantities regulate    the cusp luminosity, the disk accretion rate,   and  the  mass flow rate through the cusp i.e., mass loss accretion rate. \emph{$\Gamma$--quantities}, have general form  $\Gamma (r_\times,r_s,n)=\beta_1(n,\kappa)(W_s-W_{\times})^{\beta_2(n)}$, where  here $\beta_1(n,\kappa)=1$ and $\beta_2(n))\equiv\nu$ are  functions of the polytropic index and constant.
 \emph{$\delta$-quantities},  have  general form  $\delta={\Gamma}(r_\times,r_s,n) r_\times/\Omega(r_\times)$; where  $\Omega(r_\times)$ is the relativistic angular  frequency  at  the   tori cusp  $r_\times$ where the pressure vanishes. }\label{Fig:Plotmetripreda1}
\end{figure*}

\section*{Data availability}
There are no new data associated with this article.
No new data were generated or analysed in support of this research.


\begin{thebibliography}{99}
 \bibitem[\protect\citeauthoryear{Abramowicz}{1971}]
{M.A.Abramowicz}
Abramowicz M. A.  1971
 Acta. Astron., {{21}},  81

 \bibitem[\protect\citeauthoryear{Abramowicz}{1985}]
   {Japan}Abramowicz M. A.  1985
Astronomical Society of Japan,  37,  4,  727-734




 \bibitem[\protect\citeauthoryear{Abramowicz\&Fragile}{2013}]
  {abrafra}
  Abramowicz M.~A.\&Fragile  P.C. 2013
   Living Rev. Relativity, {16}, 1


\bibitem[\protect\citeauthoryear{Alig et al.}{2013}]
{Aligetal(2013)}	
 Alig C.,    Schartmann  M., Burkert  A. ,  Dolag K. 2013,
\apj ,  771,  2, 119

\bibitem[\protect\citeauthoryear{Aly et al.}{2015}]
{Aly:2015vqa}
Aly H.,  Dehnen W.,   Nixon C.\& King A. 2015,
Mon. Not. Roy. Astron. Soc.,  { 449}, 1,  65


\bibitem[\protect\citeauthoryear{Amaro-Seoane et al.}{2016}]
{2016A&A...591A.114A} Amaro-Seoane, P., Maureira-Fredes, C., Dotti, M., et al.\ 2016, \aap, 591, A114. 

\bibitem[\protect\citeauthoryear{Balek et al.}{1989}]
{1989BAICz..40..133B} Balek V., Bicak J., Stuchlik Z., 1989, BAICz, 40, 133

\bibitem[\protect\citeauthoryear{Blaschke\&Stuchl{\'\i}k}{2016}]
{2016PhRvD..94h6006B} Blaschke M., Stuchl{\'\i}k Z., 2016, PhRvD, 94, 086006


\bibitem[\protect\citeauthoryear{Bardeen}{1972}]{BARDEEN}
Bardeen J. M., 1972,
 Rapidly Rotating Stars, Disks,
and Black Holes
in
Black Holes
Les Astres Occlus
Les Houches,  1972
edited by C. DeWitt
and B. S. DeWitt.




    \bibitem[\protect\citeauthoryear{Bardeen\&Petterson}{1975}]
{BP75}    Bardeen J. M., Petterson J. A.,  1975, \apj, 195, L65

\bibitem[\protect\citeauthoryear{Barrabes et al.}{1995}]{Barrabes:1995ue}
Barrabes C., Boisseau B. and Israel W. 1995,
MNRAS, \textbf{276}, 432

\bibitem[\protect\citeauthoryear{Beckert\&Falcke}{2002}]{Beckert:2001az}
Beckert  T.\&Falcke H., 2002,
Astron. Astrophys. \textbf{388}, 1106


\bibitem[\protect\citeauthoryear{Bicak\&Stuchlik}{1976}]
{1976BAICz..27..129B} Bicak J.\& Stuchlik Z., 1976, BAICz, 27, 129


\bibitem[\protect\citeauthoryear{Bonnerot et al.}{2016}]
{Bonnerot:2015ara}
 Bonnerot C.,  Rossi E.M.,  Lodato G.  et al. 2016, 
MNRAS, {455}, 2,  2253
\bibitem[\protect\citeauthoryear{Carmona-Loaiza et al.}{2015}]
{Carmona-Loaiza:2015fqa}Carmona-Loaiza J.M., Colpi M., Dotti M. et al., 2015,
MNRAS,453,1608



\bibitem[\protect\citeauthoryear{Carter}{1968}]{Carter}
Carter B., 1968, Phys. Rev., 174, 1559


{ \bibitem[\protect\citeauthoryear{Chakrabarti}{1990}]{Chakrabarti0}  Chakrabarti, S. K., 1990, \mnras\ , {{245}}, 747
\bibitem[\protect\citeauthoryear{Chakrabarti}{1991}]{Chakrabarti} Chakrabarti S. K., 1991,   \mnras , {{250}},  7}

\bibitem[\protect\citeauthoryear{Christodoulou et al.}{2017}]{Christodoulou}
Christodoulou D. M.,  Laycock S. G. T.,  Kazanas D., 2017, 
 MNRAS: Letters,  470,  1,  L21--L24


\bibitem[\protect\citeauthoryear{Christodoulou\&Ruffini}{1971}]{CRR}
Christodoulou D. M.\& Ruffini R., 1971, Phys. Rev. D,4, 3552

 \bibitem[\protect\citeauthoryear{Cowperthwaite\&Reynolds.}{2012}]{Cowperthwaite}Cowperthwaite P. S.\&Christopher S. R., 2012, ApJL, 752, L21

 \bibitem[\protect\citeauthoryear{Do{\u g}an et al.}{2015}]
  {2015MNRAS.449.1251D} Do{\u g}an S.,   Nixon C.,  King A. {\it et al.}, 2015, 
MNRAS, 449, 1251

   \bibitem[\protect\citeauthoryear{Dyda et al.}{2015}]{Dyda:2014pia}
Dyda S., Lovelace R.~V.~E, et al., 2015, 
MNRAS, \textbf{446}, 613--621

    \bibitem[\protect\citeauthoryear{Ensslin}{2003}]{Ensslin:2002gn}
Ensslin T.~A., 2003,
Astron. Astrophys. \textbf{401}, 499-504

\bibitem[\protect\citeauthoryear{Evans et al.}{2010}]
{2010ApJ...710..859E} Evans, D.~A., Reeves, J.~N., Hardcastle, M.~J., et al.\ 2010, \apj, 710, 859. 

{\bibitem[\protect\citeauthoryear{Event Horizon Telescope Collaboration et al.}{2019}]{M87} Event Horizon Telescope Collaboration, Akiyama K., Alberdi A., Alef W., Asada K., Azulay R., Baczko A.-K., et al., 2019, ApJL, 875}

\bibitem[\protect\citeauthoryear{Garofalo}{2013}]{Garofalo}
 Garofalo D. 2013.,
Advances in Astronomy,  2013,  213105


\bibitem[\protect\citeauthoryear{Garofalo et al.}{2010}]{GarofaloEvans}
 Garofalo D., Evans D.A., Sambruna R. M. 2010,
MNRAS, 406, 975-986
\bibitem[\protect\citeauthoryear{Impellizzeri et al.}{2019}]{Violette}
  Impellizzeri C. M. V. et al., 2019, ApJL, 884, L28

 \bibitem[\protect\citeauthoryear{Kataoka et al.}{2007}]{Kataoka}Kataoka J. et al., 2007, PASJ, 59, 279


 \bibitem[\protect\citeauthoryear{King\&Nixon}{2018}]
{King:2018mgw}
  King A.\&Nixon C., 2018,
  \apj,  {\bf 857}, 1,  L7


\bibitem[\protect\citeauthoryear{King et al.}{2005}]
  {King2005}
 King A.~R., Lubow S.~H., Ogilvie  G.~I.,\& Pringle  J.~E. 2005
 MNRAS, { 363},  49

 { \bibitem[\protect\citeauthoryear{Koz{\l}owski et al.}{1979}]
{Koz-Jar-Abr:1978:ASTRA:}
  Koz{\l}owski M., Jaroszy{\'n}ski  M., Abramowicz  M.~A., 1998,
Astron. Astrophys., {63}, 209
}

\bibitem[\protect\citeauthoryear{Kuznetsov et al.}{1999}]{Kuznetsov}
 Kuznetsov O. A., et al., 1999, ApJ, 514, 691

 \bibitem[\protect\citeauthoryear{Kim et al.}{2016}]{Kim:2016qsr}
Kim M.~I., Christian D.~J., Garofalo D.\& D'Avanzo  J., 2016,
MNRAS \textbf{460}, 3, 3221--3231

\bibitem[\protect\citeauthoryear{Lodato\&Pringle}{2006}]
{2006MNRAS.368.1196L}   Lodato G.\& Pringle  J.~E. 2006, \mnras, \textbf{368}, 1196

\bibitem[\protect\citeauthoryear{Lovelace\&Chou}{1996}]
{Lovelace:1996kx}
    Lovelace R.~V.~E.\&Chou T.  1996,
  \apj,   { 468},  L25


  \bibitem[\protect\citeauthoryear{Lovelace et al.}{2014}]
  {Romanova}
 Lovelace R.V.E., Romanova  M.M. , Lii P. et al. 2015, 
Comp. Astroph.  Cosmology, 1-3

\bibitem[\protect\citeauthoryear{Martin et al.}{2014}]
  {Martin:2014wja}
  Martin R.~G., Nixon C., Lubow S.~H., et al. 2014, 
  Astrophys.\ J.\  {\bf 792}, L33

\bibitem[\protect\citeauthoryear{Middleton et al.}{2014}]{Middleton-miller-jones}
Middleton M. J.,Miller-Jones J. C. A., Fender R. P. 2014,
\mnras,  439,  2, 1740--1748

 \bibitem[\protect\citeauthoryear{Morningstar et al.}{2014}]
  {Morningstar}
  Morningstar  W.~R., Miller   J.~M.,  Reis R.~C.  {\it et al.} 2014,
  \apj,\  { 784}, L18

 \bibitem[\protect\citeauthoryear{Murray et al.}{1999}]{Murray} Murray J.~R., de Kool M., Li J., 1999, ApJ, 515, 738

{
\bibitem[\protect\citeauthoryear{Narayan et al.}{2022}]{2022MNRAS.511.3795N} Narayan R., Chael A., Chatterjee K., Ricarte A., Curd B., 2022, MNRAS, 511}


 \bibitem[\protect\citeauthoryear{Nealon et al.}{2015}]{Nealon:2015jya}
 Nealon  R., Price D. and Nixon C. 2015,
MNRAS,  {\bf 448}, 2,  1526

\bibitem[\protect\citeauthoryear{Nixon et al.}{2011}]{Nixonx}
Nixon C. J., Cossins P. J. , King  A. R., Pringle  J. E. 2011, 
\mnras,  412,  3,  1591--1598


\bibitem[\protect\citeauthoryear{Nixon et al.}{2012a}]
{2012MNRAS.422.2547N} Nixon, C.~J., King, A.~R.,\&Price, D.~J.\ 2012a, \mnras, 422, 2547. 



\bibitem[\protect\citeauthoryear{Nixon et al.}{2012b}]
{2012ApJ...757L..24N}  Nixon  C.~J.,   King A.~R.,\&  Price D.~J., et al.2012b, Astrophys.\ J., \textbf{757}, L24

\bibitem[\protect\citeauthoryear{Nixon et al.}{2013}]
{Nixon:2013qfa}
  Nixon C., King  A.\&Price D. 2013,
MNRAS,  { 434}, 1946






\bibitem[\protect\citeauthoryear{Paczy{\'n}ski}{1980}]
{Pac-Wii}
 Paczy{\'n}ski B, 1980,
\newblock {Acta Astron.}, {30}, 4


{\bibitem[\protect\citeauthoryear{Porth et al.}{2021}]{2021MNRAS.502.2023P} Porth O., Mizuno Y., Younsi Z., Fromm C.~M., 2021, MNRAS, 502, 2023} 

 \bibitem[\protect\citeauthoryear{Pugliese\&Montani}{2015}]{pugtot}
Pugliese  D.\& Montani G. 2015,
Phys. Rev. D, \textbf{91}, 8, 083011

  \bibitem[\protect\citeauthoryear{Pugliese\&Montani}{2018}]{Fi-Ringed}
Pugliese  D.\& Montani G. 2018,
MNRAS, \textbf{476}, 4, 4346-4361


 \bibitem[\protect\citeauthoryear{Pugliese\&Montani}{2021}]{review}
Pugliese  D.\& Montani G. 2021,
Gen. Rel. Grav., 53, 5,51





 \bibitem[\protect\citeauthoryear{Pugliese\&Quevedo}{2015}]{ergon}
Pugliese D.\&Quevedo H. 2015,
Eur. Phys. J. C, \textbf{75}, 5, 234
\bibitem[\protect\citeauthoryear{Pugliese\&Quevedo}{2018}]{observers}
Pugliese D.\&Quevedo H. 2018,
  Eur.\ Phys.\ J.\ C, {\bf 78}, 1,  69


 \bibitem[\protect\citeauthoryear{Pugliese et al.}{2011}]
  {Pu:Kerr}
Pugliese D.,  Quevedo  H. and  Ruffini R. 2011,
  \prd, {84}, 044030

\bibitem[\protect\citeauthoryear{Pugliese\&Stuchl\'{\i}k}{2015}]
  {ringed}
Pugliese D.\&Stuchl\'{\i}k Z. 2015,
Astrophys.\ J.s,  \textbf{221}, 2,  25
\bibitem[\protect\citeauthoryear{Pugliese\&Stuchl\'{\i}k}{2016}]
{open}
Pugliese D.\&Stuchl\'{\i}k Z. 2016,
   Astrophys.\ J.s,  { \textbf{223}}, 2,  27

   \bibitem[\protect\citeauthoryear{Pugliese\&Stuchl{\'{\i}}k}{2017a}]
{dsystem} Pugliese D.\&Stuchl\'{\i}k Z. 2017a,   Astrophys.\ J.s,  \textbf{229}, 2, 40


\bibitem[\protect\citeauthoryear{Pugliese\&Stuchlik}{2017b}]
 {letter}
Pugliese D.\&Stuchl\'{\i}k Z. 2017b,  Eur.\ Phys.\ J.\ \textbf{C}, {\bf 79}, 4,  288

  \bibitem[\protect\citeauthoryear{Pugliese\&Stuchlik}{2018a}]
{Multy}
Pugliese D.\&Stuchl\'{\i}k Z. 2018a,
  Class.\ Quant.\ Grav.\  {\bf 35}, 18,  185008


    \bibitem[\protect\citeauthoryear{Pugliese\&Stuchlik}{2018b}]
{long}
Pugliese D.\&Stuchl\'{\i}k Z 2018b,
 JHEAp, { 17}, 1

 \bibitem[\protect\citeauthoryear{Pugliese\&Stuchlik}{2018c}]
{proto-jets}
Pugliese D.\&Stuchl\'{\i}k Z. 2018c
  Class.\ Quant.\ Grav.,\  {\bf 35},  10,  105005



  \bibitem[\protect\citeauthoryear{Pugliese\&Stuchlik}{2020a}]{mnras2}  Pugliese D.\& Stuchlik Z. 2020a, MNRAS., 493, 3, 4229--4255


\bibitem[\protect\citeauthoryear{Pugliese\&Stuchlik}{2020b}]{cqg2020}
Pugliese D.\&  Stuchlik Z., 2020b,
Class.Quant.Grav., 37, 19, 195025



\bibitem[\protect\citeauthoryear{Pugliese\&Stuchlik}{2021a}]{ella-correlation}
Pugliese D.\&Stuchl\'{\i}k Z. 2021a,
 Class. Quant. Grav., 38, 14, 145014

 \bibitem[\protect\citeauthoryear{Pugliese\&Stuchlik}{2021b}]{ella-jet} Pugliese D.\&Stuchl\'{\i}k Z. 2021b
PASJ,
 73,  5, 1333-1366



%
\bibitem[\protect\citeauthoryear{Pugliese\&Stuchlik}{2021c}]{dragged}Pugliese D.\&Stuchl\'{\i}k Z. 2021c
PASJ,73
 6, 1497-1539

 \bibitem[\protect\citeauthoryear{Pugliese\&Stuchl{\'{\i}}k}{2022}]
{new}Pugliese D.\&Stuchl\'{\i}k Z.  2022, \emph{in preparation}


\bibitem[\protect\citeauthoryear{Rao\&Vadawale}{2012}]{Rao}  Rao A.\&Vadawale S.V.,  2012, ApJL, 757, L12
\bibitem[\protect\citeauthoryear{Reis et al.}{2013}]{Reis} Reis R. C. et al., 2013, ApJ, 778, 155



 \bibitem[\protect\citeauthoryear{Sadowski et al.}{2016}]{sadowtree}
 Sadowski, A., Lasota, J.P., Abramowicz, M.A.\&Narayan, R.
2016, MNRAS, 456, 4, 391

 \bibitem[\protect\citeauthoryear{Scheuerl\&Feiler}{1996}]
{Feiler}
Scheuerl P. A. O.\& Feiler R. 1996,
MNRAS, \textbf{282}, 291-294

\bibitem[\protect\citeauthoryear{Stuchlik}{1980}]
{1980BAICz..31..129S} Stuchlik Z., 1980, BAICz, 31, 129%

\bibitem[\protect\citeauthoryear{Stuchl{\'\i}k}{2005}]
 {2005MPLA...20..561S} Stuchl{\'\i}k Z., 2005, MPLA, 20, 561


 \bibitem[\protect\citeauthoryear{Stuchl{\'\i}k\&Kolo{\v{s}}}{2016}]{2016EPJC...76...32S} Stuchl{\'\i}k Z., Kolo{\v{s}} M., 2016, EPJC, 76, 32
\bibitem[\protect\citeauthoryear{Stuchl{\'\i}k et al.}{2020}]
{2020Univ....6...26S} Stuchl{\'\i}k Z., Kolo{\v{s}} M., Kov{\'a}{\v{r}} J., Slan{\'y} P., Tursunov A., 2020, Univ, 6, 26

    \bibitem[\protect\citeauthoryear{Stuchl{\'\i}k, Kolo{\v{s}},\&Tursunov}{2021}]{2021Univ....7..416S} Stuchl{\'\i}k Z., Kolo{\v{s}} M., Tursunov A., 2021, Univ, 7, 416


\bibitem[\protect\citeauthoryear{Stuchl{\'\i}k\&Schee}{2012}]
{2012CQGra..29f5002S} Stuchl{\'\i}k Z., Schee J., 2012, CQGra, 29, 065002

\bibitem[\protect\citeauthoryear{Stuchl{\'\i}k\&Schee}{2013}]
{2013CQGra..30g5012S} Stuchl{\'\i}k Z., Schee J., 2013, CQGra  30, 075012
\bibitem[\protect\citeauthoryear{Stuchl{\'\i}k, Hled{\'\i}k,\&Truparov{\'a}}{2011}]
{2011CQGra..28o5017S} Stuchl{\'\i}k Z., Hled{\'\i}k S., Truparov{\'a} K., 2011, CQGra, 28, 155017



\bibitem[\protect\citeauthoryear{Stuchl{\'\i}k, Slan{\'y},\&Kov{\'a}{\v{r}}}{2009}]
{2009CQGra..26u5013S} Stuchl{\'\i}k Z., Slan{\'y} P., Kov{\'a}{\v{r}} J., 2009, CQGra, 26, 215013




{\bibitem[\protect\citeauthoryear{Tejeda et al.}{2017}]{2017MNRAS.469.4483T} Tejeda E., Gafton E., Rosswog S., Miller J.~C., 2017, MNRAS, 469, 4483 } 

\bibitem[\protect\citeauthoryear{Volonteri}{2010}]
{2010ASPC..427....3V} Volonteri, M.\ 2010, Accretion and Ejection in AGN: a Global View, 427, 3
\bibitem[\protect\citeauthoryear{Volonteri et al.}{2003}]
{Volonteri:2002vz}
Volonteri   M., Haardt  F.,\& Madau P., 2003
 \apj,{ 582}  559

 {\bibitem[\protect\citeauthoryear{Wong et al.}{2021}]{Wongetal}  Wong G. N. et al., 2021, The Astrophysical Journal, 914:55} 

\bibitem[\protect\citeauthoryear{Zanotti\&Pugliese}{2014}]
{zanotti}
  Zanotti O.\&Pugliese D. 2015,
 Gen.\ Rel.\ Grav.,  { 47}, 4,  44


\bibitem[\protect\citeauthoryear{Zhang et al.}{1997}]{Cui}Zhang S. N., Cui W., Chen W., 1997, ApJ, 482, L155



{\bibitem[\protect\citeauthoryear{Zhang et al.}{2015}]{Zhangetal}Zhang W. et al., 2015,The Astrophysical Journal, 807:89} %








\end{thebibliography}
\end{document}